\documentclass{aa}
\usepackage{graphicx,supertabular}
\usepackage{txfonts}
\begin{document}
   \title{Towards a library of synthetic galaxy spectra and preliminary results of classification and parametrization of unresolved galaxies for Gaia - II}
   \titlerunning{A library of synthetic galaxy spectra for Gaia - II}

   \author{P. Tsalmantza\inst{1,2}
          \and M. Kontizas\inst{2}
          \and B. Rocca-Volmerange\inst{3,4}
          \and C. A. L. Bailer-Jones\inst{1}
          \and E. Kontizas\inst{5}
          \and I. Bellas-Velidis\inst{5}
          \and E. Livanou\inst{2}
          \and R. Korakitis\inst{6}
          \and A. Dapergolas\inst{5}
          \and A. Vallenari\inst{7}
          \and M. Fioc\inst{3,8}}

    \offprints{P. Tsalmantza\\
    \email{vivitsal@mpia-hd.mpg.de}}

    \institute{Max-Planck-Institut f\"ur Astronomie, K\"onigstuhl 17, 69117 Heidelberg, Germany
         \and
               Department of Astrophysics Astronomy \& Mechanics, Faculty
               of Physics, University of Athens, GR-15783 Athens, Greece                 
         \and
              Institut d'Astrophysique de Paris, 98bis Bd Arago, 75014 Paris, France
         \and
              Universit\'e de Paris-Sud XI, I.A.S., 91405 Orsay Cedex, France              
         \and
              IAA, National Observatory of Athens, P.O. Box 20048, GR-118 10 Athens, Greece
         \and
              Dionysos Satellite Observatory, National Technical University of Athens, 15780 Athens, Greece
         \and 
              INAF, Padova Observatory, Vicolo dell'Osservatorio 5, 35122 Padova, Italy
         \and
              Universit\'e Pierre et Marie Curie, 4 place Jussieu, 75005 Paris, France}

\date{Received date / accepted}

% \abstract{}{}{}{}{}
% 5 {} token are mandatory

  \abstract
   {} %leave it empty if necessary
   {This paper is the second in a series, implementing a classification system for 
     Gaia observations of unresolved 
     galaxies. Our goals are to determine spectral classes and
     estimate intrinsic astrophysical parameters via synthetic templates.
     Here we describe (1) a new extended library of synthetic galaxy spectra, (2) 
     its comparison with various observations, and (3) first results of classification 
     and parametrization experiments using simulated Gaia spectrophotometry of this library.}
   {Using the P\'EGASE.2 code, based on galaxy evolution models that take
     account of metallicity evolution, extinction correction, and emission lines
     (with stellar spectra based on the BaSeL library), we improved our first 
     library and extended it to cover the domain of most of the SDSS catalogue. 
     Our classification and regression models were Support Vector Machines (SVMs).}
   {We produce an extended library of 28\,885 synthetic galaxy spectra at zero 
     redshift covering four general Hubble types of galaxies, over the wavelength 
     range between 250 and 1050\,nm at a sampling of 1\,nm or less. The library 
     is also produced for 4 random values of redshift in the range of 0-0.2.
     It is computed on a random grid of four key astrophysical parameters 
     (infall timescale and 3 parameters defining the SFR) and, depending on the 
     galaxy type, on two values of the age of the galaxy.
     The synthetic library was compared and found to be in good agreement with various observations.
     The first results from the SVM classifiers and parametrizers are
     promising, indicating that Hubble types can be reliably predicted and
     several parameters estimated with low bias and variance.}
   {}
   

\maketitle

\section{Introduction}
The ESA satellite mission Gaia (e.g., Perryman et al. \cite{perryman}, 
Turon et al. \cite{turon}, Bailer-Jones \cite{bailer1}) will complete 
observations of the entire sky, detecting any point source brighter 
than 20th magnitude, including several million unresolved galaxies. 
During its five years of operation, Gaia will observe every source 70 
times providing astrometry as well as low and high resolution spectroscopy 
for the wavelength ranges 330--1050\,nm and 847--874\,nm, respectively. 
Our primary goal is to use low resolution spectroscopic observations to 
classify and determine the main astrophysical parameters of all the 
unresolved galaxies that Gaia will observe. To proceed with this task, 
we produced a library of synthetic spectra of galaxies that was used to 
simulate Gaia observations and to train classification and parametrization 
algorithms.

The first library of synthetic spectra of galaxies for Gaia purposes 
(Tsalmantza et al. \cite{tsalmantza}) was produced with P\'EGASE.2 
code\footnote{http://www.iap.fr/pegase} (Fioc \& Rocca-Volmerange 1997). 
The galaxy evolution model P\'EGASE.2 was developed principally to model 
the spectral evolution of galaxies. It is based on the stellar evolutionary 
tracks of the Padova group, extended to the thermally pulsating asymptotic 
giant branch (AGB) and post-AGB phases (Groenewegen \& de Jong \cite{groenewegen}). 
These tracks cover all the masses, metallicities, and phases of interest to galaxy 
spectral synthesis. P\'EGASE.2 uses the BaSeL 2.2 library of stellar spectra and 
can synthesize low resolution (R$\approx$200) ultraviolet to near-infrared spectra 
of Hubble sequence galaxies, as well as starburst galaxies. For a given 
evolutionary scenario (typically characterized by a star formation law, 
an initial mass function, and possibly, infall or galactic winds), 
the code consistently calculates the spectral energy distribution (SED), 
star formation rate, and metallicity as a function of time. The nebular 
component (continuum and lines) produced by HII regions is calculated 
and added to the stellar component. Depending on the geometry of the 
galaxy (disk or spheroidal), the attenuation of the spectrum by dust 
is then computed using a radiative transfer code (which takes account 
of the scattering).

Assuming a star formation rate proportional to the gas mass, the IMF 
of Rana \& Basu (\cite{rana}), and varying the values of the infall 
input parameters, galactic winds and SFR, eight synthetic spectra 
corresponding to different typical types of Hubble sequence galaxies 
(E, S0, Sa, Sb, Sbc, Sc, Sd, and Im) were produced using P\'EGASE.2 
(Fioc \cite{fioc97}; Fioc \& Rocca-Volmerange \cite{fioc99}; Le Borgne 
\& Rocca-Volmerange, \cite{le}). By expanding the range of the input 
parameters values of these eight typical models and applying selection 
criteria for each type, we produced our first library of synthetic galaxy 
spectra (Tsalmantza et al. \cite{tsalmantza}). This library consists of 
888 spectra produced on a regular grid of input parameters values and 
2709 spectra produced on a random grid. These spectra correspond to 
seven spectral types of galaxies:  E-S0, Sa, Sb, Sbc, Sc, Sd, and Im. 
For only E-S0 galaxies did we model galactic winds, 
as was the case for the original P\'EGASE.2 models. The part of the 
library constructed by the regular grid was also produced for 4 
values of redshift: 0.05, 0.1, 0.15, and 0.2.

\begin{figure}
\centering
\includegraphics[width=6cm,angle=-90]{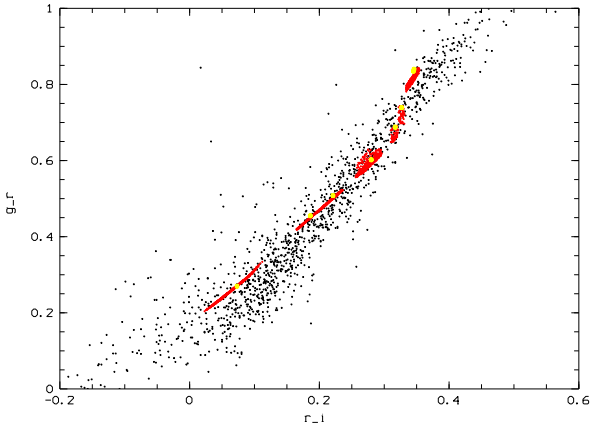}
\caption{The first library of synthetic galaxy spectra. The SDSS galaxies, 
the galaxies produced in the first library, and the typical synthetic 
spectra of P\'EGASE.2 are presented with black, red, and yellow dots, respectively.}
\label{f1}
\end{figure}

This first library of synthetic galaxy spectra at zero redshift was 
compared with the SDSS data (DR4) of galaxies. Although the photometry 
produced by the synthetic spectra was in very good agreement with the 
observational data, only a narrow locus of the SDSS colour-colour 
diagram was covered (Fig. \ref{f1}). For the classification and 
parametrization tasks of Gaia, the production of a large variety of 
galaxies is mandatory, to interpret all observational data. To accomplish 
this, we attempted to cover most of the SDSS colour-colour diagram in the 
second library presented here.

For the extension of the first library of synthetic spectra of galaxies, 
we had to overcome two main problems when comparing with SDSS data (Fig. \ref{f1}):
i) the spread in the blue part of the diagram, where true data 
have a large variance, while all the synthetic irregular galaxies are 
distributed along a line;
and ii) the systematic deviation between synthetic and true data in 
the red part of the diagram, where early-type galaxies are located.

In Sect. 2, where we describe our method to produce our second library 
of synthetic galaxy spectra, we develop and apply solutions to these 
two problems (Sect. 2.1 and 2.2). In Sect. 3, we check our library in 
other colours and in Sect. 4 we present our final library produced at 
a random grid of physical parameters. In sect. 5 we compare our library 
with the Kennicutt Atlas of galaxies. The simulated Gaia spectra for 
the final library are described in Sect. 6, while in Sect. 7 we present 
the classification and parametrization models used and their first 
results for these data. A brief discussion follows in Sect. 8.

\section{The second library of the synthetic spectra of galaxies}

\subsection{The blue part of the colour-colour diagram--developing scenarios for quenched star-forming galaxies}

\begin{figure}
\centering
\includegraphics[width=6cm,angle=-90]{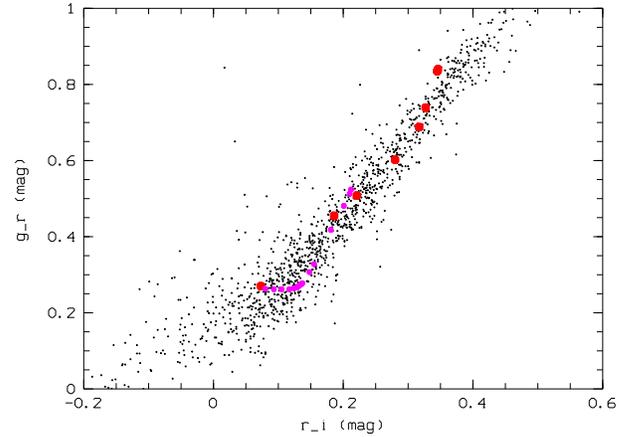}
\caption{ Models of Im galaxies with SFR stopping at 1Myr to 2Gyr ago 
(magenta). Black dots are  the SDSS galaxies and red the 8 typical 
synthetic spectra of P\'EGASE.2.}
\label{f4}
\end{figure}

In the first library of synthetic spectra of galaxies (Fig. \ref{f1}) 
the blue part of the SDSS colour-colour diagram (r-i$<$0.15 mag) is 
covered only by irregular (Im) galaxies. However, starburst galaxies 
could also have such blue colours. In the models of P\'EGASE.2 used to 
produce starburst galaxies (Le Borgne \& Rocca-Volmerange, \cite{le}), 
the age of the galaxy can vary from 1\,Myr to 2\,Gyr, while the SFR is 
given by a delta function lasting for only 1\,Myr. To use models with 
more realistic values of parameters, we investigated various scenarios. 
The one providing the most comprehensive coverage of the blue part of the 
SDSS colour-colour diagram was based on models of irregular galaxies in 
which star formation had stopped at a certain time in the past instead of 
continuing until the present. Using the original model of Im galaxies and 
stopping star formation at various ages from 1\,Myr to 2\,Gyr before the 
present (BP), we produced some examples of these models. Their synthesized 
colours are presented in Fig. \ref{f4} where we see that the spread in the 
properties of the SDSS galaxies in the blue part of the diagram can be 
covered by applying this approach to all the irregular galaxies in our library.

We are able to reproduce the properties of galaxies with such blue 
colours by assuming that the SF in the irregular models stops at a 
certain time (see last row of Table \ref{tab1}, where $p_{3}$ varies 
from 1\,Myr to 250\,Myr BP in the produced spectra). In this way, we 
can include the bulk of the flux produced by supergiants and AGB stars 
(which have very high masses and evolve rapidly). Assuming that the star 
formation stopped even earlier would create galaxies with too red colours 
(i.e., redder than the ones corresponding to the other types of galaxies 
included in our library). In the sections that follow, we refer to the 
galaxies produced by this model as quenched star-forming galaxies.

This model is clearly more realistic than one with a delta function star 
formation history (SFH). Blue galaxies produced in this way have an age 
of 9\,Gyr and not just a few Myr. In their quiescent phase, this 
population of quenched star formation resemble periodic bursting dwarfs, 
the properties of which are indeed required to fit the UV galaxy counts 
(Fioc \& Rocca-Volmerange \cite{fioc99b}). A further analysis of the SEDs 
in the far-UV will be required to confirm this result.

The SFH used to reproduce the properties of quenched star-forming galaxies 
in the new version of our library is given in Table \ref{tab1}. Using 10 
different values for $p_{3}$, we produced \mbox{1584x10=15\,840} synthetic 
spectra for quenched star-forming galaxies. In Table \ref{t1}, the range of 
the input parameters is given, while Fig. \ref{f6} shows our results 
(magenta points). It is obvious that most of the blue part of the SDSS 
colour-colour diagram is now covered.

\subsection{The red part of the colour-colour diagram -- adopting an exponential SFR law for early-type galaxies}

In the first library, we assumed a star-formation rate that is proportional 
to the gas mass  \mbox{(SFR=$(M_{gas}^{p_{1}})/p_{2}$)} and the presence 
of infall and galactic winds to reproduce spectra of early-type galaxies. 
As described in the introduction and as can be seen in Fig. \ref{f1}, 
there is a small deviation between the predicted properties of this type 
of galaxy produced in the first library and those observed for red galaxies 
of SDSS. To solve this problem, we tested several methods. The most successful 
was to increase the amplitude of the initial starburst and decrease its duration. 
To achieve this, we adopted an exponential SFR for early-type galaxies. Using 
this scenario, one cannot include infall since the presence of infall means 
that the mass of the gas is zero at t=0. Galactic winds are also not included 
in this model.

To test this model, we initially used a small set of values for the parameters 
$p_{1}$ and $p_{2}$ of the exponential \mbox{SFR=($p_{2}$exp(-t/$p_{1}$)/$p_{1}$)} 
and created models using all the combinations of the following parameters values:

$p_{1}$: 50,  100, 250, 500, 1000, 1500, 2000 \& 2500\,Myr

$p_{2}$: 0.5, 0.6, 0.65, 0.7, 0.75 \& 1 $M_{\odot}$

\begin{figure}
\centering
\includegraphics[width=6cm,angle=-90]{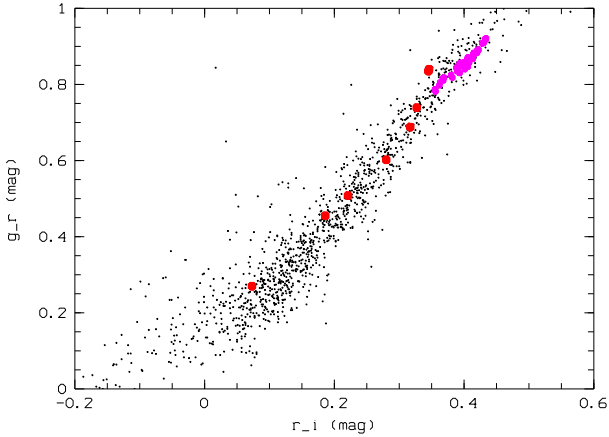}
\caption{Models of E galaxies produced assuming exponential 
star formation rate (magenta). Black dots are SDSS galaxies 
and red the 8 typical synthetic spectra of P\'EGASE.2.}
\label{f5}
\end{figure}

The synthesized SDSS photometry of these models is shown in Fig. \ref{f5}.
By increasing $p_{2}$, we produce galaxies with redder colours while 
ensuring that the influence of $p_{1}$ is weaker (since it appears both in 
the order of the exponential and in the denominator of the SFR).
In Fig. \ref{f5}, it is clear that the differences between the red parts of 
the spectrum of synthetic and true galaxies have decreased compared to the 
first library. For this reason we decided to assume this scenario 
(Table \ref{tab1}) when modelling the new library of synthetic galaxy spectra 
of E and S0 galaxies. Initially, we produced 2015 synthetic spectra based on 
a regular grid of input parameters (Table \ref{t1}). The simulated colours are 
presented in Fig. \ref{f6} (red points). In this plot, we can see that there 
is now a closer agreement between the synthetic and observed data than possible 
with the first library.

\subsection{Spiral and Irregular galaxies}

\begin{figure}
\includegraphics[width=6cm,angle=-90]{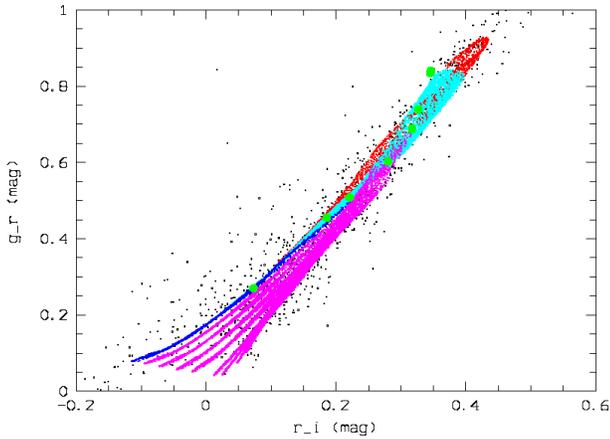}
\caption{Models of irregular (blue), quenched star-forming galaxies 
(magenta), spirals (light blue), and early-type galaxies (red). Black 
dots are SDSS galaxies and green the 8 typical synthetic spectra of P\'EGASE.2.}
\label{f6}
\end{figure}

For the case of spiral and irregular galaxies, the scenarios used in the 
first library were adopted here (table \ref{tab1}), while their input 
parameter values are extended to a wider range (Table \ref{t1}). Gas infall 
was taken into account, while the age of the galaxies was kept at 9\,Gyr and 
13\,Gyr for irregular and spiral galaxies, respectively. In Fig. \ref{f6}, 
the colour-colour diagram for the 9590 model spiral galaxies (light blue points) 
and the 1584 model irregular galaxies (blue points) is presented. In this plot, 
we can see that the properties of the new synthetic spectra of spiral galaxies 
produced here covers a larger part of the SDSS colour-colour diagram than 
the first library (Fig. \ref{f1}).

\begin{table}[h]
 \caption {Models of SF assumed in the new library.}
 \begin{tabular}{c c}
 \hline\hline
  
Galaxy type         & SFR           \\
                    &                          \\
 \hline
\hline
Early-type galaxies & $p_{2}$exp(-t/$p_{1}$)/$p_{1}$                        \\
\hline
Spiral galaxies     & $(M_{gas}^{p_{1}})/p_{2}$                   \\
\hline
Irregular galaxies  & $(M_{gas}^{p_{1}})/p_{2}$                   \\
\hline
Quenched star-forming galaxies & $(M_{gas}^{p_{1}})/p_{2}$ for $t<t_{f}-p_{3}$ \\
                               & 0 for $t>t_{f}-p_{3}$ where \\
                               & ($t_{f}$=9\,Gyr, the age of the galaxy) \\
\hline
\end{tabular}
\label{tab1}
\end{table}

\begin{table}[h]
 \caption {Input parameters for the galaxy scenarios in the new library.}
 \begin{tabular}{c c}
 \hline\hline
  
parameter           & range of value           \\
                    &                          \\
 \hline
\hline
Early-type galaxies &                          \\
\hline
$p_{1}$             & 10-30 000 (Myr)           \\
$p_{2}$             & 0.2-1.5 ($M_{\odot}$)    \\
age                 & 13 (Gyr)                 \\
\hline
Spiral galaxies     &                          \\
\hline
$p_{1}$             & 0.3-2.4                  \\
$p_{2}$             & 5-30 000 (Myr/$M_{\odot}$)\\
infall timescale    & 5-16 000 (Myr)            \\
age                 & 13 (Gyr)                 \\
\hline
Irregular galaxies  &                          \\
\hline
$p_{1}$             & 0.6-3.9                  \\
$p_{2}$             & 4000-70 000 (Myr/$M_{\odot}$) \\
infall timescale    & 5000-30 000 (Myr)         \\
age                 & 9 (Gyr)                  \\
\hline
Quenched star-forming galaxies &                          \\
\hline
$p_{1}$             & 0.6-3.9                  \\
$p_{2}$             & 4000-70 000 (Myr/$M_{\odot}$) \\
$p_{3}$             & 1-250 (Myr)              \\
infall timescale    & 5000-30 000 (Myr)         \\
age                 & 9 (Gyr)                  \\
\hline
\end{tabular}
\label{t1}
\end{table}

Synthesized colours of all 29\,029 synthetic spectra in the new 
library are superimposed on the SDSS observed data in Fig. \ref{f6}. 
The coverage is much improved, although a problem remains: the 
boundaries of the areas corresponding to each of the 4 scenarios 
are not clear. To solve this problem, we used UBV photometry as we 
describe in the next section.

\section{Criteria describing the spectral type based on UBV photometry}

It is known that the \mbox{U-B} versus \mbox{B-V} colour-colour diagram 
can provide a means to spectrally classifying of galaxies. To define 
the parameter-space corresponding to each of the 4 scenarios defined 
in Sect. 2.3, we compare our synthetic UBV colours with observational 
values for galaxies of known spectral type.

\begin{figure}
\centering
\includegraphics[width=6cm,angle=-90]{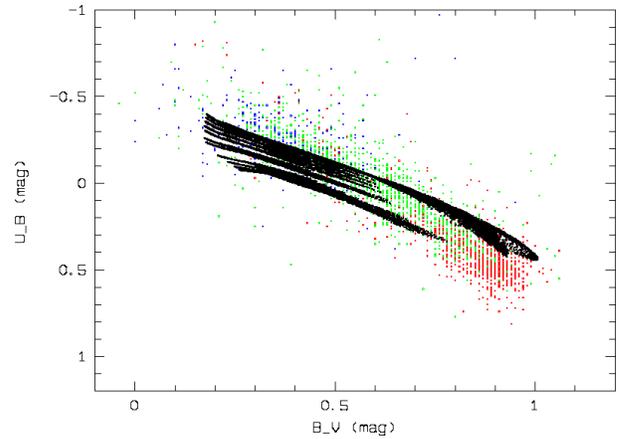}
\caption{Early-type, spiral, and irregular (red, green, and blue) 
galaxies derived from the LEDA catalog. Black dots are all galaxies 
(including quenched star-forming galaxies) produced by P\'EGASE.2.}
\label{f7a}
\end{figure}

\begin{figure}
\centering
\includegraphics[width=6cm,angle=-90]{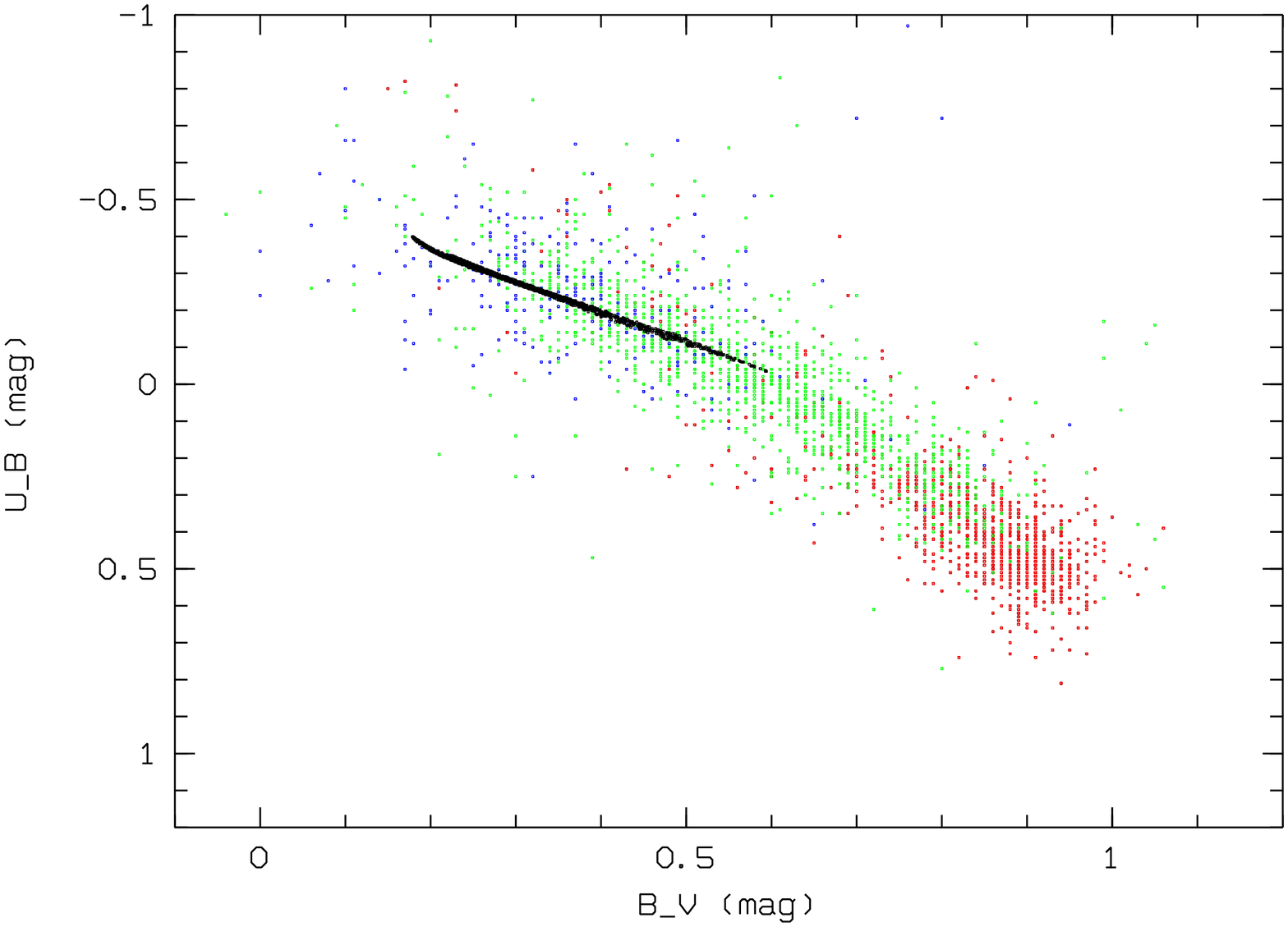}
\caption{As in Fig. \ref{f7a}, but now the black dots are the 
properties of irregular galaxies produced by P\'EGASE.2.}
\label{f7b}
\end{figure}

\begin{figure}
\centering
\includegraphics[width=6cm,angle=-90]{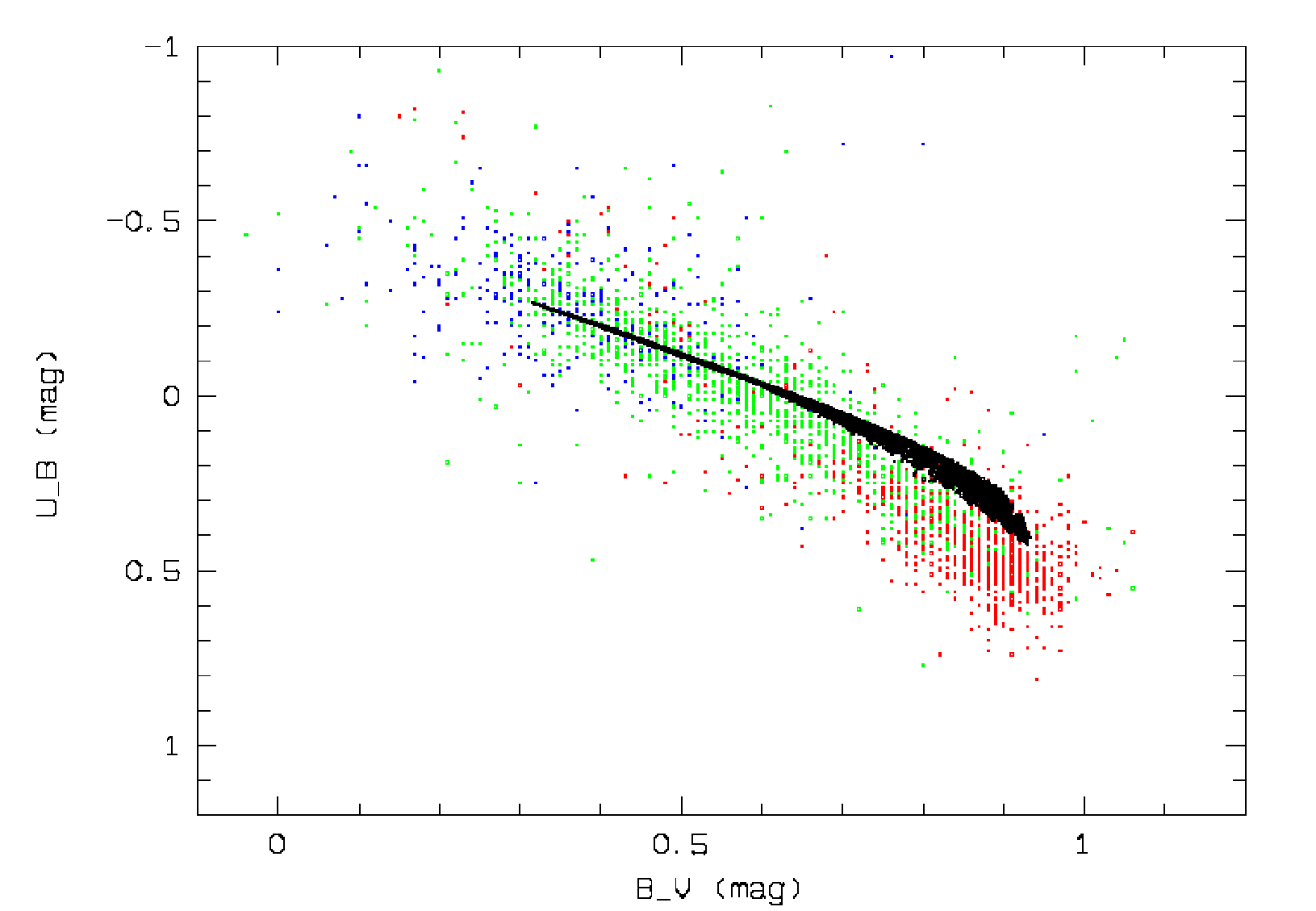}
\caption{As in Fig. \ref{f7a}, but now the black dots are the 
properties of spiral galaxies produced by P\'EGASE.2.}
\label{f7c}
\end{figure}

\begin{figure}
\centering
\includegraphics[width=6cm,angle=-90]{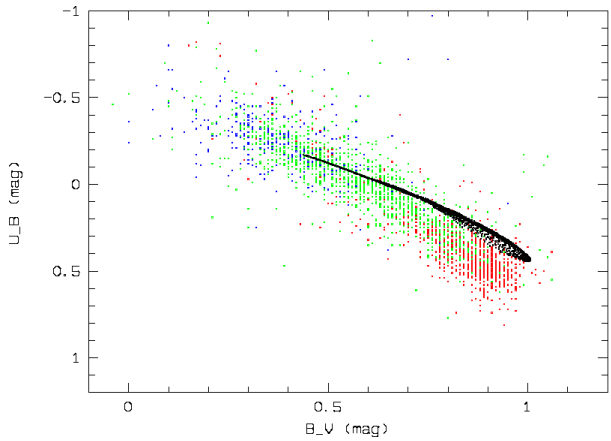}
\caption{As in Fig. \ref{f7a}, but now the black dots are 
the properties of early-type galaxies produced by P\'EGASE.2}
\label{f7d}
\end{figure}

The observational data used here were taken from the LEDA 
catalog\footnote{http://leda.univ-lyon1.fr/} (Paturel et al., \cite{paturel}). 
This catalog contains 2672 galaxies with estimated numeric photometric 
type (T) corresponding to the RC3 catalog and calculated total 
apparent \mbox{U-B} and \mbox{B-V} colours. Those colours are corrected 
for Galactic extinction, inclination, and redshift. In Figs. \ref{f7a} - \ref{f7d}, 
we present the \mbox{U-B} versus \mbox{B-V} colour-colour diagram 
for all the synthetic spectra produced here (black dots) plotted 
over the galaxies of the LEDA catalog. According to this catalog, 
galaxies with $T<0.5$ are considered to be early-type galaxies, 
with $0.5<T<8.5$ spirals and with $T>8.5$ irregulars. These type 
of galaxies are presented in Figs. \ref{f7a} - \ref{f7d}. The spread of 
the real data in the colour-colour diagram is larger than the one of 
the synthetic spectra, most probably because of errors in the calculations 
of UBV colours in the LEDA catalog and limitations of the P\'EGASE model.

\begin{figure}
\centering
\includegraphics[width=6cm,angle=-90]{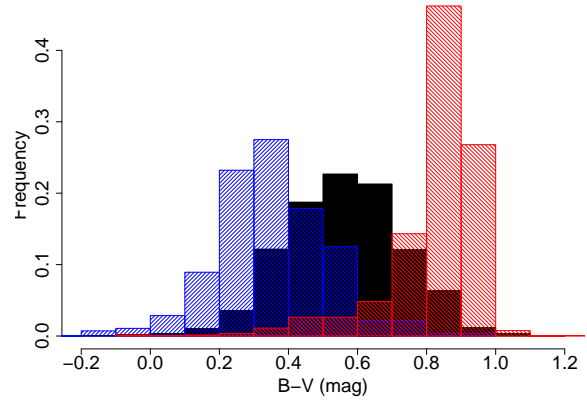}
\caption{Distribution (normalized) of B-V colours for early-type, 
spiral, and irregular (red, black, and blue respectively) galaxies 
derived from the LEDA catalog.}
\label{f8}
\end{figure}

\begin{figure}
\centering
\includegraphics[width=6cm,angle=-90]{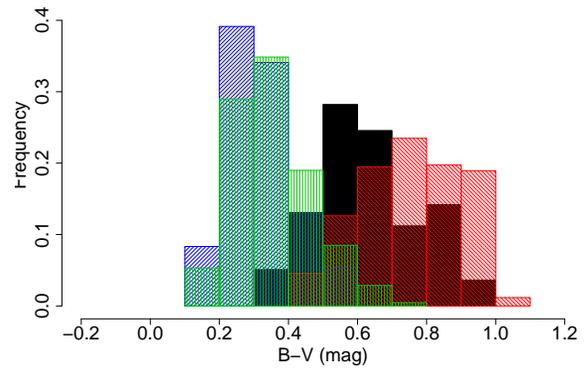}
\caption{Distribution (normalized) of B-V colours for early-type, 
spiral, irregular and quenched star-forming (red, black, blue and 
green respectively) galaxies produced by P\'EGASE.2 code.}
\label{f9}
\end{figure}

The synthetic and observational data are in good agreement, even 
though a small difference in the slope is observed. This could be 
explained by considering that the \mbox{U-B} colour is the one with 
the largest errors in the simulated photometry (Yi, \cite{yi}). For 
that reason, we based our selection criteria only on the \mbox{B-V} 
colour. In Figs. \ref{f8} and \ref{f9}, we present the normalized to 
the total number of galaxies \mbox{B-V} distributions for the 
observational LEDA catalog and synthetic spectra, respectively, 
for the three types of galaxies. In Fig. \ref{f9}, we included 
quenched star-forming galaxies (green) even though they are not 
included in the LEDA catalog. Since this type of galaxy is produced 
by models of irregular galaxies we exclude some of them because of 
the selection criteria applied to the irregular galaxies.

Comparing these two histograms, we can see that for each type 
of galaxy the peaks of the histograms are approximately coincident 
for both observational and synthetic spectra. Using the 
distribution of the observational data, we decided to keep in our 
library early-type galaxies with $B-V>0.6$, spirals with $0.3<B-V<0.9$, 
and irregulars with $B-V<0.6$. In all cases, more than 90\% of real 
galaxies of each type remain after applying the above selection 
criteria. In the case of synthetic spectra, these criteria affect 
mainly spiral and early-type galaxies where the number of galaxies 
is reduced after their application. However, the range of parameters 
remains the same in all cases and only some combinations are removed 
from our original sample.

\section{The random library of synthetic galaxy spectra}
\begin{figure}
\centering
\includegraphics[width=6cm,angle=-90]{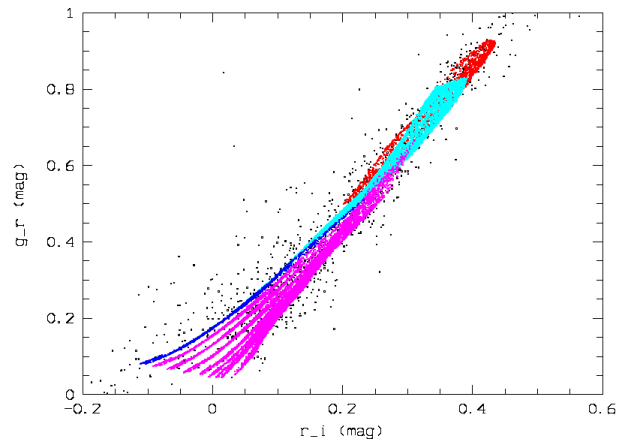}
\caption{Random models of irregular (blue), quenched star-forming 
galaxies (magenta), spirals (light blue), and early-type galaxies 
(red). Black dots are SDSS galaxies and green the 8 typical synthetic 
spectra of P\'EGASE.2.}
\label{f10}
\end{figure}

The library of synthetic spectra of galaxies described in Sect. 2 was 
produced by using a regular grid of input paramaters of P\'EGASE. To 
achieve optimal training and assessment of the classification and 
parametrization algorithms, one should use data produced by a random 
grid of parameters. For that reason, we used the range of parameters 
given in Table \ref{t1} to compile 30\,500 random scenarios of galaxies. 
To create spectra using P\'EGASE.2, we used GRID-technology provided by 
SEE-GRID (South Eastern Europe). We applied the B-V criteria described in 
the previous section to the resulting spectra. By applying this procedure, 
we produced 2816 early-type galaxies, 10\,569 spirals, 1500 irregulars, 
and 14\,000 quenched star-forming galaxies. The derived library of 28\,885 
synthetic galaxy spectra is presented in Fig. \ref{f10} where the colours 
of those spectra are plotted over the SDSS data. The new synthesized colours 
are in very good agreement with the SDSS observations.

Each spectrum in this library was simulated at four random values of 
redshift, each lying within four intervals from 0 to 0.05, 0.05 to 0.1, 
0.1 to 0.15, and 0.15 to 0.2. The final library now includes 144\,425 
synthetic spectra produced by a random grid of parameters. 

The spectra were linearly interpolated to produce a wavelength sampling of 
1nm or shorter to be used for simulations of Gaia observations.

\section{Comparison of the second library with Kennicutt's atlas}
Even though the second library of synthetic spectra has been compared and 
found to be in good agreement with both SDSS and LEDA photometric observations, 
we also used the library to fit real spectra of galaxies in order to examine 
in more detail its ability to represent reality and to classify galaxies 
according to their Hubble type. We first compared the library with a small 
amount of observational spectra of high S/N and known Hubble type. A catalog 
meeting these criteria is Kennicutt's atlas of galaxies (Kennicutt, \cite{kennicutt}). 
This is a spectophotometric atlas containing spectra of 55 nearby normal 
and peculiar galaxies, most covering the spectral range of 365-710\,nm at 
a resolution of 0.2\,nm. The spectra were normalized with respect to the 
flux at 550\,nm. Since in P\'EGASE spectra, the flux at 550\,nm is not 
provided, our spectra were normalized to the mean flux between 549 and 551\,nm.

Even though the galaxies observed in this catalog are nearby galaxies, we 
first needed to transform them to the rest frame. This was achieved by 
keeping the energy in each spectral bin constant, while relabeling the 
wavelength axis. After this step, we had to rebin the spectra to ensure 
that their resolution was equal to that of P\'EGASE.2 (2\,nm). We also 
ensured that all the observational spectra had a common spectral range, 
namely 371 to 679\,nm (155 data points).

To fit these spectra to our new library, we used two different methods. 
In the first method, we did not account for the wavelength ranges including 
the seven most important emission lines, while in the second method we did. 
These seven emission lines are the H$\alpha$ + [NII] blend (654.8 and 658.3\,nm 
(the second line is not included in the P\'EGASE data)), H$\beta$ (486.1\,nm), 
[OII] (372.7\,nm), [OIII] (500.7\,nm), and [SII] (671.7 and 673.1\,nm). By 
excluding the three regions 370-380\,nm, 480-510\,nm, and 650-680\,nm, these 
lines were in most cases excluded. This left us with 123 data points in each 
spectrum. In the second method, we decreased the spectral resolution of both 
the synthetic and observational spectra in the wavelength ranges contatining 
those lines, and replaced the fluxes with broadband values, we then had 126 
data points in each spectrum. We followed this procedure because P\'EGASE 
predicts only the total energy of each emission line and we have no 
information about the line's shape to compare with when fitting the 
observational spectra. In both cases, we performed a $\chi^{2}$--fitting 
of the Kennicutt galaxy spectra with our library and in this way checked 
the ability of the synthetic spectra to reproduce and classify the 
observational ones. We should also mention that the spectrophotometric 
calibration of Kennicutt's atlas has an error of 10\%, which of course 
affects our results.

When we excluded the above emission lines from our comparison, 17 of 55 
galaxies were classified correctly, while when we included them, the 
$\chi^{2}$--fitting gave 2 additional correct results (19 in total). 
These results are very good if we bear in mind that of the 55 galaxies 
included in the Kennicutt atlas only 28 correspond to typical Hubble types. 
In addition to the 9 misclassifications that occur, two may be caused by 
other effects and not by problems with our library. More specifically, 
NGC6217 is an Sc strongly interacting galaxy and therefore the error 
in its classification as an irregular galaxy is not very important. 
This is also the case for the misclassification of the irregular 
galaxy NGC1569 as an early-type or spiral galaxy (when lines were 
excluded or included, respectively), since its spectrum was affected 
by high Galactic reddening. Of the remaining 27 galaxies in 
Kennicutt's atlas that do not correspond to Hubble types, 8 are 
starburst galaxies, 4 are extreme emission-line galaxies, 7 are 
Seyfert galaxies, and 4 are peculiar and merger galaxies. These 
galaxy types are not included in our library. However, since the 
quenched star-forming galaxies that we produced cover the same 
part of the SDSS colour-colour diagram as the starburst galaxies 
(Sect. 2.1), we are interested in fitting the starbust galaxy spectra. 
The Kennicutt's atlas includes four galaxies undergoing global bursts of 
star formation and four galaxies that are nuclear starburst galaxies. 
The results of fitting these galaxy spectra with those included in our 
library showed that none of these galaxies were classified as quenched 
star-forming galaxy. This could be because our method does not place 
much importance on the emission lines, which are the most significant 
features in the spectra of this galaxy type. However, a closer comparison 
between the model quenched star-forming galaxies included in our library 
and that of observational starbursts showed that the emission lines of 
this type of synthetic spectra are not as strong as in the observational 
spectra.

The results of the classification are presented in Table \ref{t2} for 
when the emission lines are taken into consideration, while the fitting 
of the spectra when excluding and including the emission lines is 
presented in Appendix A. In all cases, the fitting of the continua is 
very good and in most cases it is also good when we included in the 
comparison the emission lines. We note that the continua fitting does 
not deteriorate in most cases when the emission lines are included. 
In most cases, the difference in the scaling of the flux axis highlights 
the details of the continua most clearly.

After deciding the best-fit model spectrum for each galaxy we can extract 
all the other information provided by P\'EGASE for each spectrum, such as 
SFR, metallicity history, mass of both stars and gas, and luminosity of the galaxy. 
                                                                               
\begin{table}[h]
 \begin{center}
 \caption {Classification results for galaxy spectra in the Kennicutt's atlas.}       
 \begin{tabular}{c c c c c}          
 \hline\hline                        
Kennicutt/P\'EGASE                & Early-type & Spiral     & Irregular \\
 \hline
Early-type                      & 4          & 4          & 0         \\
Spiral                          & 1          & 12         & 3         \\
Irregular                       & 1          & 0          & 3         \\
 \hline
Starburst with global burst     & 0          & 1          & 3         \\
Nucleus starburst               & 1          & 2          & 1         \\
\hline
\end{tabular}
\label{t2}
\end{center}
{\small \textsc{Note.} Rows indicate the true class and columns the ones 
predicted by the fitting with the synthetic spectra of our library. 
For these results we have included the emission lines in the $x^{2}$--fitting.} 
\end{table}

\section{Simulated Gaia spectra}
The Gaia spectrophotometer is a slitless prism spectrograph comprising blue
and red channels (called BP and RP respectively) that operate over the
wavelength ranges 330--680\,nm and 640-1050\,nm respectively. Each 
of BP and RP is simulated with 48 pixels,
whereby the dispersion varies from 3--29\,nm/pix and 6--15\,nm/pix
respectively. The 144 425 synthetic spectra of galaxies produced here were 
simulated for BP and RP Gaia spectra during cycle 3 of Gaia simulations. 
Additionally they were reddened by applying to each of them a random 
value of extinction in our Galaxy (Av in the range 0-10). The simulated 
spectra are given for three values of G-band magnitude (G=15, G=18.5, 
and G=20). Randomly sampled noise, including the source Poisson noise, 
background Poisson noise, and CCD readout noise, was added to all spectra. 
The final library contains 1\,733\,100 simulated Gaia spectra. In the 
sections that follow, we present the results of the classification and 
parametrization of spectra with a noise characteristic of G=18.5\,mag.

\section{Classification \& parametrization}

As in Paper I, we have used Support Vector Machine
classifiers (SVMs) (C-classification) to determine spectral types
of the simulated spectra and regression SVMs ($\epsilon$-regression)
when estimating their astrophysical parameters. For a more detailed
description of SVMs and references, we refer to Paper I, while a more 
general aspect of the Gaia classification scheme can be 
found in Bailer-Jones et al. (\cite{bailer}).

\begin{figure}
\centering
\includegraphics[width=6cm,angle=-90]{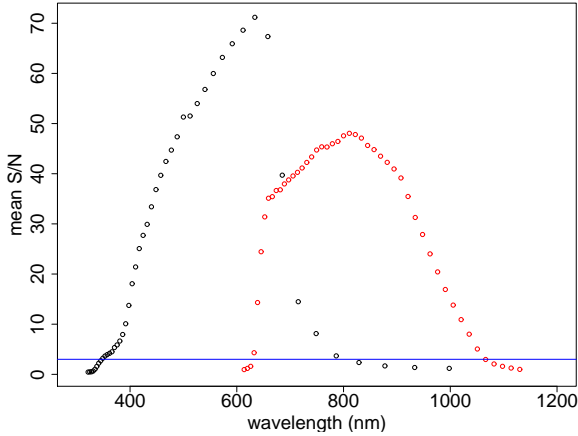}
\caption{The mean S/N spectrum of the simulated galaxy spectra used 
in the classification tests. Only points with SNR$>$3 (horizontal line) 
were used.}
\label{f10b}
\end{figure}

Throughout this section, we consider truncated spectra, retaining only
77 of the 96 data points of the simulated BP and RP spectra
corresponding to the wavelength range 321.43-998.02\,nm and 
613.78-1130.25\,nm for the two photometers, respectively. The data of
the 19 pixels were excluded based on their very low values of SNR
($<$3) in the mean spectrum of galaxies with zero redshift and
reddening (Fig \ref{f10b}). The rejection of these pixels from the
data improves the performance of the SVM. Each of the remaining 77 pixels were
standardized, that is scaled to have zero mean and unit variance
across the whole sample, before being used by SVM.

For all the classification and regression tests performed here, the
 samples used in each case were randomly split into two sets. The
 first one was used for training the SVMs and the second for testing
 their performance. For the tuning of SVMs, i.e., the selection of the
 optimal values of the internal parameters used by the method (C and
 g, and in the case of regression $\epsilon$ also), we used two
 different schemes (four-fold cross validation or a fixed scheme), 
 depending on the amount of data in the training set. In Table \ref{t9},
 we present the total number of spectra used in each test as well as
 the number of spectra in the training and test sets and the scheme
 used for the tuning of SVMs in each case.

\begin{table*}
 \centering
 \caption {The number of spectra and the tuning scheme used 
for the various SVM tests.}
 \begin{tabular}{l c c c c}          
 \hline\hline                        
  
Test performed & Total number & Number of spectra & Number of spectra & Scheme used for the \\
 & of spectra & in the training set & in the test set & tuning of SVMs \\
 \hline
\textit{Galaxies without reddening at zero redshift} & & & & \\
 \hline
Classification of galaxy types & 28 885 & 4885 & 24 000 & 4-fold cross validation \\
Regression of input APs for early-type galaxy models & 2816 & 704 & 2112 & 4-fold cross validation \\
Regression of input APs for spiral galaxy models & 10 569 & 2643 & 7926 & fixed (fix=4) \\
Regression of input APs for irregular galaxy models & 1500 & 750 & 750 & 4-fold cross validation \\
Regression of input APs for quenched star &  &  &  &  \\
forming galaxy models & 14 000 & 3500 & 10 500 & fixed (fix=4) \\
Regression of output APs for all galaxy models & 28885 & 4885 & 24000 & fixed (fix=4) \\
\hline
\textit{Galaxies with redshift, without reddening} & & & & \\
\hline
Classification of galaxy types & 144 425 & 14 440 & 129 985 & fixed (fix=4) \\
Regression of redshift for all galaxy models & 144 425 & 14 440 & 129 985 & fixed (fix=4) \\
\hline
\textit{Galaxies with redshift and reddening} & & & & \\
\hline
Regression of redshift for all galaxy models & 144 425 & 14 440 & 129 985 & fixed (fix=4) \\
Regression of reddening for all galaxy models & 144 425 & 14 440 & 129 985 & fixed (fix=4) \\
\hline
\end{tabular}
\label{t9}
\end{table*}

\subsection{Galaxies without reddening at zero redshift at G=18.5} 
This subset of the library comprises 28 885 galaxies produced by a
random grid of parameters in P\'EGASE.2 with zero reddening and
redshift at G=18.5\,mag. For these data, we performed both classification of the
galaxy type and regression of the input parameters of P\'EGASE
(i.e., the parameters included in the SFR in each case and the infall
timescale if present) as well as regression of the most significant
output parameters of the model.

\subsubsection{Classification}
Using the first subset presented in Table \ref{t9}, we trained the SVMs
to classify the data into the four different galaxy types that it
contains. The number of support vectors was 438 and the results for
the training and test set are given in the Tables \ref{t3}-\ref{t4}.

\begin{table}
      
 \begin{center} 
 \caption {Galaxy classification with the SVM for the training set.}                                   
 \begin{tabular}{l | c c c c}          
 \hline\hline                        
  
Type & E   & S    & Im  & QSFG   \\
 \hline
E    & 484 & 25   & 0   & 0    \\
S    & 5   & 1820 & 0   & 0    \\
Im   & 0   & 0    & 255 & 0    \\ 
QSFG & 0   & 0    & 0   & 2296 \\

\hline
\end{tabular}
\label{t3}
\end{center}
 \begin{center}
 \caption {As Table~\ref{t4} but for the test set.}
 \begin{tabular}{l | c c c c}          
 \hline\hline                        
  
Type & E    & S    & Im   & QSFG   \\
 \hline
E    & 2033 & 274  & 0    & 0    \\
S    & 215  & 8434 & 95   & 0    \\
Im   & 0    & 91   & 1153 & 1    \\ 
QSFG   & 3    & 1    & 9    & 11691\\

\hline
\end{tabular}
\label{t4}
\end{center}
{\small \textsc{Note.} The confusion matrices for galaxies at 
z=0 and G=18.5\,mag. Columns indicate the true class and rows 
the predicted ones. The labels E, S, Im, and QSFG correspond 
to early-type, spiral, irregular, and quenched star-forming 
galaxies, respectively.}
\end{table}

From these tables, we see that the results are very good. The
number of missclassifications is 30 for the training set and 689 for the testing set, 
corresponding to errors of 0.6\% and 2.9\%, respectively.

Because of the large differences between the two spectral libraries used
in this work and the work presented in Paper I, the results are not
directly comparable. More specifically, the library used here includes
four galaxy types instead of the seven that were included in the previous
version. In Paper I, we applied strict selection criteria to the
spectra to determine the galaxy type as accurately as possible. 
This led to artificial gaps in the colour-colour space,
which of course made the classification process easier. Finally, noise was not included in 
the data in Paper I, in contrast to what we wrote 
there, because of an error in our software. The library 
in the present paper - which now includes the correct values of 
noise - is also more comprehensive, so the results shown here supersede 
those in Paper I. All these factors have a
large impact on the results with SVMs.

\subsubsection{Regression of input parameters of P\'EGASE}
\begin{figure*}[t]
  \setlength{\unitlength}{1cm}
\begin{picture}(18,20)
\put(0,20){\includegraphics{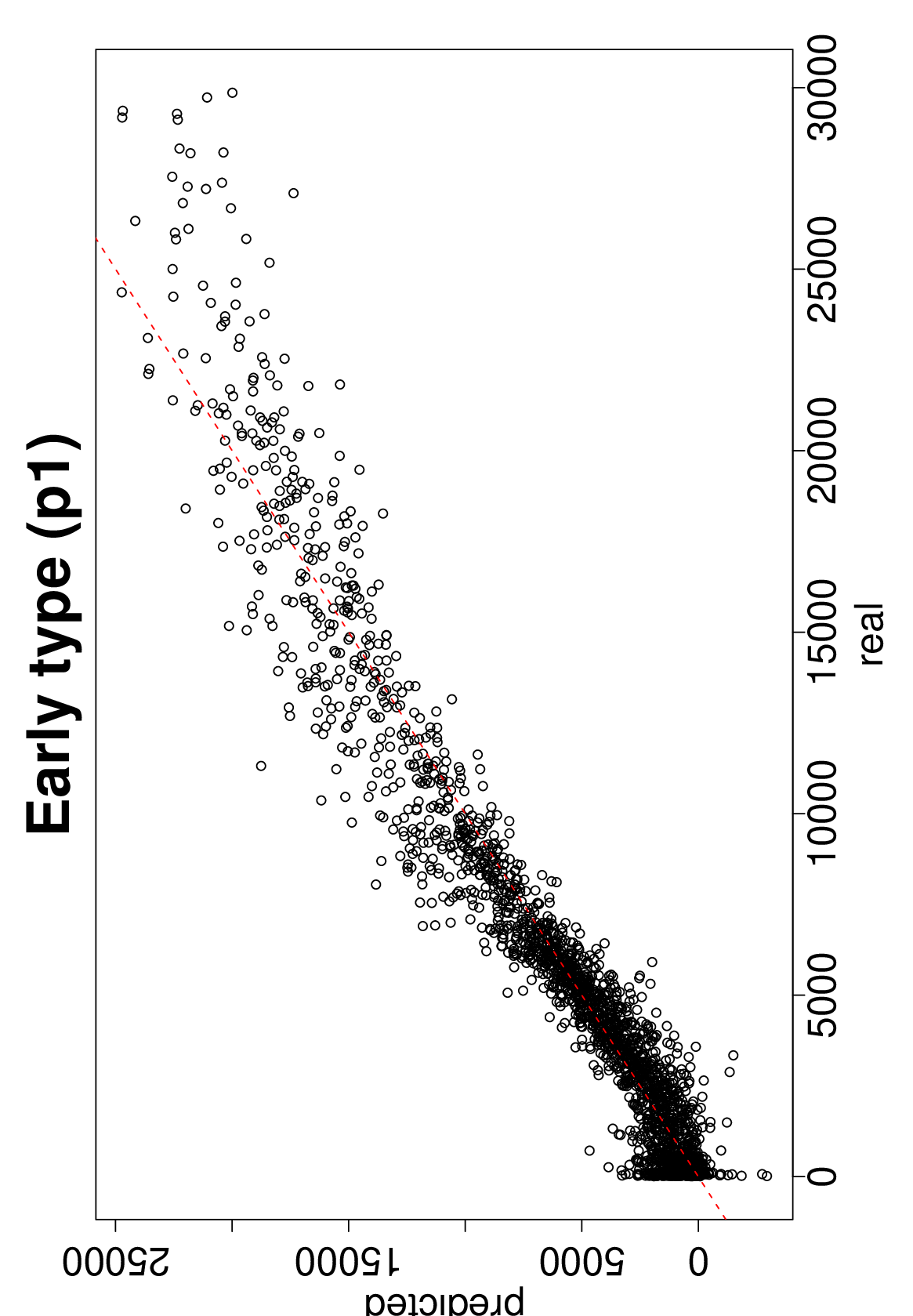}}
\put(0,15){\includegraphics{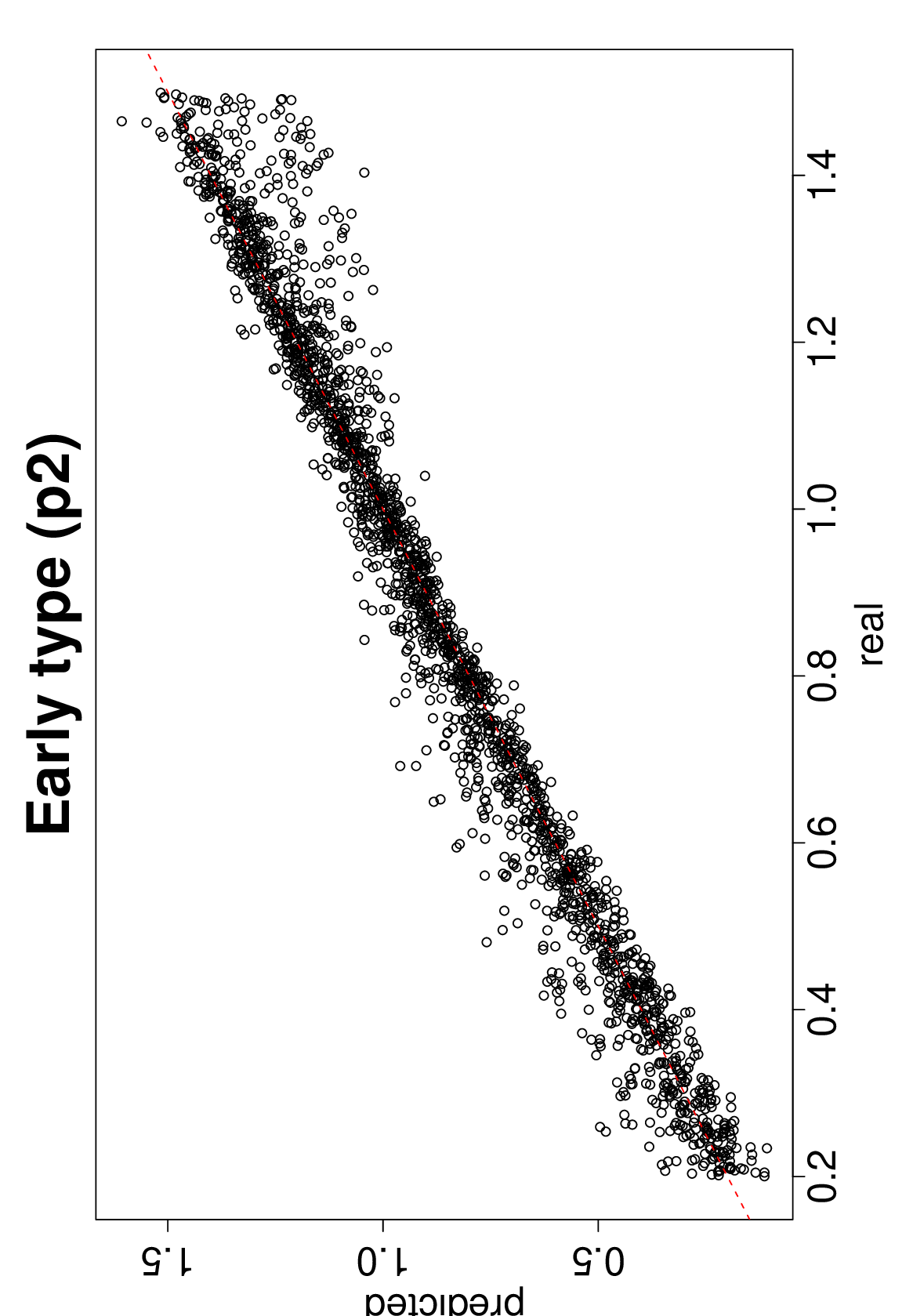}}
\put(0,10){\includegraphics{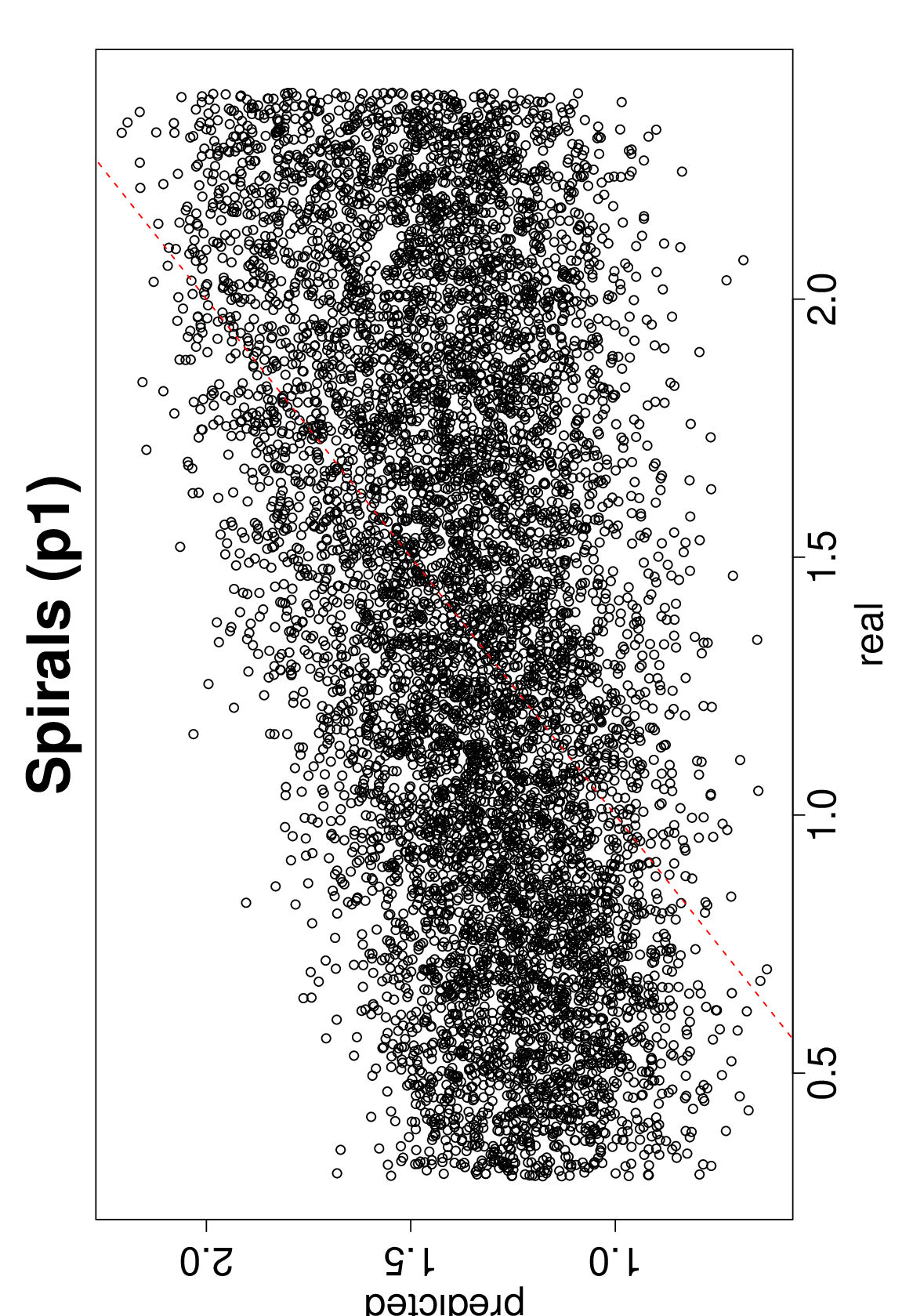}}
\put(0,5){\includegraphics{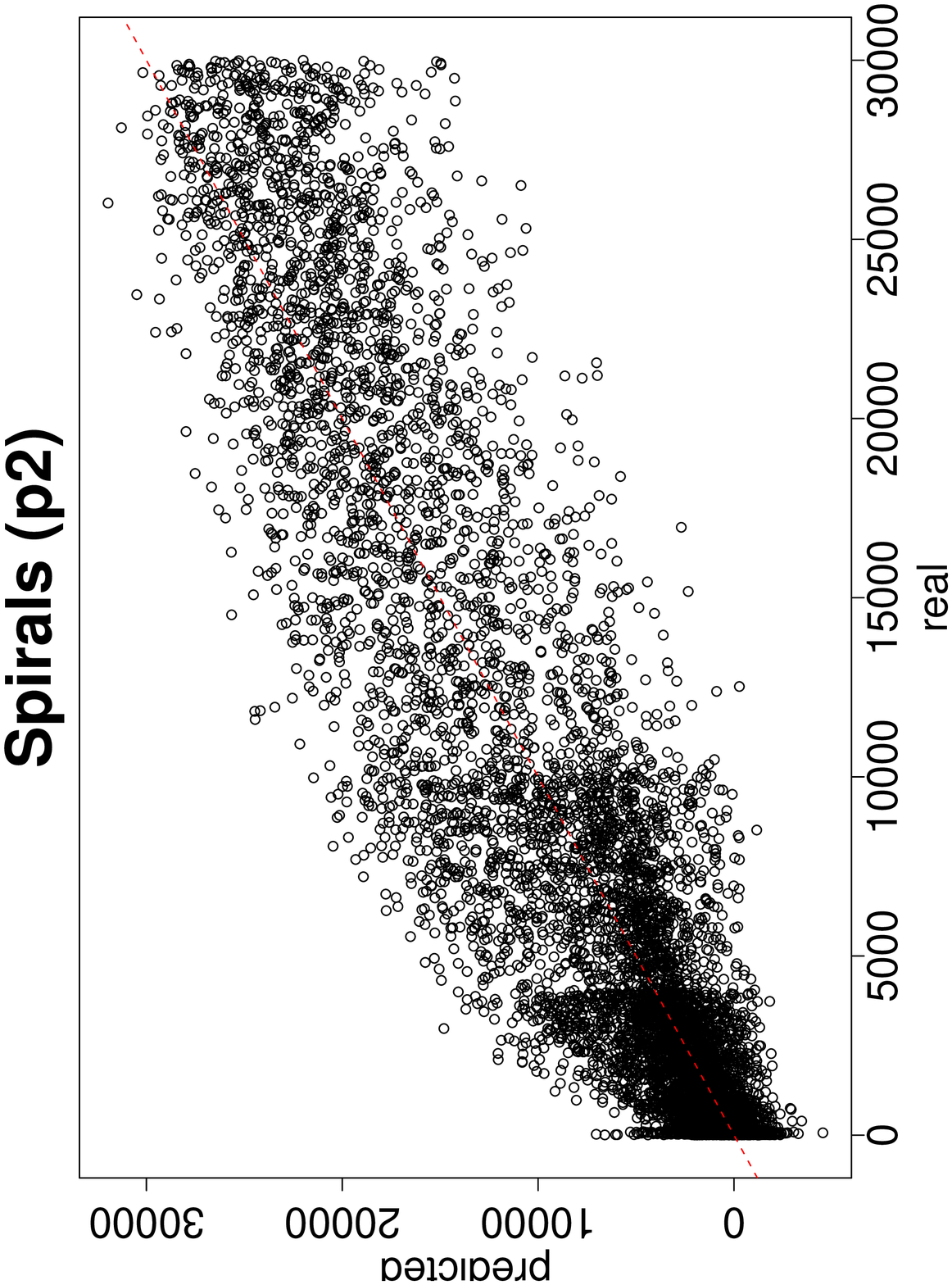}}
\put(6,20){\includegraphics{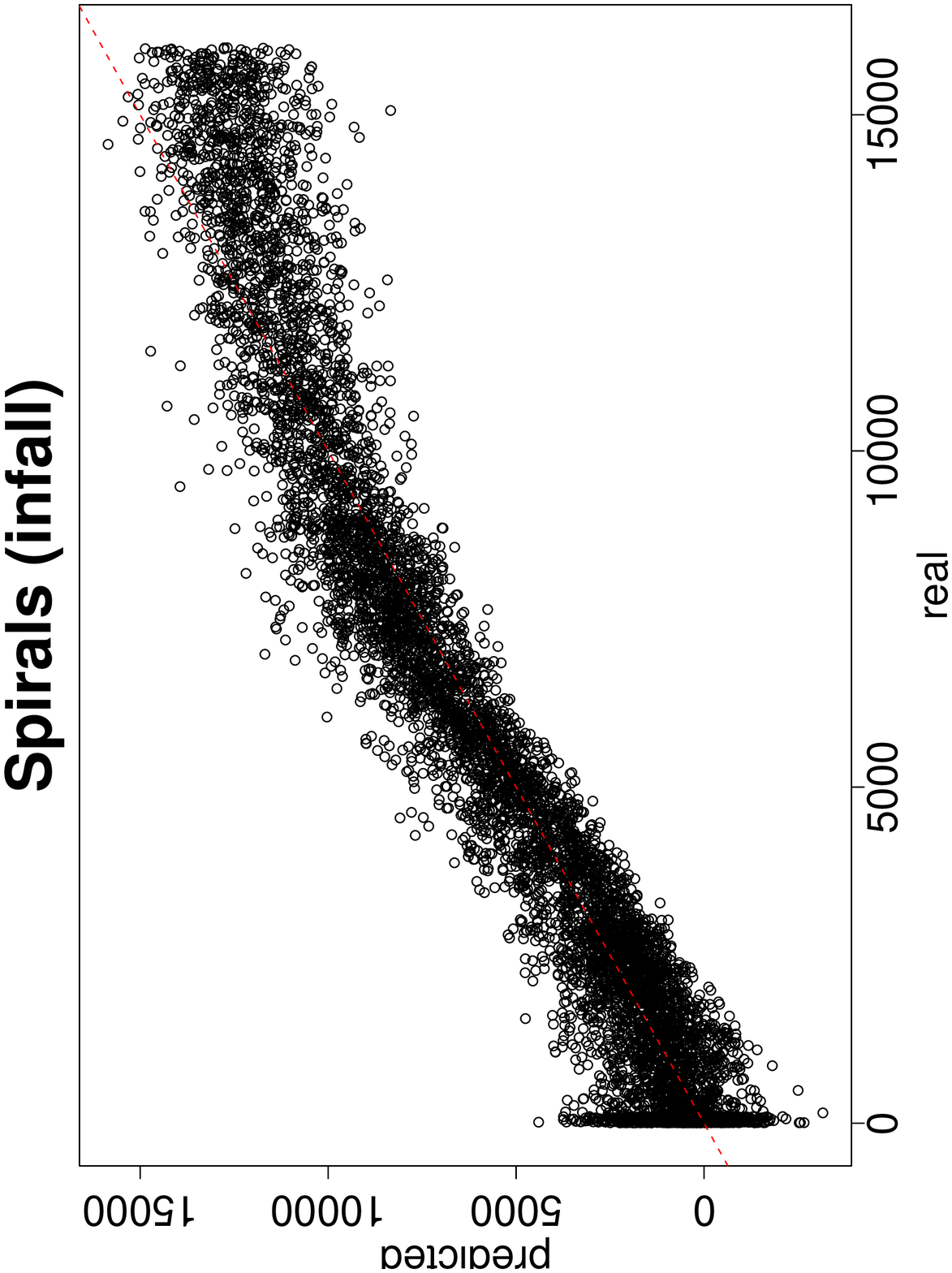}}
\put(6,15){\includegraphics{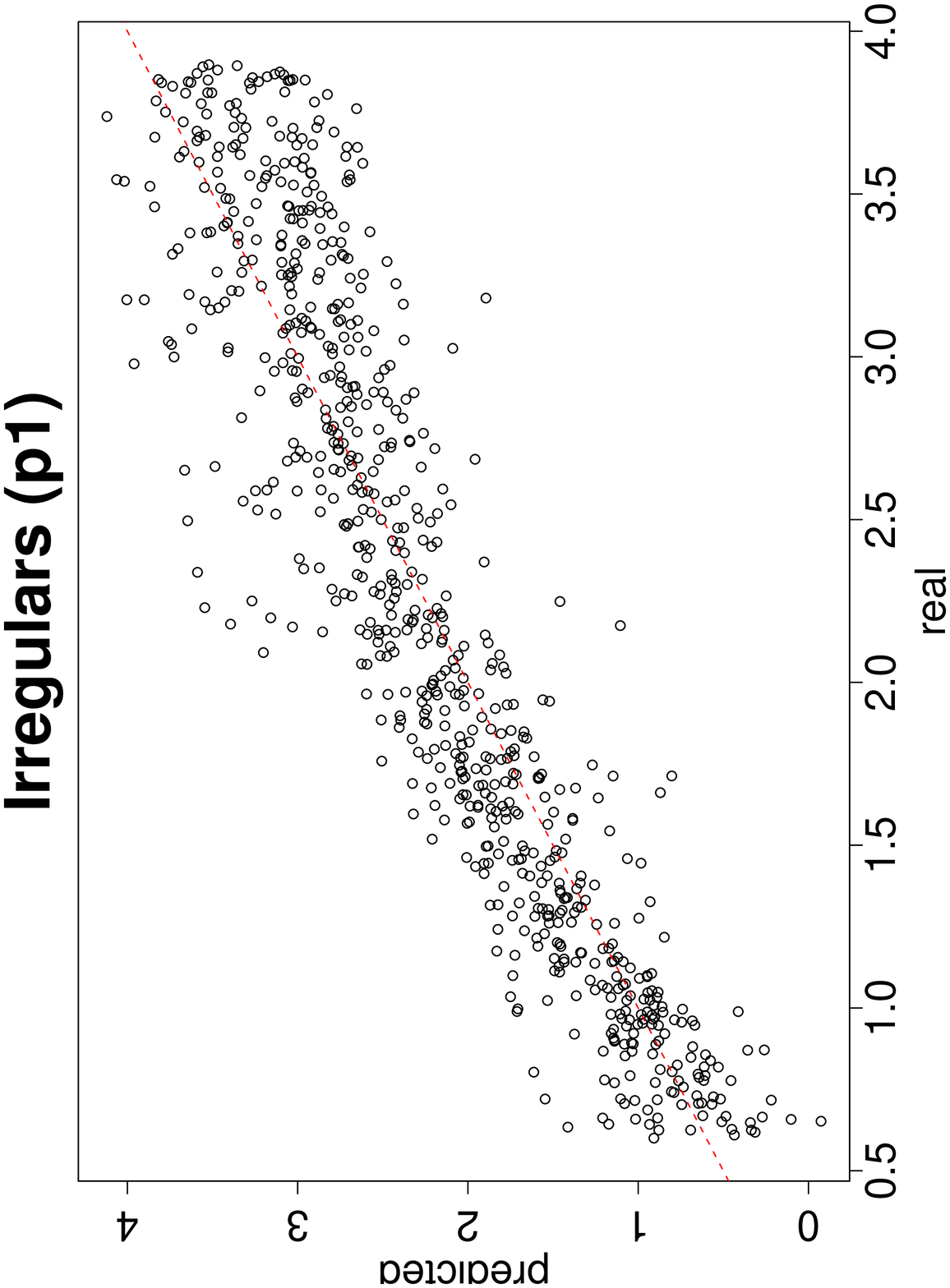}}
\put(6,10){\includegraphics{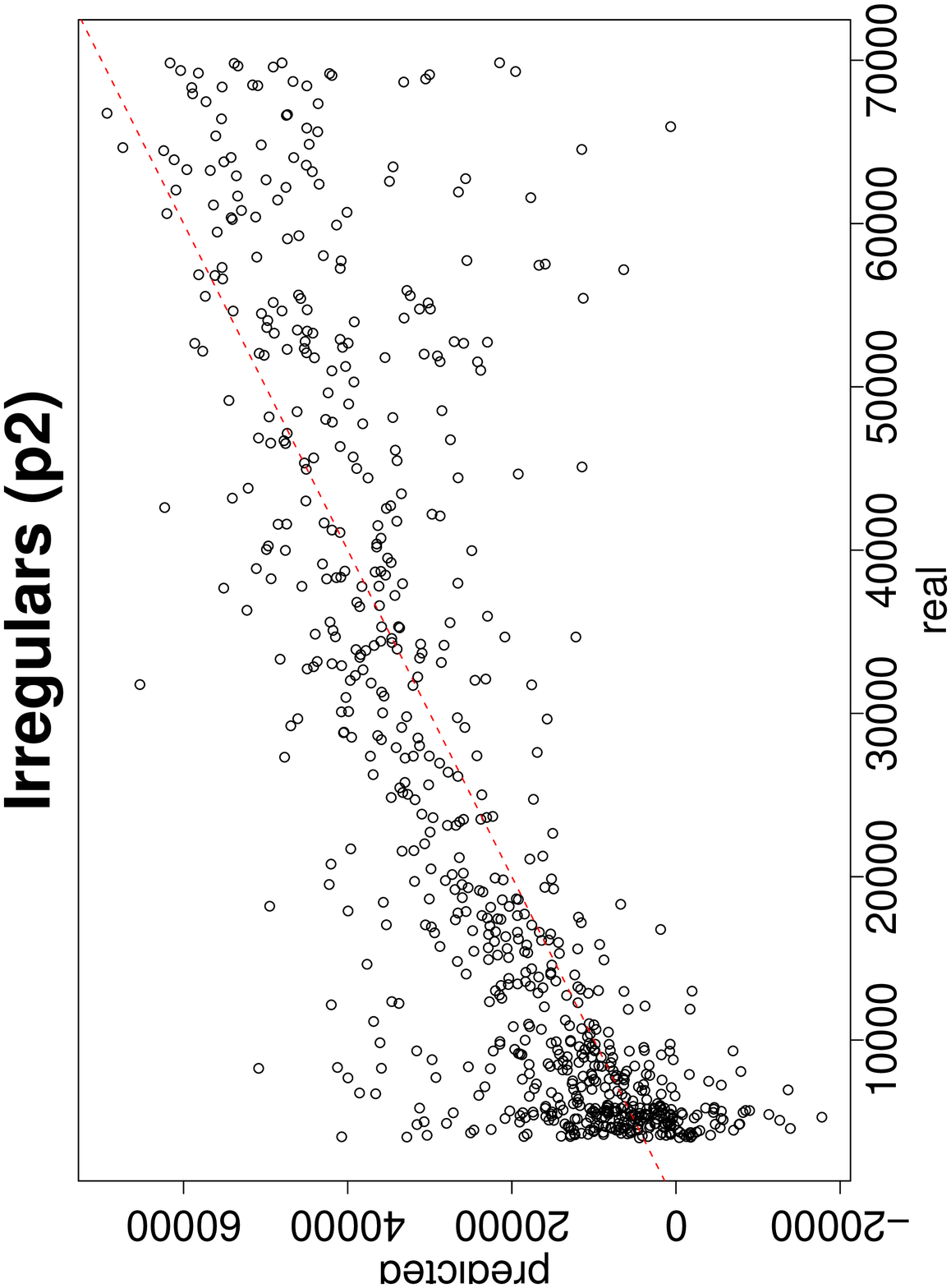}}
\put(6,5){\includegraphics{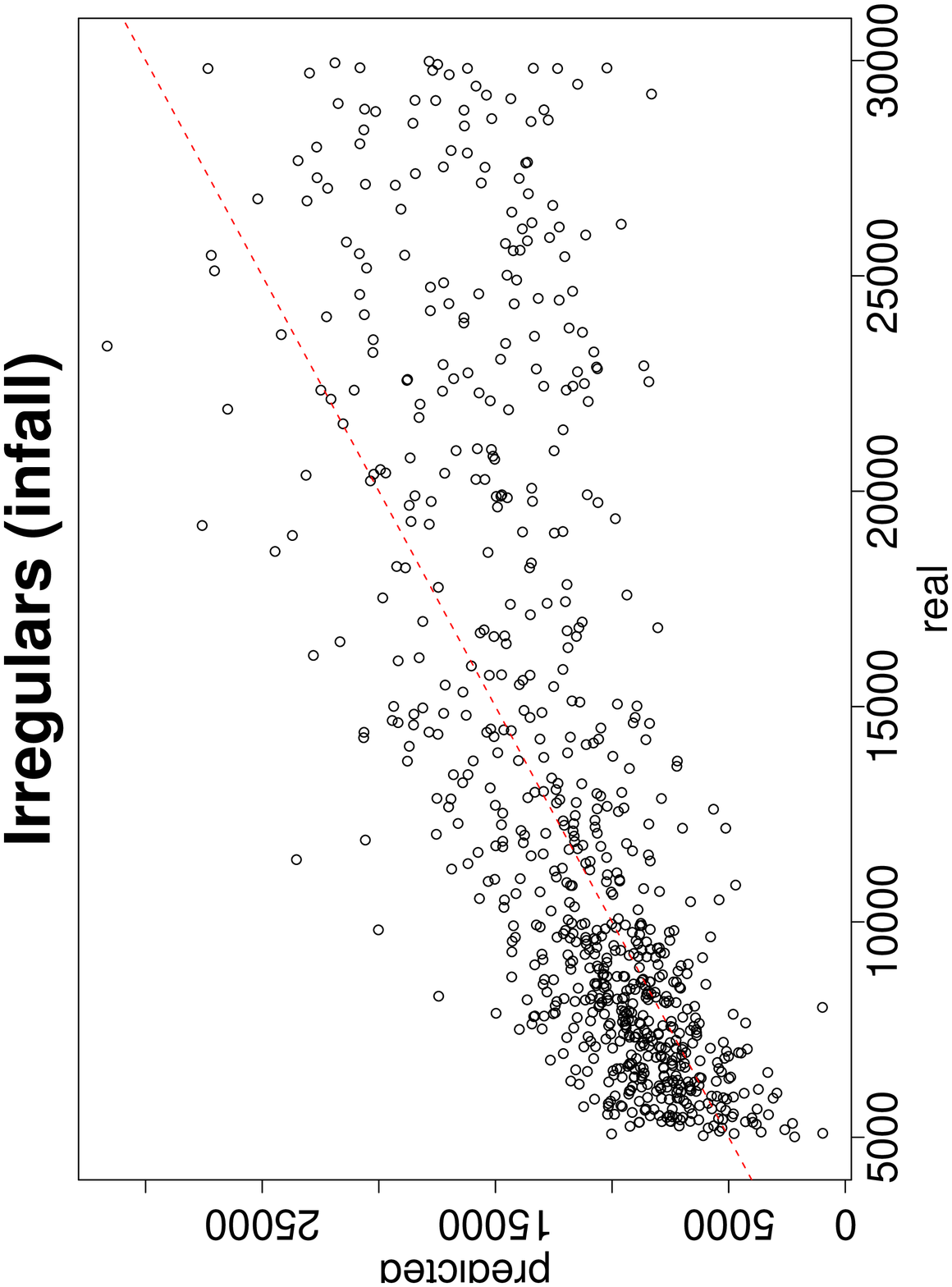}}
\put(12,20){\includegraphics{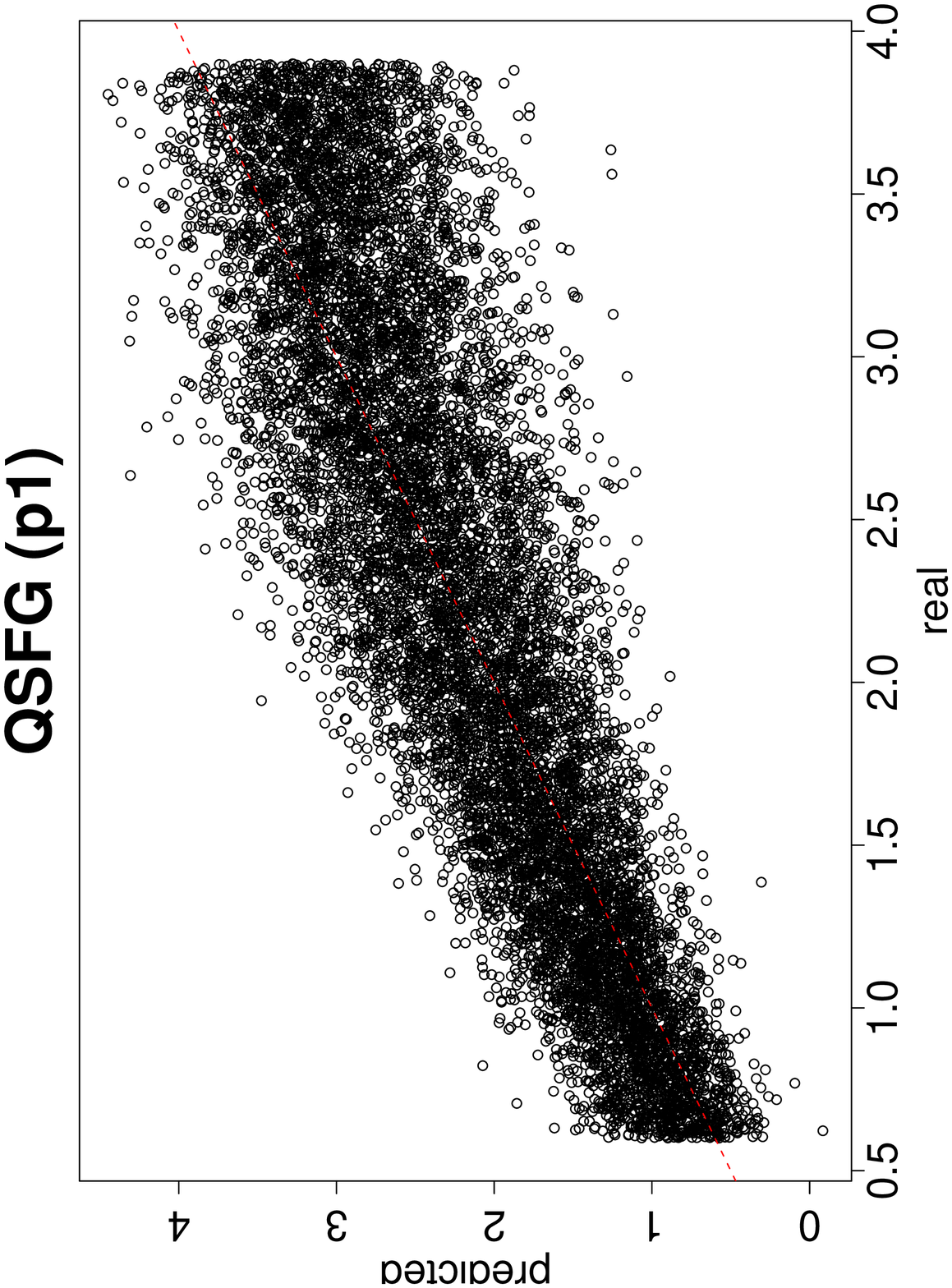}}
\put(12,15){\includegraphics{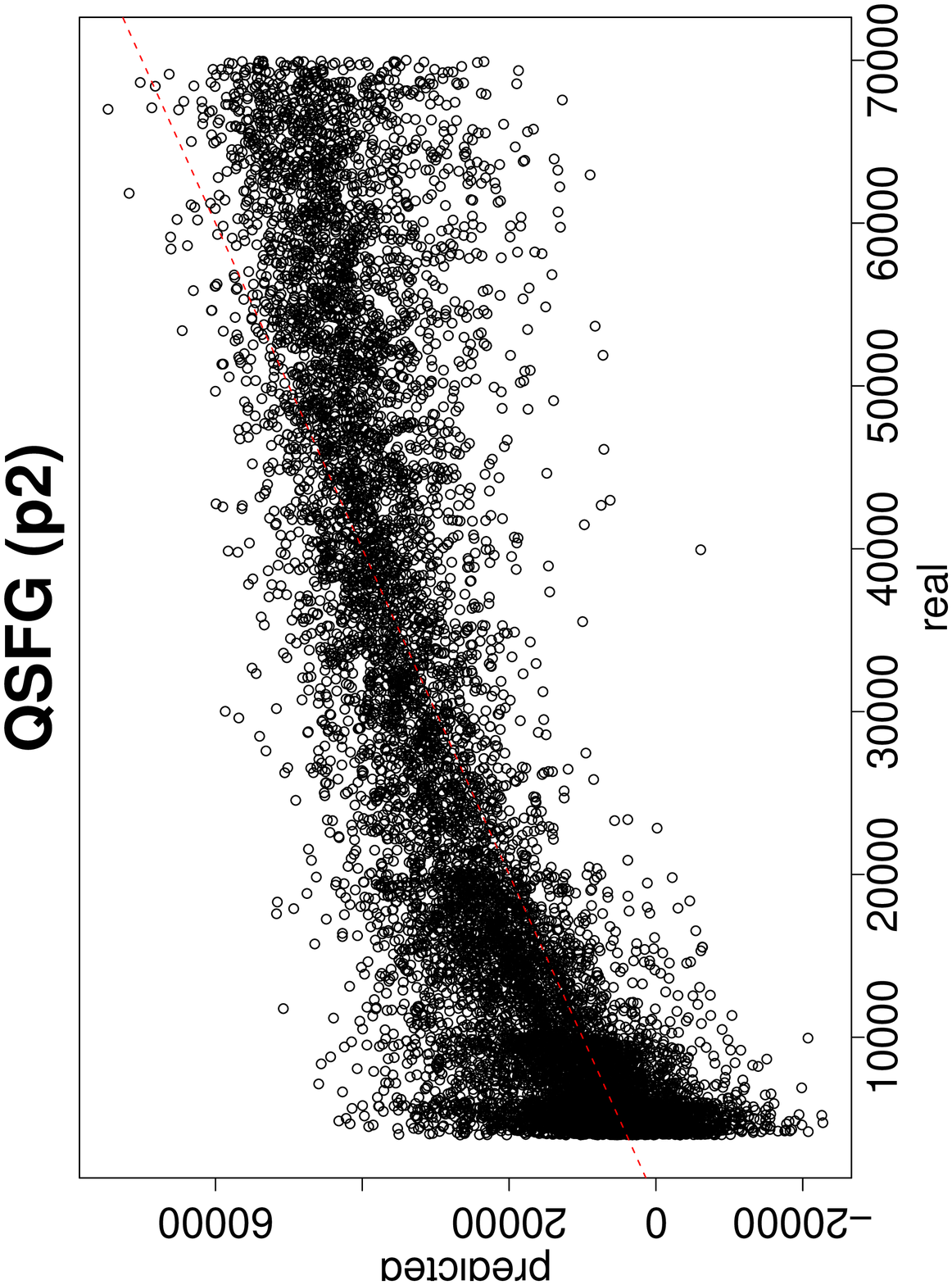}}
\put(12,10){\includegraphics{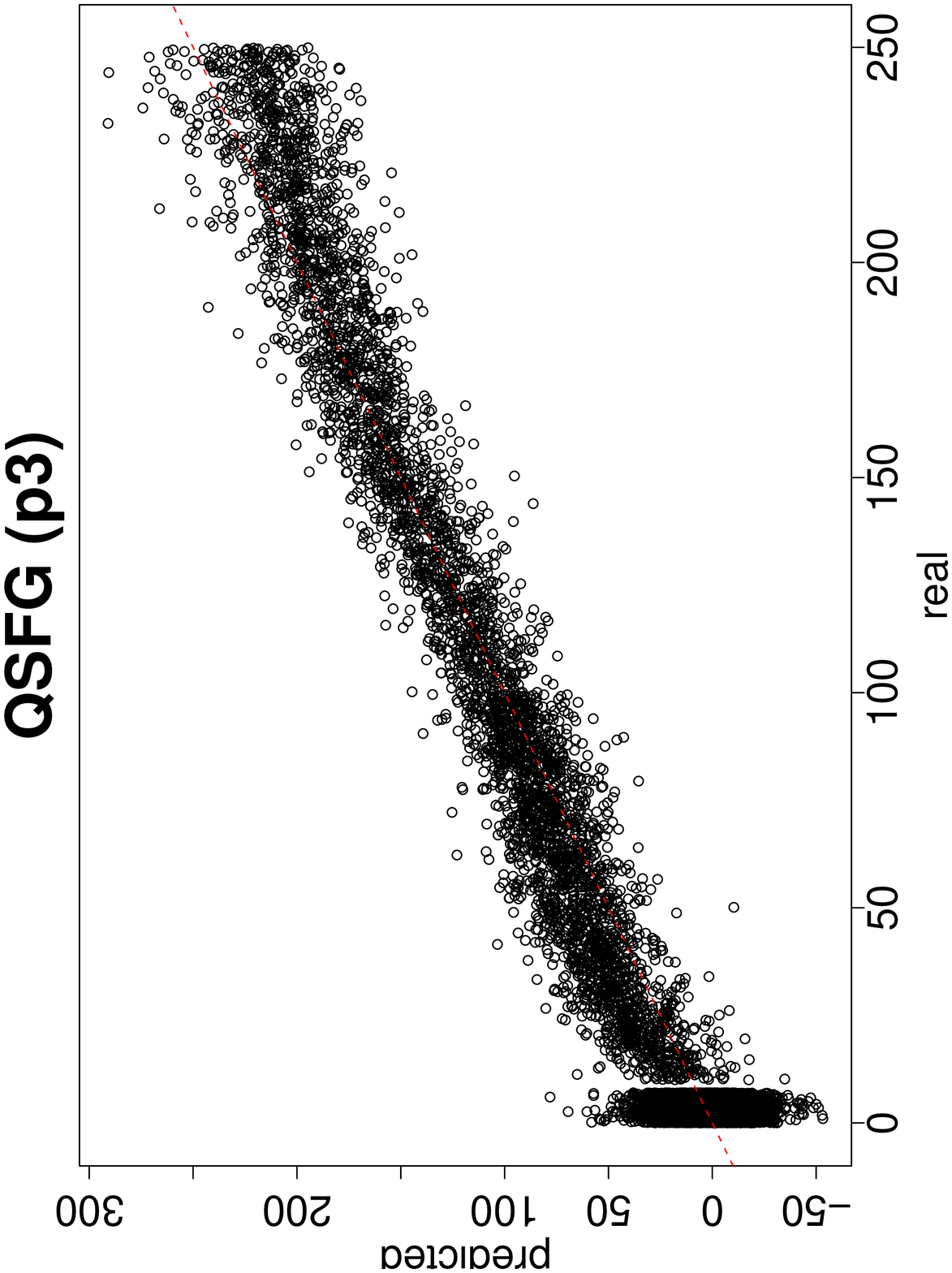}}
\put(12,5){\includegraphics{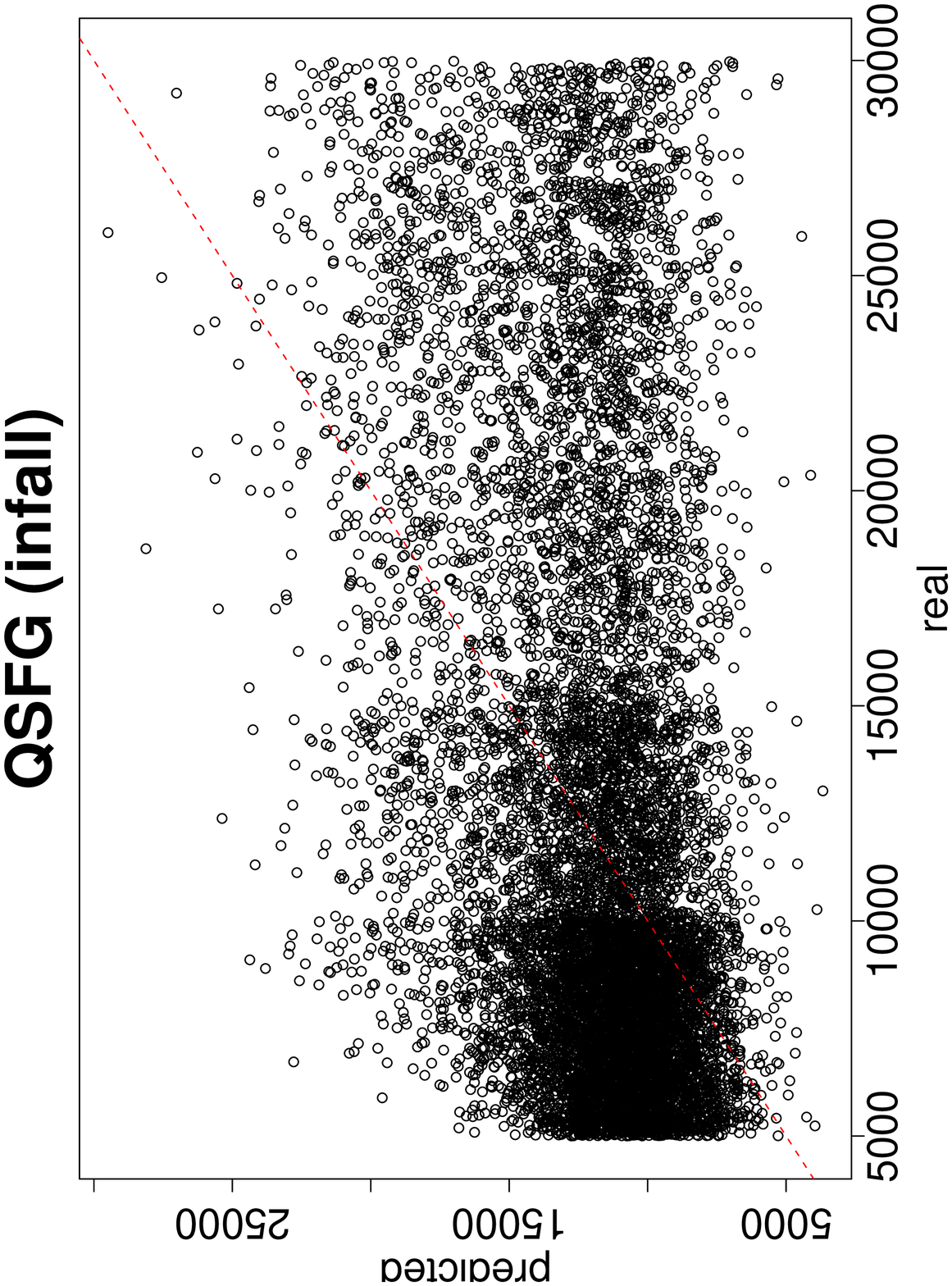}}
\end{picture}
\caption{
  Galaxy parameter estimation performance. For each of the input APs we plot
  the predicted vs.\ true AP values for the test set. The red line indicates
  the line of perfect estimation. The summary errors are given in
  Table~\ref{t5}.}
\label{f11}
\end{figure*}

As described in the introduction, the models used to produce 
the second library (Table \ref{tab1}) are not the same for each galaxy
type and therefore the parameters included are different (Table \ref{t1}). For this
reason, we decided to perform the parametrization independently for
each galaxy type. In the future, we plan to perform first a
classification of the galaxy type and then to have four parametrizers
(one for each type) that would be selected based on the type of
galaxy extracted. Here we perform the regression tests assuming the
classification was 100\% correct. In the paragraphs that follow we
discuss the results, presented in Fig. \ref{f11} and Table \ref{t5}, 
for each type separately.

i) \textbf{Early-type galaxies.} For the spectra of early-type galaxies, 
we performed the regression for the $p_{1}$ and $p_{2}$ parameters 
using the data described in Table \ref{t9} and obtained quite good results.
Comparing these results with those for
models of other galaxy types we see that the estimation of the
parameters for these models is much more precise. This is possibly a result of
the simpler model used to produce this type of galaxy. Using an exponential 
SF law characterized by only 2
parameters, we produced spectra characterized by fewer
degeneracies and therefore easier to parametrize.

ii) \textbf{Spiral galaxies.}  For spiral galaxies the parametrization was
performed on three parameters (2 for SFR and 1 for infall
timescale). The results of the regression are quite
poor, especially for the case of the $p_{1}$ parameter. This is 
partly because the SVM tuning was performed with a less detailed
scheme, but mainly to the higher complexity of the model used to
produce the spectra of this galaxy type.

iii) \textbf{Irregular galaxies.}  As in the case of spiral galaxies, 
the regression for spectra of irregular galaxies was performed for the same three
parameters with SVM. Once again the results are not very
good as can be seen from the large scatter 
in the resulting plots of the real against the
predicted values. The results are similar to the case
of spiral galaxies, which was expected since the models producing
these two types of galaxies are the same and they differ only in the
values of the model parameters. The results between these two types
seem to differ a lot only in the case of the infall timescale
parameter, where the very sparse grid for higher values of infall
timescale leads to a poor training of the SVMs.

iv) \textbf{Quenched star-forming galaxies.} To perform regression for the four
parameters of the models of quenched star-forming galaxies, 
we used the sample and the tuning scheme
presented in Table \ref{t9}. For the parameters in common 
between the models used to produce both irregular and quenched star-forming
galaxies, we can see that the performance of the SVMs in estimating them is
very similar and very poor. The only parameter of this galaxy type 
extracted with quite good accuracy is $p_{3}$, which seems
to have a large and direct impact on the galaxy spectra.

\begin{table*}
 \begin{center}
 \caption {Summary of the performance of the SVM regression 
models for predicting the input APs of the galaxy models.}
 \begin{tabular}{l c c c}          
 \hline\hline                        
  
Astrophysical parameter                & mean(real-predicted)/mean(real) & sd(real-predicted)/mean(real) & SVs   \\
 \hline
                                       & Early-type galaxies             &                               &       \\
 \hline
$p_{1}$                                & -0.008                          & 0.250                         & 674   \\
$p_{2}$                                & -0.001                          & 0.076                         & 646   \\
\hline
                                       & Spiral galaxies                 &                               &       \\
\hline
$p_{1}$                                & 0.018                           & 0.384                         & 2440  \\
$p_{2}$                                & 0.010                           & 0.451                         & 1866  \\
infall timescale                       & 0.009                           & 0.234                         & 1786  \\
\hline
                                       & Irregular galaxies              &                               &       \\
 \hline
$p_{1}$                                & 0.004                           & 0.174                         & 647   \\
$p_{2}$                                & 0.030                           & 0.519                         & 736   \\
infall timescale                       & 0.089                           & 0.393                         & 655   \\
\hline
                   & Quenched star-forming galaxies              &                               &       \\
\hline
$p_{1}$                                & 0.014                           & 0.195                         & 3116  \\
$p_{2}$                                & 0.044                           & 0.479                         & 3127  \\
$p_{3}$                                & 0.003                           & 0.335                         & 3077  \\
infall timescale                       & 0.038                           & 0.524                         & 3466  \\
\hline
\end{tabular}
\label{t5}
\end{center}
{\small \textsc{Note.} The sample is for zero redshift and interstellar extinction ($A_{v}$). The second and third columns list the mean and RMS errors, respectively. The final column gives the number of support vectors in the SVM model.}
\end{table*}

\subsubsection{Regression of output parameters of P\'EGASE}
\begin{figure*}[t]
  \setlength{\unitlength}{1cm}
\begin{picture}(18,15)
\put(0,15){\includegraphics{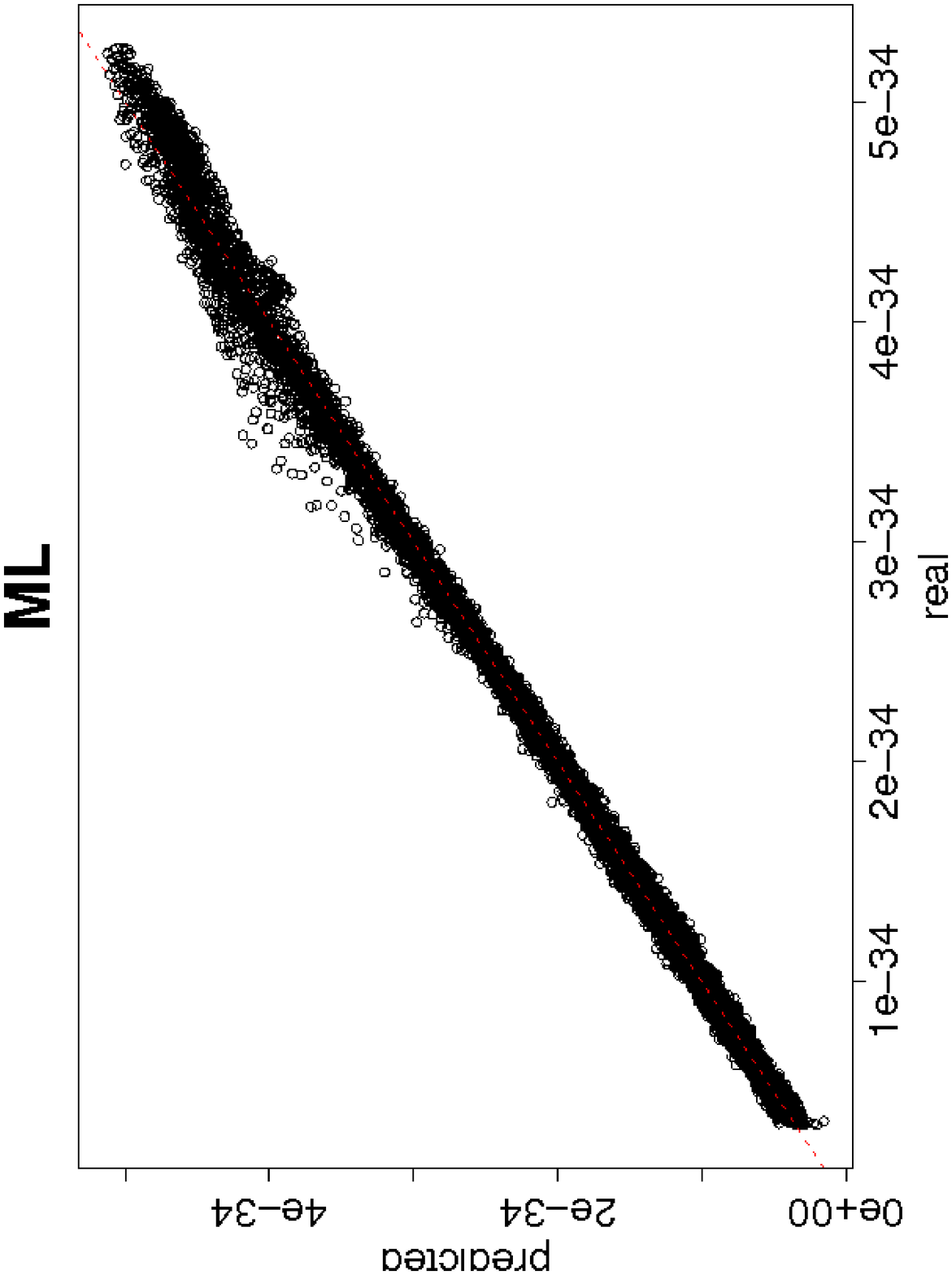}}
\put(0,10){\includegraphics{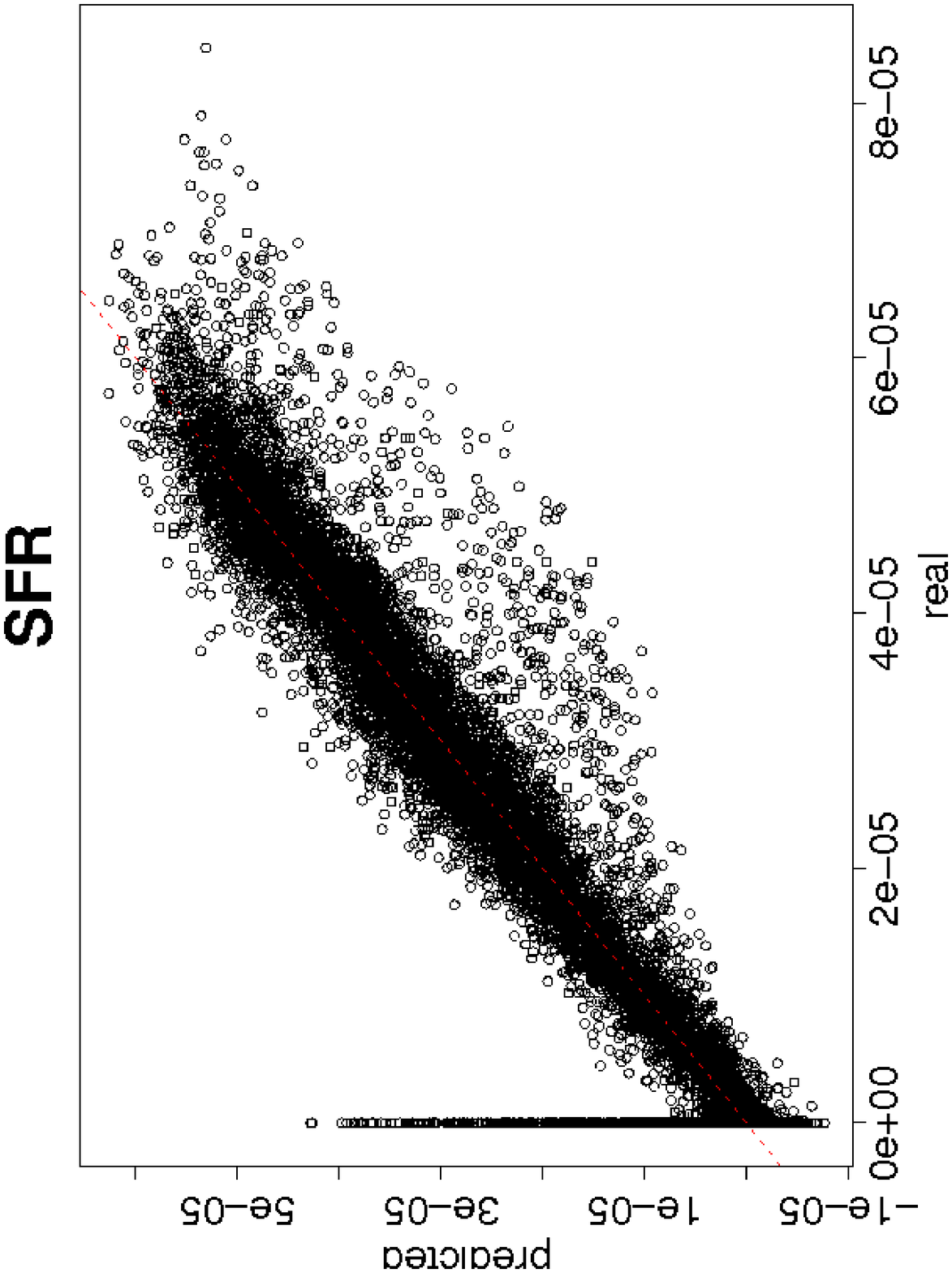}}
\put(0,5){\includegraphics{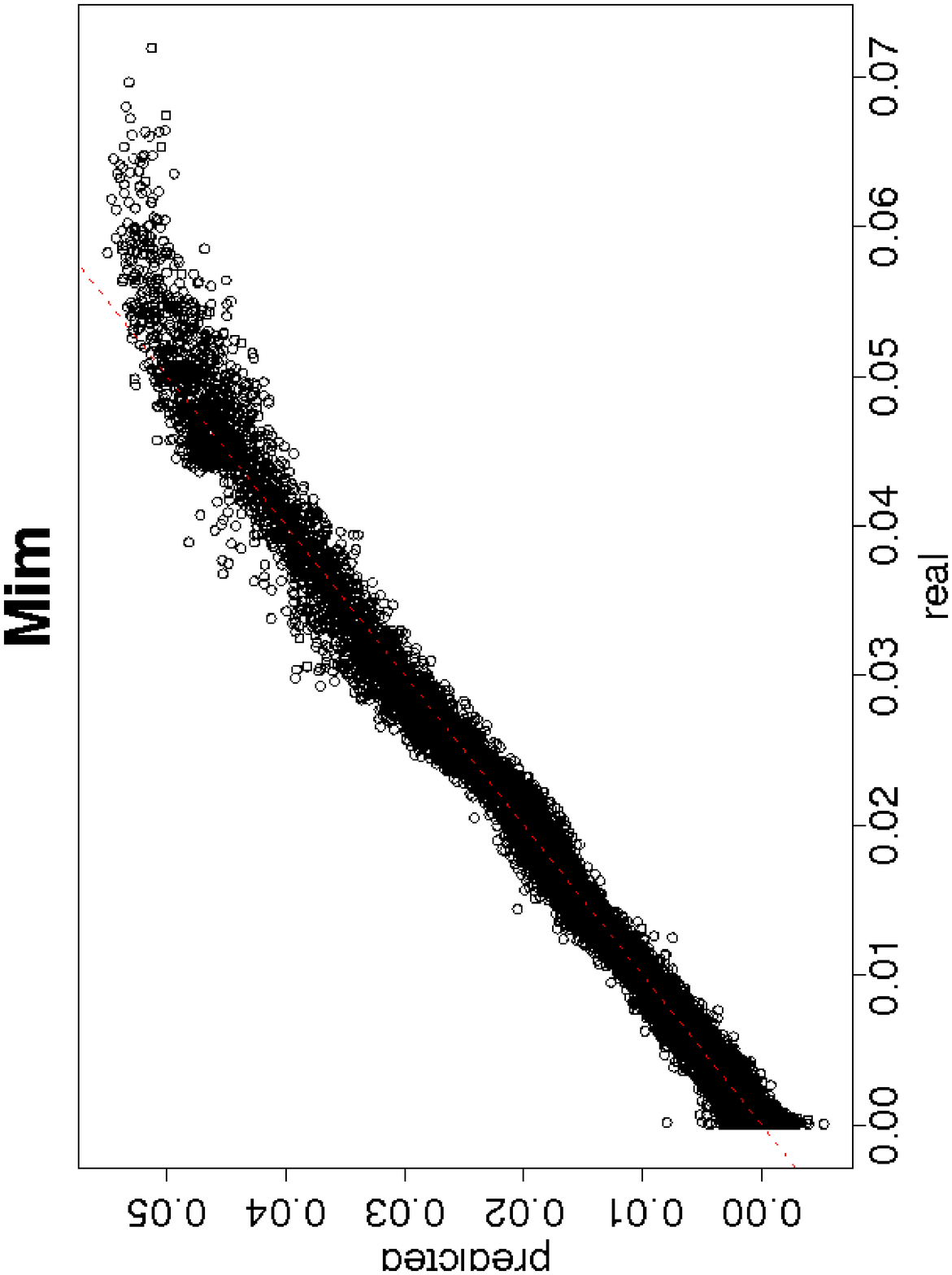}}
\put(6,15){\includegraphics{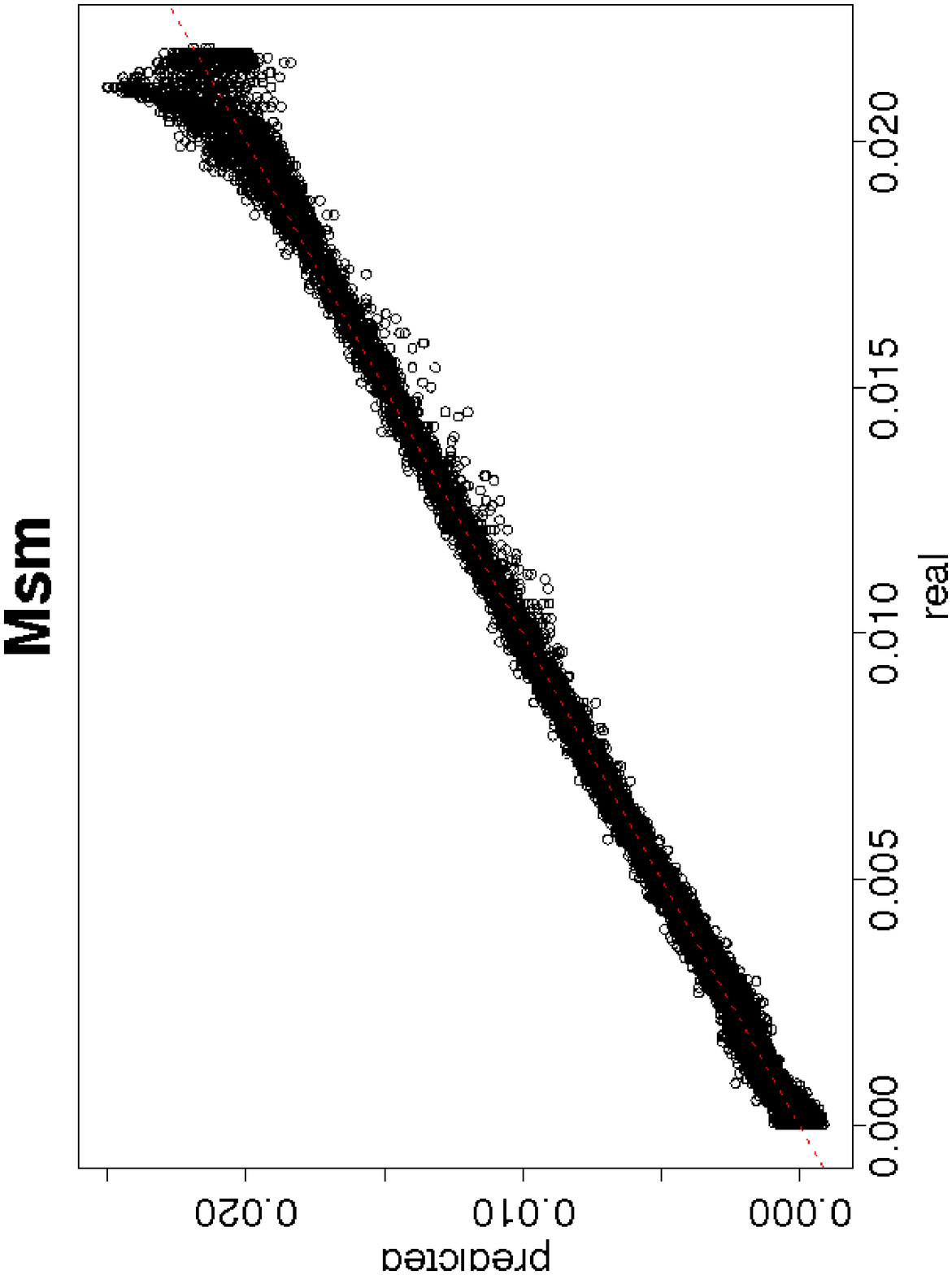}}
\put(6,10){\includegraphics{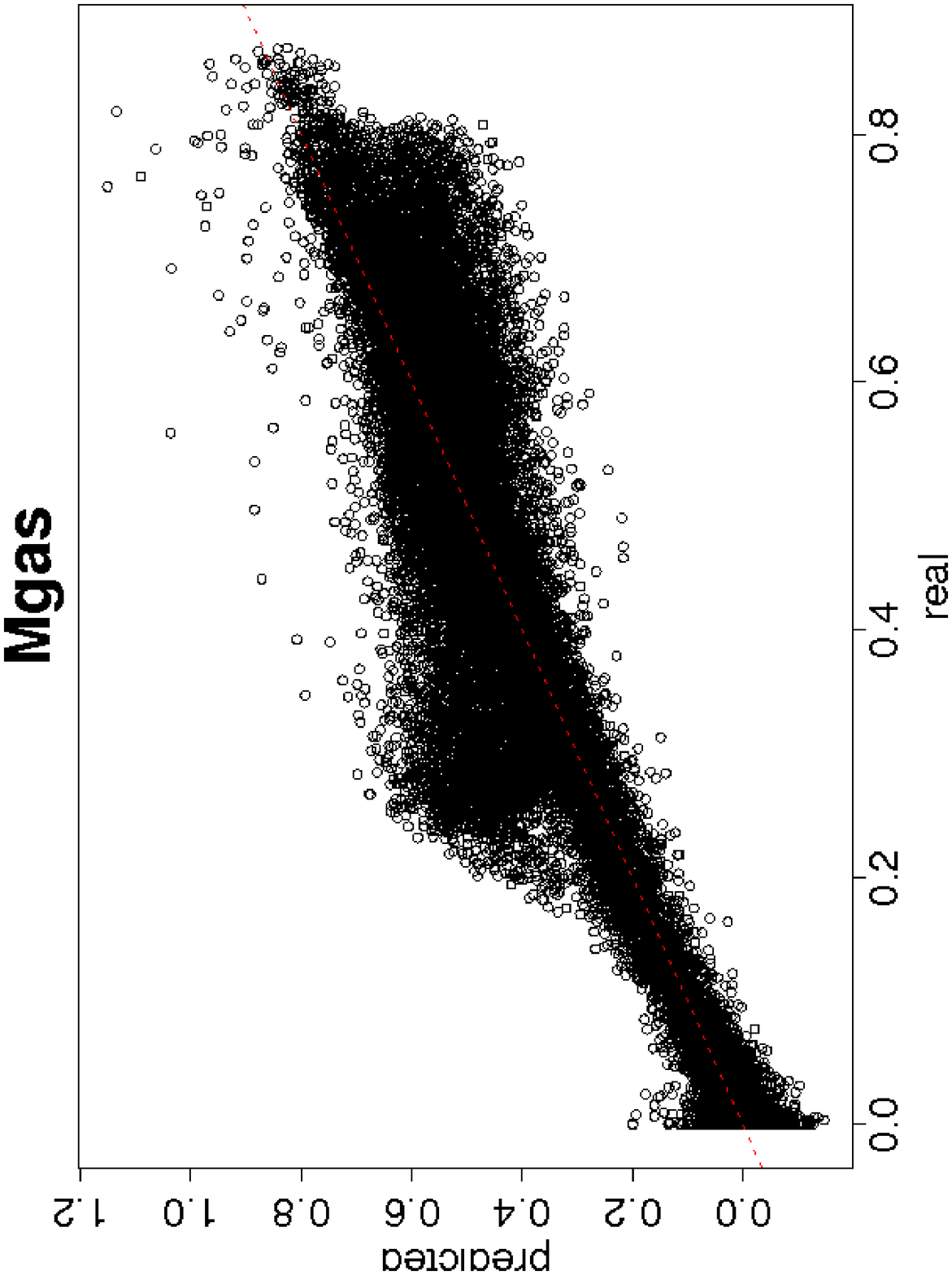}}
\put(6,5){\includegraphics{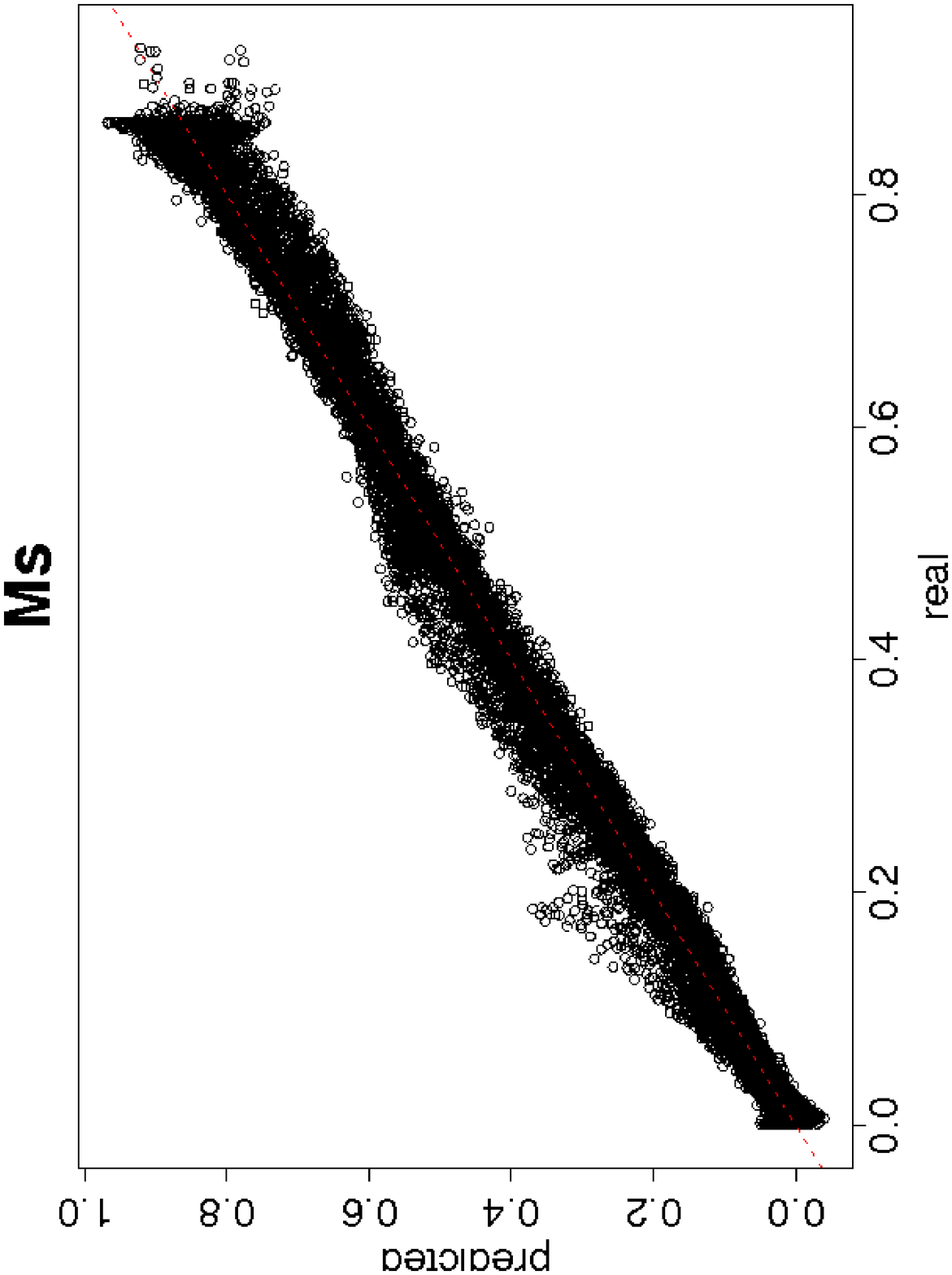}}
\put(12,15){\includegraphics{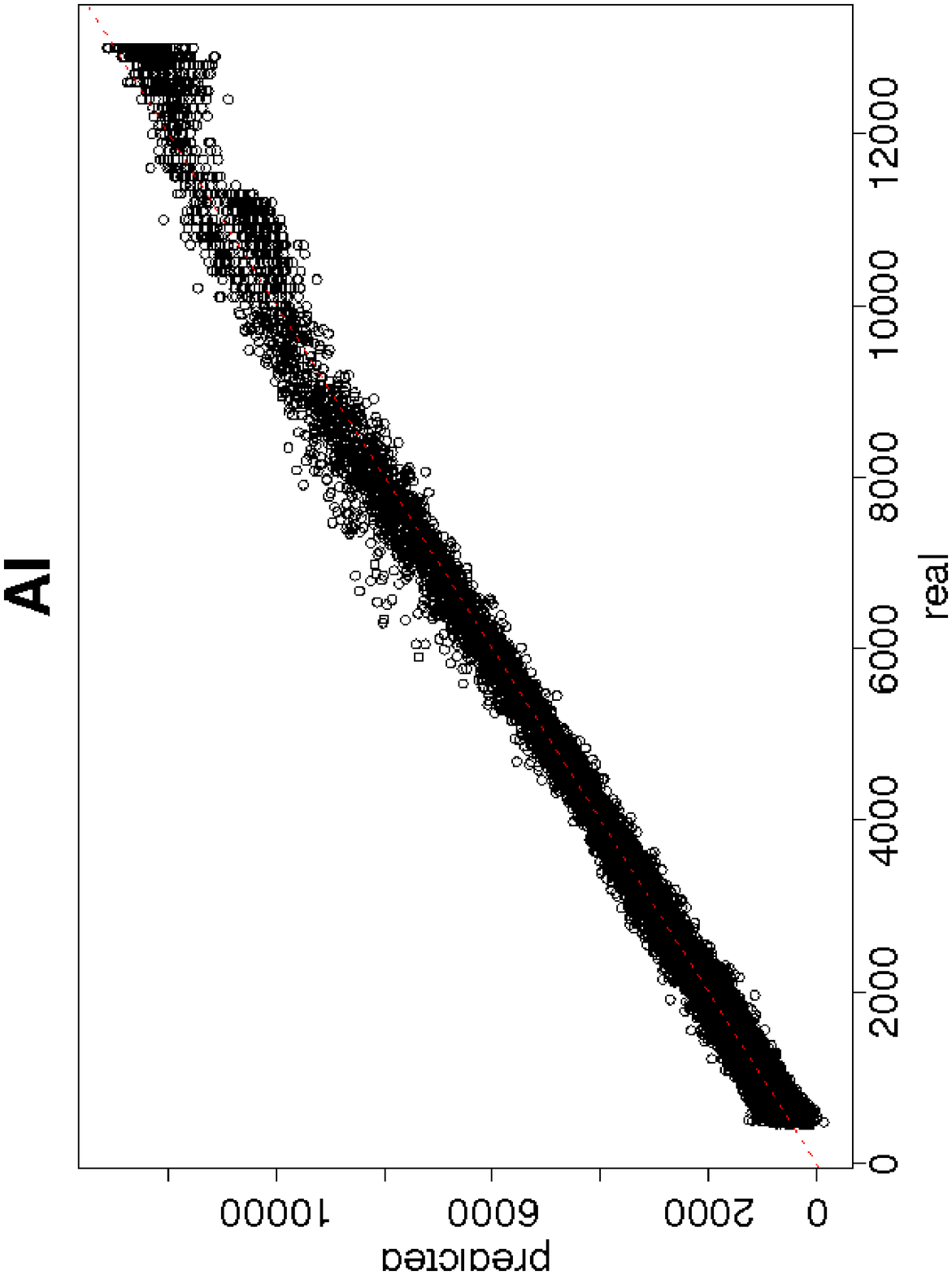}}
\put(12,10){\includegraphics{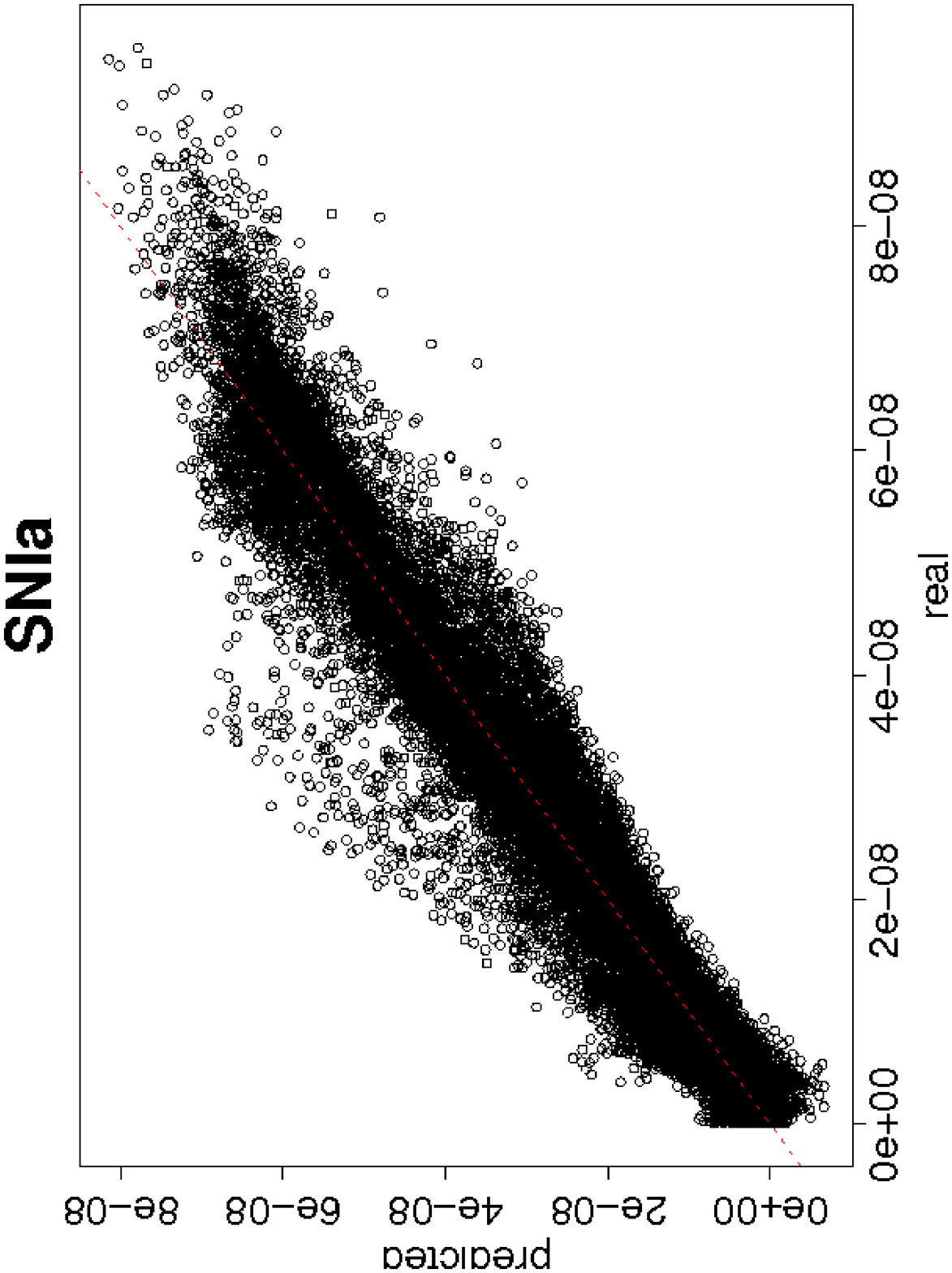}}
\put(12,5){\includegraphics{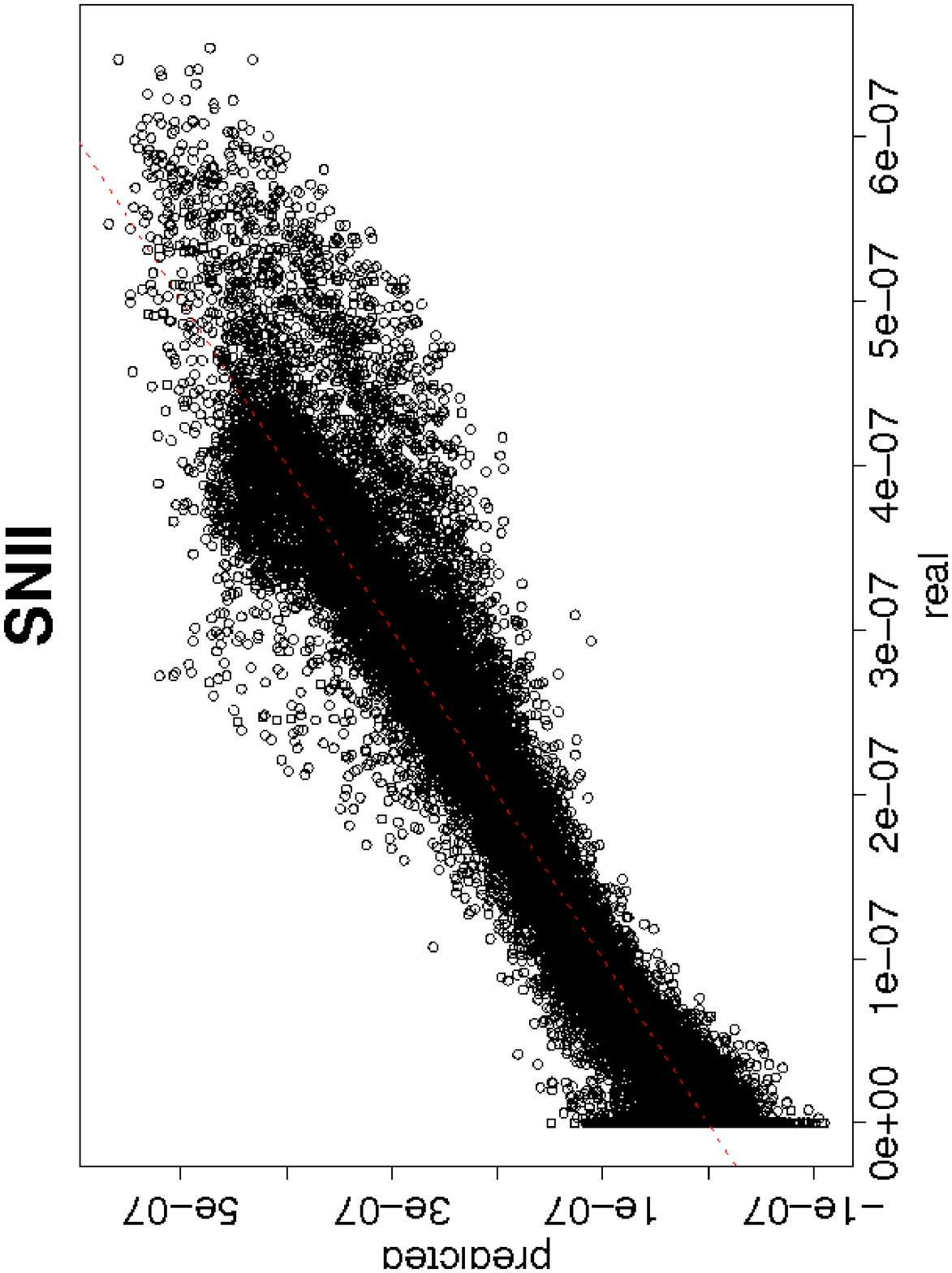}}
\end{picture}
\caption{
  Galaxy parameter estimation performance. For each of the output APs we plot
  the predicted vs.\ true AP values for the test set. The red line indicates
  the line of perfect estimation. The summary errors are given in
  Table~\ref{t11}.}
\label{f16}
\end{figure*}

The regression of the galaxy parameters was also performed for the
  nine most significant output parameters of P\'EGASE as in Paper
  I. We expect these parameters to be more strongly and directly
  related to the galaxy spectra and therefore easier to extract with
  the SVMs. Since these parameters are common for all the galaxy types,
  we performed regression for the whole sample of our simulated
  spectra. We can indeed estimate them more accurately 
  (see Table \ref{t9} and Fig. \ref{f16}) than the input parameters. 
  The largest errors appear in the cases
  of SNIa and SNII rate as well as in the case of the remaining mass
  of the gas in the galaxy and the current SFR. For the two first
  parameters, this was expected since they are more related to the
  emission lines of the spectrum, to which Gaia observations will not
  be very sensitive because of their low resolution. For the case of the
  gas mass, the problem is caused mainly due to irregular and
  quenched star-forming galaxies. Even though these types of galaxies are produced
  with very similar models to the ones used for spiral galaxies, the
  range of the input parameters used for them is very different. In
  particular, the values of the infall timescale are much higher in the
  case of starbust and irregular galaxies, especially compared to
  their age (see table \ref{t1}). This leads to degeneracies in the
  produced spectra. For example, galaxies that include a significant gas
  component and are dominated by a young stellar population might have
  the same amount of gas as galaxies with older stellar populations
  that are expected to have depleted their gas component, but may
  have ongoing gas infall. For the case of the current SFR,
  we observe that for many galaxies with a zero SFR, the estimated
  value was quite high. This is mainly a problem caused by the
  quenched star-forming galaxies for which the SFR is 
  currently zero but in many cases the SF
  stopped only very recently (e.g., 1\,Myr ago). This scenario would lead to
  a spectrum that is very similar to that of an irregular 
  galaxy (i.e., with a high current SFR) and
  it is therefore difficult for the SVMs to identify.

For all the other parameters, the results are very accurate. These
results indicate that we will be able to predict most of the
astrophysical parameters that characterize the galaxy spectra with
quite good accuracy, at least for G$\leq$18.5.

\begin{table*}
 \centering
 \caption {Similar to Table \ref{t5} but for the prediction of the output APs of the galaxy models.}
 \begin{tabular}{l c c c}          
 \hline\hline                        
  
Astrophysical parameter                                  & mean(real-predicted)/mean(real) & sd(real-predicted)/mean(real) & SVs   \\
 \hline
mass to light ratio (M/L)                                &  4.79e-4             & 3.97e-2            & 4566  \\
normalized star formation rate (SFR)                     & -7.69e-3             & 3.05e-1            & 4419  \\
metallicity of interstellar medium (Mim)                 &  4.52e-3             & 4.99e-1            & 3617  \\ 
metallicity of stars averaged on mass (Msm)              &  2.73e-3             & 6.98e-2            & 4796  \\
normalized mass of gas (Mgas)                            & -2.72e-3             & 2.25e-1            & 4499  \\ 
normalized mass in stars (Ms)                            & -2.57e-3             & 9.03e-2            & 4583  \\
mean age of stars averaged on bolometric luminosity (Al) & -5.49e-5             & 8.25e-2            & 4768  \\
normalized SNIa rate (SNIa)                              & -1.17e-3             & 1.85e-1            & 2966  \\
normalized SNII rate (SNII)                              &  1.70e-2             & 2.47e-1            & 3660  \\
\hline
\end{tabular}
\label{t11}
\end{table*}

\subsection{Galaxies with redshift, without reddening at G=18.5} 

We present both the classifications of galaxy type and
the regressions of redshift for 144 425 simulated spectra of
random redshift, without reddening at G=18.5. In both cases, 14 440
spectra were used for training and 129 985 for testing. Only a coarse search for the 
optimal SVM hyperparameters was performed.

\begin{table}
      
 \begin{center} 
 \caption {Galaxy classification with the SVM for the training set.}                                   
 \begin{tabular}{l | c c c c}          
 \hline\hline                        
  
Type     & E    & S    & Im  & QSFG   \\
 \hline
E        & 1131 & 276  & 0   & 1    \\
S        & 41   & 5238 & 10  & 3    \\
Im       & 0    & 63   & 701 & 3    \\ 
QSFG     & 0    & 53   & 4   & 6916 \\

\hline
\end{tabular}
\label{t6}
 \centering
 \caption {As Table~\ref{t4} but for the test set.}
 \begin{tabular}{l | c c c c}          
 \hline\hline                        
  
Type     & E     & S     & Im   & QSFG   \\
 \hline
E        & 9343  & 3319  & 0    & 10    \\
S        & 666   & 46283 & 455  & 149   \\
Im       & 0     & 1022  & 5516 & 195   \\ 
QSFG     & 0     & 792   & 205  & 62030 \\

\hline
\end{tabular} 
\label{t7}
\end{center}
{\small \textsc{Note.} The confusion matrices for galaxies 
at $z\geq{0}$. Columns indicate the true class and the rows 
the predicted ones. The labels E, S, Im, and QSFG represent 
early-type, spiral, irregular, and quenched star-forming galaxies, 
respectively.}
\end{table}

In Tables \ref{t6} and \ref{t7} (confusion matrices), 
we can see that the misclassifications
numbered 454 for the training and 6813 for the testing set
correspond to errors of 3.1\% and 5.2\% respectively. Comparing
the results with those in Sect. 7.1.1 (2.9\% for
the testing set), we can see that although the error is small it
is still two times larger. This result agrees with those in our previous
paper (Tsalmantza et al. \cite{tsalmantza}) for the tests of the first
library, where we concluded that the redshift is a parameter that
should be estimated in advance of the classification and
the estimation of the other parameters.

\begin{figure}
\centering
\includegraphics[width=6cm,angle=-90]{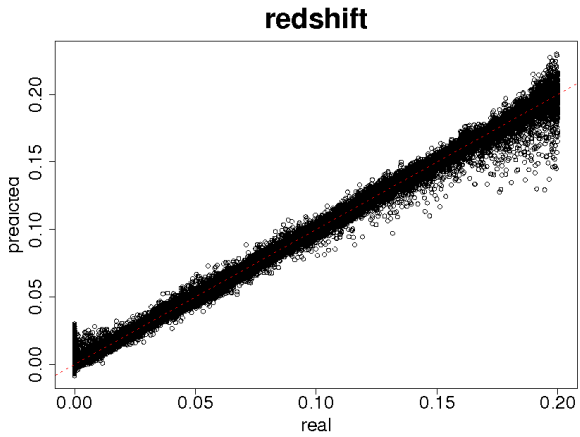}
\caption{Galaxy redshift estimation performance. We plot
  the predicted versus true z values for the test set. The red line indicates
  the line of perfect estimation. The summary errors are given in
  Table~\ref{t8}.}
\label{f13}
\end{figure}

If we follow this classification scheme, the accuracy
in the performance of regression for the redshift parameter will be very
important to all the results we will extract from galaxy observations
with Gaia. Using the same subsets for training and testing SVMs 
and following the same procedure for tuning as in
the classification we extracted the values of redshift. 
The results are very good and they 
are presented in Table \ref{t8} and Fig. \ref{f13}. This is very 
promising for the performance of the classification and 
parametrization of the Gaia galaxy observations.

\begin{table*}
 \begin{center}
 \caption {Summary of the performance of the SVM regression models for predicting the z.}
 \begin{tabular}{l c c c}          
 \hline\hline                        
  
Astrophysical parameter                & mean(real-predicted)/mean(real) & sd(real-predicted)/mean(real) & SVs   \\
 \hline
 z                                     & -3.61e-5                        & 0.070                         & 2160   \\
\hline
 \end{tabular}
 \label{t8}
 \end{center}
{\small \textsc{Note.} The sample is for zero interstellar 
extinction ($A_{v}$) and five random values of redshift 
in the range 0-0.2. The second and third columns list the 
mean and RMS errors respectively. The final column gives 
the number of support vectors in the SVM model.}
\end{table*}

\subsection{Galaxies with reddening, redshift at G=18.5} 
We used the 144 425 galaxy spectra that include the effects of
reddening to perform a regression analysis of the Av and z parameters. The number of 
spectra used as a training and testing set, as well as the scheme used 
for the tuning of the SVMs, is given once again in Table \ref{t9}.

\begin{figure}
\centering
\includegraphics[width=6cm,angle=-90]{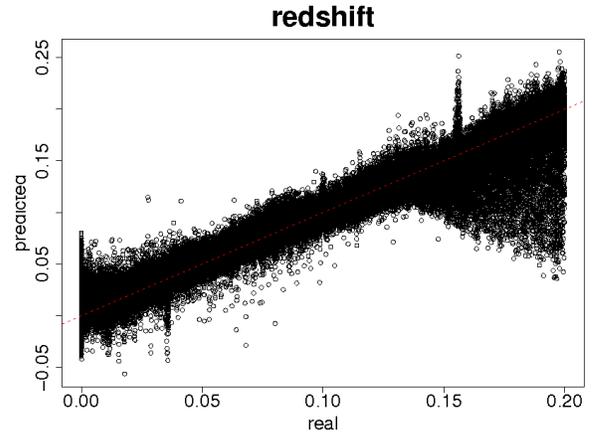}
\caption{Galaxy redshift estimation performance. We plot
  the predicted versus true z values for the test set. The red line indicates
  the line of perfect estimation. The summary errors are given in
  Table~\ref{t15}.}
\label{f17}
\end{figure}

As expected, the results of the regression analysis for the redshift parameter 
(Fig. \ref{f17} and Table \ref{t15}) are worse
than in the case where no reddening was included in the data, the
problem becoming increasingly obvious towards high redshift. This
indicates that we should estimate the reddening values before performing the
regression of the redshift.

\begin{table*}
 \centering
 \caption {As table \ref{t8} for the z and $A_{v}$ parameters 
for galaxy spectra that include redshift and reddening.}
 \begin{tabular}{l c c c}          
 \hline\hline                        
  
Astrophysical parameter                & mean(real-predicted)/mean(real) & sd(real-predicted)/mean(real) & SVs   \\
 \hline
 z                                     & 0.012                           & 0.190                         & 7291  \\
 Av                                    & 0.002                           & 0.100                         & 5478  \\
\hline
\end{tabular}
\label{t15}
\end{table*}

\begin{figure}
\centering
\includegraphics[width=6cm,angle=-90]{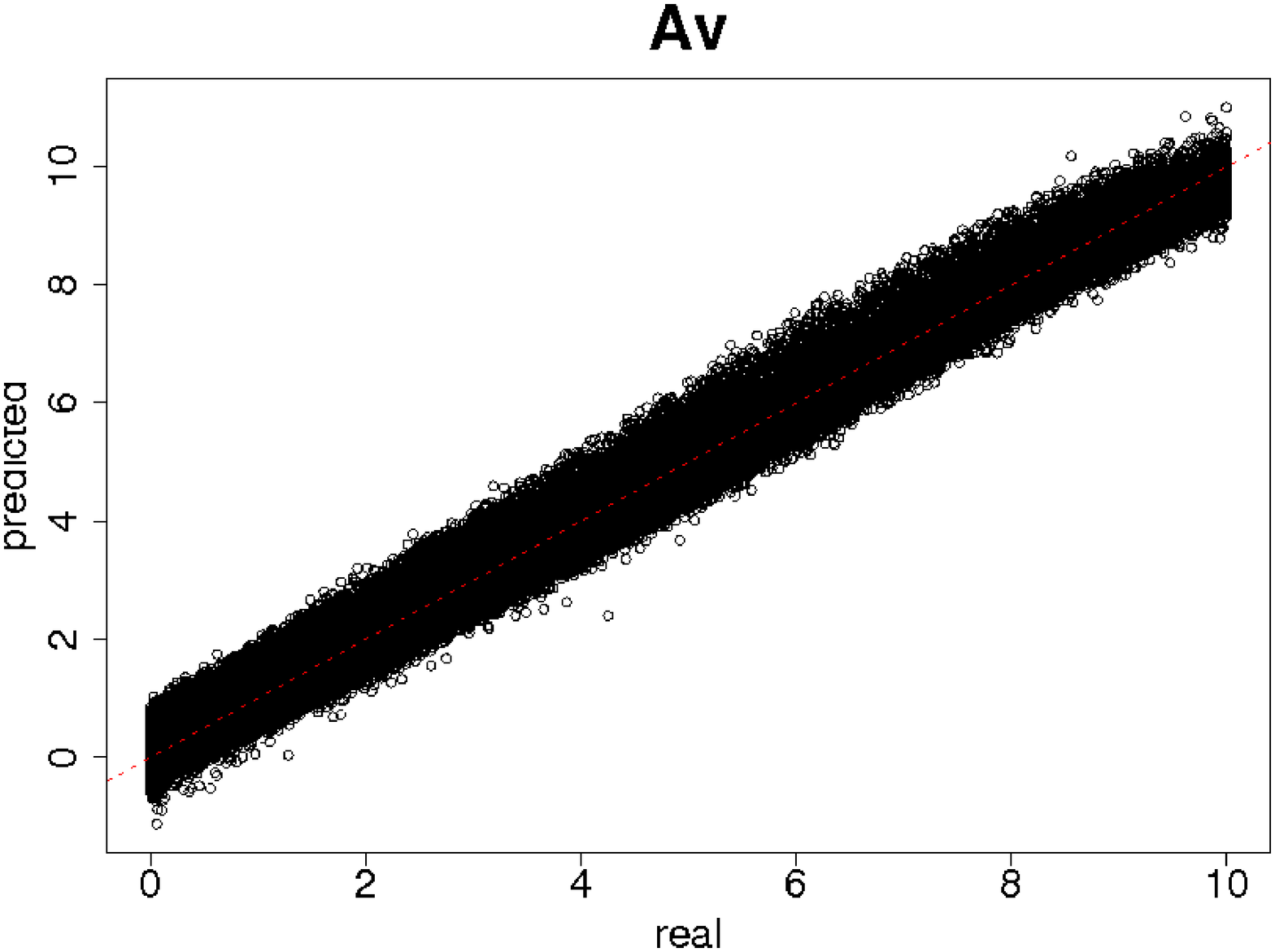}
\caption{Galactic-interstellar extinction estimation performance. We plot
  the predicted versus true Av values for the test set. The red line indicates
  the line of perfect estimation. The summary errors are given in
  Table~\ref{t15}.}
\label{f15}
\end{figure}

The results of regression for the Av parameter are once again very
good (Fig. \ref{f15} and table \ref{t15}). These results are promising, 
because they indicate that we should be able to 
estimate the effects of extinction with reasonable
accuracy before extracting the values of the redshift and performing
the rest of our classification scheme. We plan to compare the Av 
estimated in this way with that estimated from Gaia data of nearby 
stars.

\section{Discussion and conclusion}

The first results of the SVM classification and parametrization of the
second library of synthetic galaxy spectra are very good for the 
classification of galaxy types and regression of most of the output
parameters of the model, redshift, and reddening. However, we 
emphasize that although the regression results seem to be quite accurate 
for most output parameters estimated here (e.g., current 
SFR, metallicity, stellar mass), they might include large errors because of 
discrepancies between models and reality. All the parameter values 
used here to train the SVM classifiers are model dependent. The models used
are a simplification of the complex structure and evolution of galaxies and 
therefore cannot lead to accurate predictions of their parameters. These 
models also are unable to simulate the complete range of detail 
occuring in the universe. For these reasons, the results presented here should 
be used for statistical studies of the main galaxy properties in the local 
universe and not as absolute values for each individual galaxy.
In contrast to the output parameters, the results are very 
poor for the majority of the astrophysical parameters used to produce 
each type of galaxy, implying that Gaia will not be able to provide 
accurate measurements for input galaxy parameters. However, this is 
something that we should investigate further (i.e., using different 
parametrization methods) to check whether these results can be improved.

We have used the P\'EGASE.2 galaxy evolution model and
observational data from SDSS to solve problems with our first
library and extend the library to cover the large majority of
observational data parameter space. In this way, 
an extended library of 28 885 synthetic galaxy spectra was 
created at zero redshift and reproduced in addition for 4
random values of redshift. The whole library was produced for a
random grid of the astrophysical parameters used by P\'EGASE.2 models.
The models used in P\'EGASE.2 to create early-type galaxies was
changed and an exponential model for their SFR was adopted. Models for
quenched star-forming galaxies which were not included in the first library were
also added. In the case of irregular and spiral galaxies we extended
the range of input parameter values. The resulting library
includes four general Hubble types instead of seven that were included
in the first library and covers almost all the variance in the SDSS
photometric observations. To investigate the range of
input parameters in the models for each type we made use of
photometric data (SDSS and Paturel et al. \cite{paturel}). Even
though the comparison of our library with colour observations provides
good results, it is possible that some combinations or values of
input parameters produce spectra that do not correspond to
realistic galaxy spectra of those types. As an example, we propose
that values of the $p_{1}$ parameter of the early-type galaxies as
high as 30\,Gyr might lead to unrealistic spectra of early-type
galaxies. These values were kept because of the good agreement with the
photometric observations but they might be excluded or characterized
as spectra of a different galaxy type in future versions of our
library if they are found not to match real spectra. For this purpose,
we intend to compare the second library of synthetic spectra of
galaxies with a larger sample of observational spectra from SDSS.

The second library produced here was compared with other observations, 
both photometric (Paturel et al. \cite{paturel}) and spectroscopic 
(Kennicutt \cite{kennicutt}), and found to be in good agreement 
with them. The only problem appears in the case of quenched star-forming 
galaxies where the synthetic spectra of this type do not seem 
to fit very well any type of observed spectra in the Kennicutt 
Atlas (Fig. \ref{a5}). Additionally, even though the SDSS colours 
of this type of galaxy are very similar to those of starburst galaxies, 
the comparison of the spectra of these two types showed that they do not 
reproduce the strong emission lines present in the observational data. 
To solve this problem we intend to produce starburst galaxies using a 
new version of the P\'EGASE model that includes new models for all the 
mechanisms that are important to this type of galaxy. In future versions 
of our synthetic library, we intend to investigate the role of a wider 
range of astrophysical parameters in the models used in P\'EGASE. 
For example, we need to investigate the range of galaxy age parameter, 
which has a great impact on the output spectra. Here, it was simply kept 
constant at 9 or 13\,Gyr depending on the galaxy type.

For the task of classification and parametrization of unresolved 
galaxies with Gaia, we will also construct a semi-empirical library 
of galaxy spectra. This library will include observational spectra
from SDSS, which will be extended to the wavelength range of Gaia 
by our synthetic spectra. The advantage of this library is that it 
provides a set of real observed spectra, with the corresponding 
astrophysical parameters, as defined by comparing each spectrum 
with the synthetic spectra. A library of observed galaxy spectra 
combined with the already produced synthetic libraries will check 
and improve our classification system and test the reliability and 
completeness of our libraries.

\section{Acknowledgments}
The authors would like to thank the Greek General Secretariat
of Research and Technology (GSRT) for financial support and the ELKE of UOA. 
P. Tsalmantza would also like to thank the Institut d'Astrophysique de Paris (IAP) 
for their support and hospitality.

This work makes use of Gaia simulated observations and we
thank the members of the Gaia DPAC Coordination Unit
2, in particular Paola Sartoretti and Yago Isasi, for their
work. These data simulations were done with the MareNostrum
supercomputer at the Barcelona Supercomputing Center
- Centro Nacional de Supercomputacion (The Spanish
National Supercomputing Center).

Funding for the Sloan Digital Sky Survey (SDSS) has been provided 
by the Alfred P. Sloan Foundation, the Participating 
Institutions, the National Aeronautics and Space Administration, 
the National Science Foundation, the U.S. Department of 
Energy, the Japanese Monbukagakusho, and the Max Planck Society. 
The SDSS Web site is http://www.sdss.org/. 
The SDSS is managed by the Astrophysical Research Consortium (ARC) 
for the Participating Institutions. The Participating 
Institutions are The University of Chicago, Fermilab, the Institute 
for Advanced Study, the Japan Participation Group, 
The Johns Hopkins University, the Korean Scientist Group, Los Alamos 
National Laboratory, the Max-Planck-Institute for 
Astronomy (MPIA), the Max-Planck-Institute for Astrophysics (MPA), 
New Mexico State University, University of Pittsburgh, 
University of Portsmouth, Princeton University, the United States 
Naval Observatory, and the University of Washington.

\appendix

\section{Comparison of the second library with Kennicutt's atlas}
We present the results of the $\chi^{2}$-fitting of the galaxy spectra 
in Kennicutt's atlas (black lines) with the ones in our library (red lines). 
The results are presented in 5 blocks, each one containing the real 
spectra of a particular galaxy type. The synthetic spectrum plotted 
over the real one corresponds to the one with the smallest 
$\chi^{2}$-difference. For every spectrum, we present both 
results extracted when the $\chi^{2}$-fitting was performed 
by excluding or including the areas with the strongest emission 
lines. The vertical green lines indicate those spectral areas 
(from the beginning until the first line, between the two lines 
in the middle and from the last line until the end of the spectrum). 
When the strongest emission lines were included in the comparison, 
the resolution in those three areas was decreased to only one point 
for both sets of spectra.

\begin{figure*}[h]
\center
\includegraphics[angle=-90,width=0.225\columnwidth]{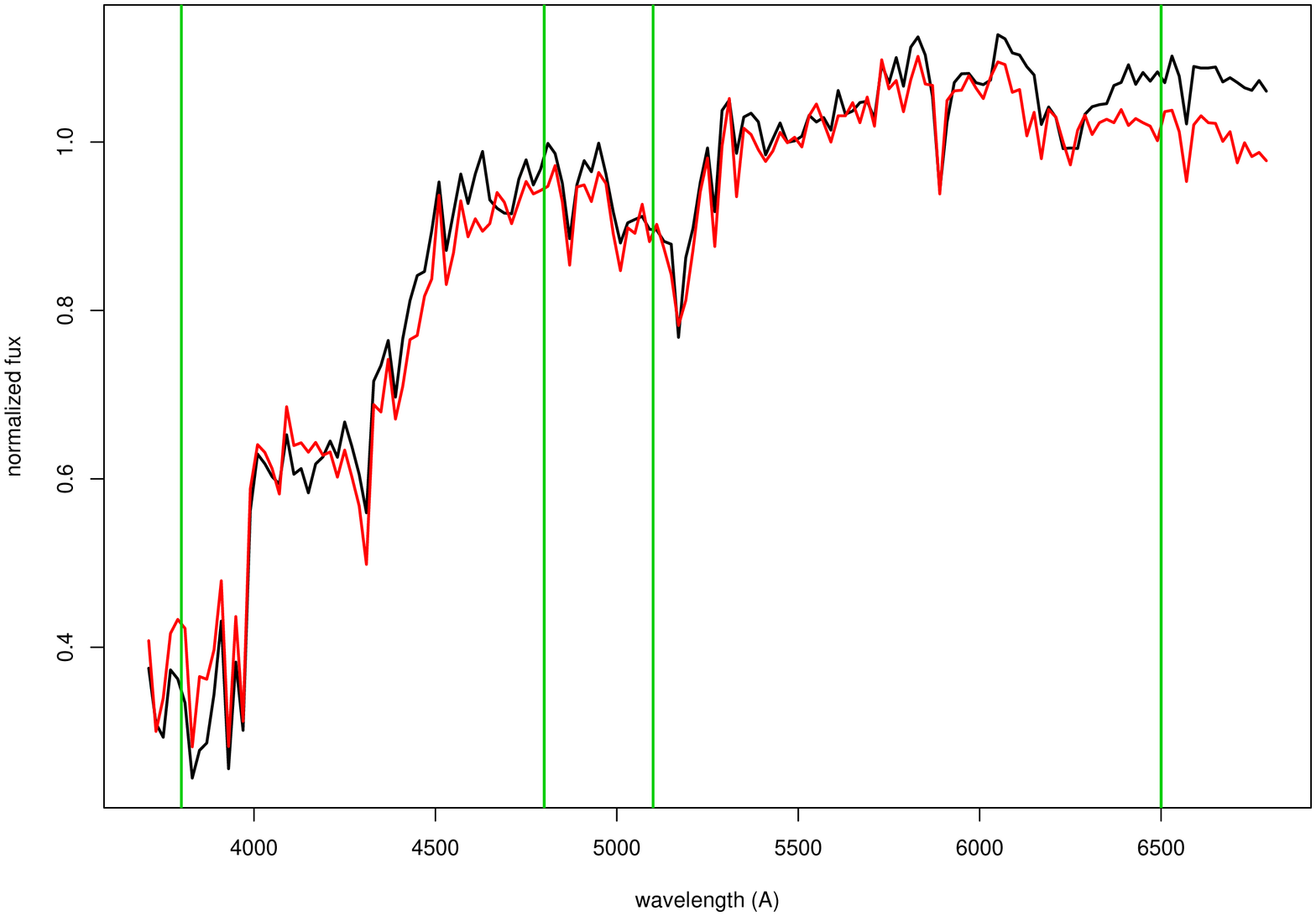}
\includegraphics[angle=-90,width=0.225\columnwidth]{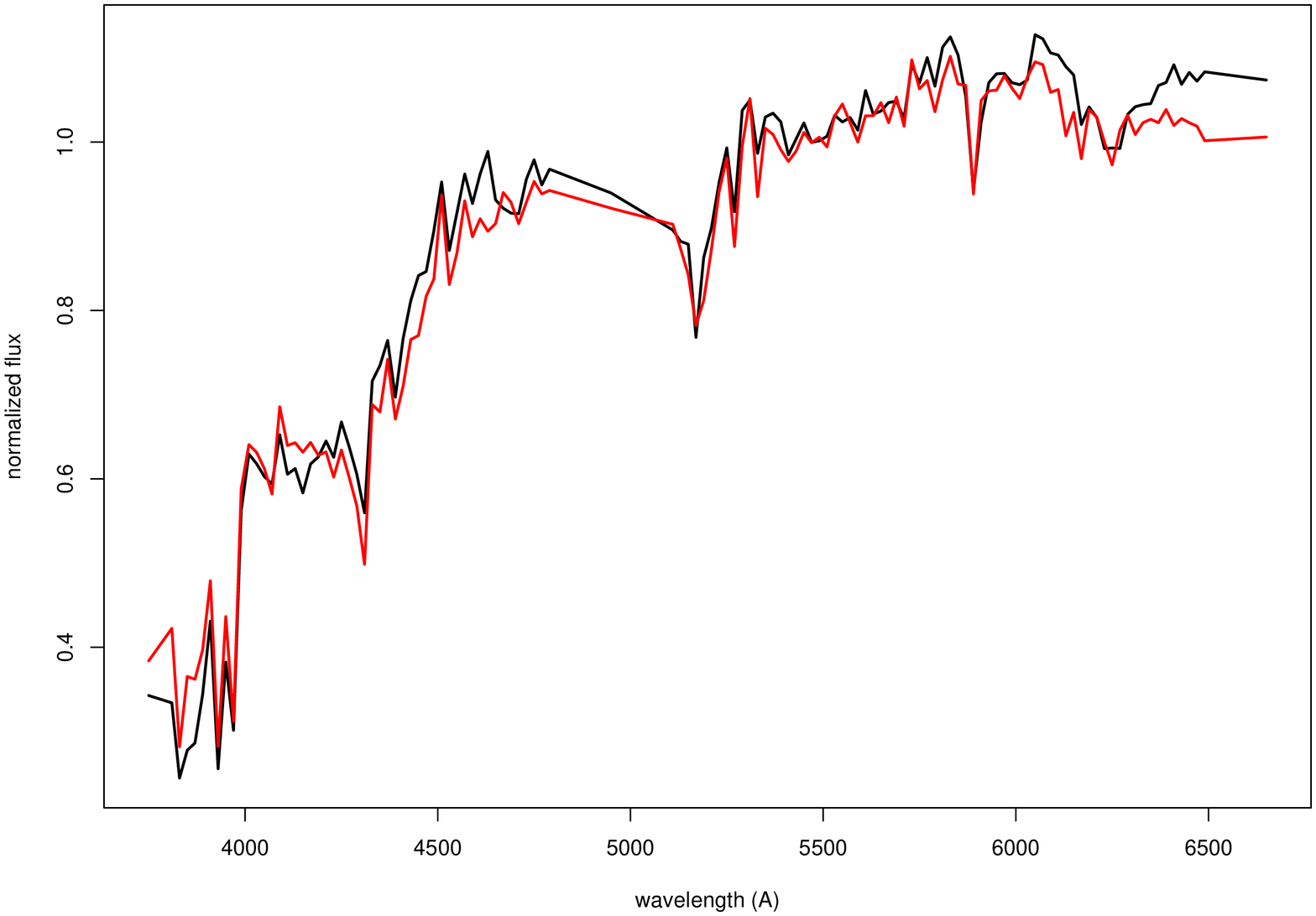}
\includegraphics[angle=-90,width=0.225\columnwidth]{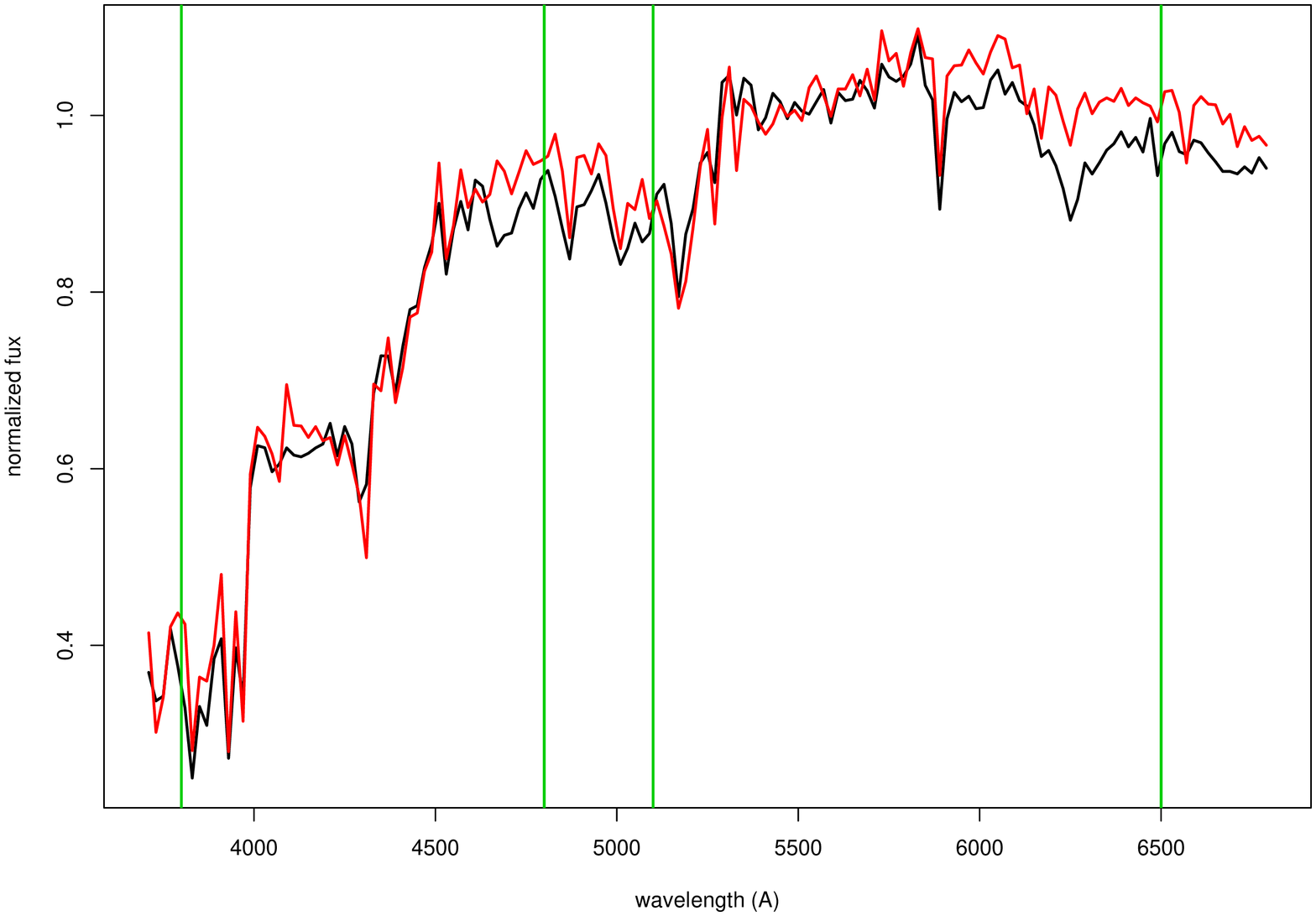}
\includegraphics[angle=-90,width=0.225\columnwidth]{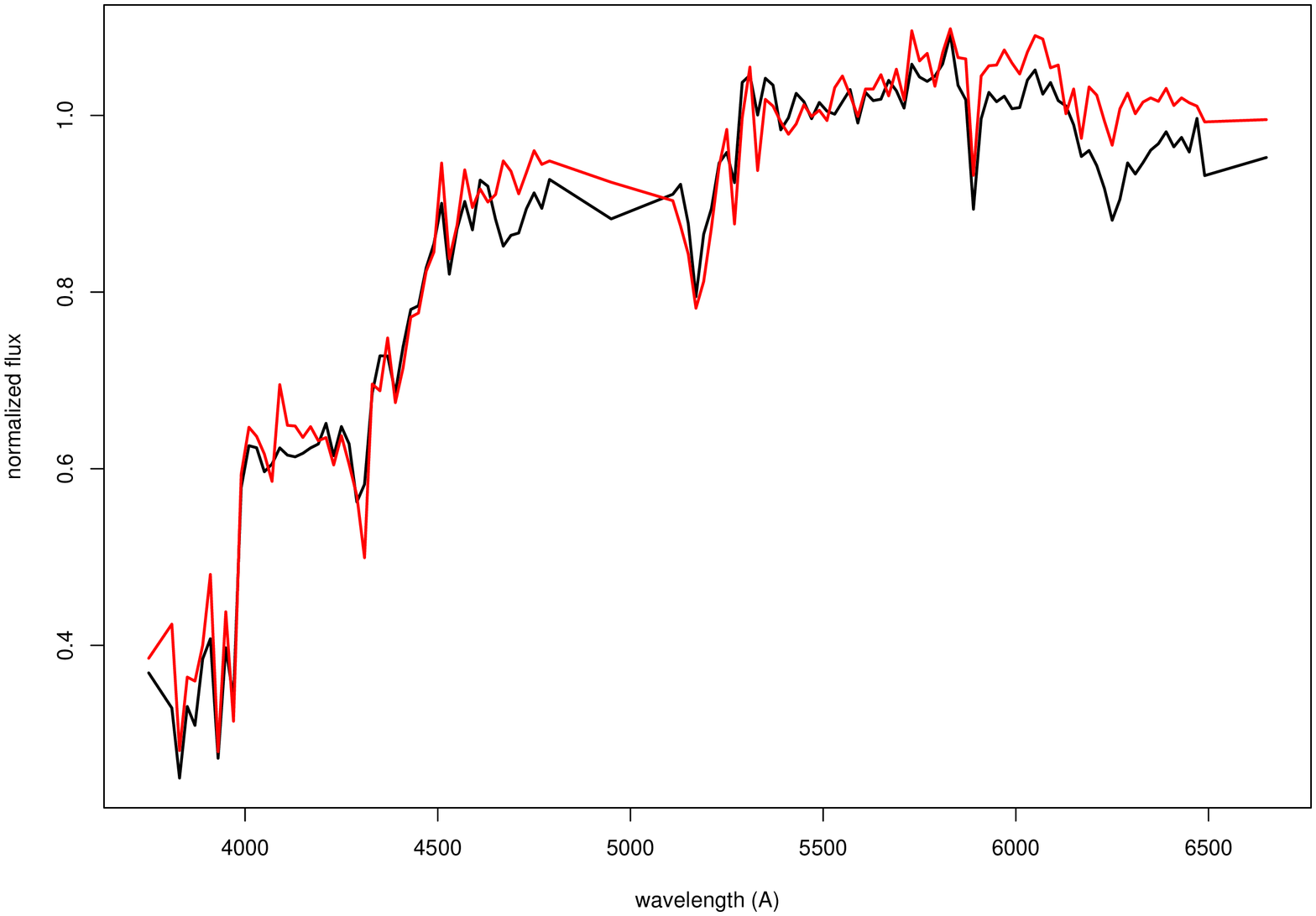}
\includegraphics[angle=-90,width=0.225\columnwidth]{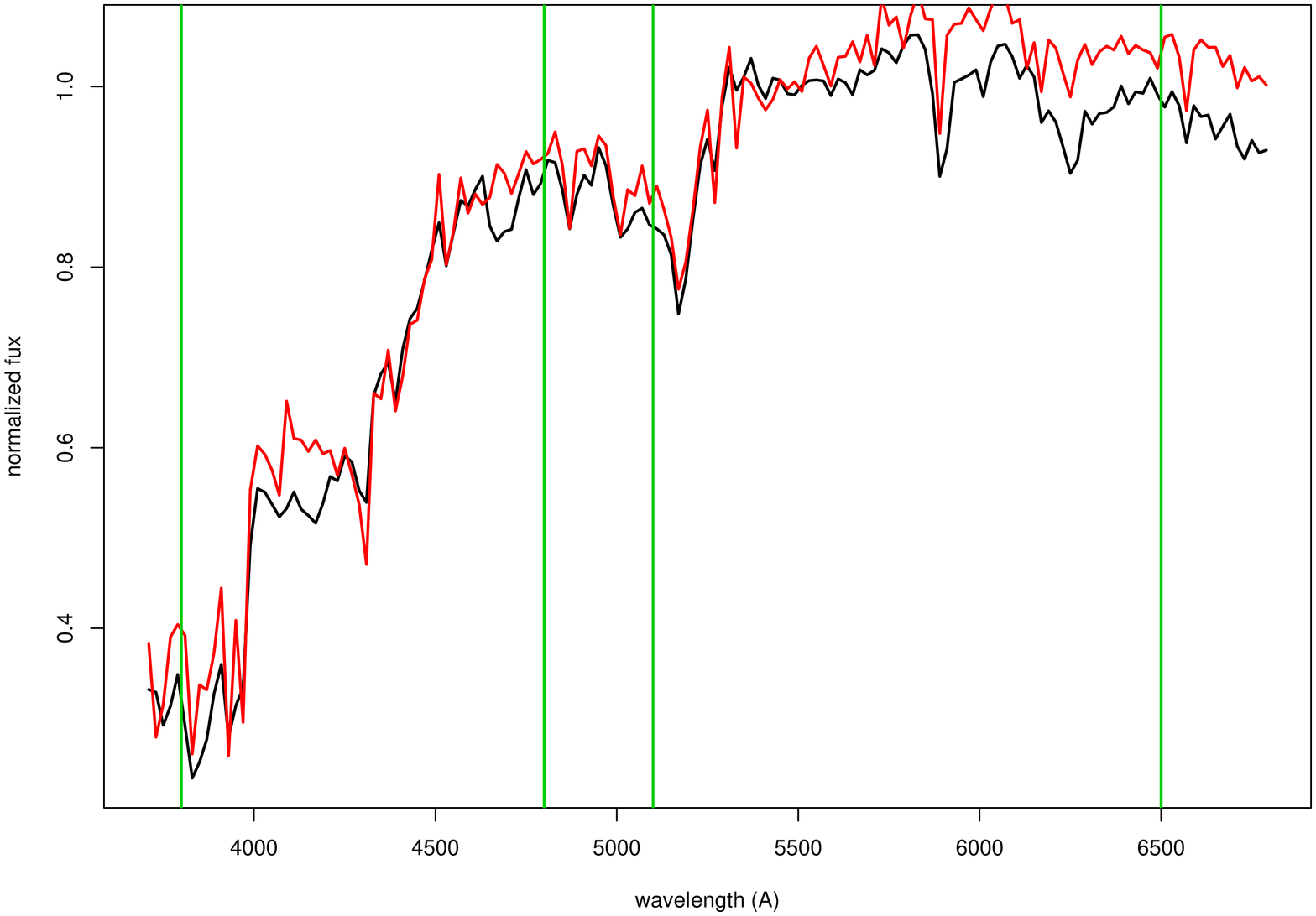}
\includegraphics[angle=-90,width=0.225\columnwidth]{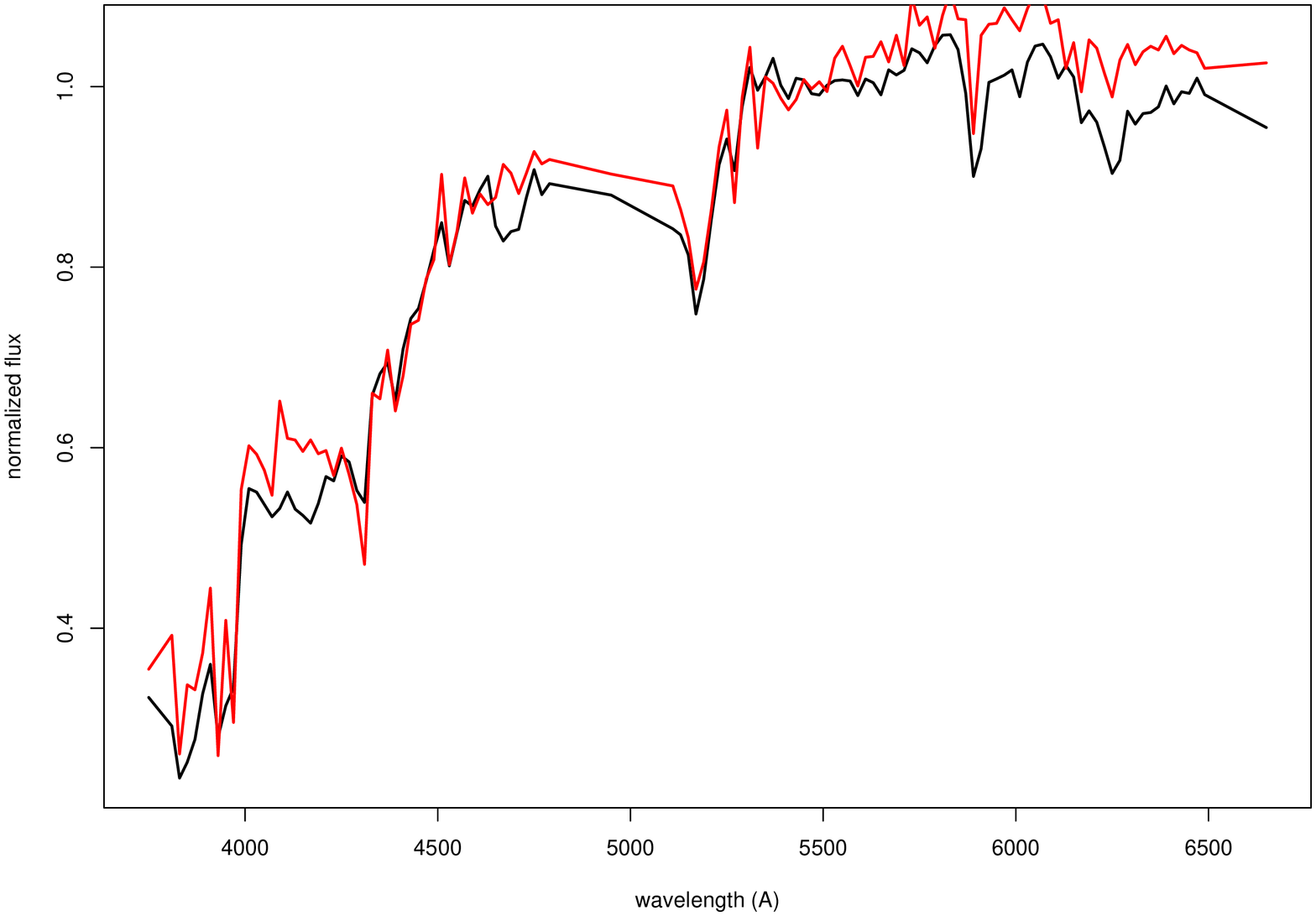}
\includegraphics[angle=-90,width=0.225\columnwidth]{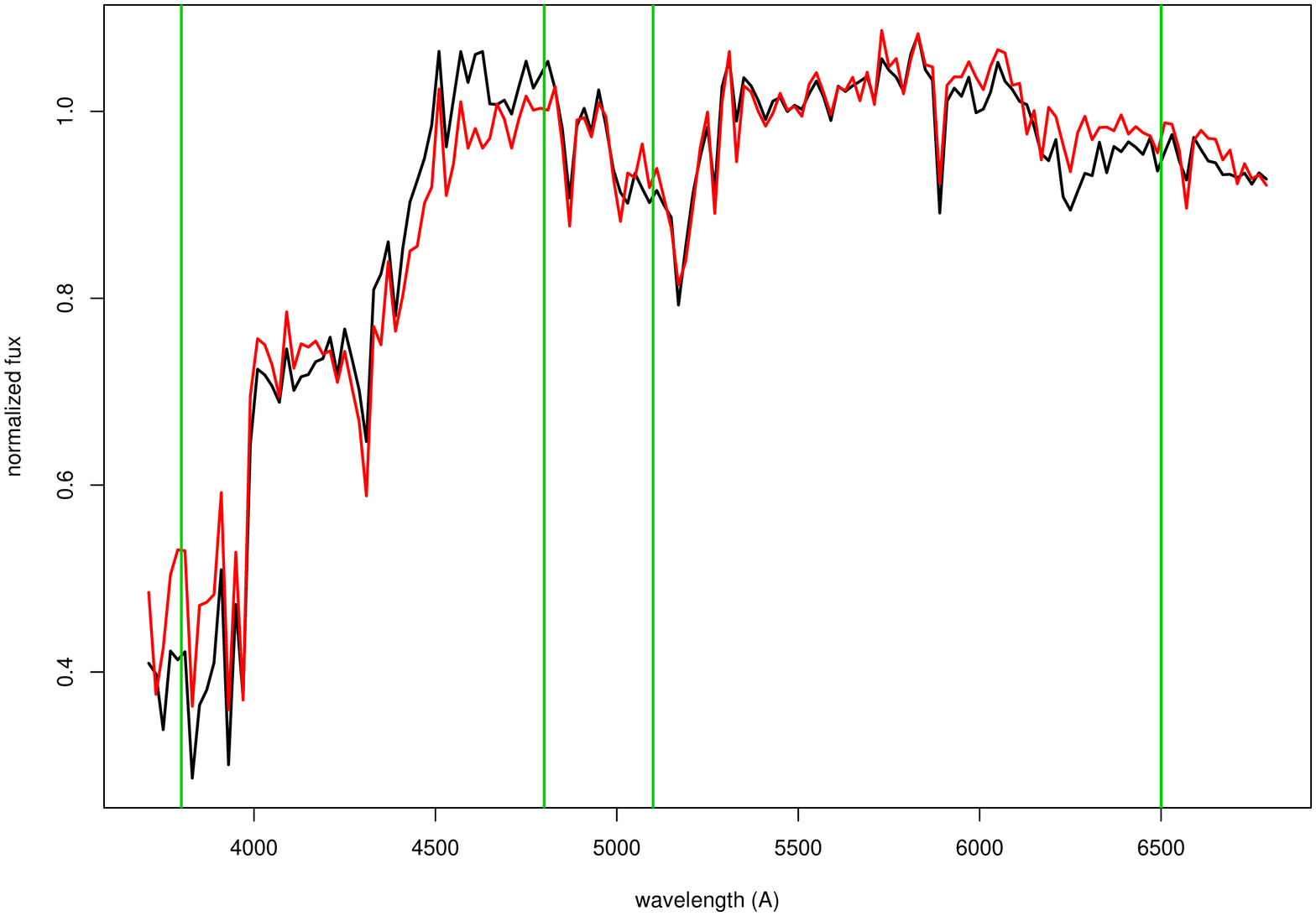}
\includegraphics[angle=-90,width=0.225\columnwidth]{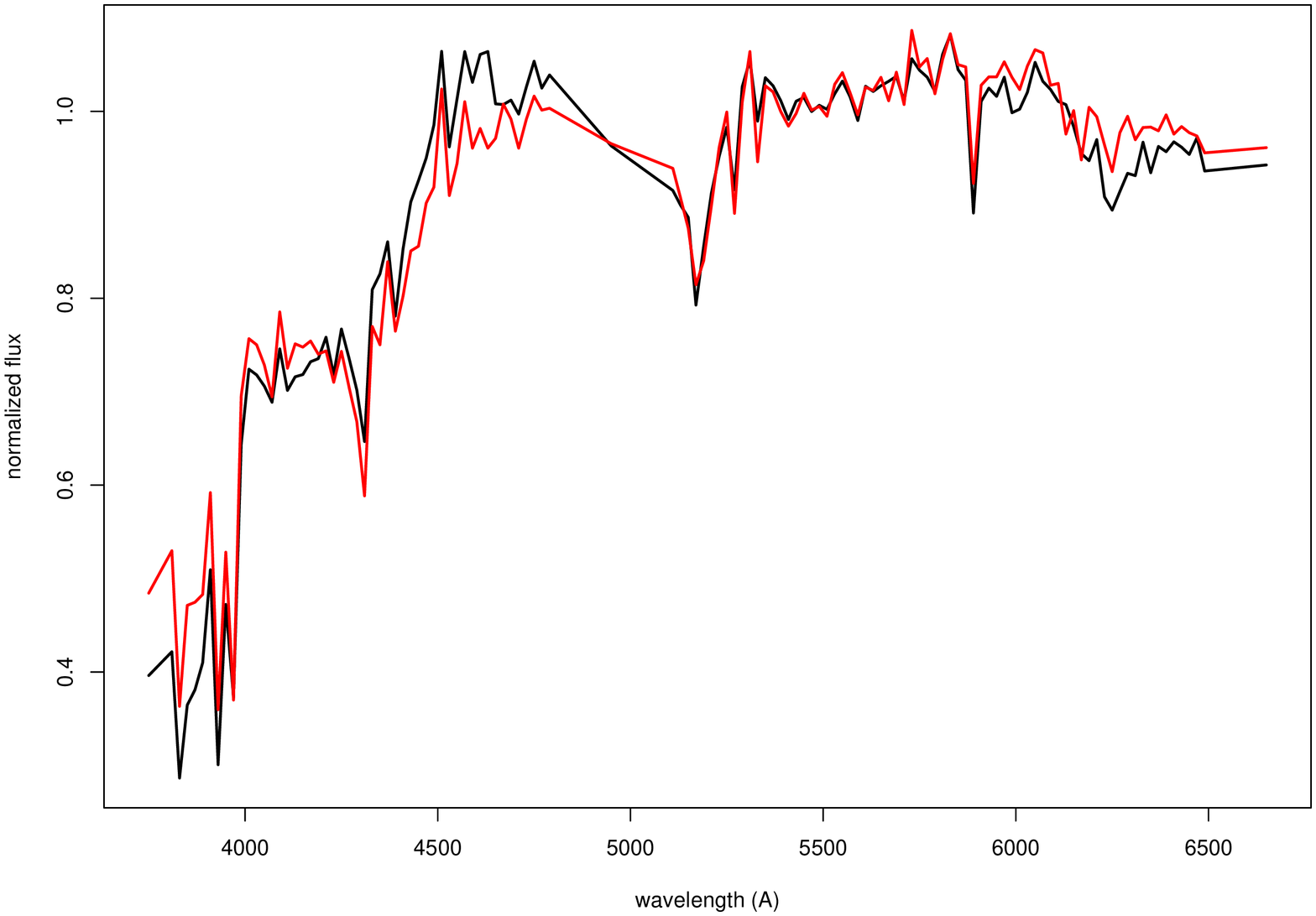}
\includegraphics[angle=-90,width=0.225\columnwidth]{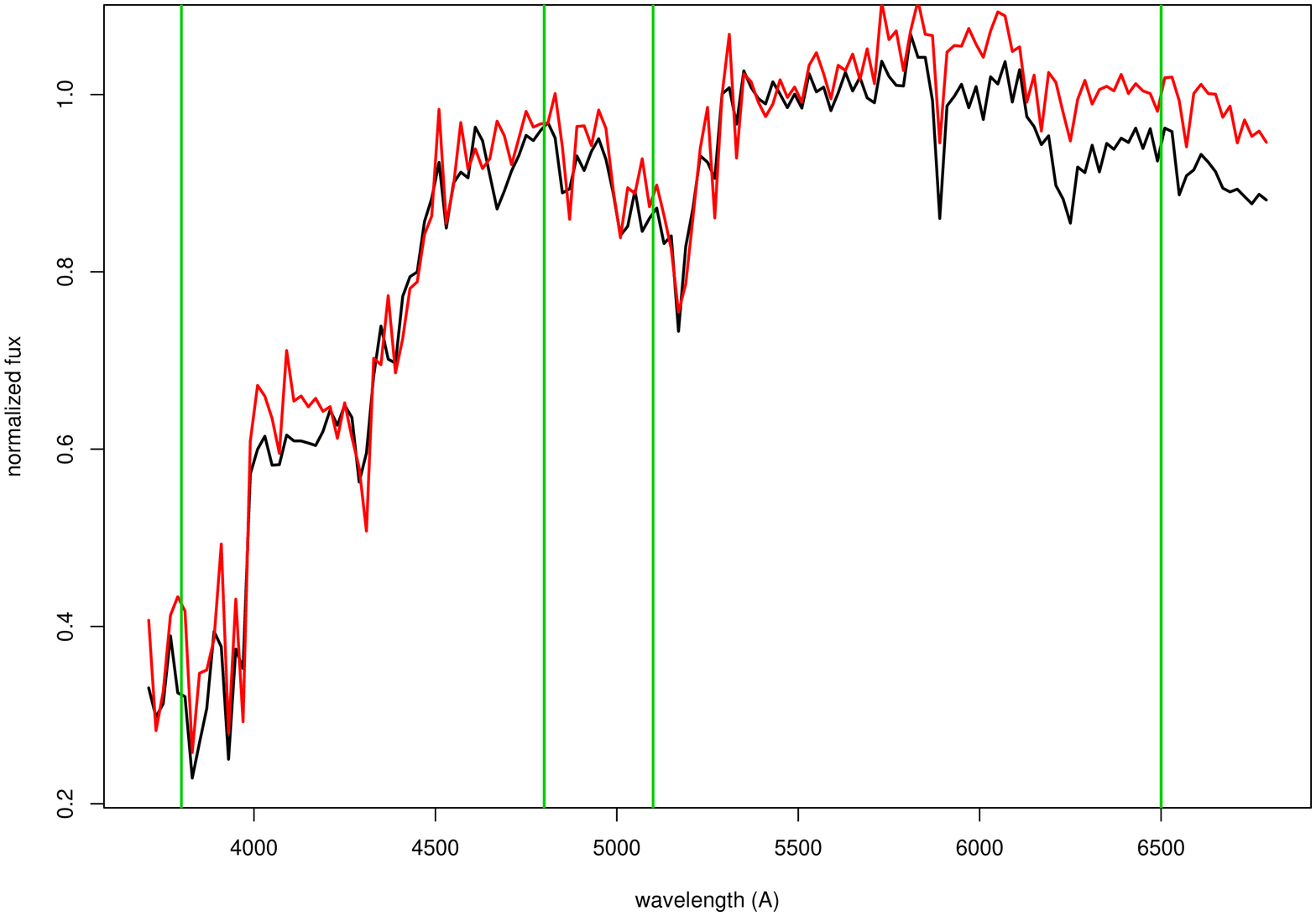}
\includegraphics[angle=-90,width=0.225\columnwidth]{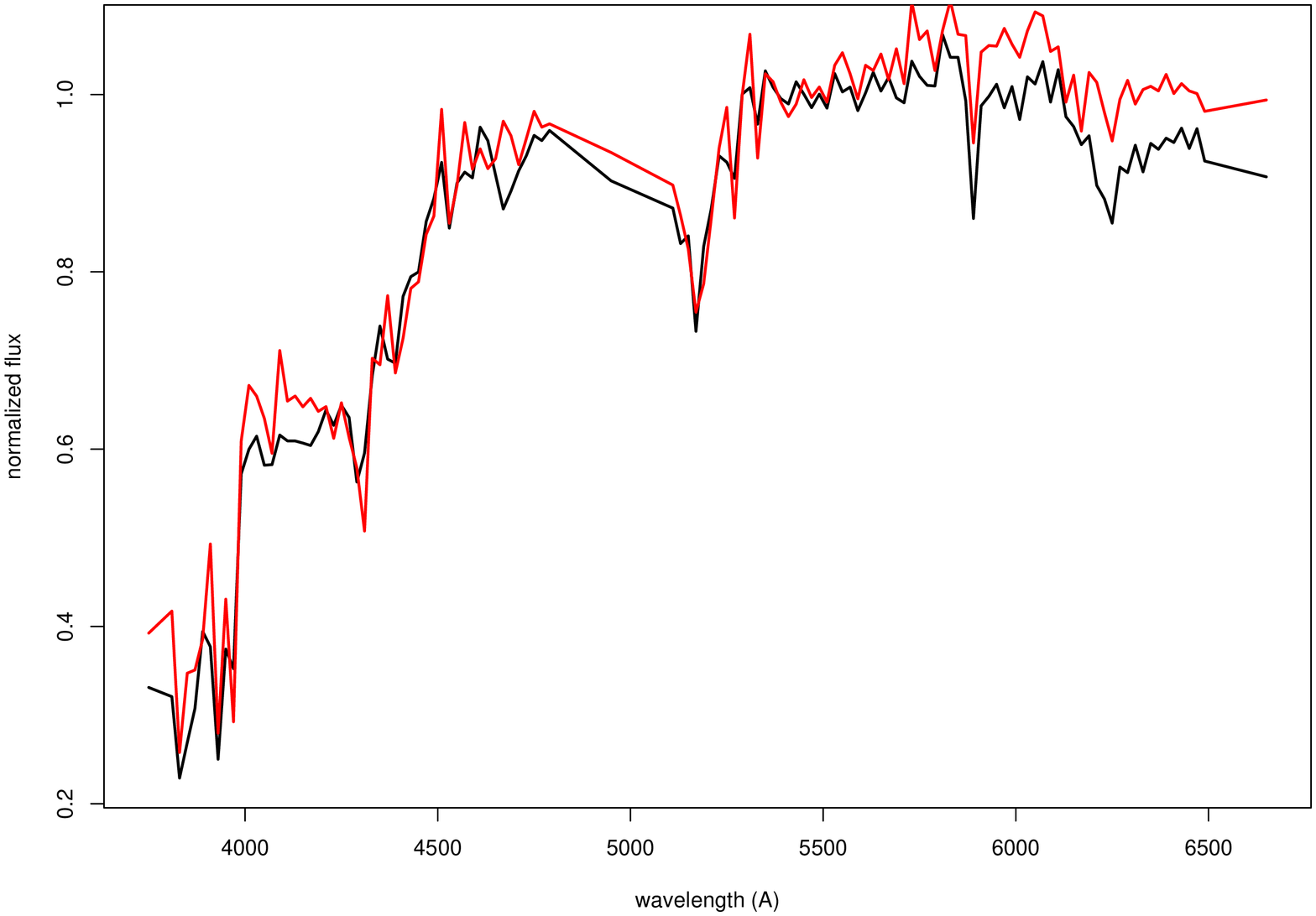}
\includegraphics[angle=-90,width=0.225\columnwidth]{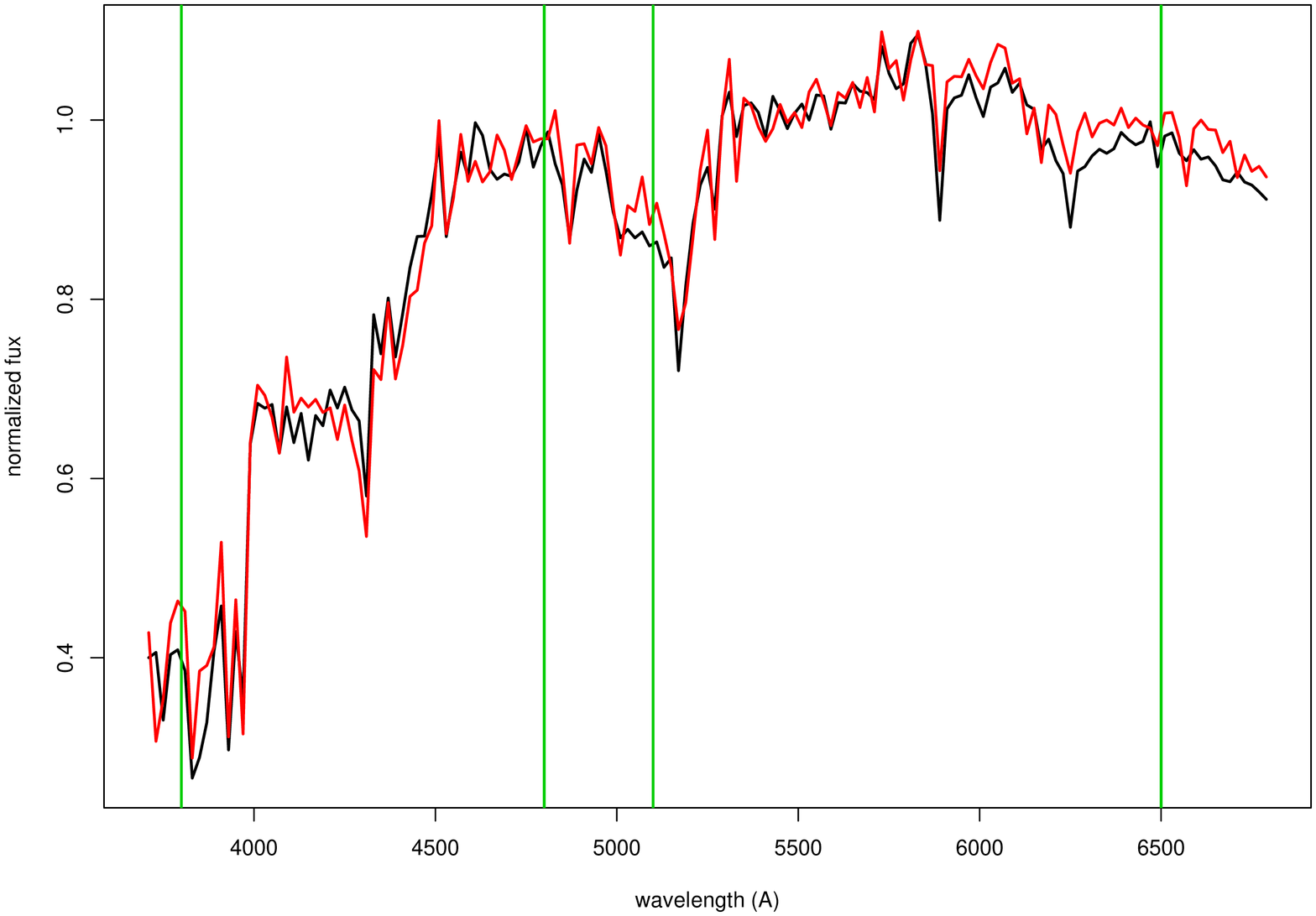}
\includegraphics[angle=-90,width=0.225\columnwidth]{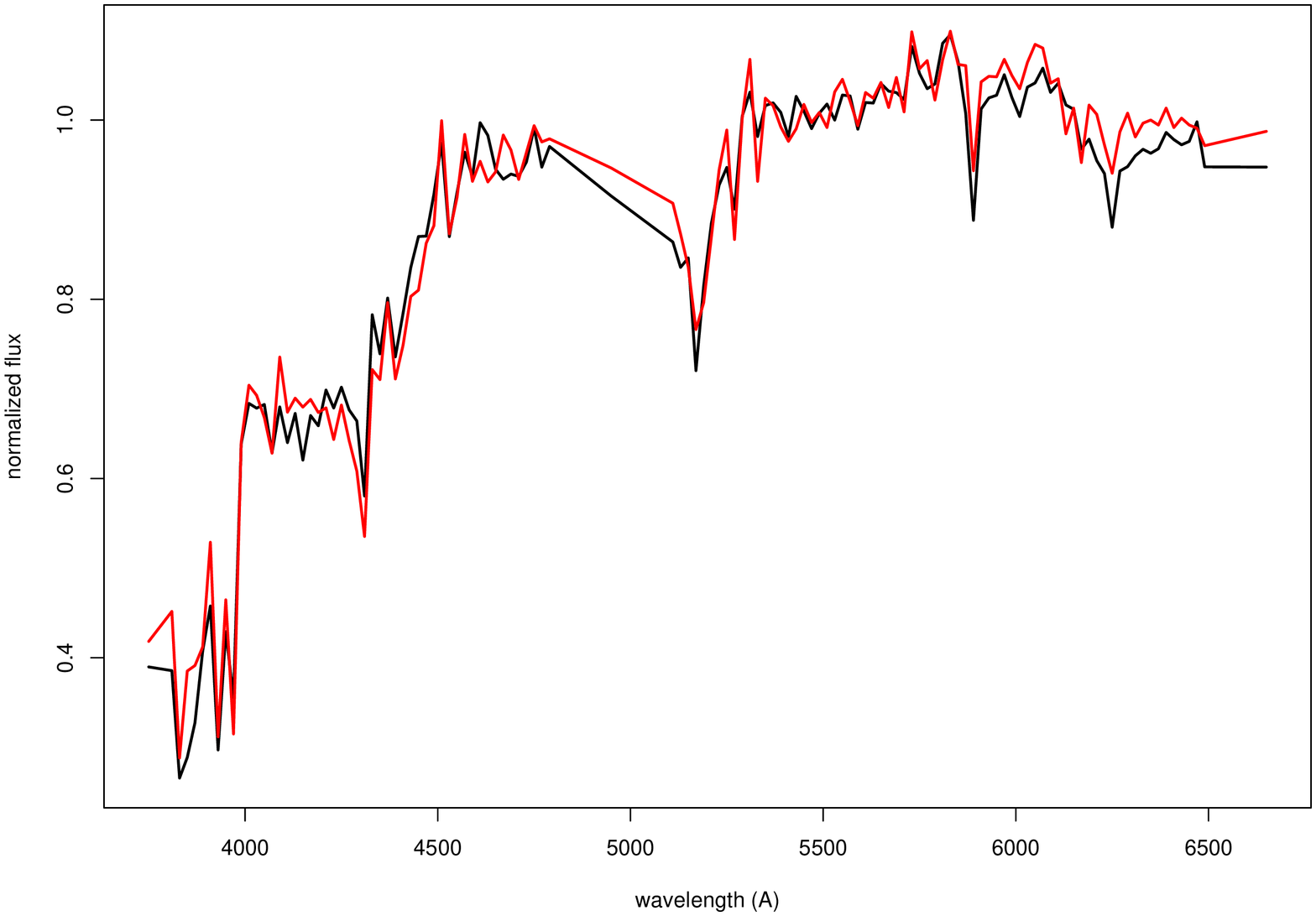}
\includegraphics[angle=-90,width=0.225\columnwidth]{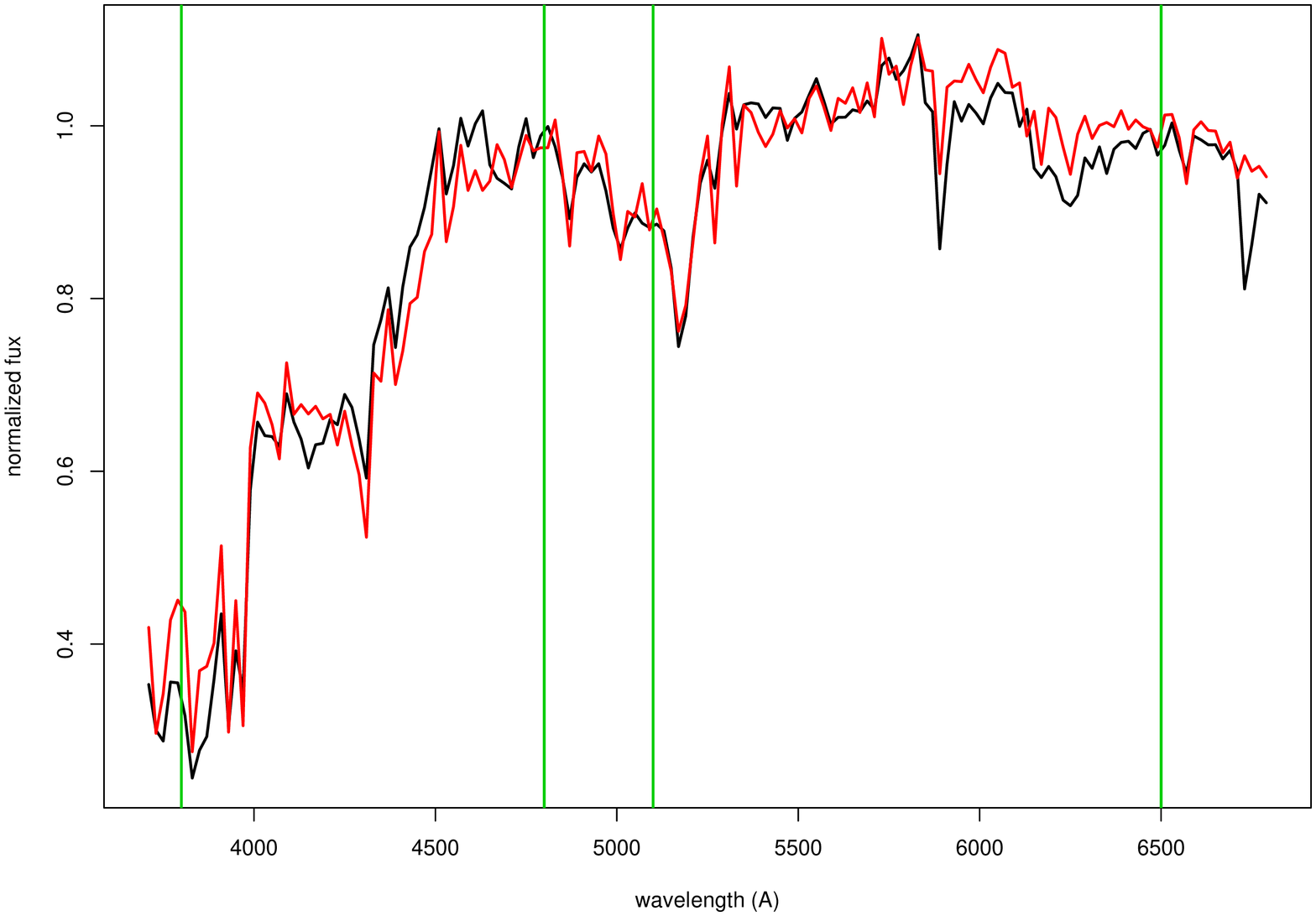}
\includegraphics[angle=-90,width=0.225\columnwidth]{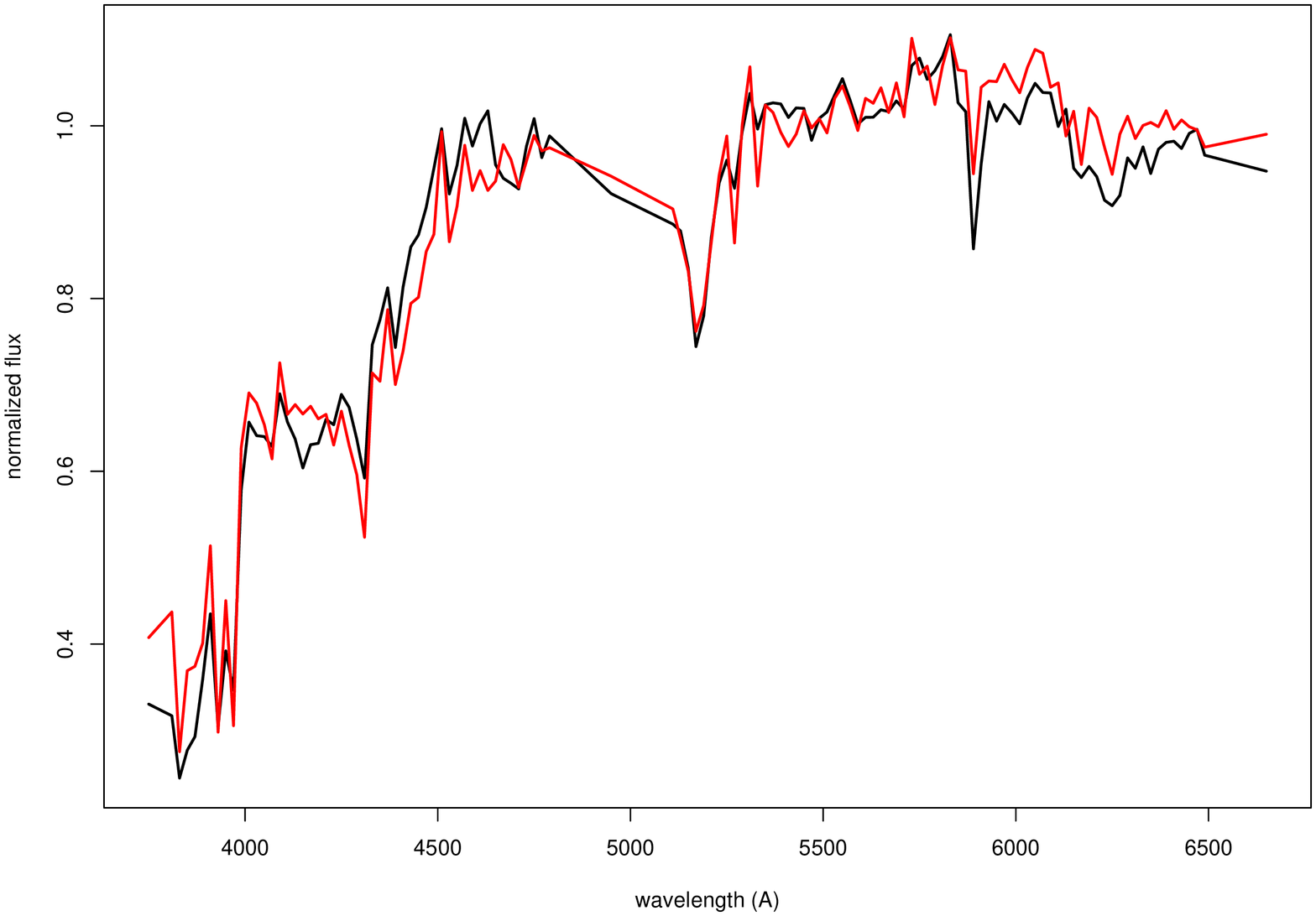}
\includegraphics[angle=-90,width=0.225\columnwidth]{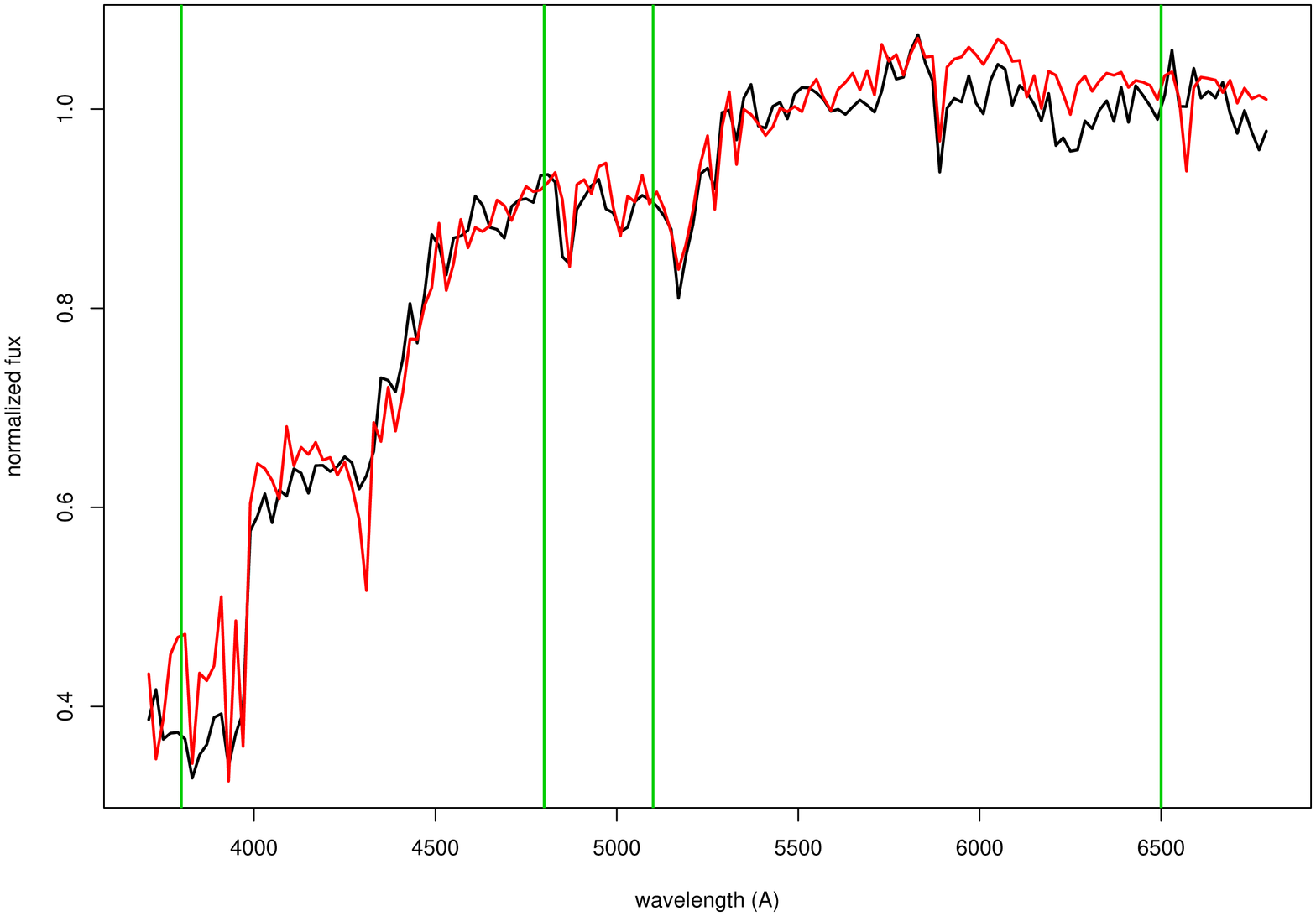}
\includegraphics[angle=-90,width=0.225\columnwidth]{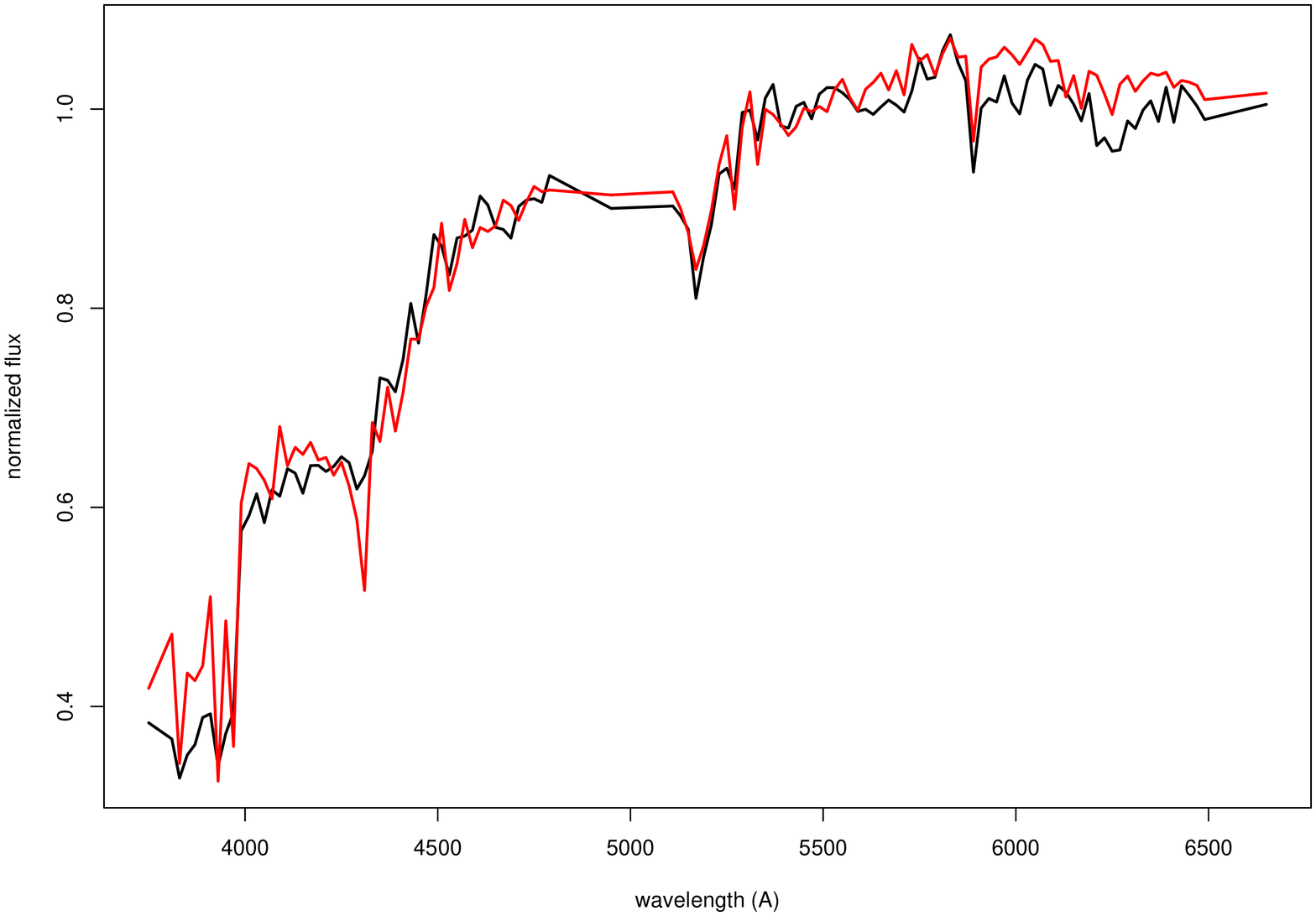}
\caption{Results of the $\chi^{2}$-fitting of early-type galaxy spectra 
in the Kennicutt's Atlas (black lines) with our library (red lines). 
The results are presented in rows. Every two plots give the results 
for one observed galaxy when the $\chi^{2}$-fitting was performed by 
excluding or including the areas with the strongest emission lines 
respectively. The y-axis corresponds to the normalized fluxes, while 
the x-axis presents the wavelengths and extends from 350 to 700\,nm. 
The galaxies presented here are: \textbf{first row:} NGC3379, NGC3245, 
NGC4472, and NGC3941, \textbf{second row:} NGC4648, NGC4889, NGC4262, 
and NGC5866.}
\label{a1}
\end{figure*}

\begin{figure*}[h]
\center
\includegraphics[angle=-90,width=0.225\columnwidth]{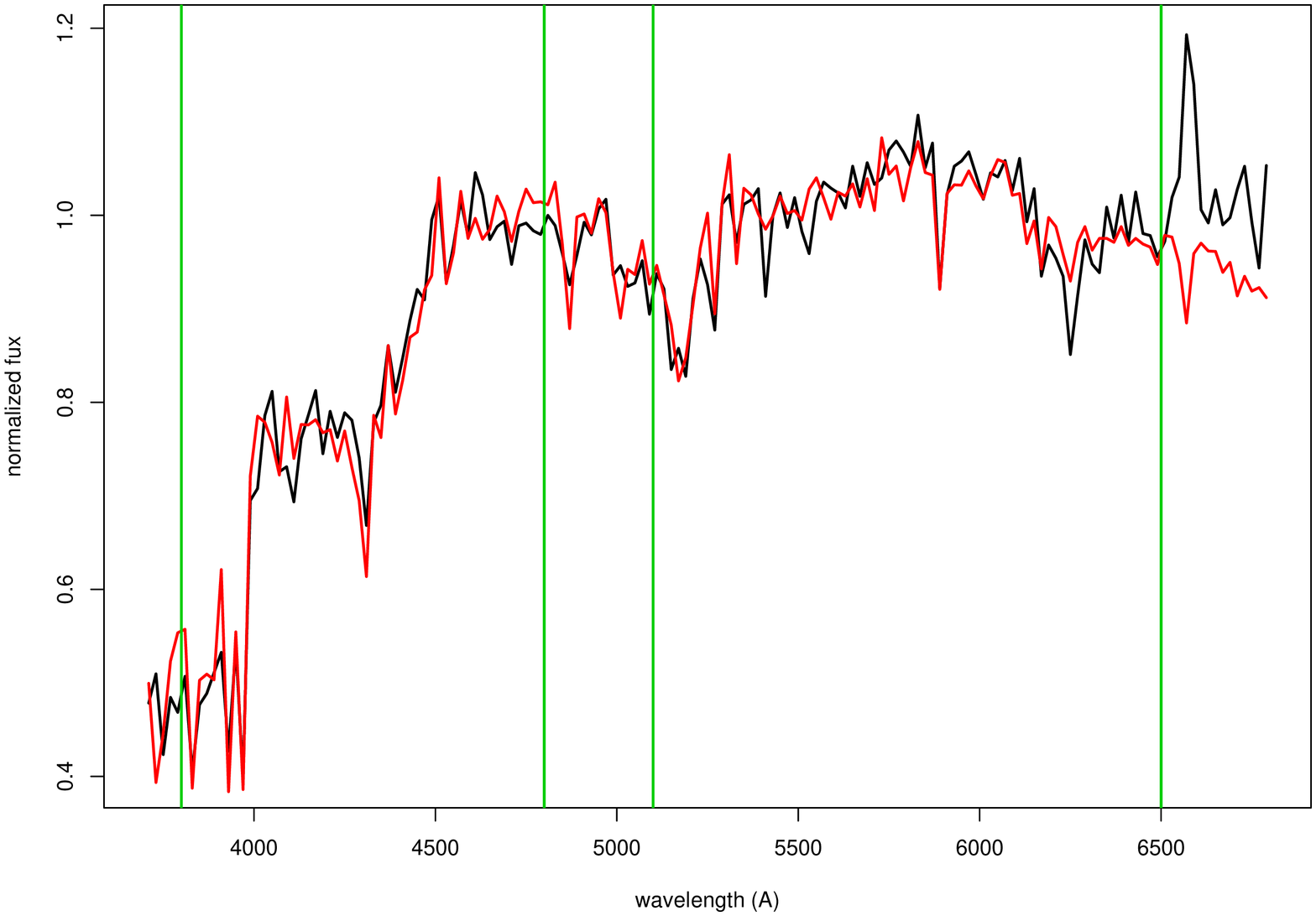}
\includegraphics[angle=-90,width=0.225\columnwidth]{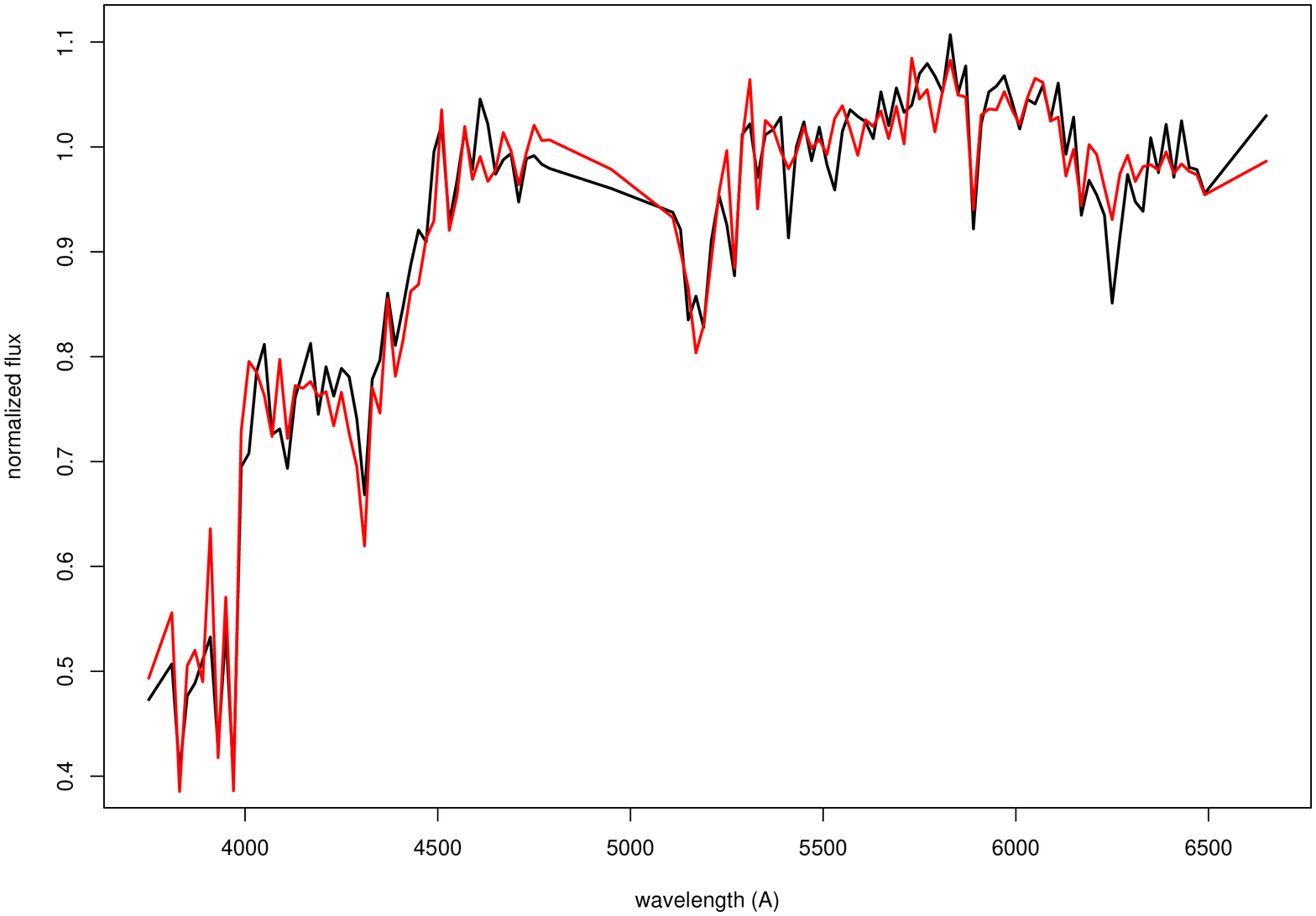}
\includegraphics[angle=-90,width=0.225\columnwidth]{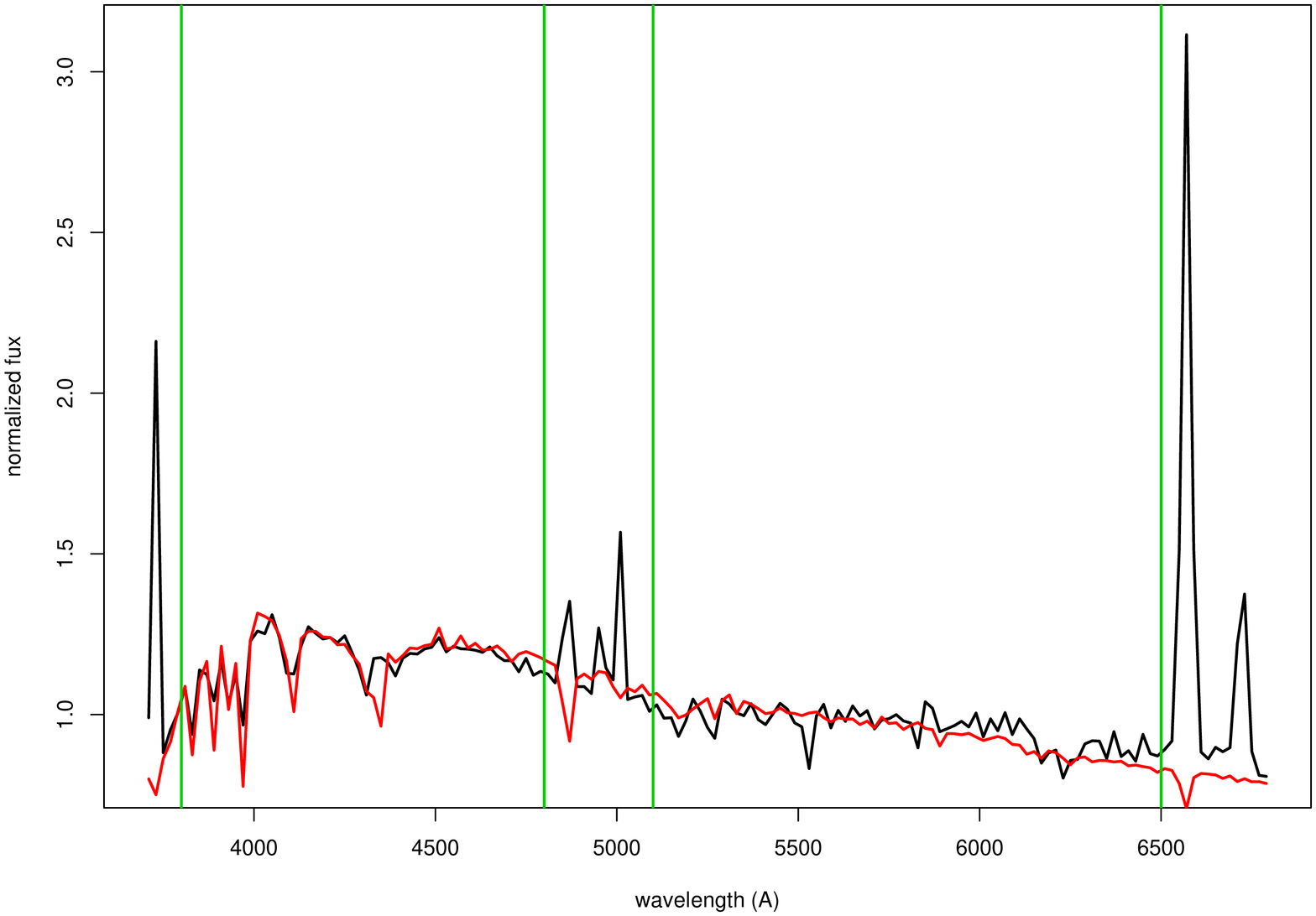}
\includegraphics[angle=-90,width=0.225\columnwidth]{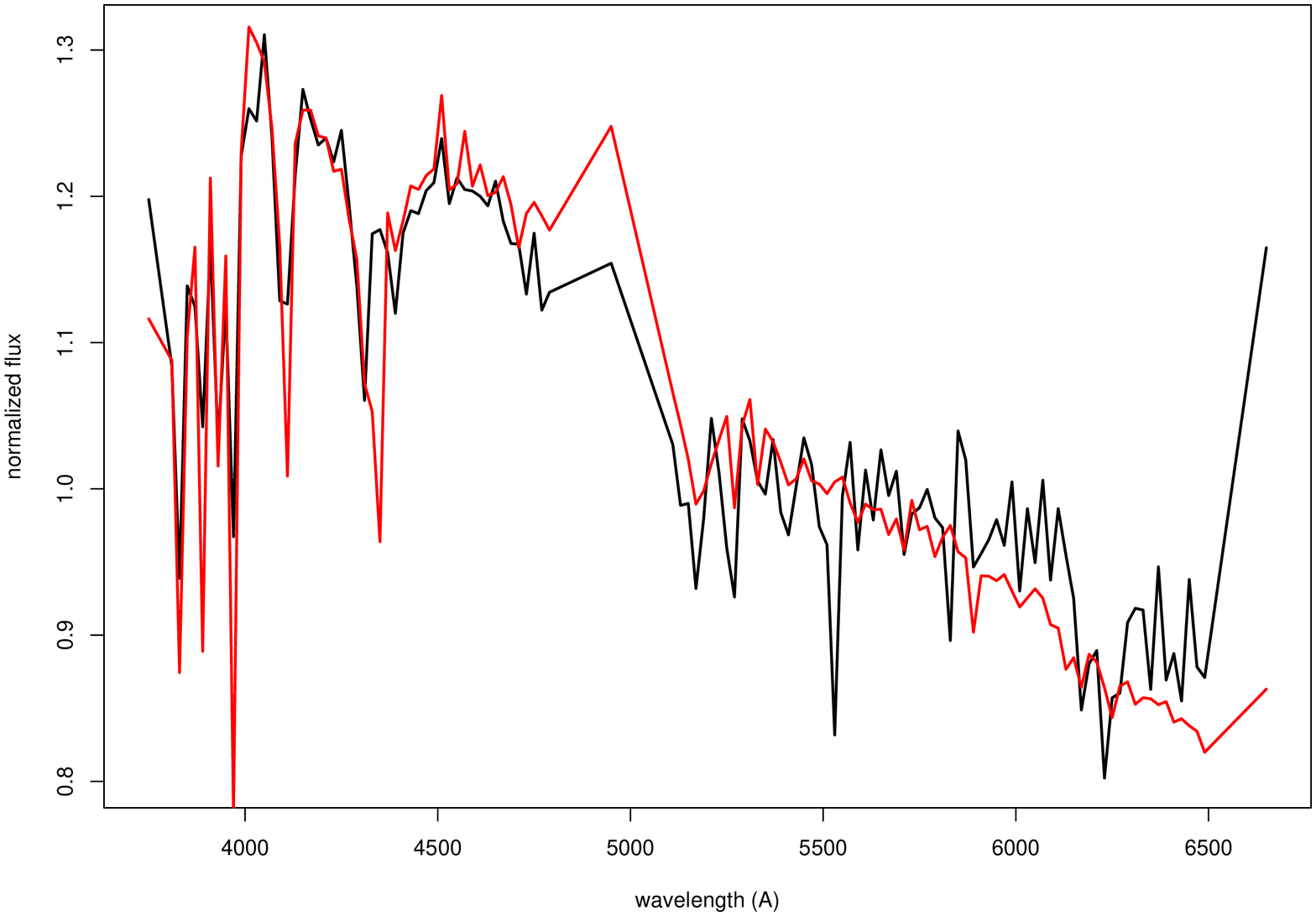}
\includegraphics[angle=-90,width=0.225\columnwidth]{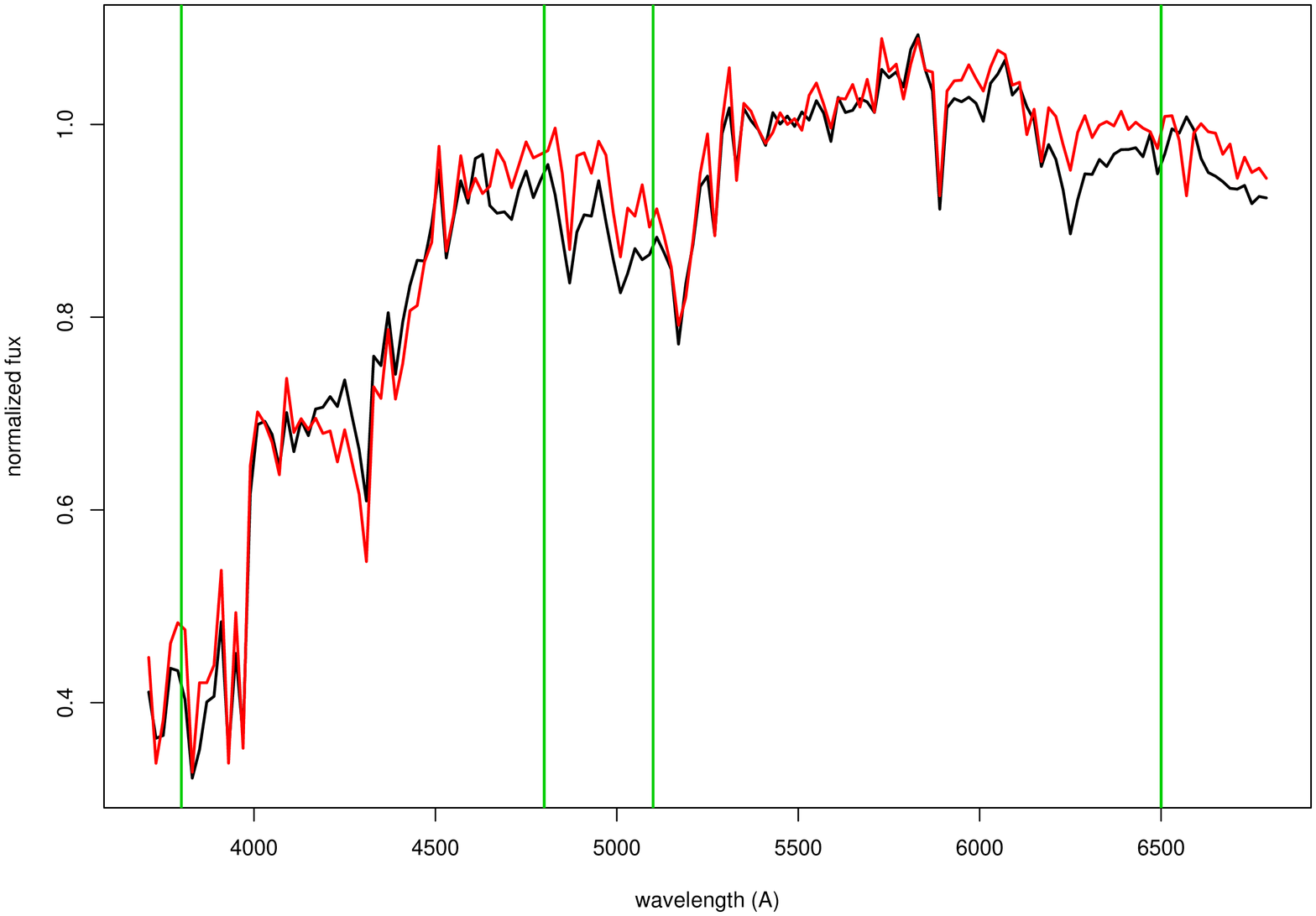}
\includegraphics[angle=-90,width=0.225\columnwidth]{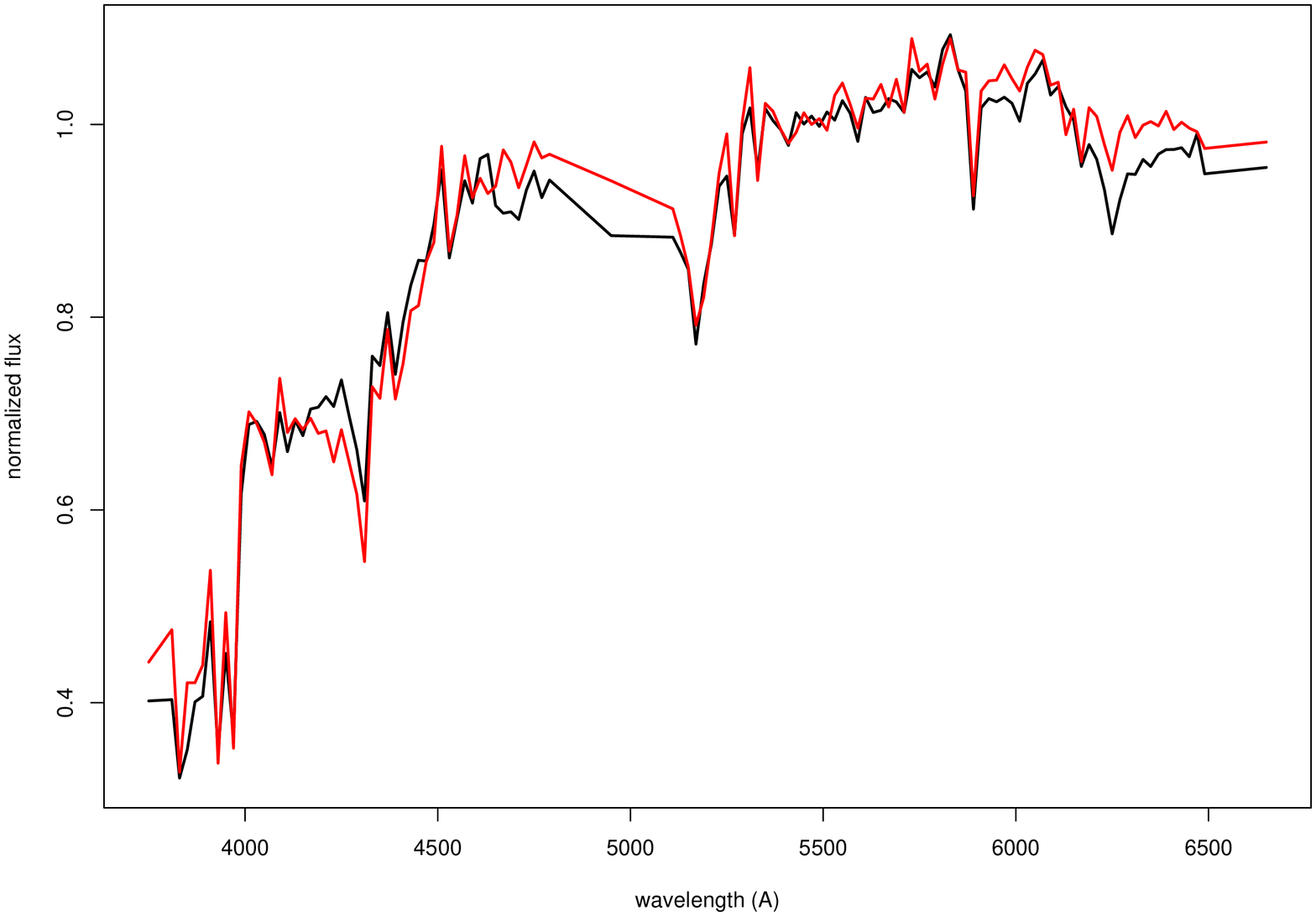}
\includegraphics[angle=-90,width=0.225\columnwidth]{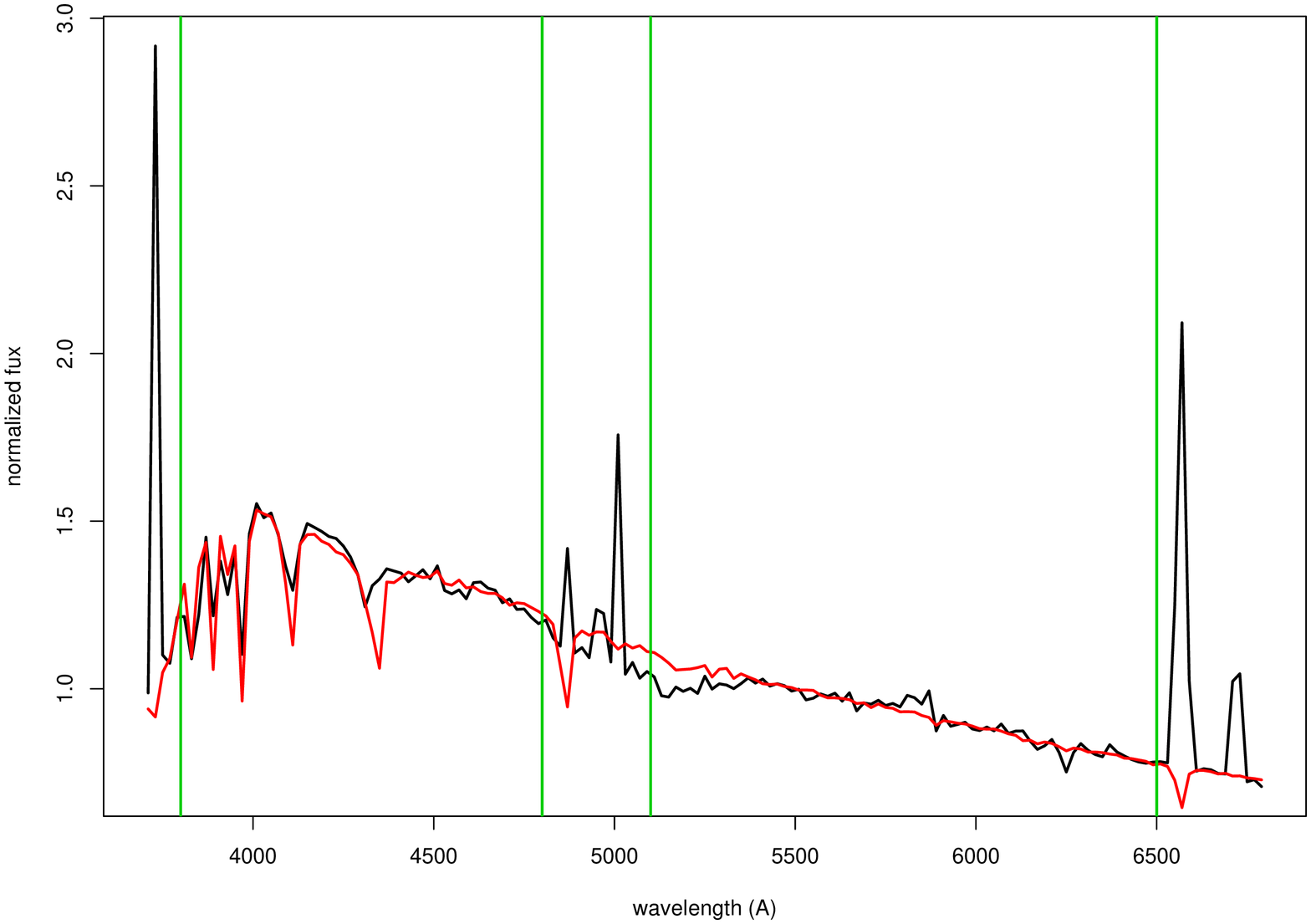}
\includegraphics[angle=-90,width=0.225\columnwidth]{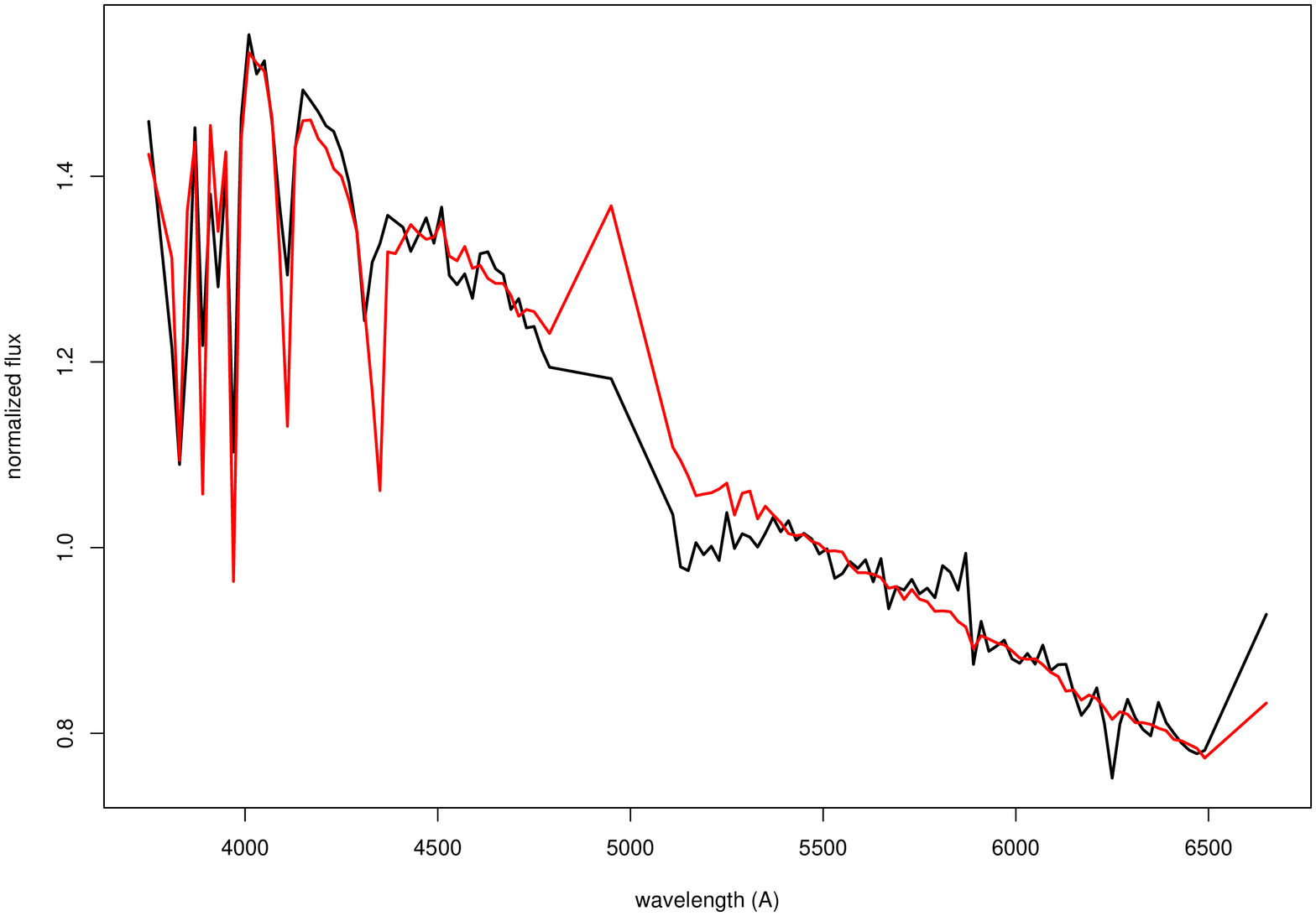}
\includegraphics[angle=-90,width=0.225\columnwidth]{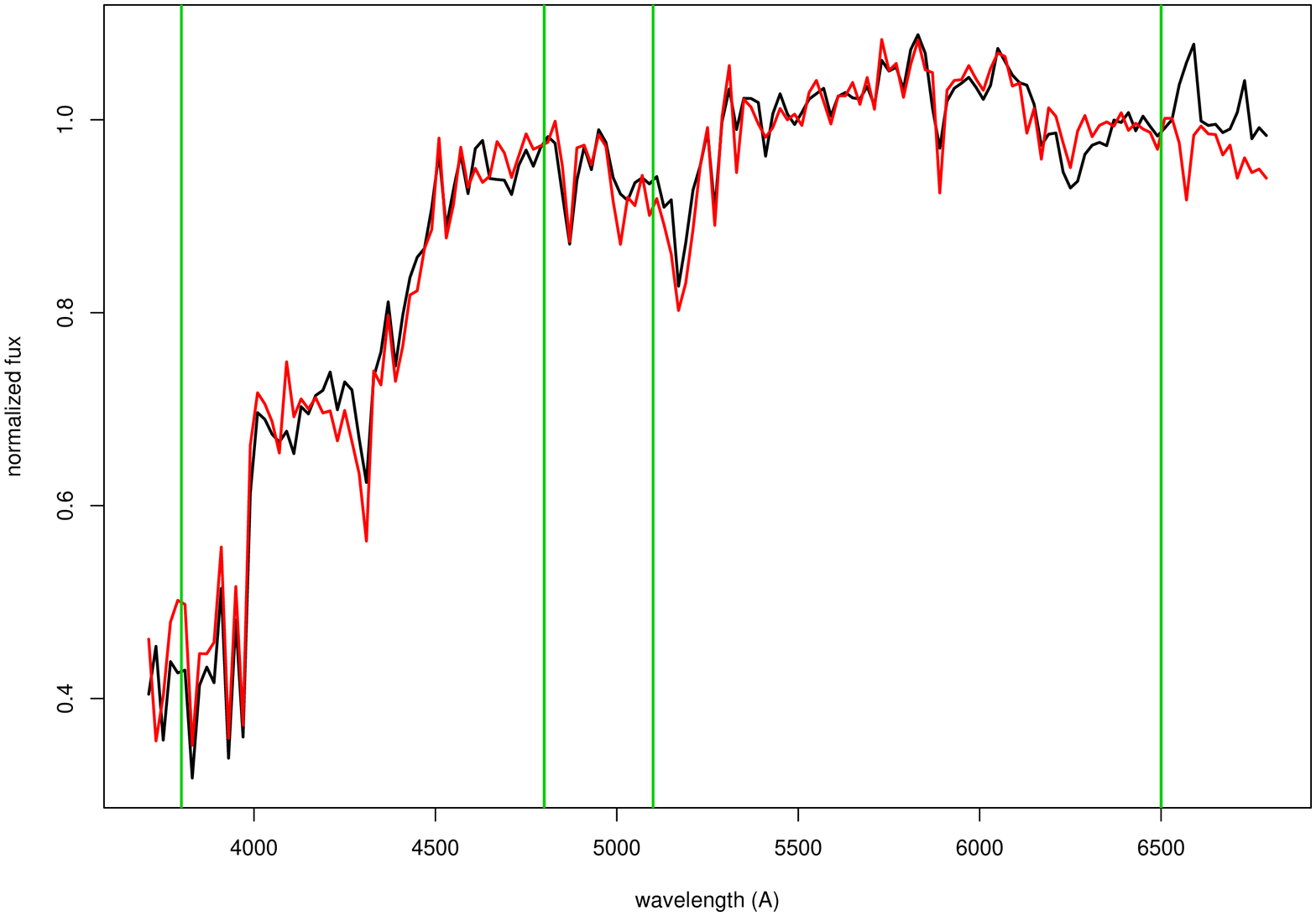}
\includegraphics[angle=-90,width=0.225\columnwidth]{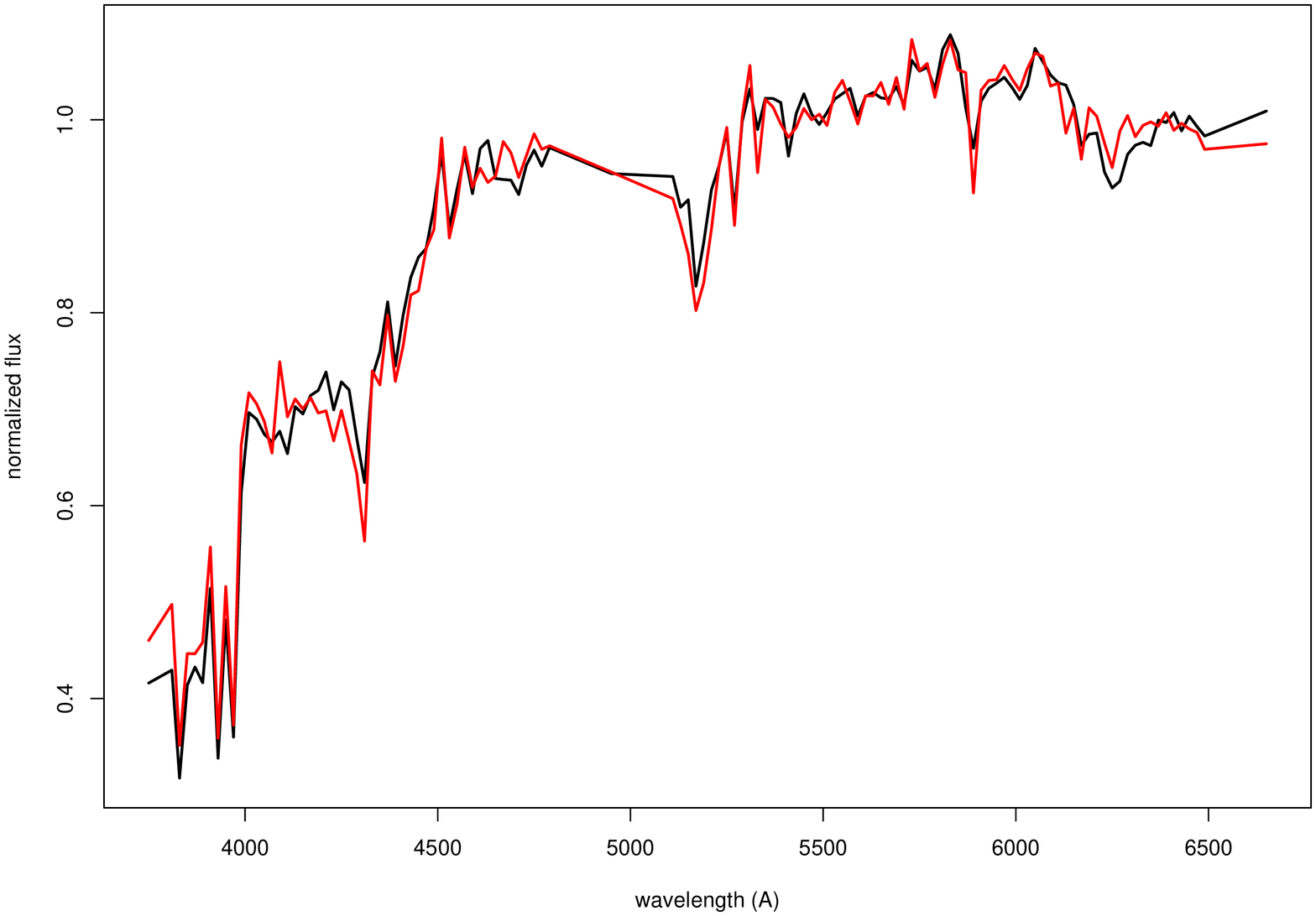}
\includegraphics[angle=-90,width=0.225\columnwidth]{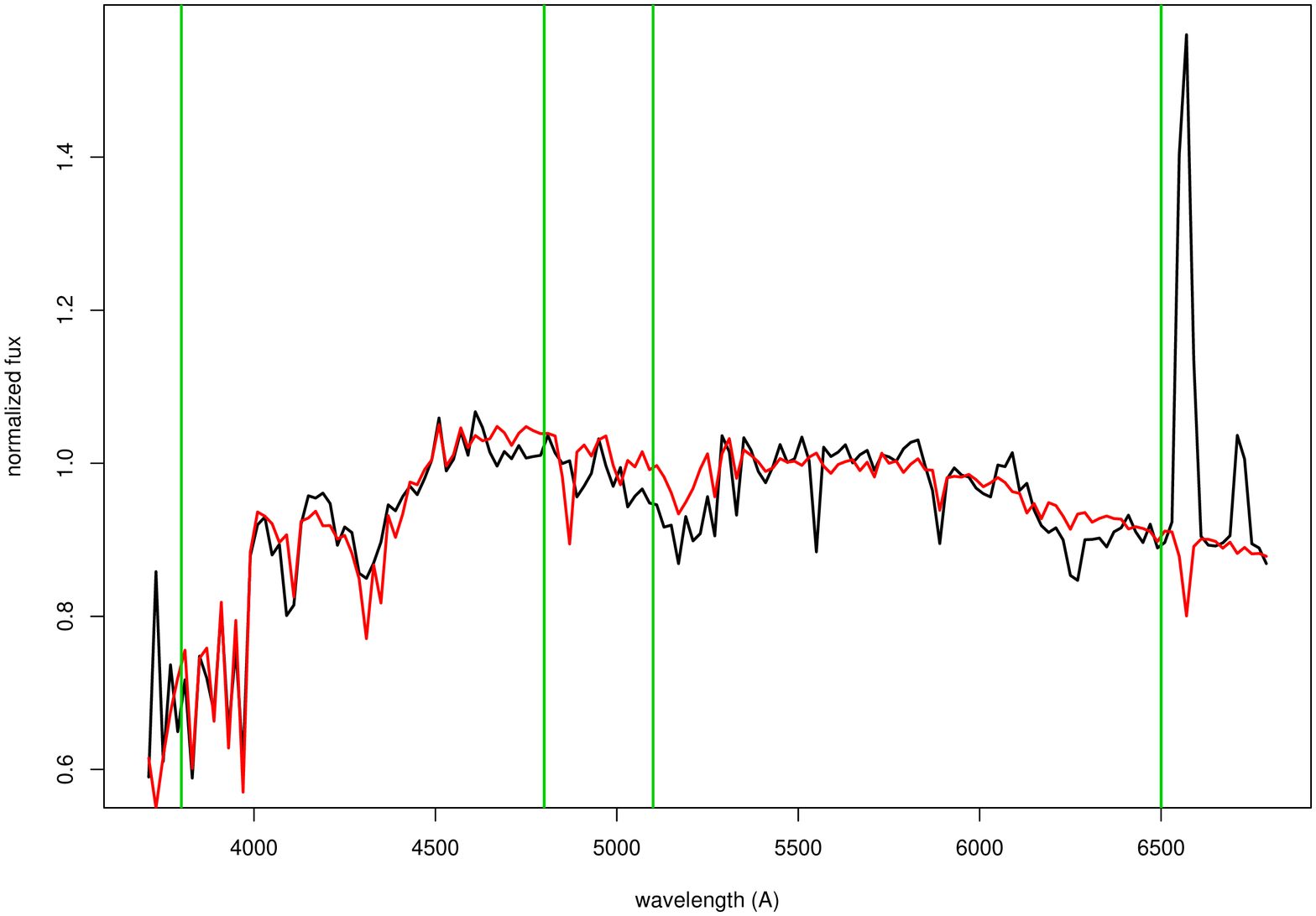}
\includegraphics[angle=-90,width=0.225\columnwidth]{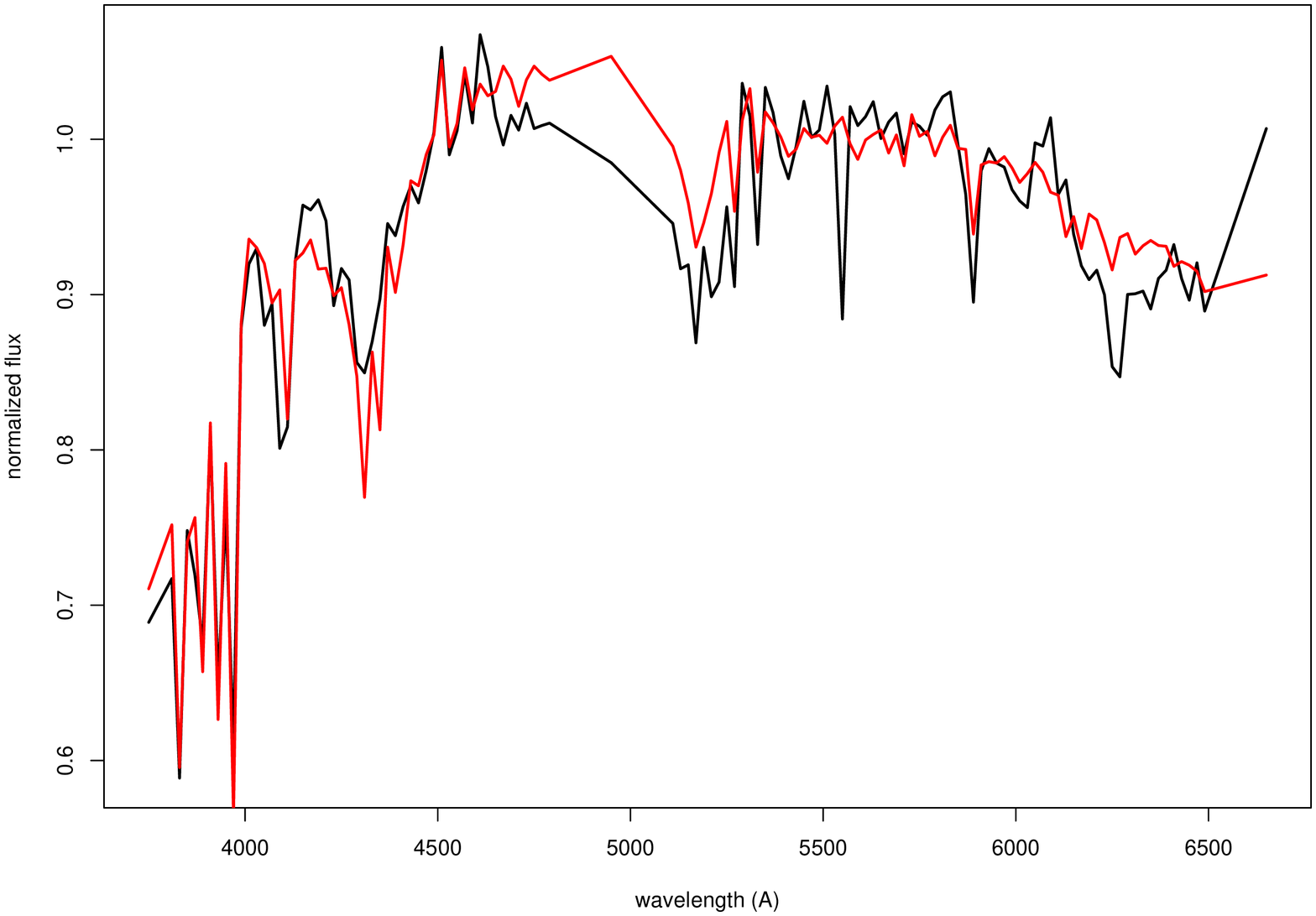}
\includegraphics[angle=-90,width=0.225\columnwidth]{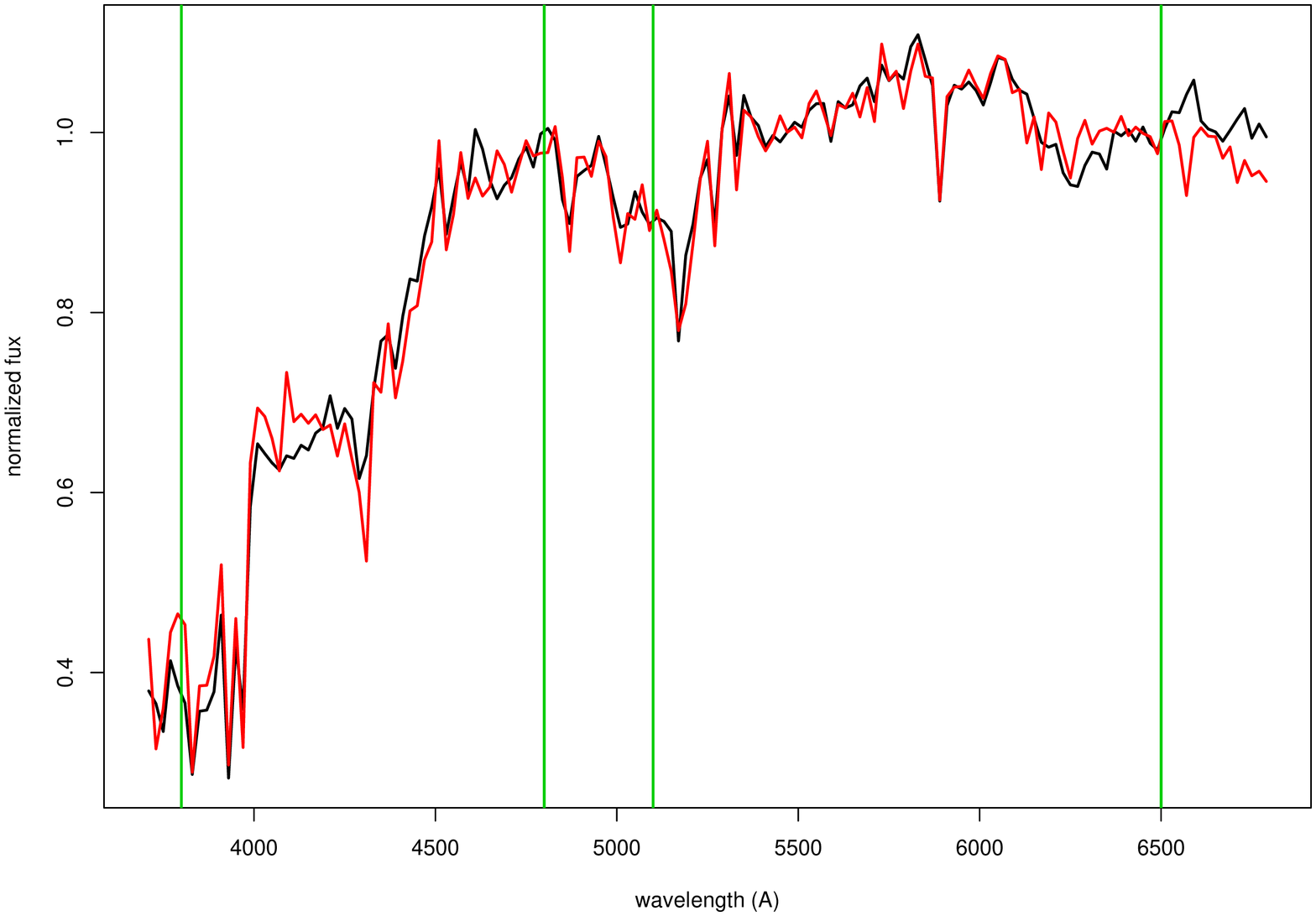}
\includegraphics[angle=-90,width=0.225\columnwidth]{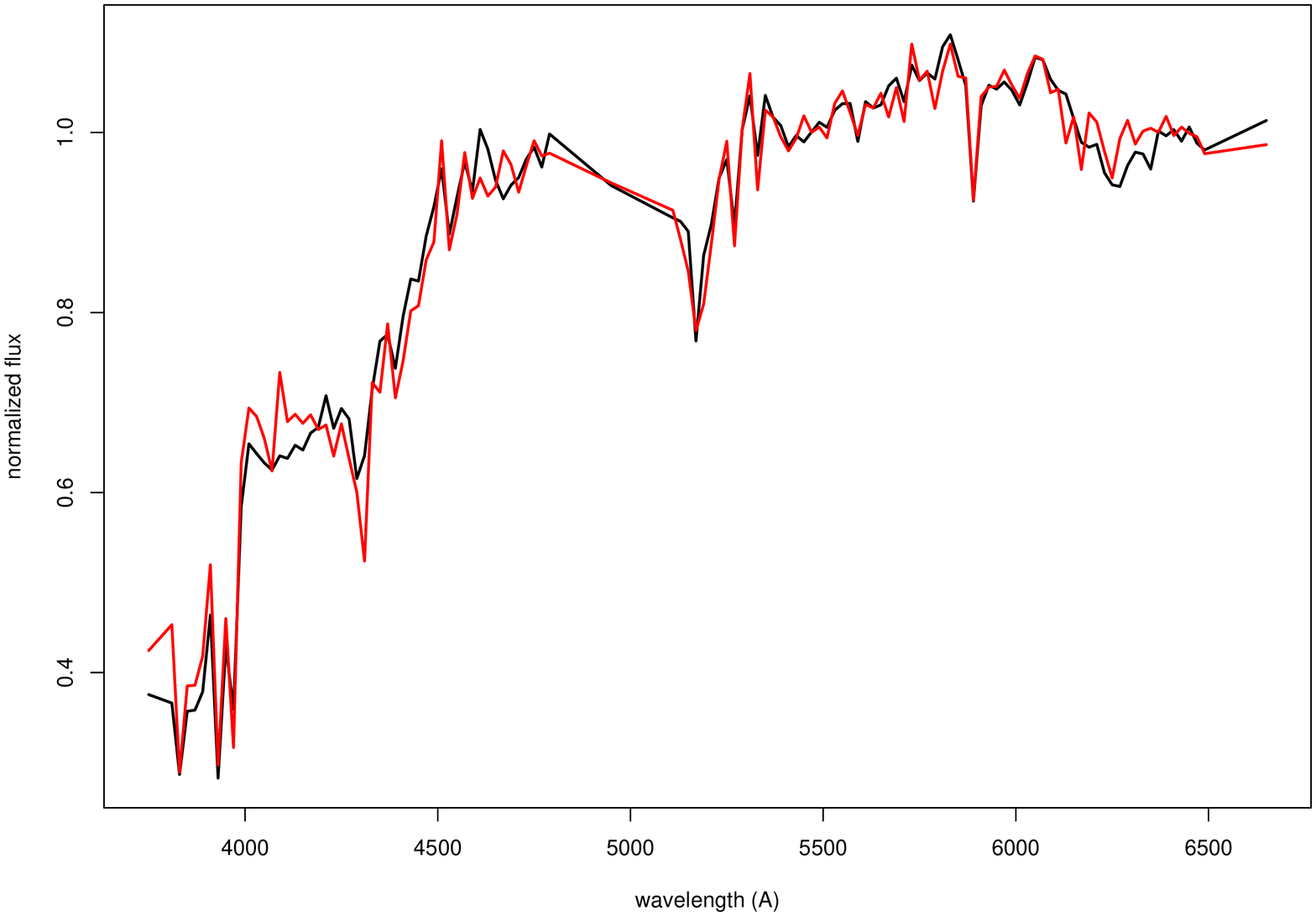}
\includegraphics[angle=-90,width=0.225\columnwidth]{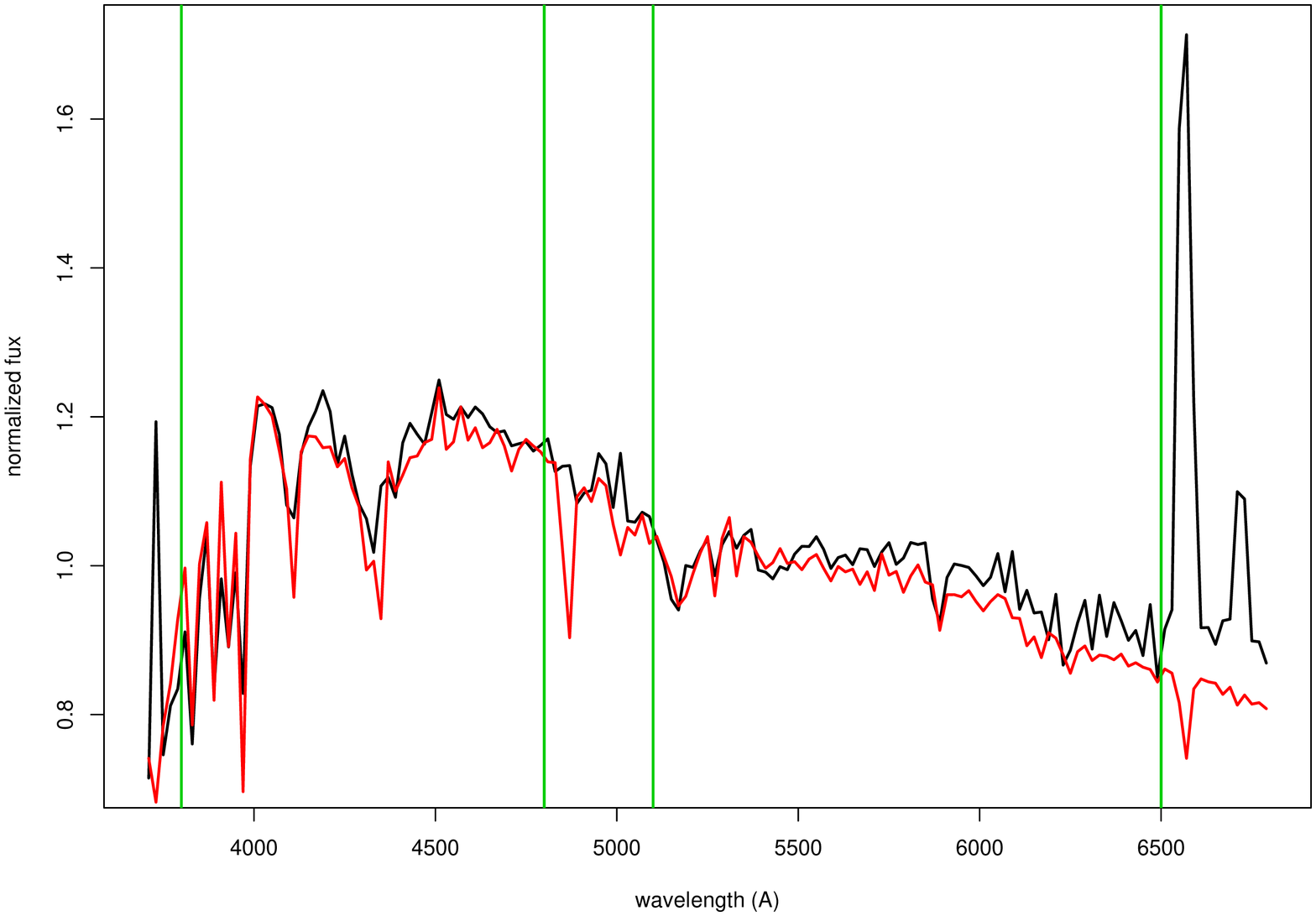}
\includegraphics[angle=-90,width=0.225\columnwidth]{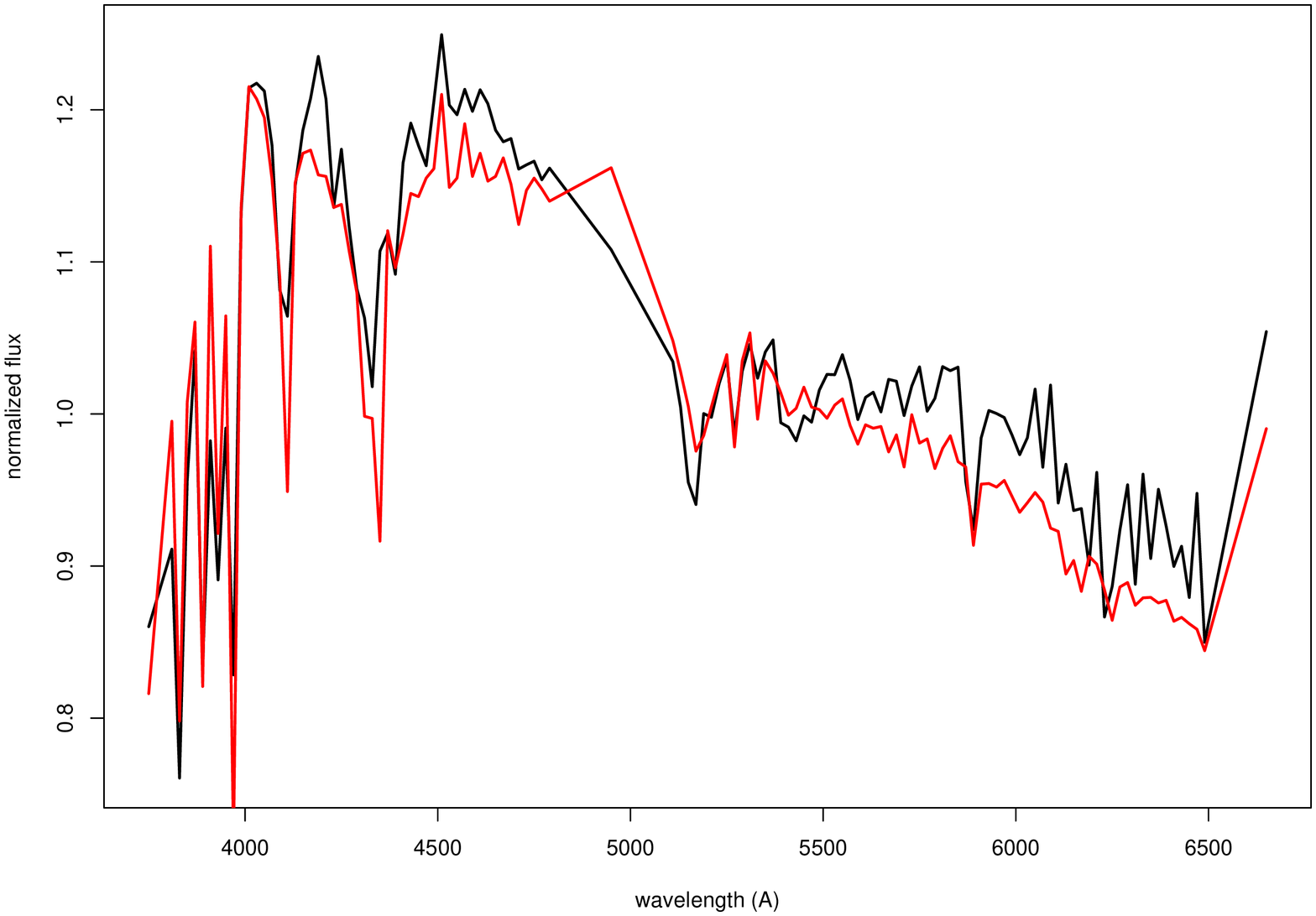}
\includegraphics[angle=-90,width=0.225\columnwidth]{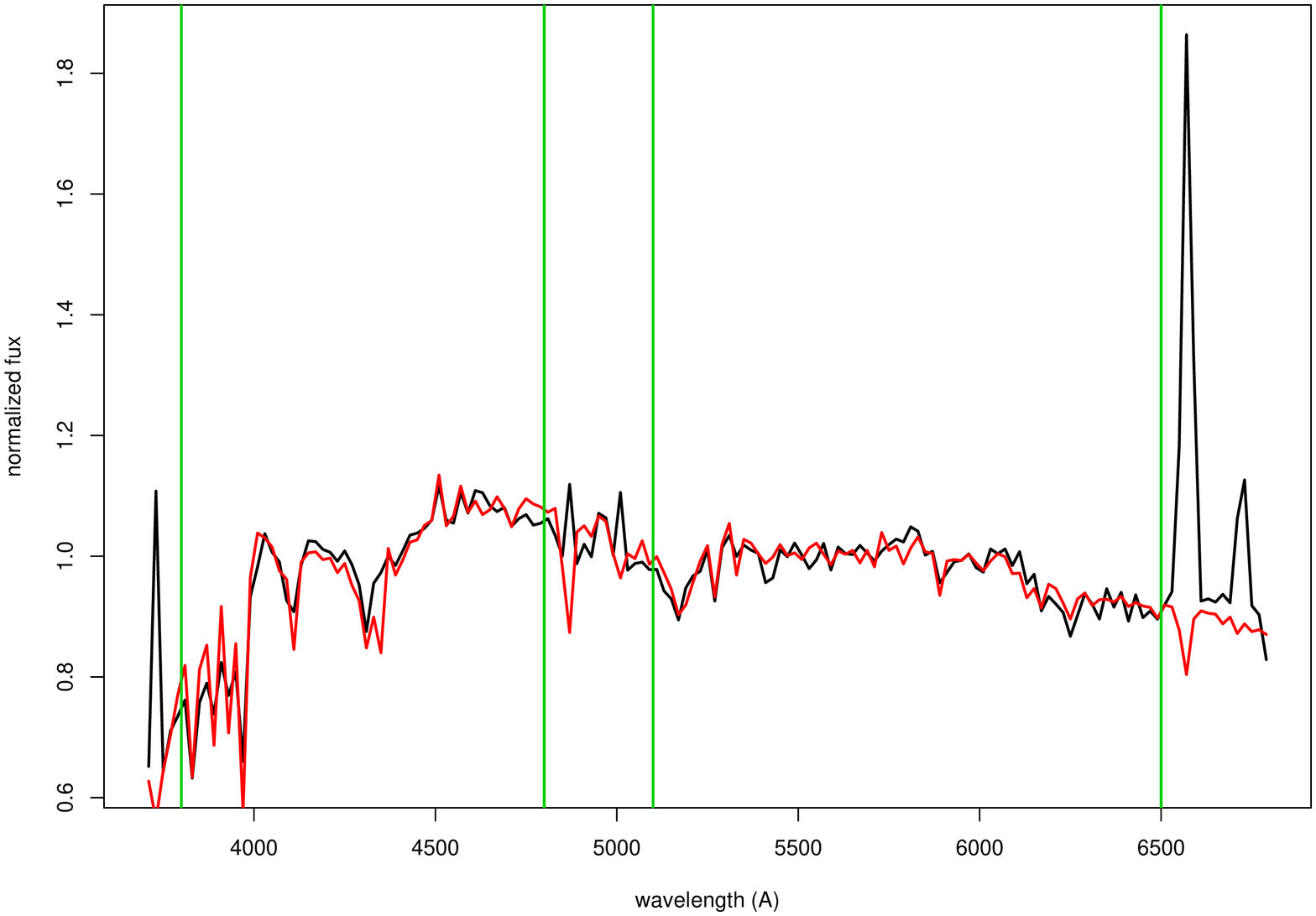}
\includegraphics[angle=-90,width=0.225\columnwidth]{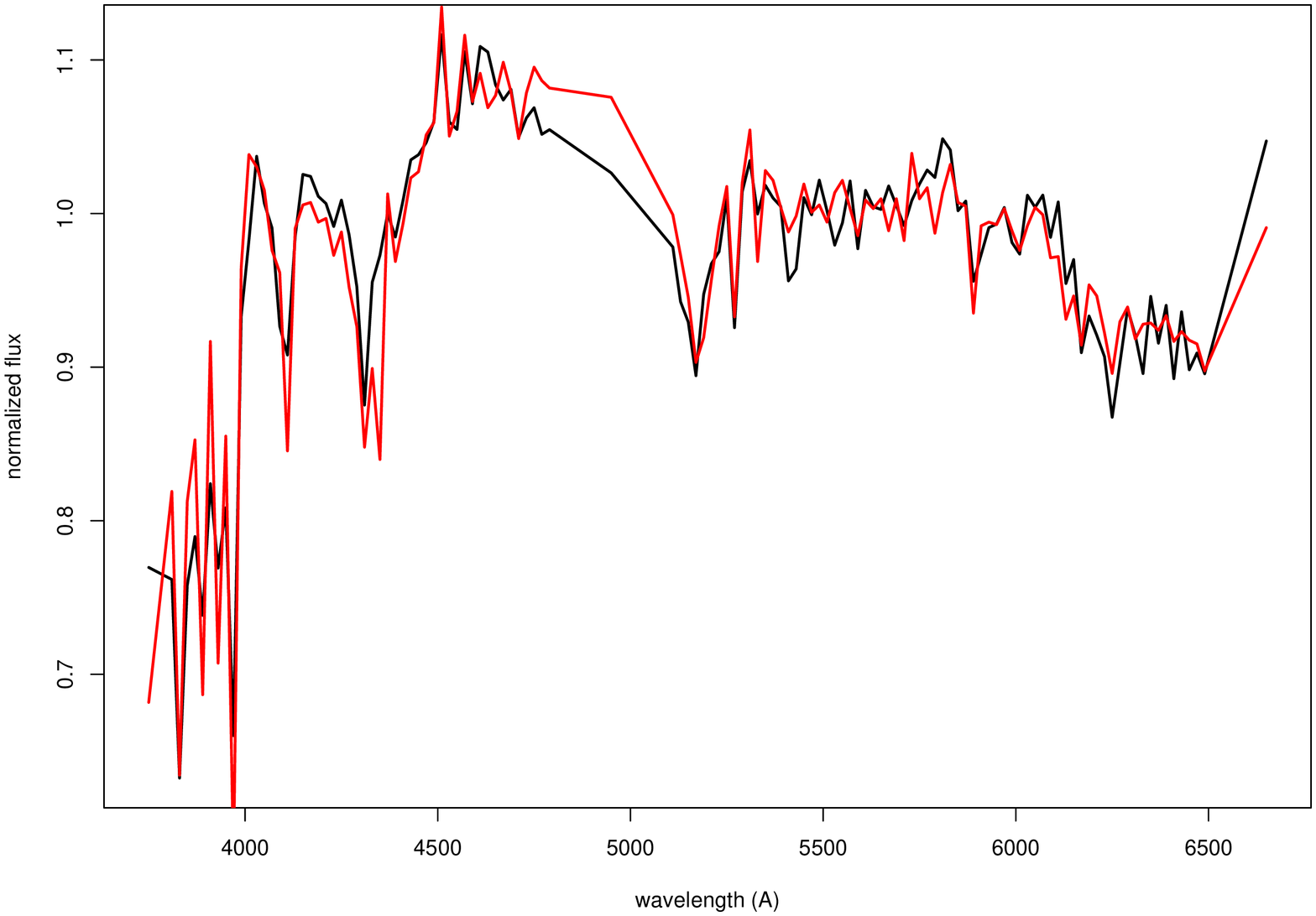}
\includegraphics[angle=-90,width=0.225\columnwidth]{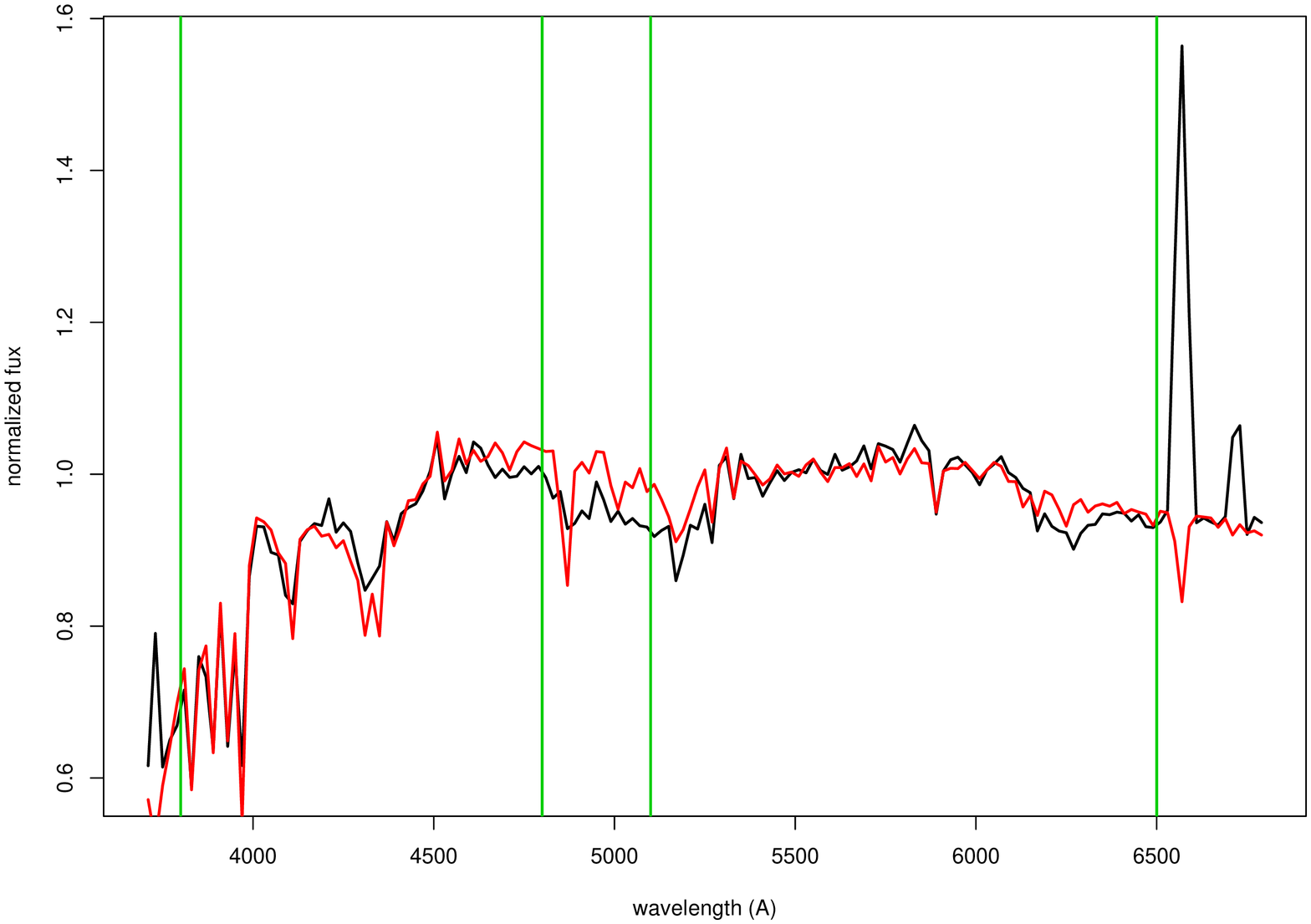}
\includegraphics[angle=-90,width=0.225\columnwidth]{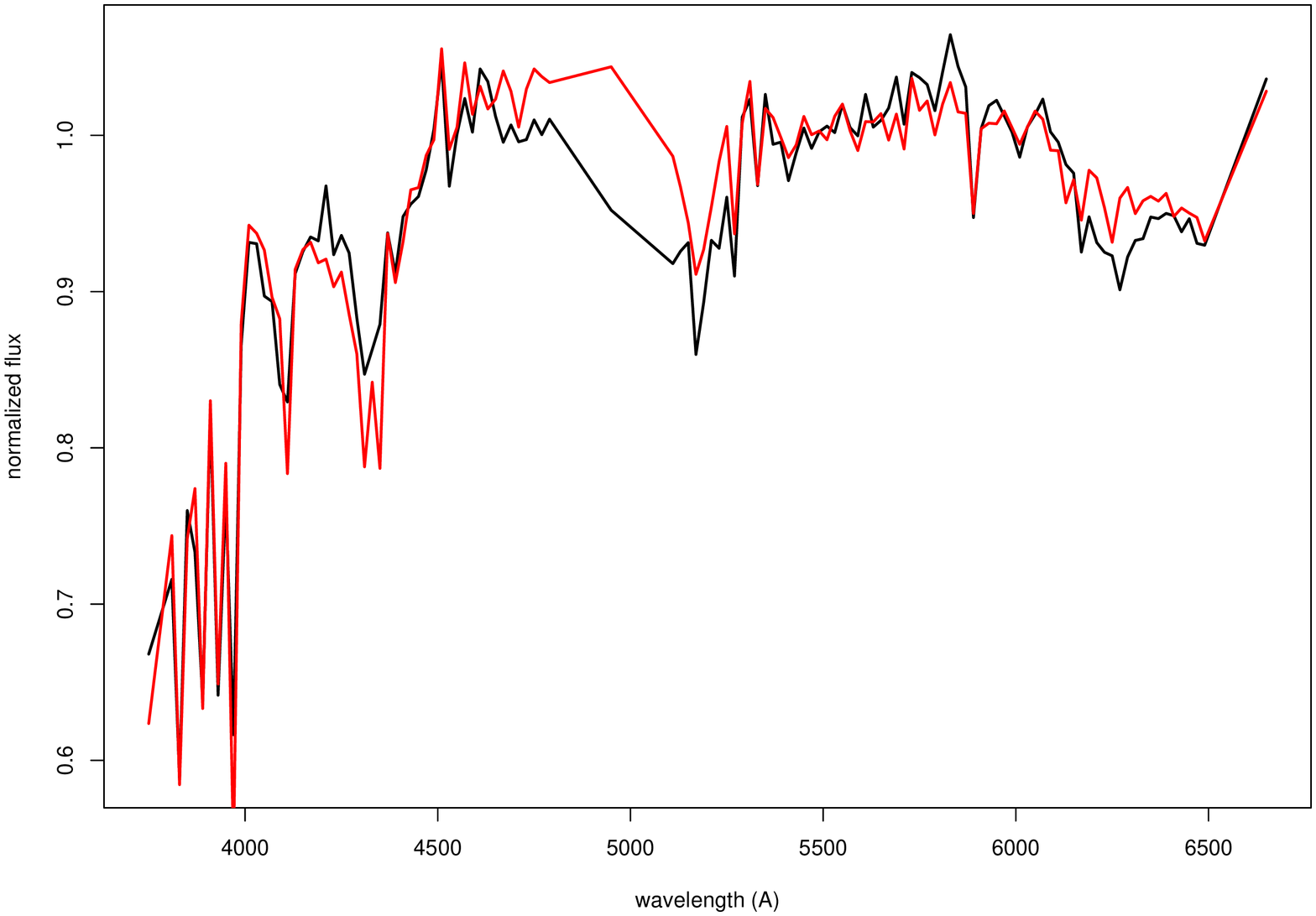}
\includegraphics[angle=-90,width=0.225\columnwidth]{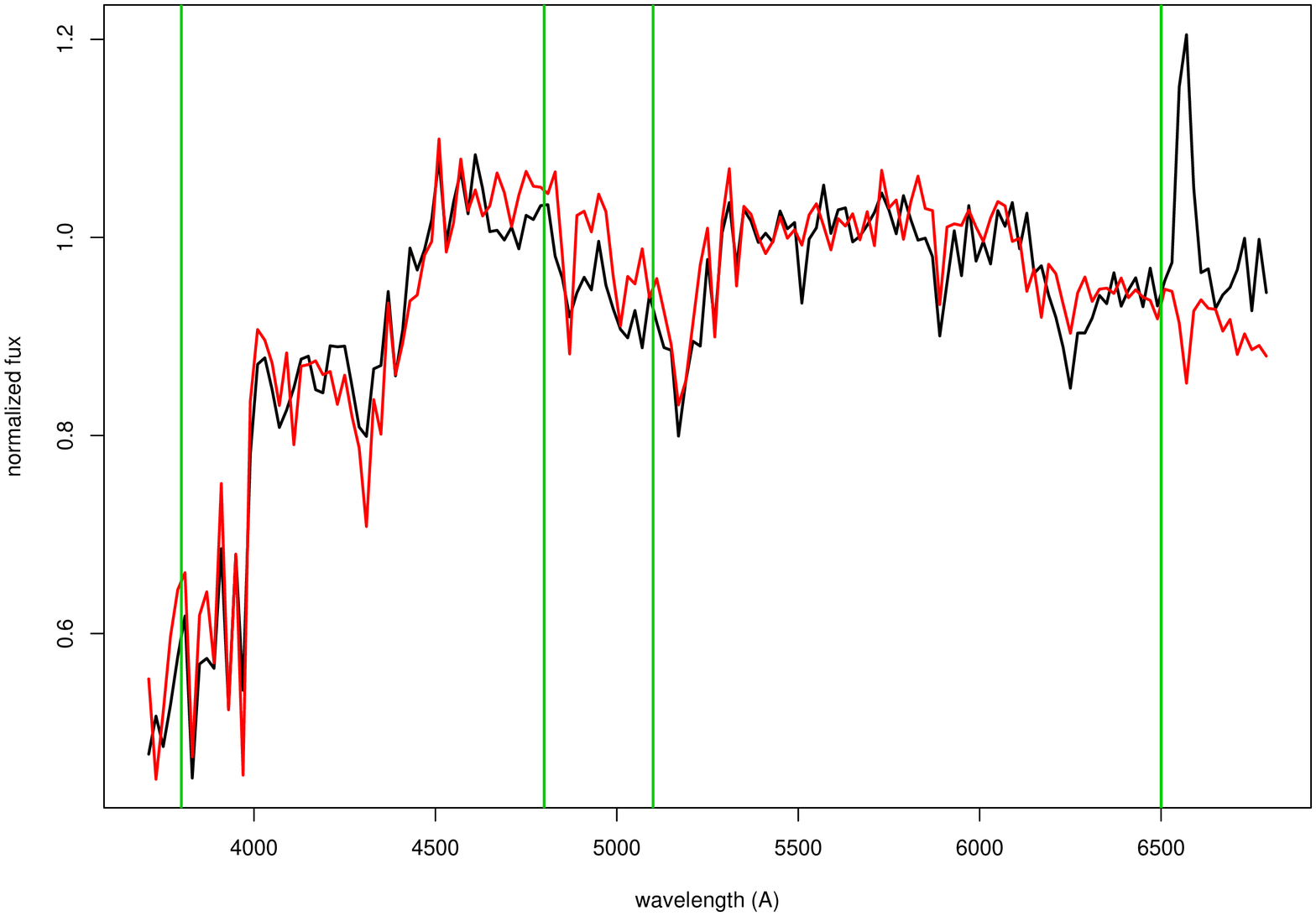}
\includegraphics[angle=-90,width=0.225\columnwidth]{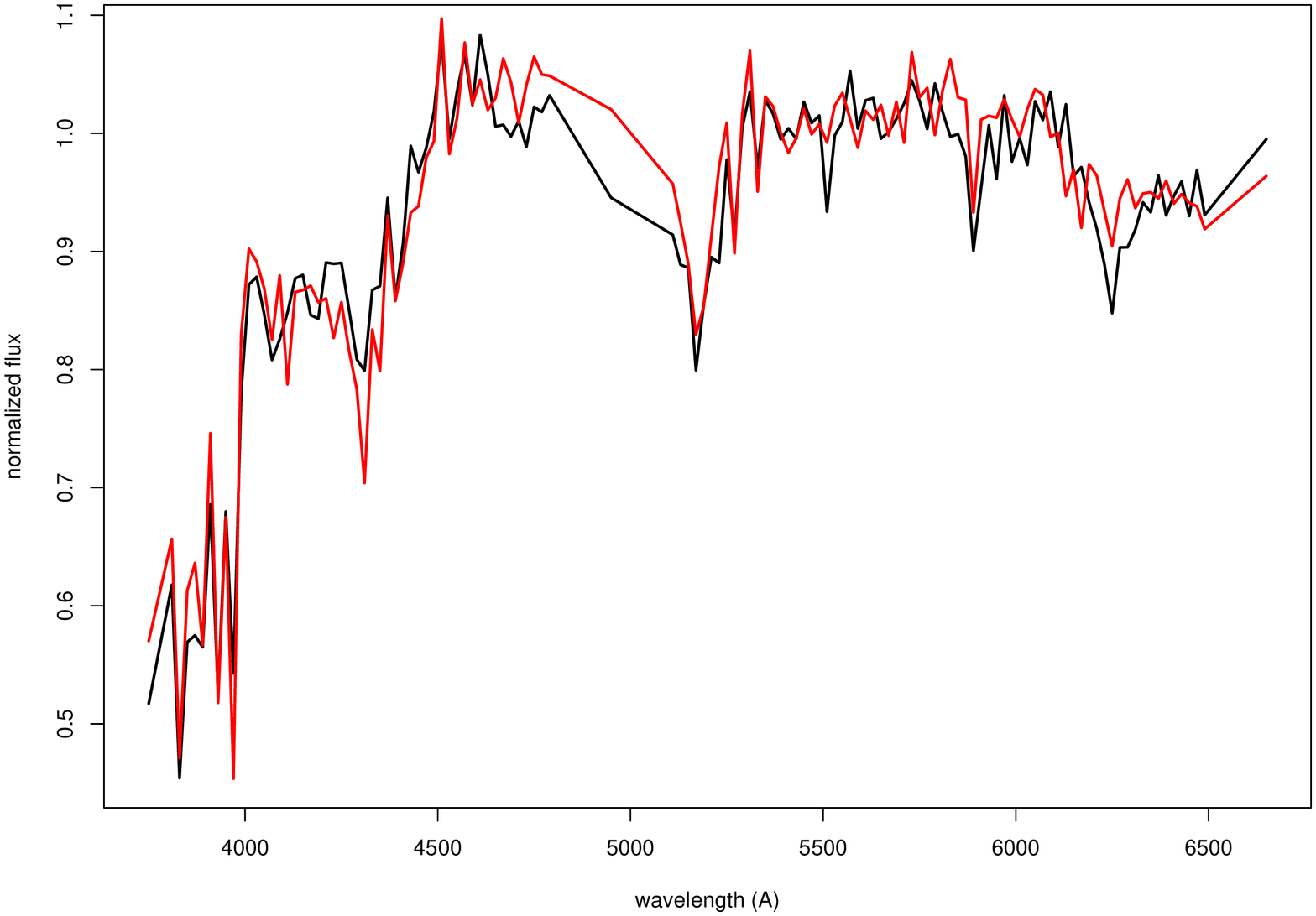}
\includegraphics[angle=-90,width=0.225\columnwidth]{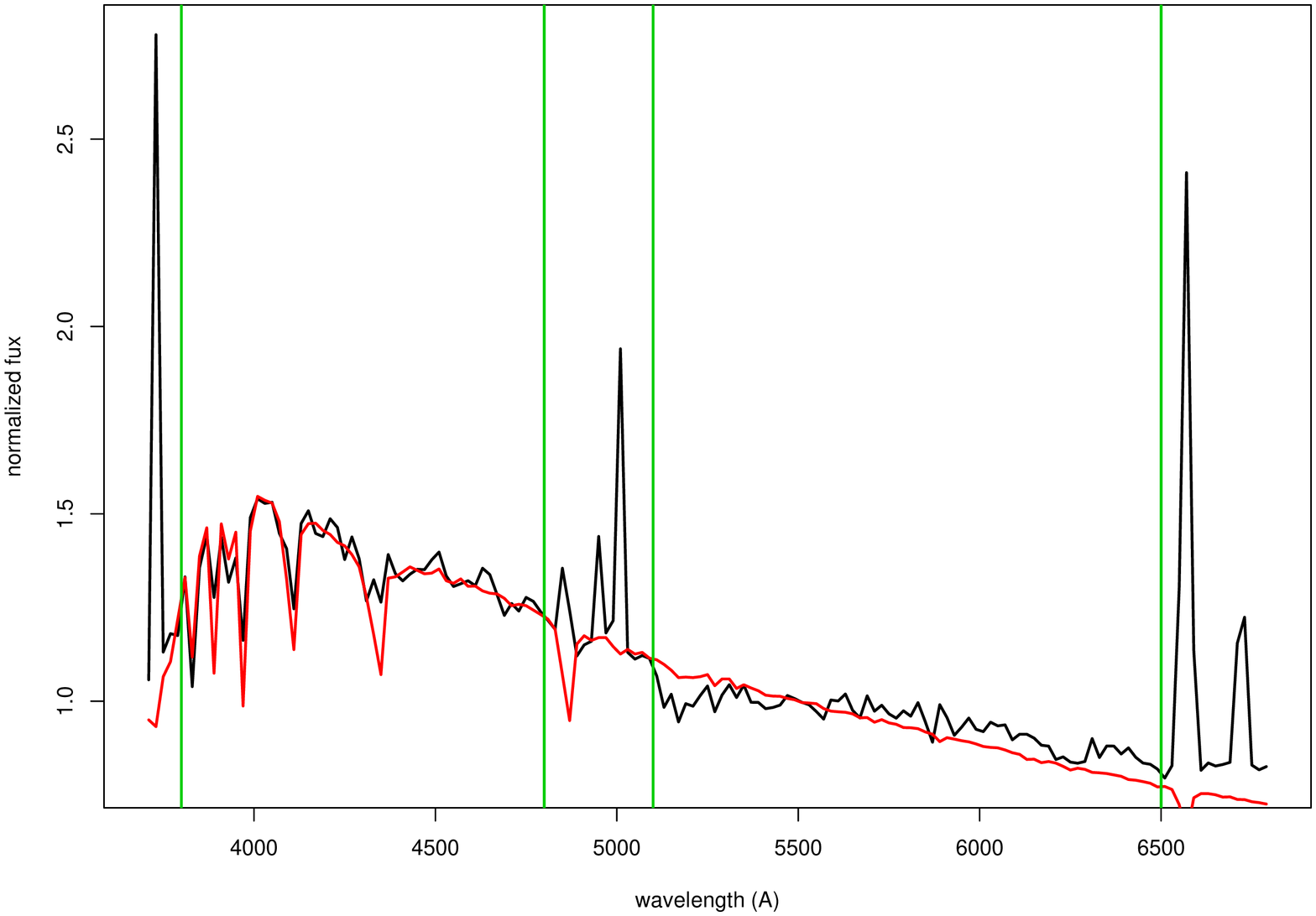}
\includegraphics[angle=-90,width=0.225\columnwidth]{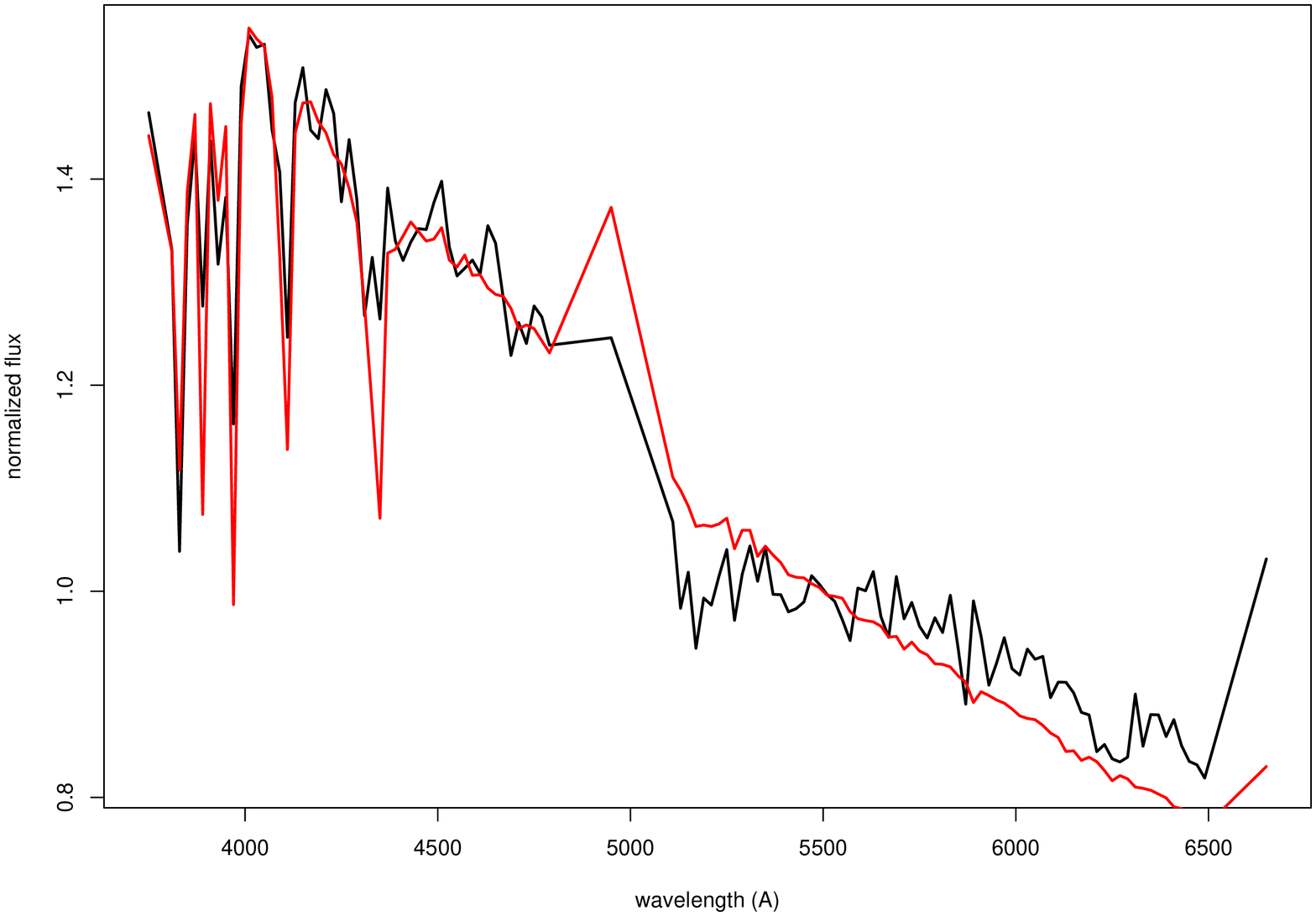}
\includegraphics[angle=-90,width=0.225\columnwidth]{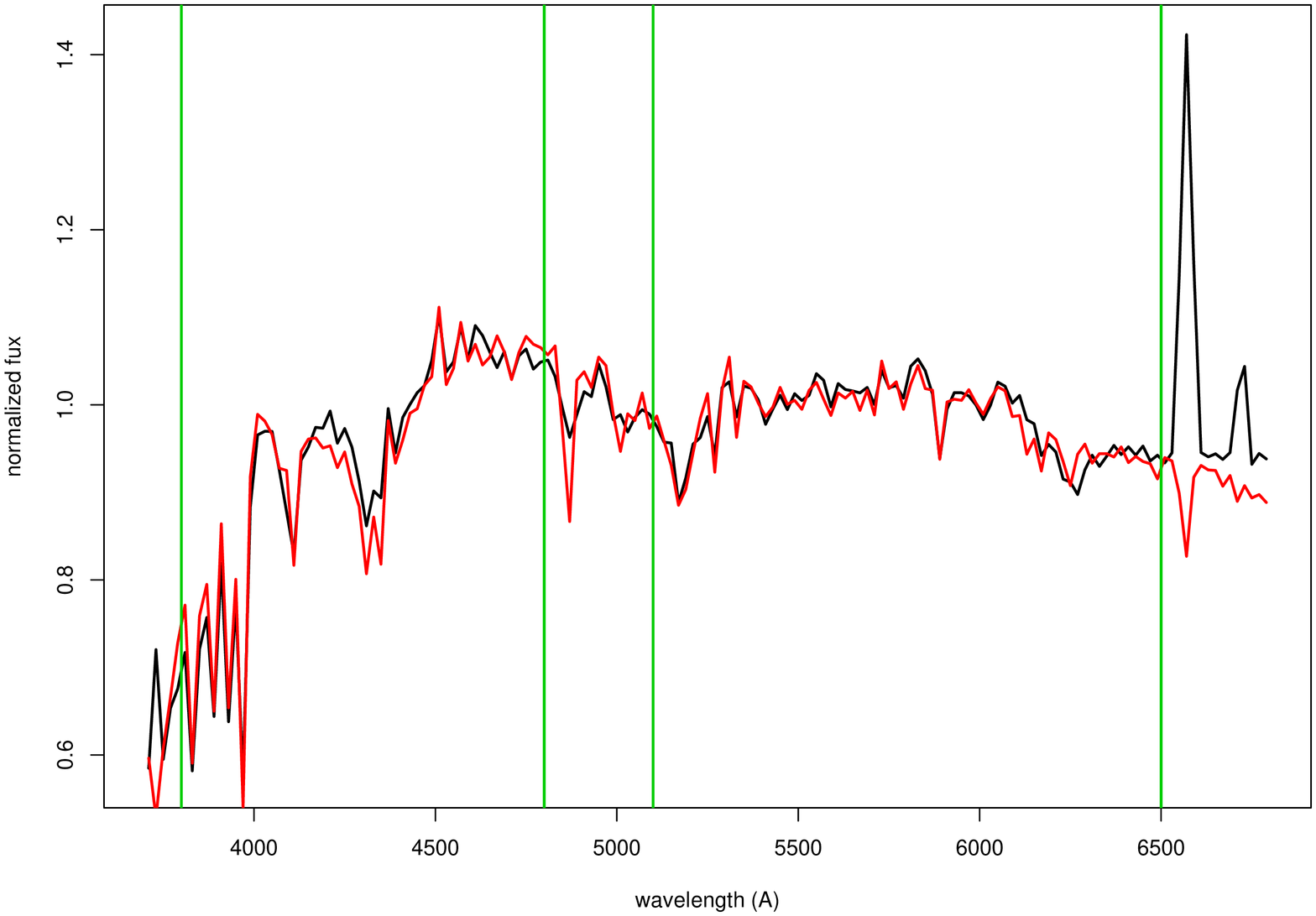}
\includegraphics[angle=-90,width=0.225\columnwidth]{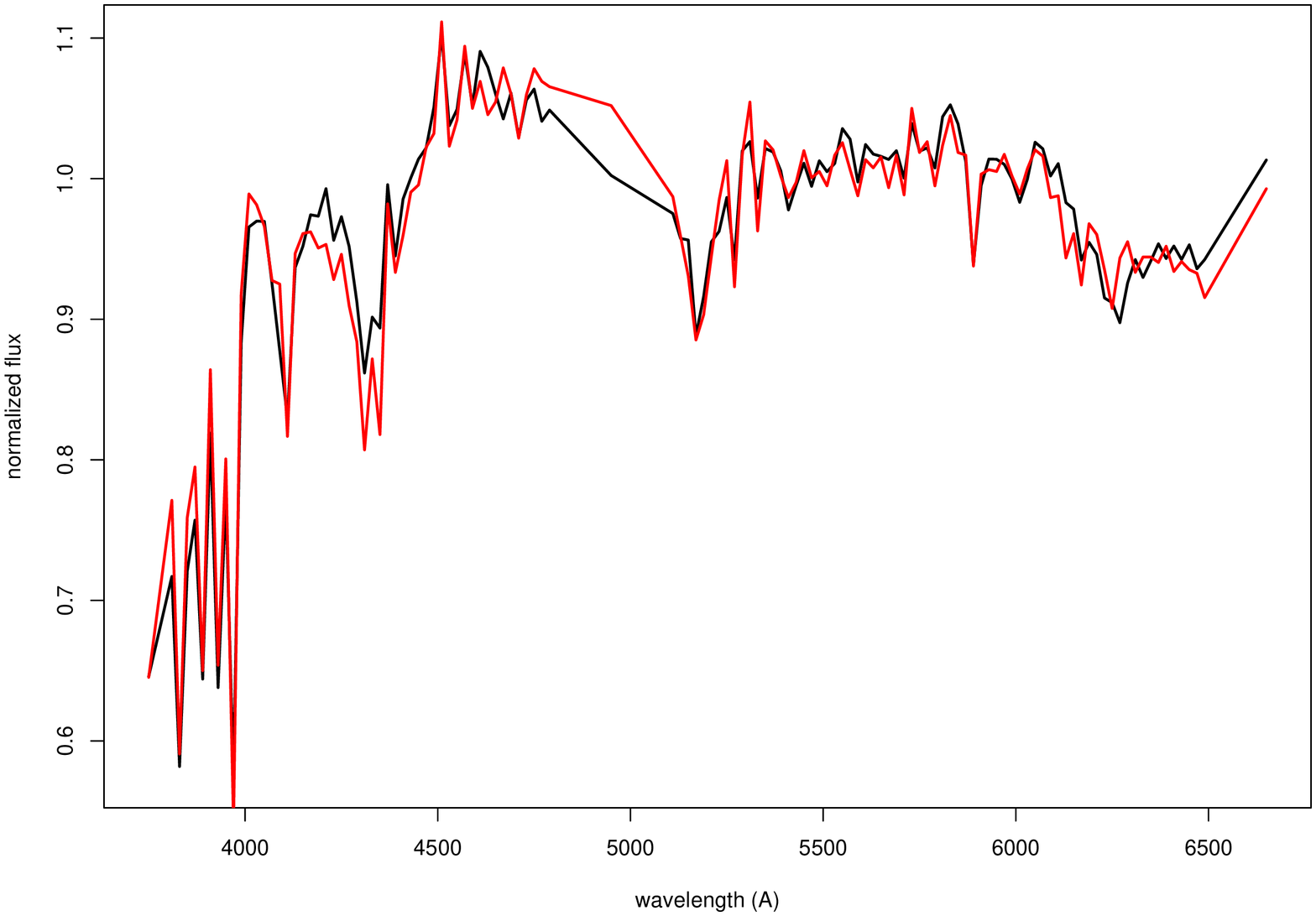}
\includegraphics[angle=-90,width=0.225\columnwidth]{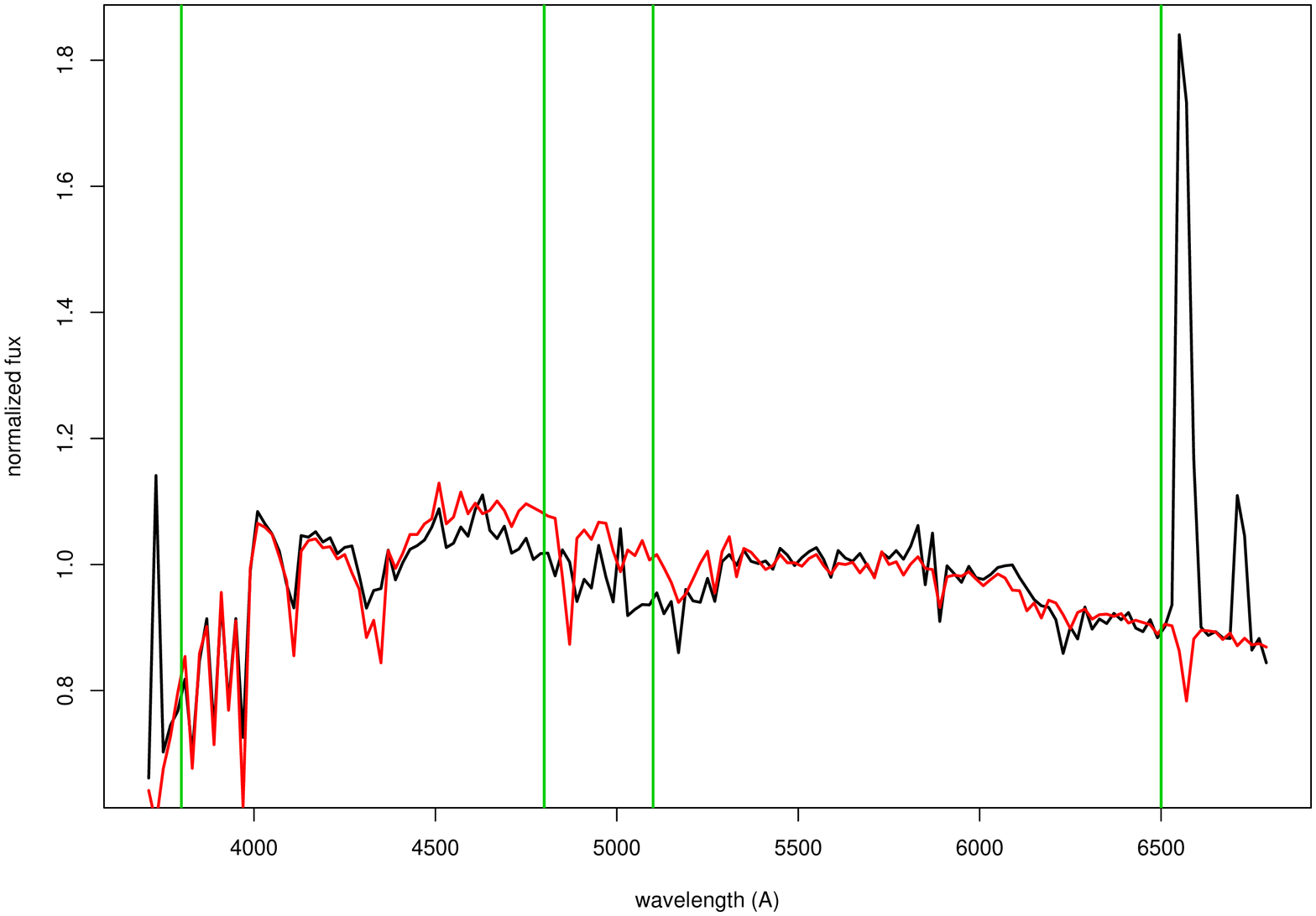}
\includegraphics[angle=-90,width=0.225\columnwidth]{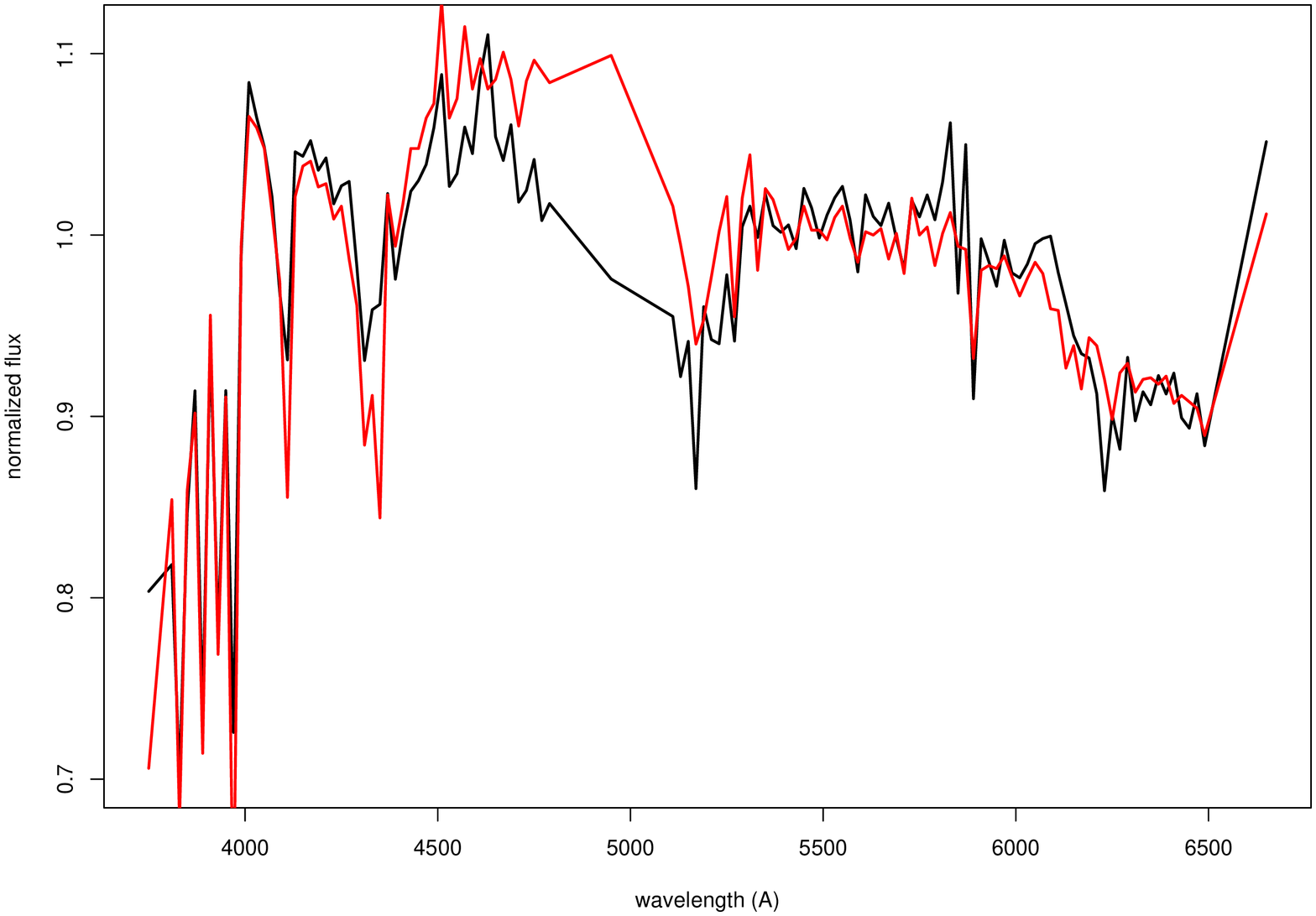}
\includegraphics[angle=-90,width=0.225\columnwidth]{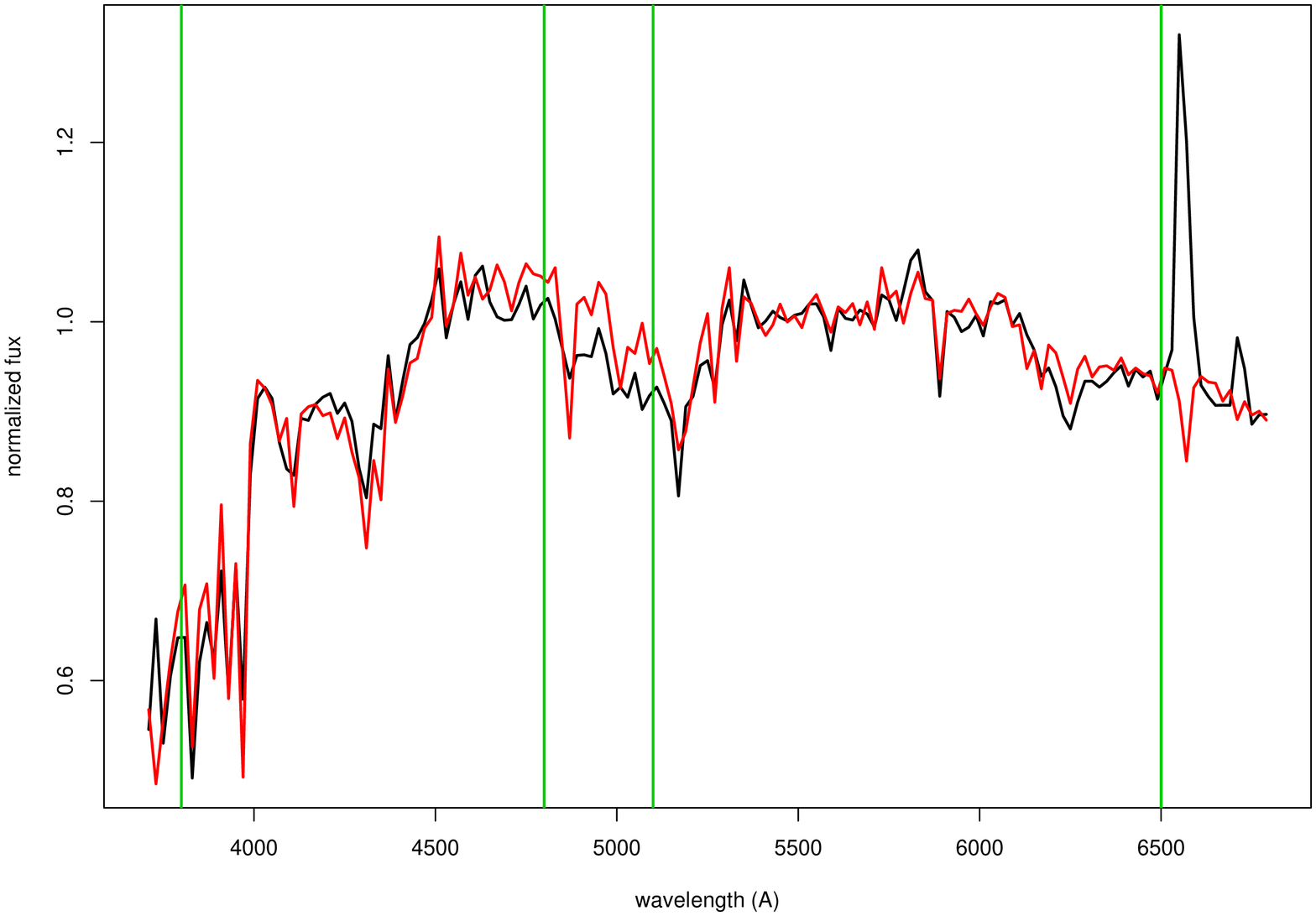}
\includegraphics[angle=-90,width=0.225\columnwidth]{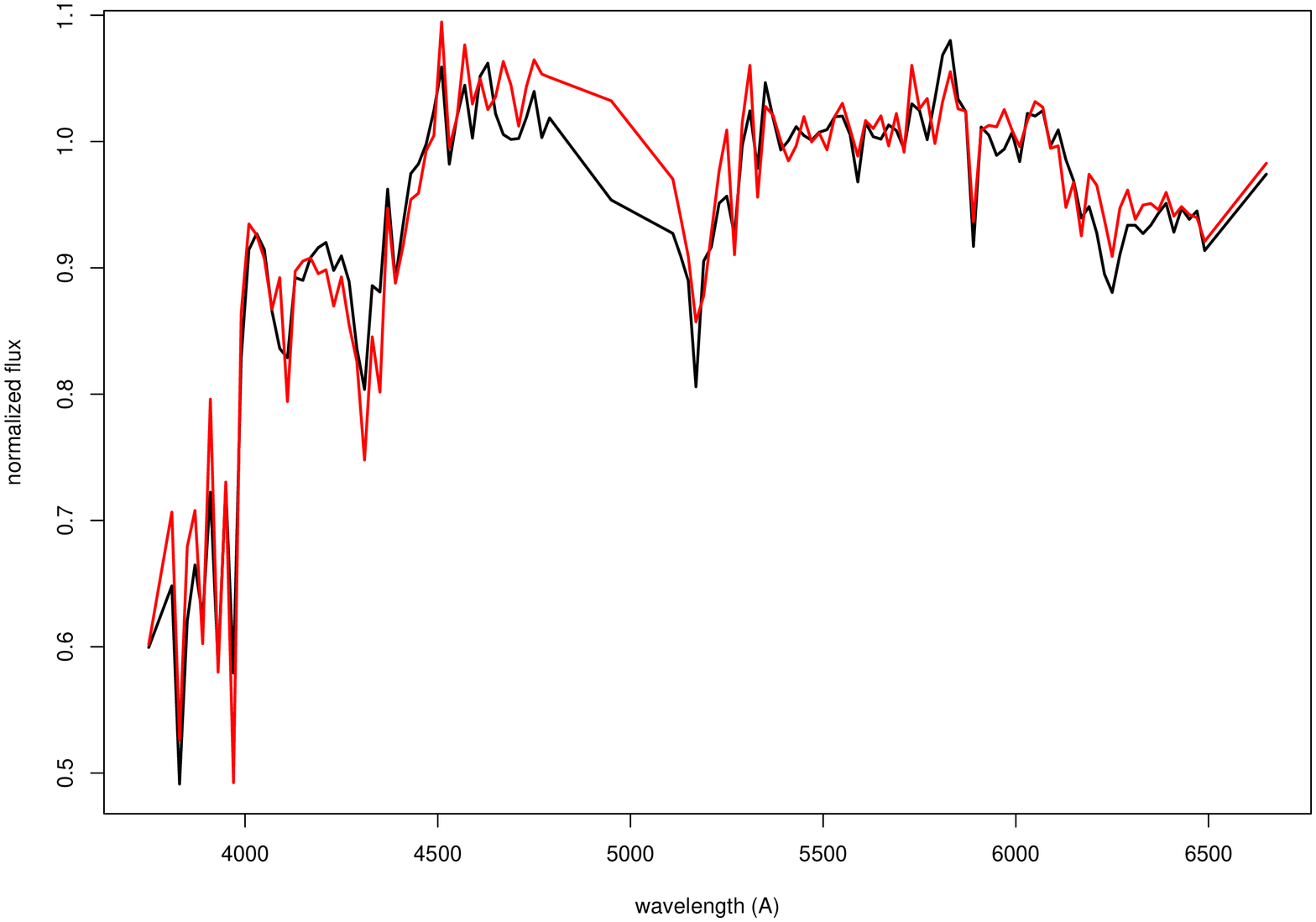}
\includegraphics[angle=-90,width=0.225\columnwidth]{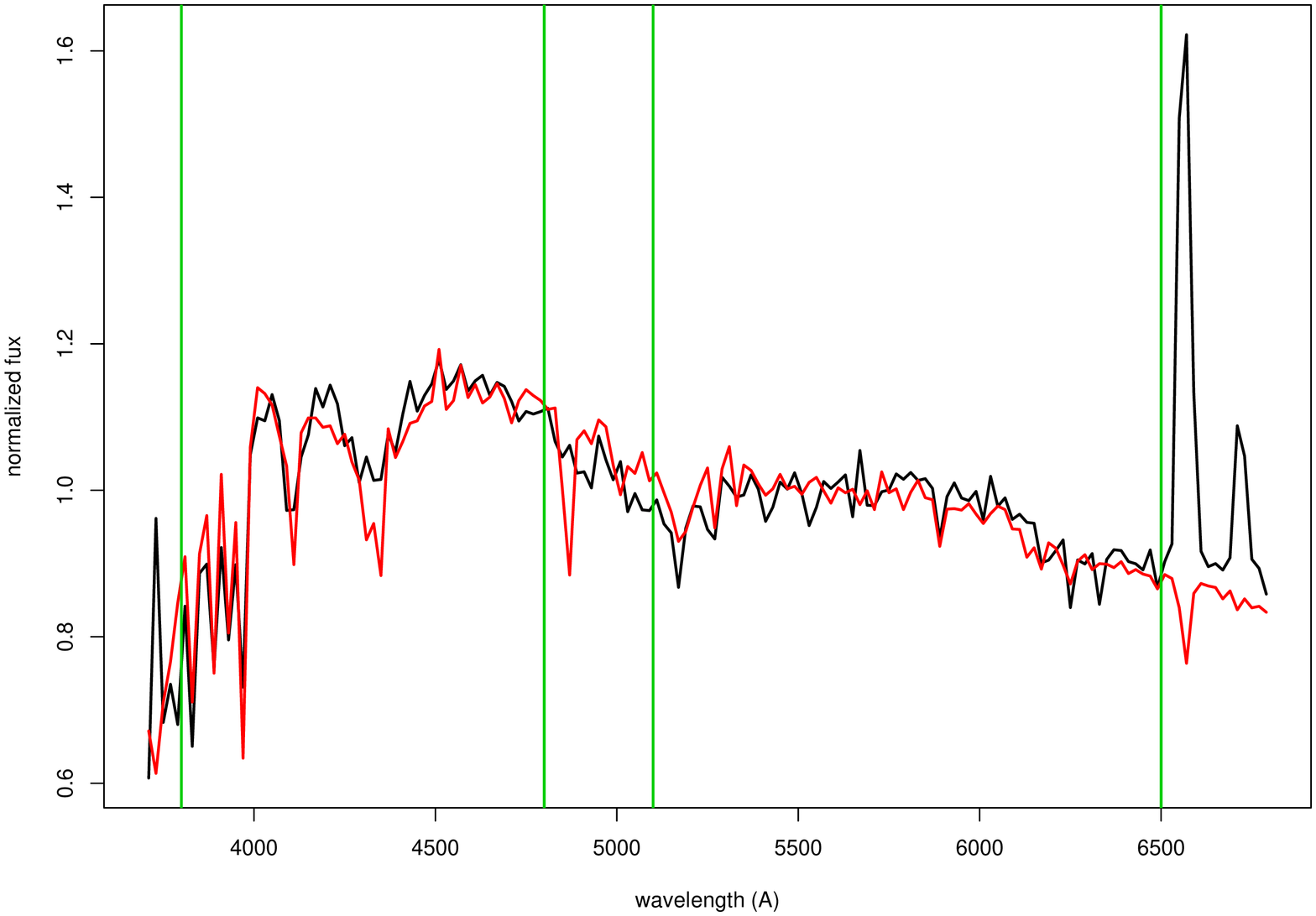}
\includegraphics[angle=-90,width=0.225\columnwidth]{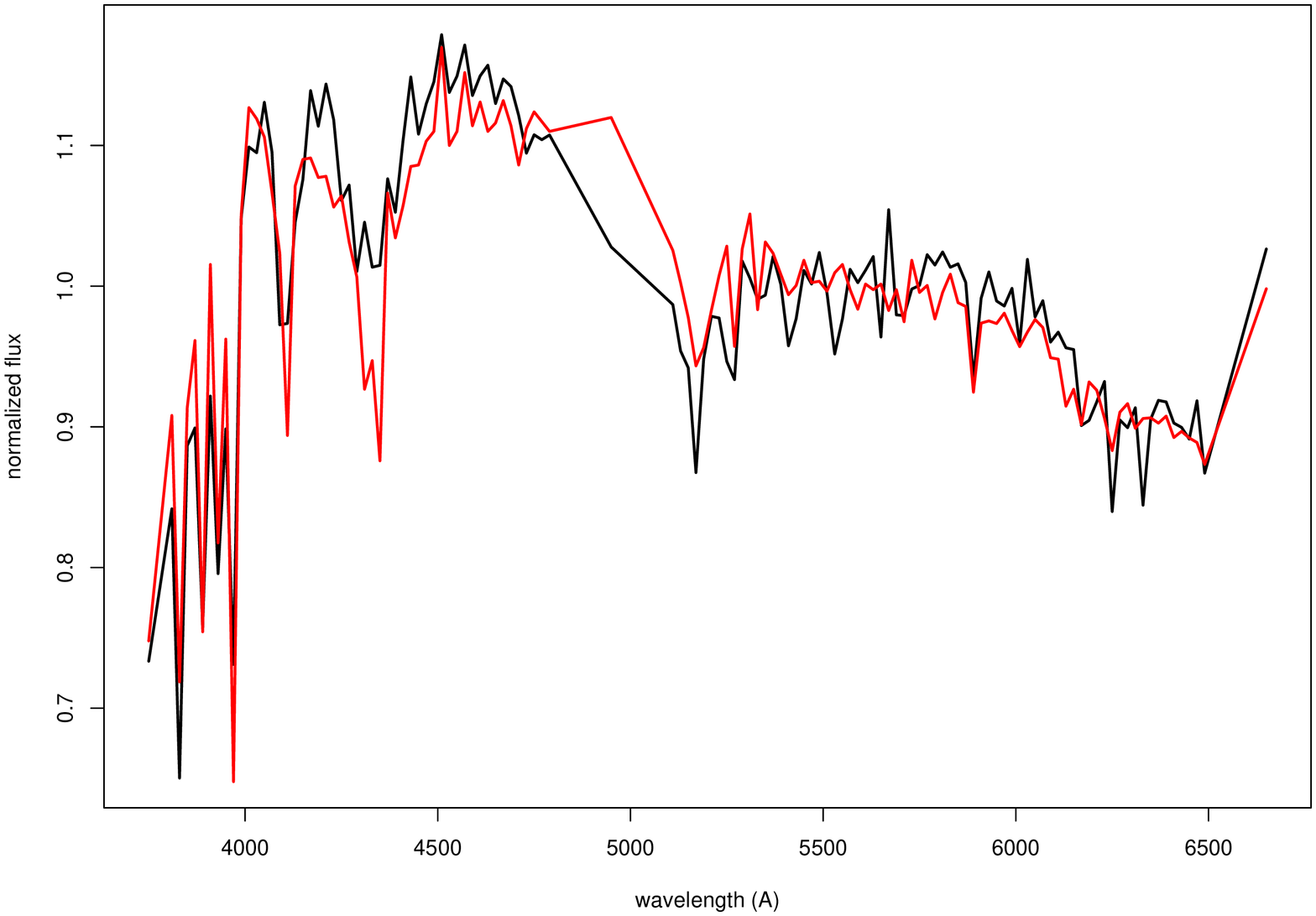}
\caption{The same with Fig \ref{a1} for the spiral galaxies in Kennicutt's 
atlas. \textbf{first row:} NGC1357, NGC2276, NGC2775, and NGC4775, 
\textbf{second row:} NGC5248, NGC3368, NGC3623, and NGC6217, 
\textbf{third row:} NGC1832, NGC2903, NGC3147, and NGC4631, 
\textbf{fourth row:} NGC3627, NGC6181, NGC4750, and NGC6643.}
\label{a2}
\end{figure*}

\begin{figure*}[h]
\center
\includegraphics[angle=-90,width=0.225\columnwidth]{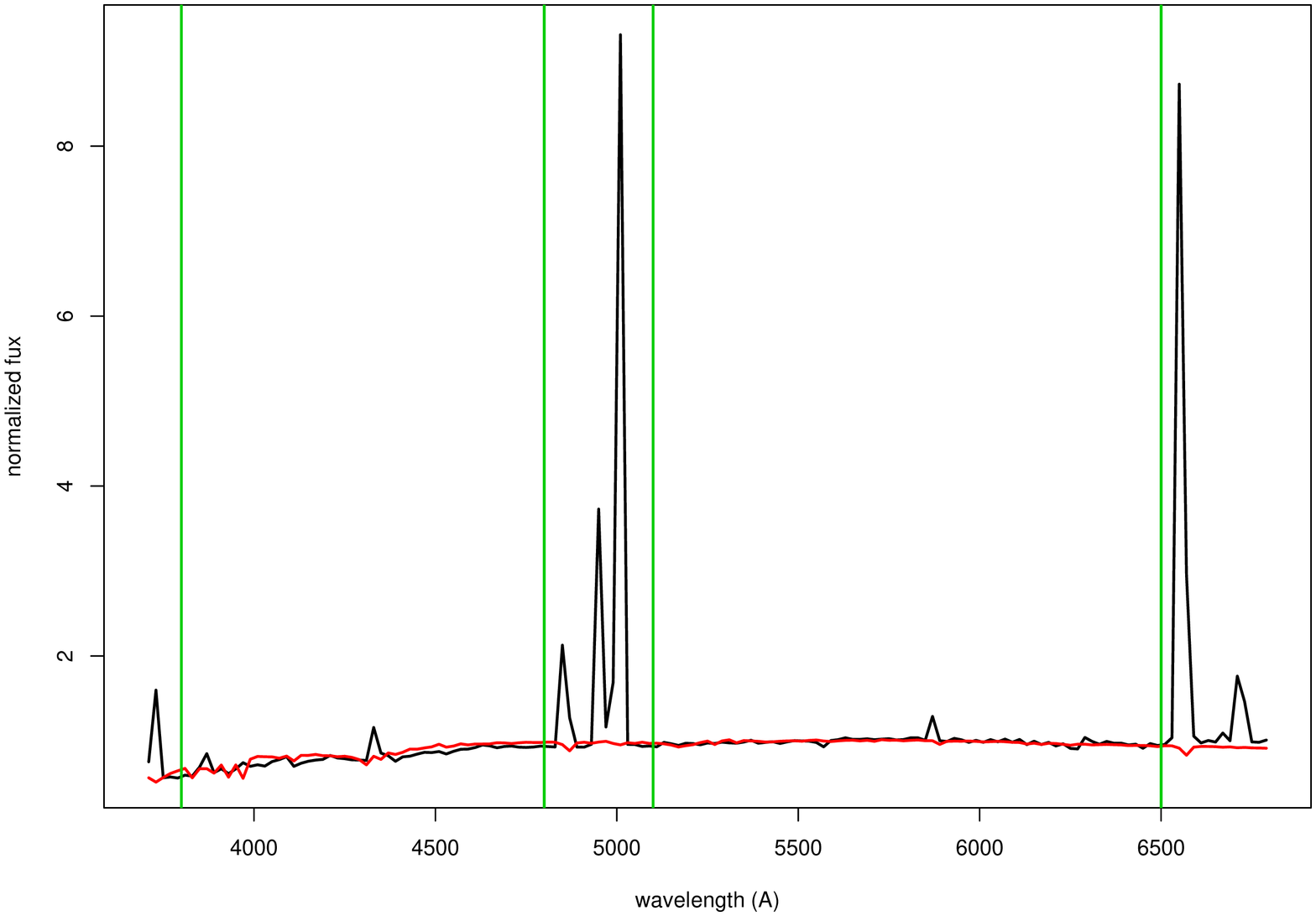}
\includegraphics[angle=-90,width=0.225\columnwidth]{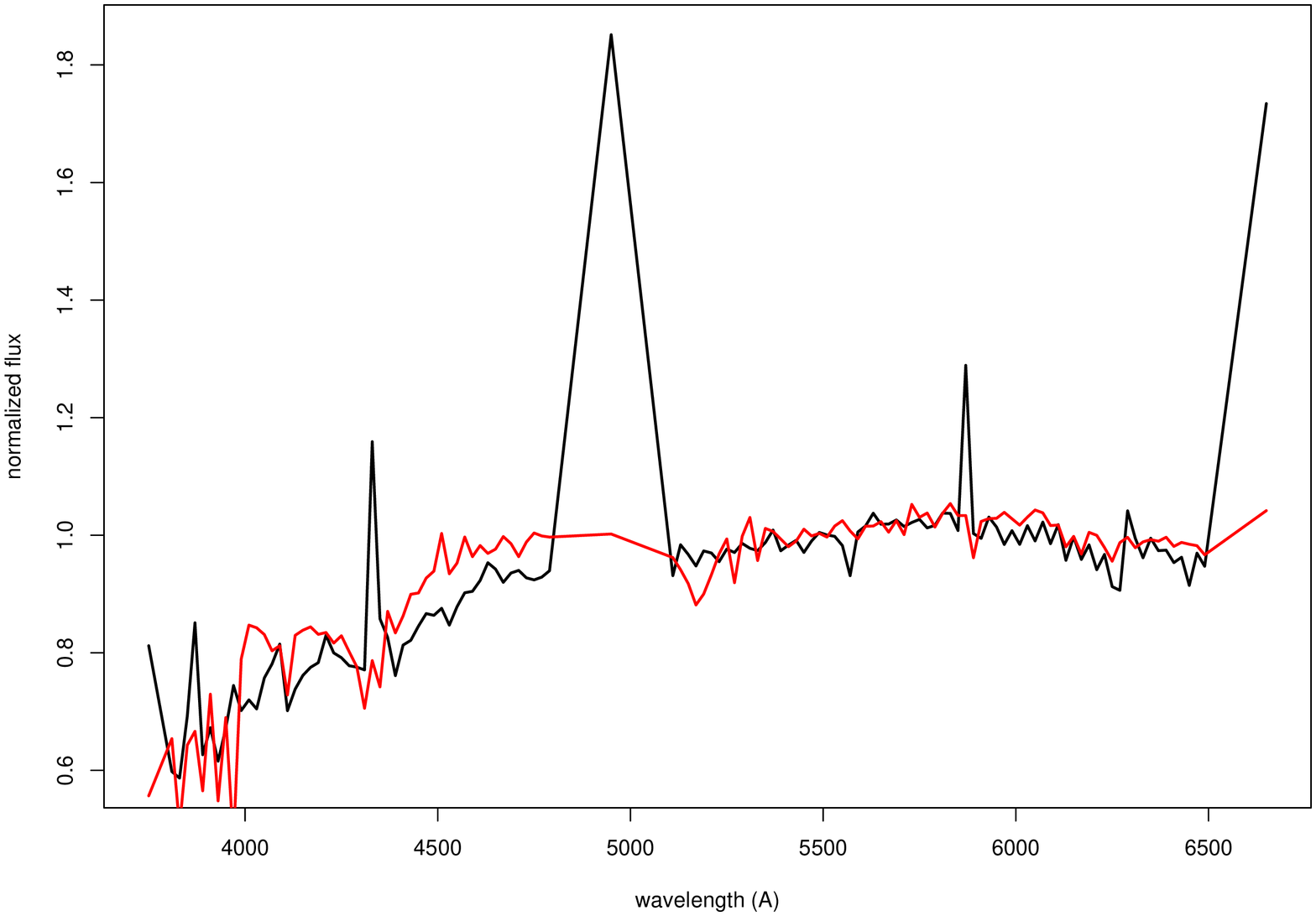}
\includegraphics[angle=-90,width=0.225\columnwidth]{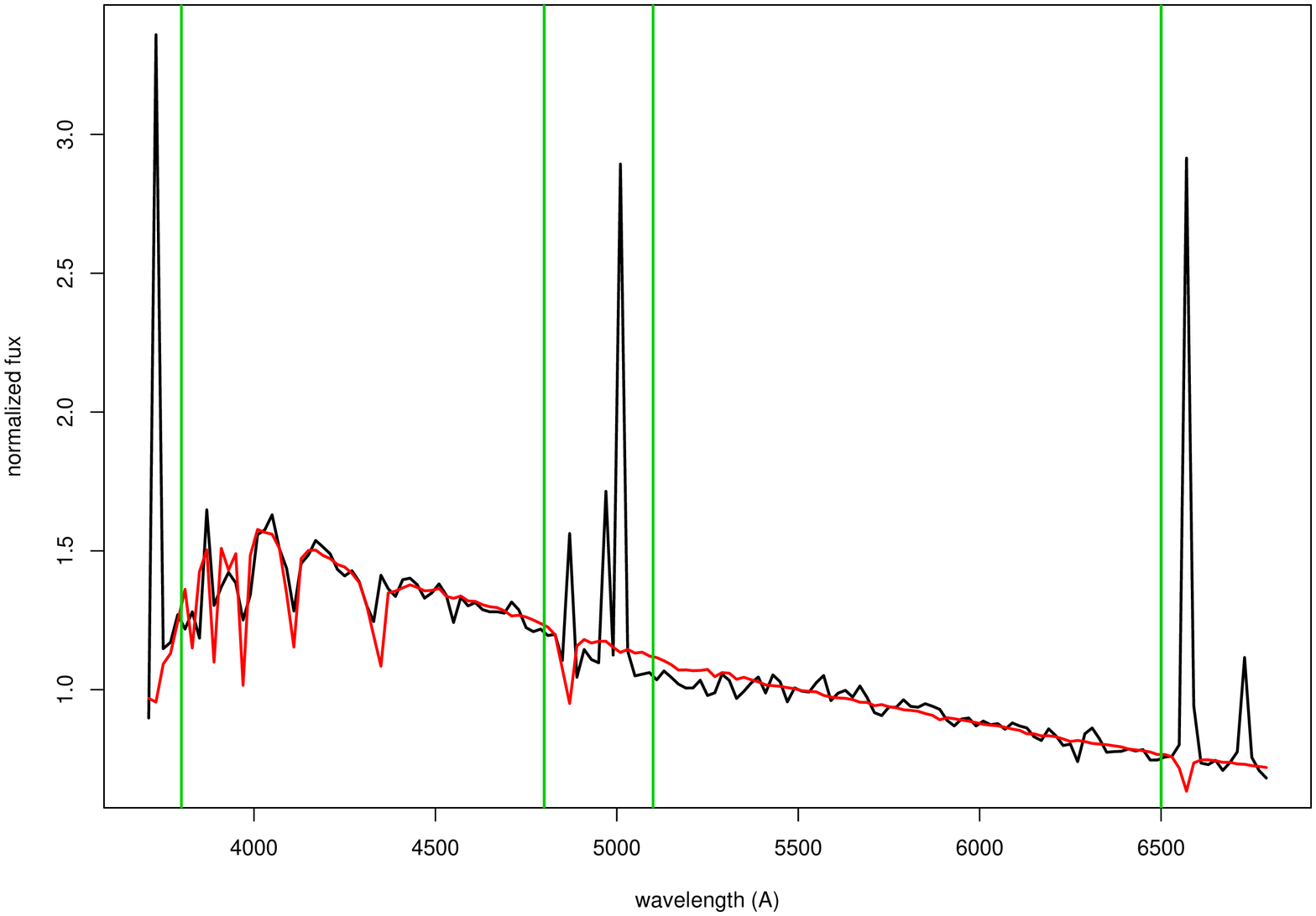}
\includegraphics[angle=-90,width=0.225\columnwidth]{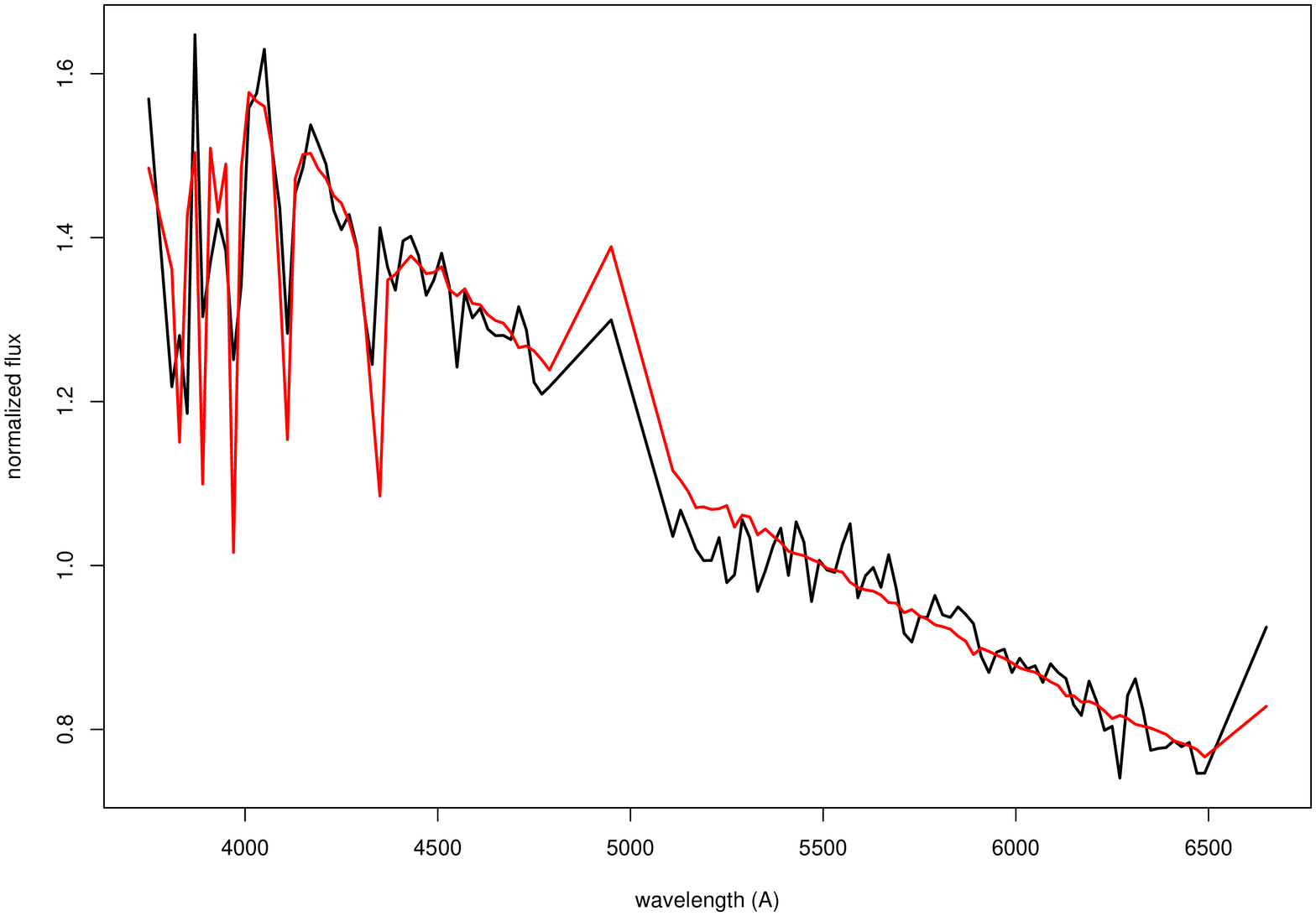}
\includegraphics[angle=-90,width=0.225\columnwidth]{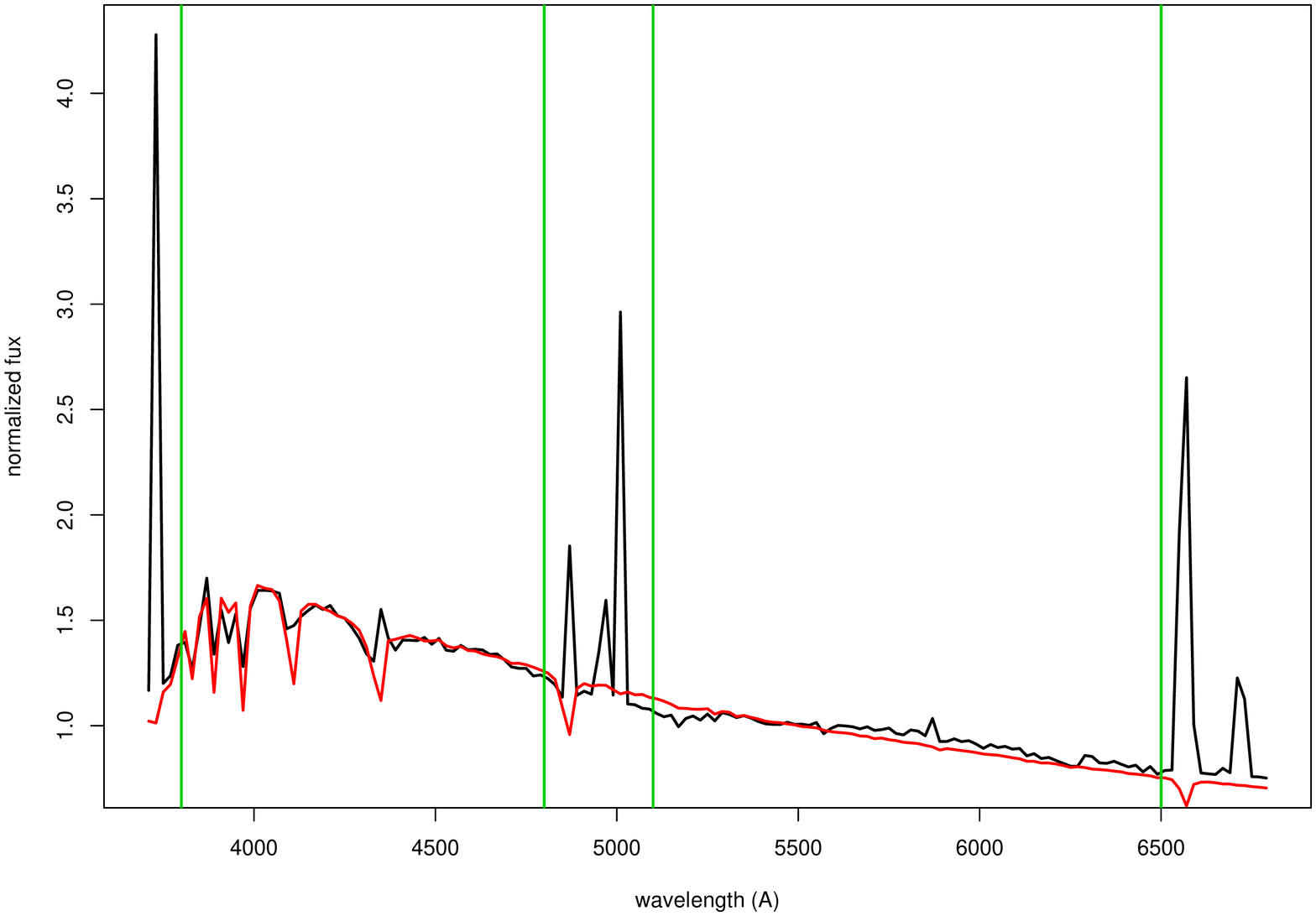}
\includegraphics[angle=-90,width=0.225\columnwidth]{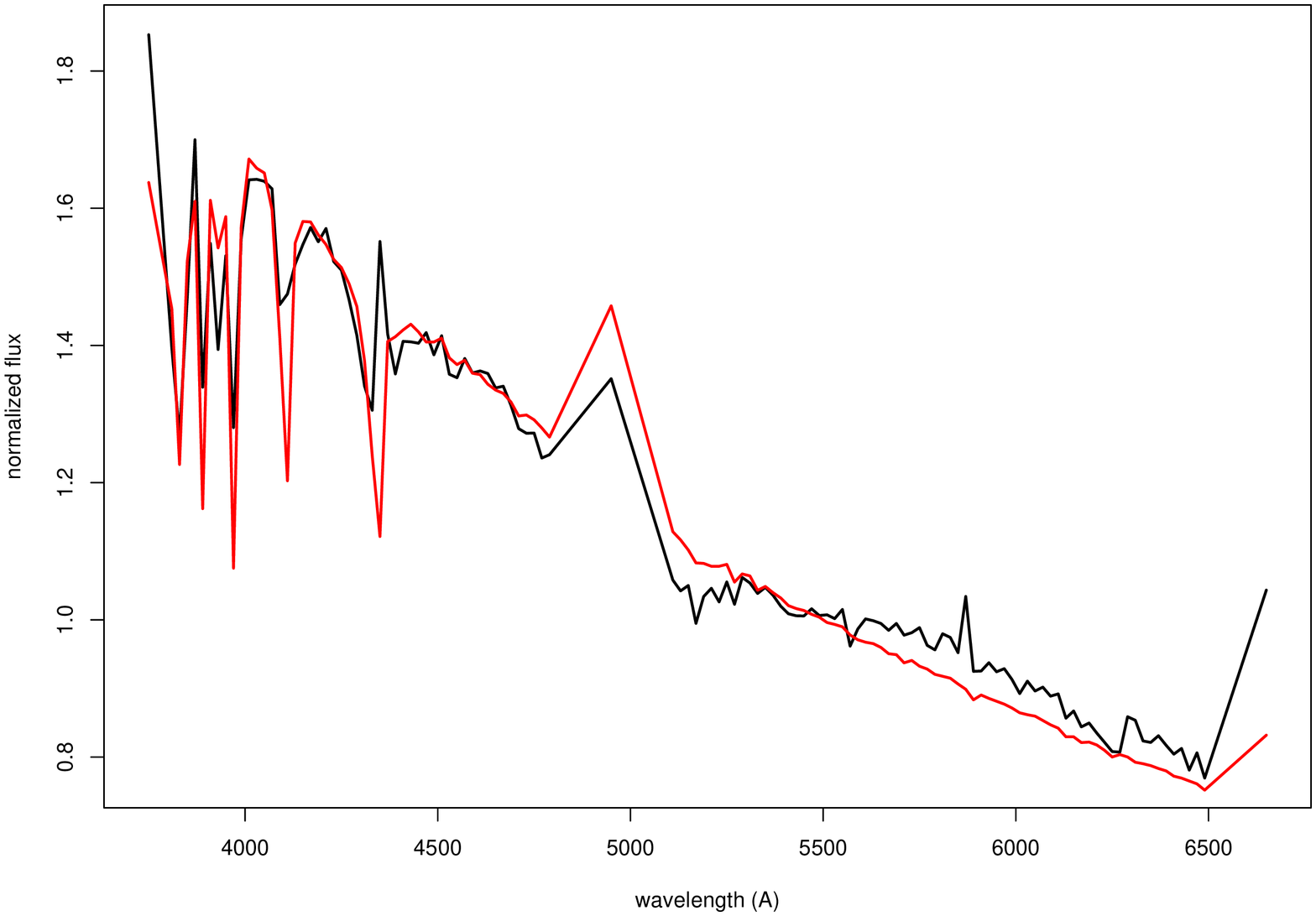}
\includegraphics[angle=-90,width=0.225\columnwidth]{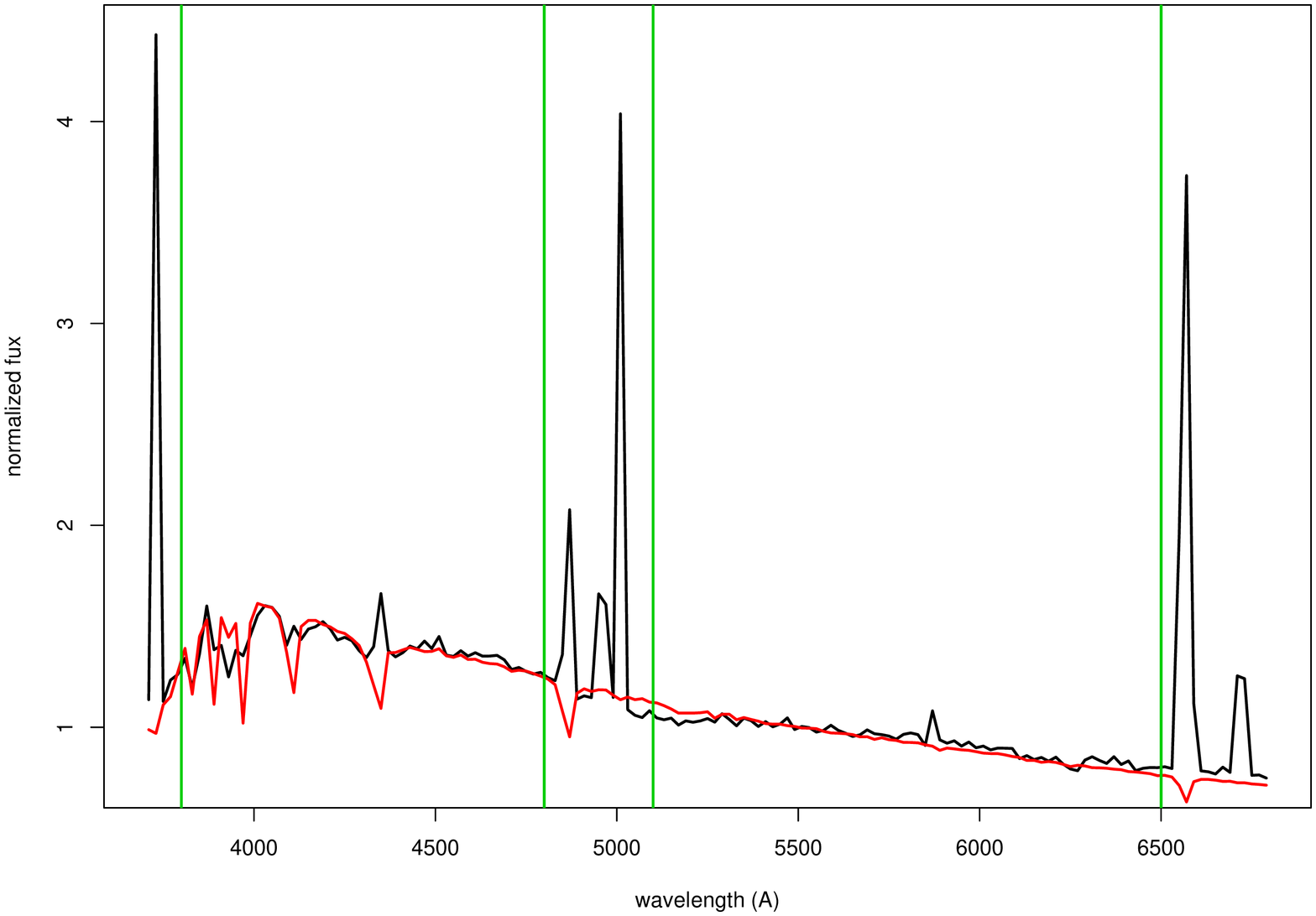}
\includegraphics[angle=-90,width=0.225\columnwidth]{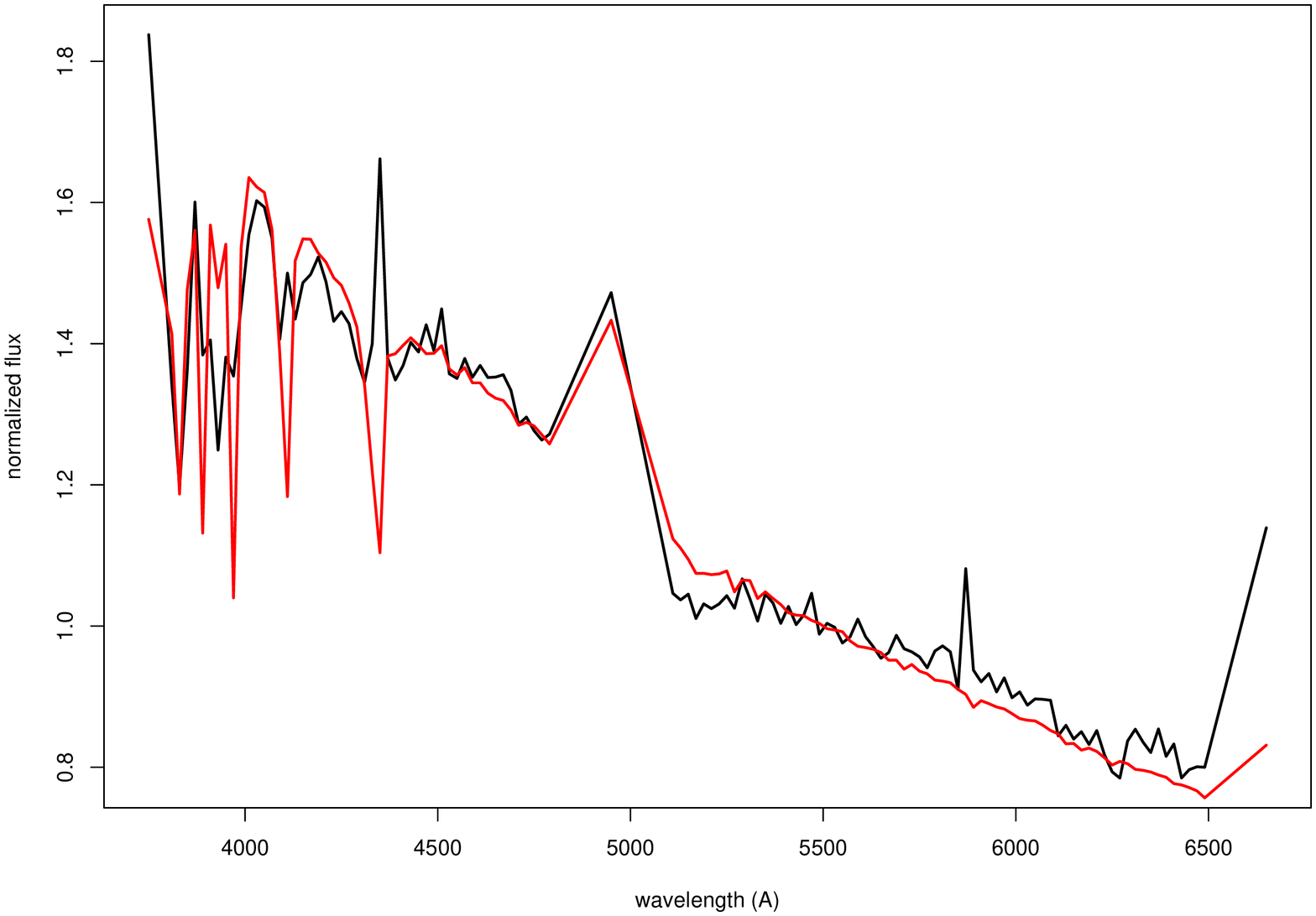}
\caption{The same with Fig \ref{a1} for the magelanic type 
irregular galaxies in Kennicutt's atlas. Here we present: 
NGC1569, NGC4485, NGC4449, and NGC4670.}
\label{a3}
\end{figure*}

\begin{figure*}[h]
\center
\includegraphics[angle=-90,width=0.225\columnwidth]{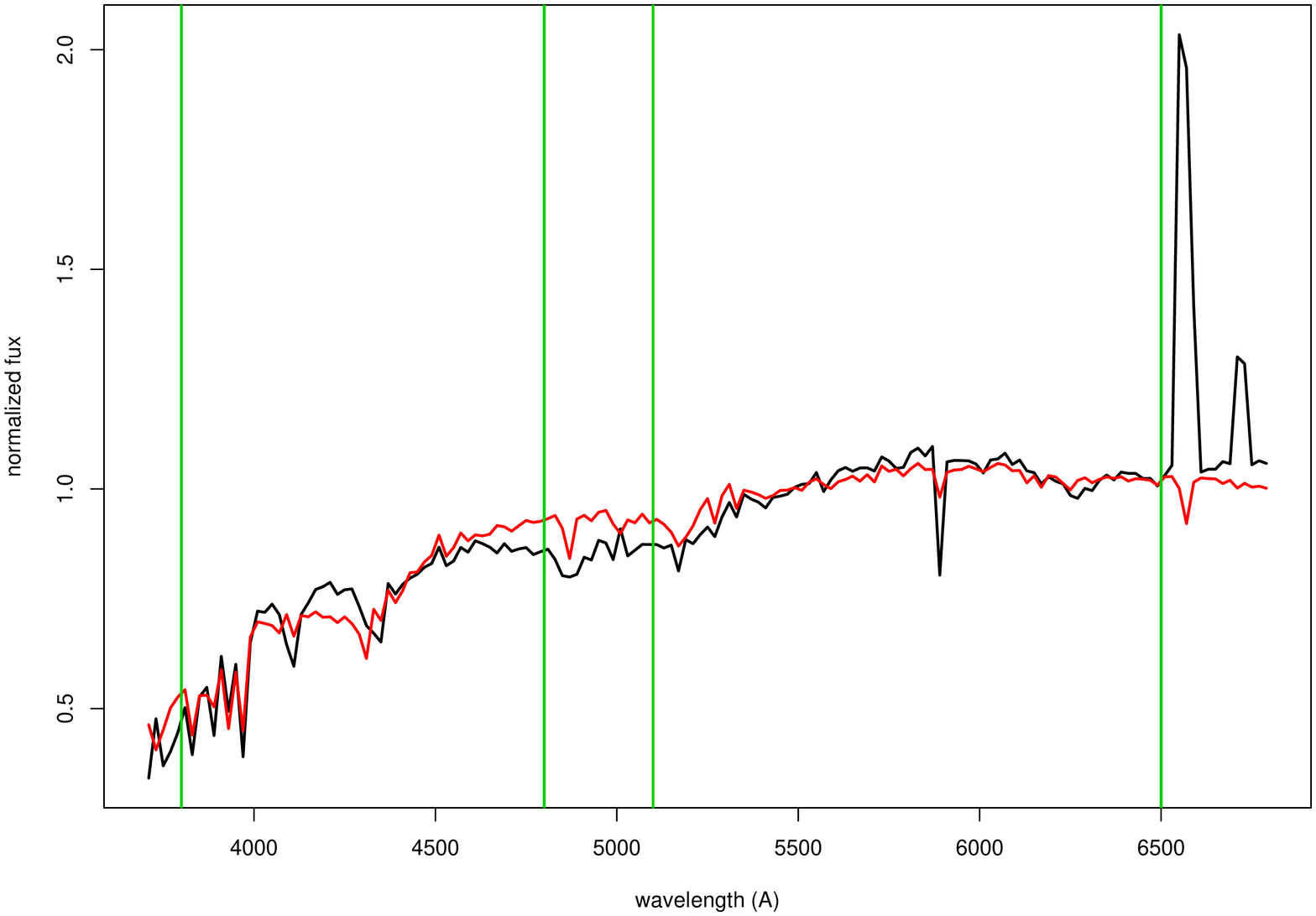}
\includegraphics[angle=-90,width=0.225\columnwidth]{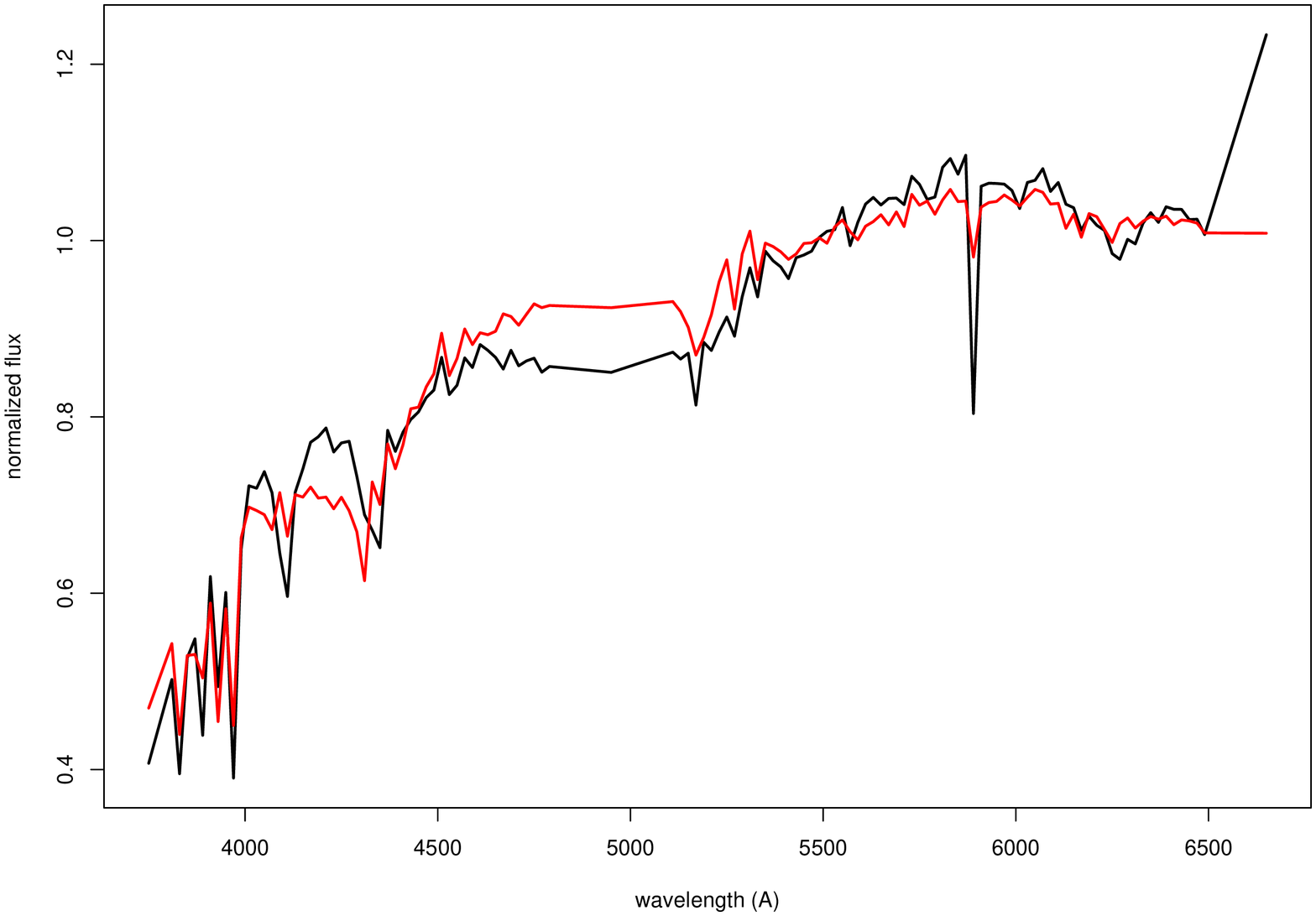}
\includegraphics[angle=-90,width=0.225\columnwidth]{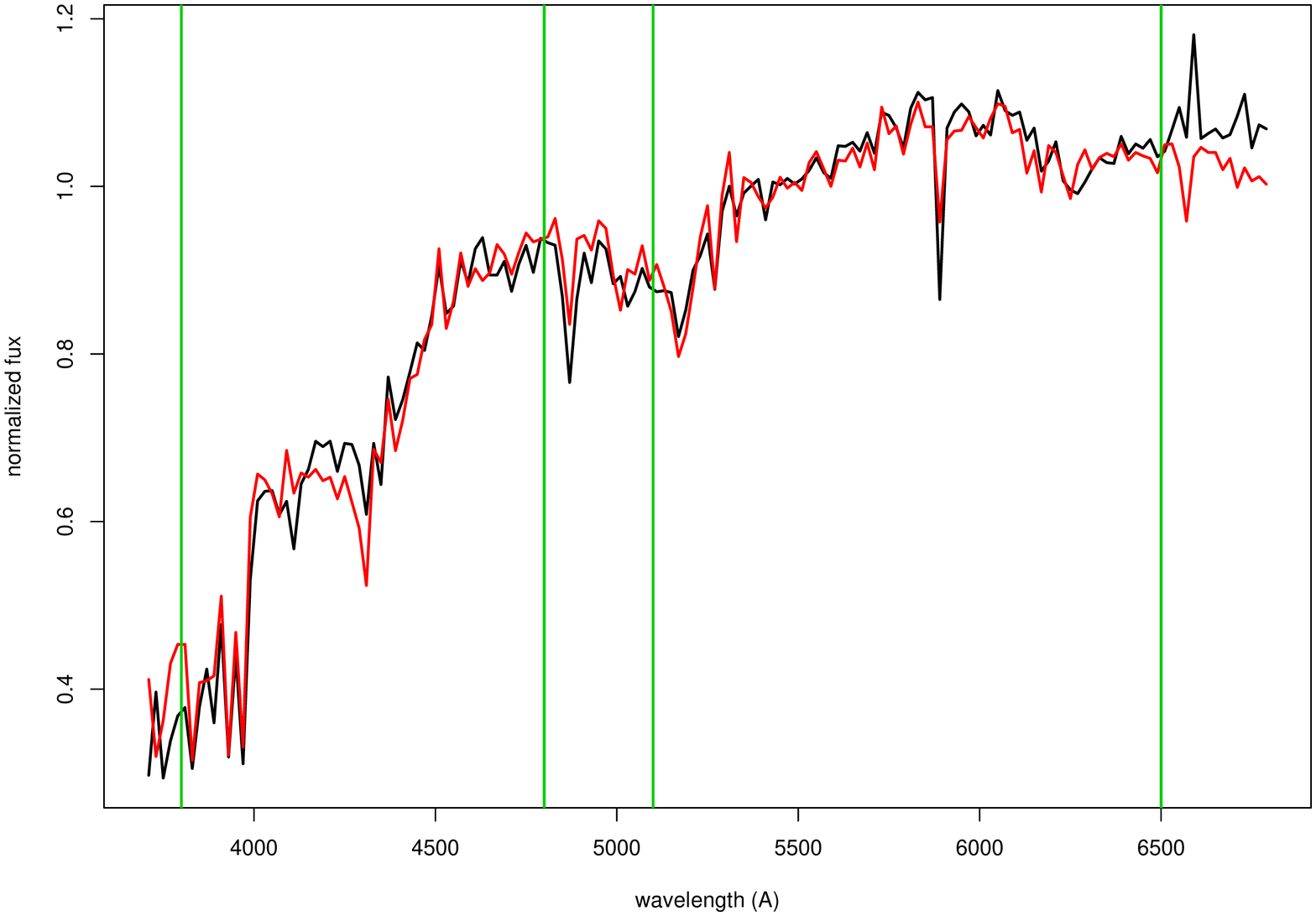}
\includegraphics[angle=-90,width=0.225\columnwidth]{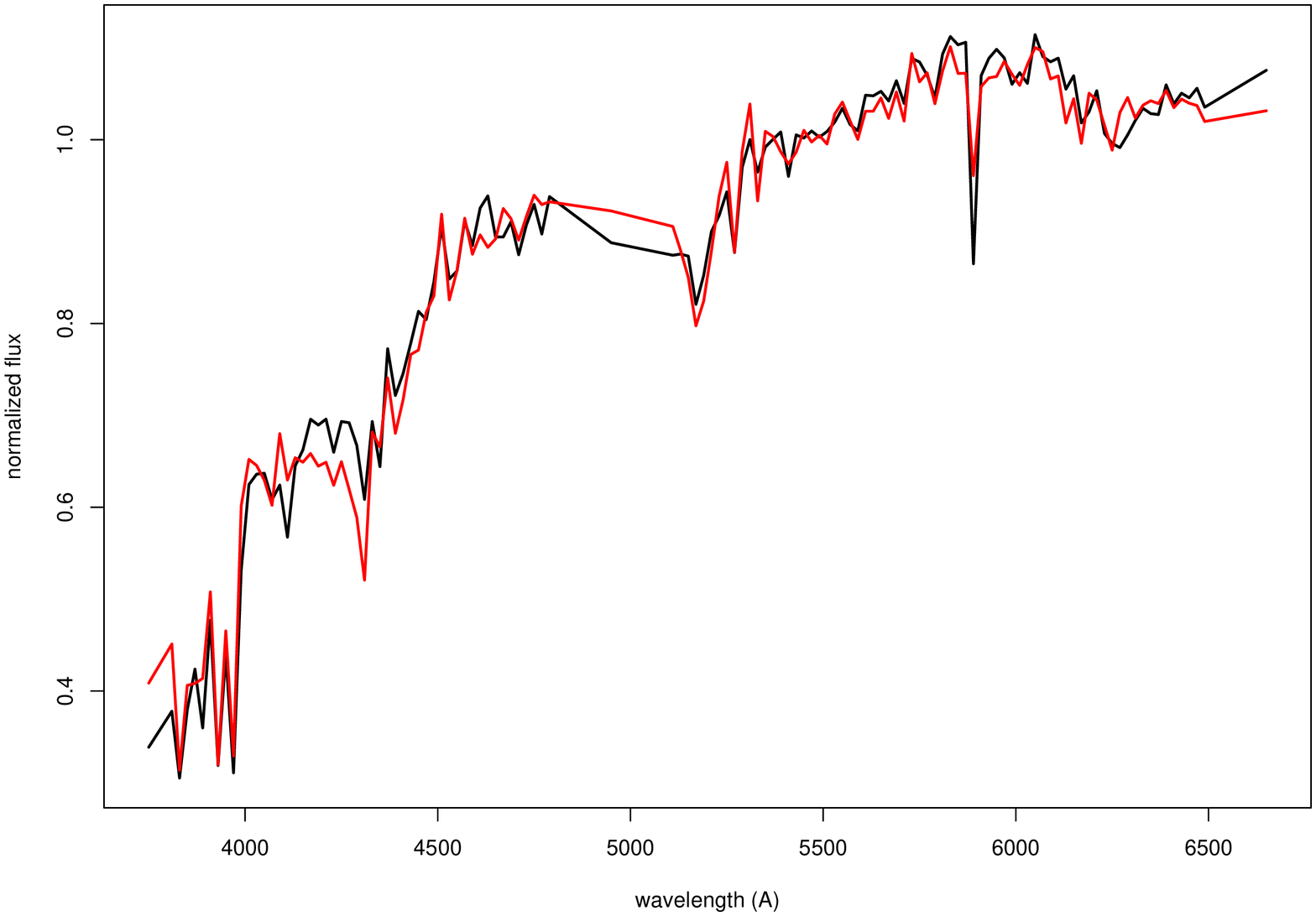}
\includegraphics[angle=-90,width=0.225\columnwidth]{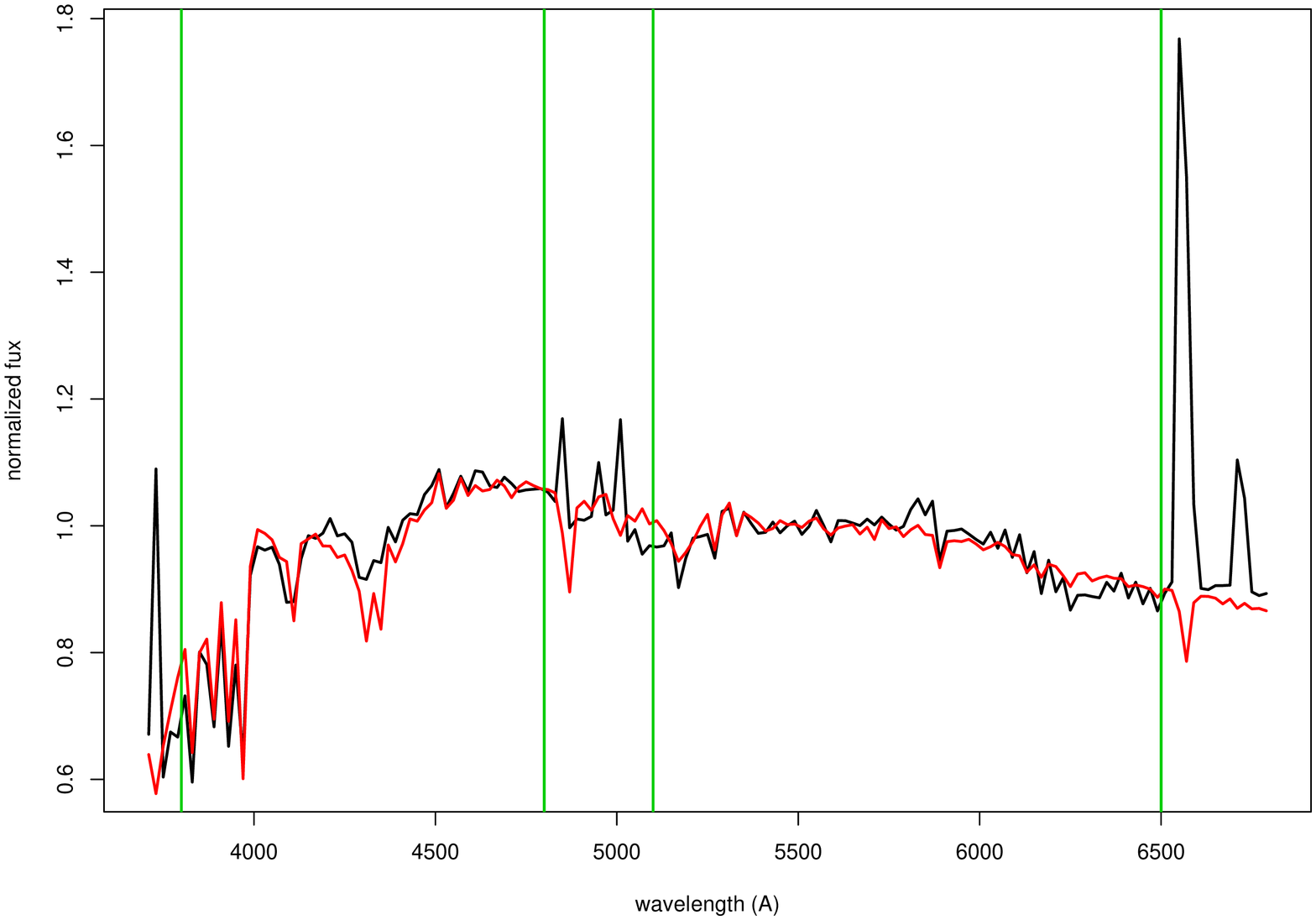}
\includegraphics[angle=-90,width=0.225\columnwidth]{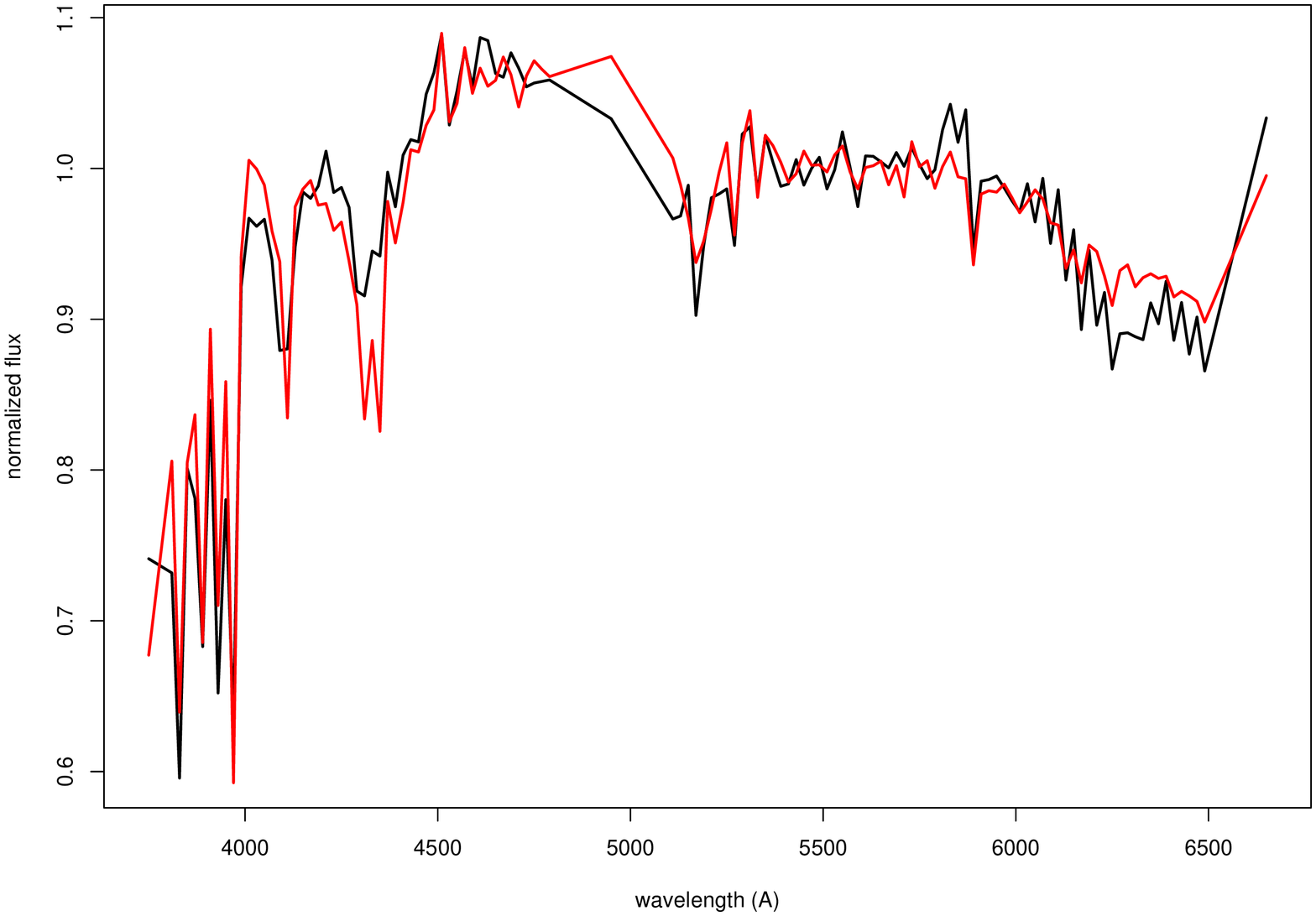}
\includegraphics[angle=-90,width=0.225\columnwidth]{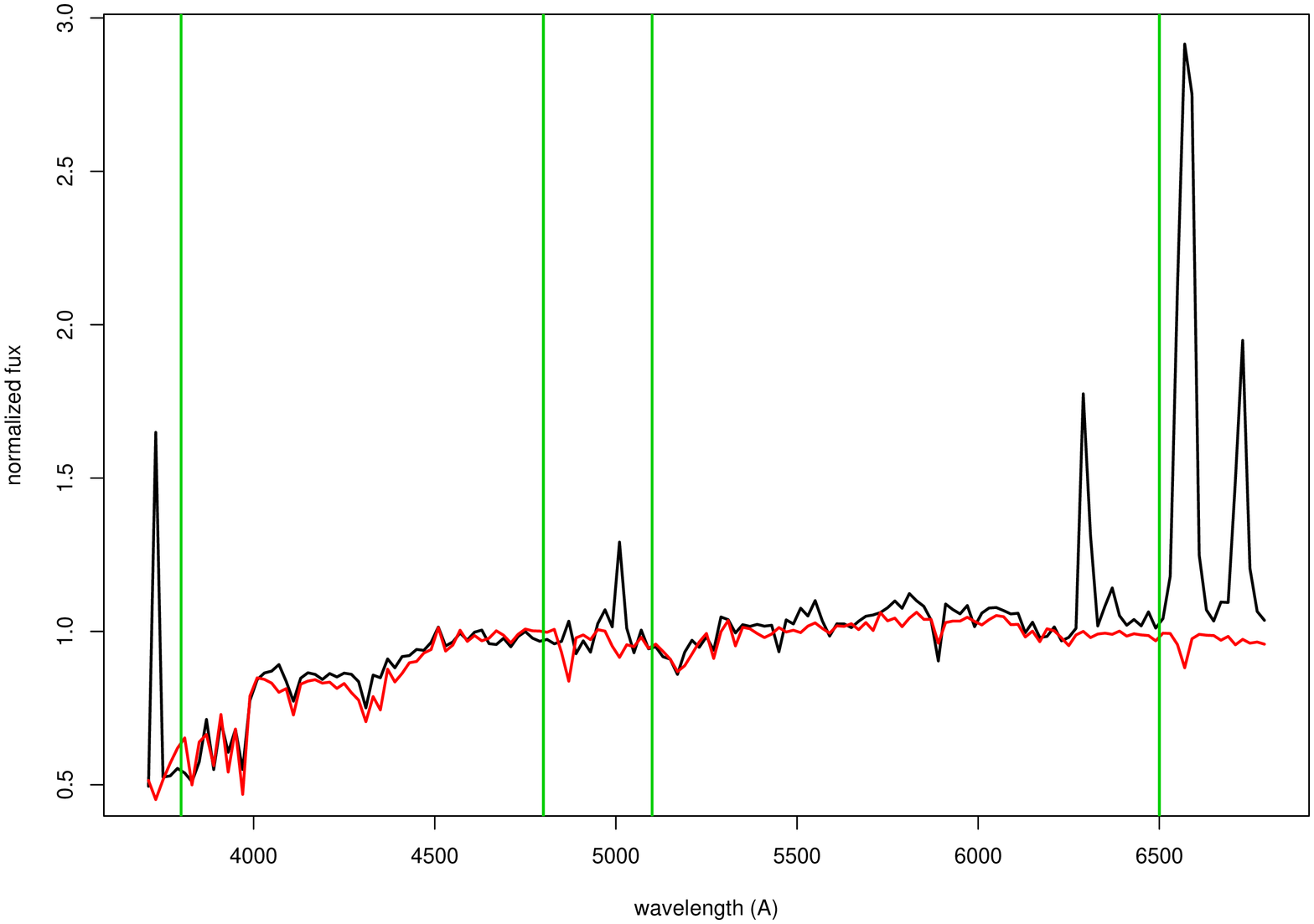}
\includegraphics[angle=-90,width=0.225\columnwidth]{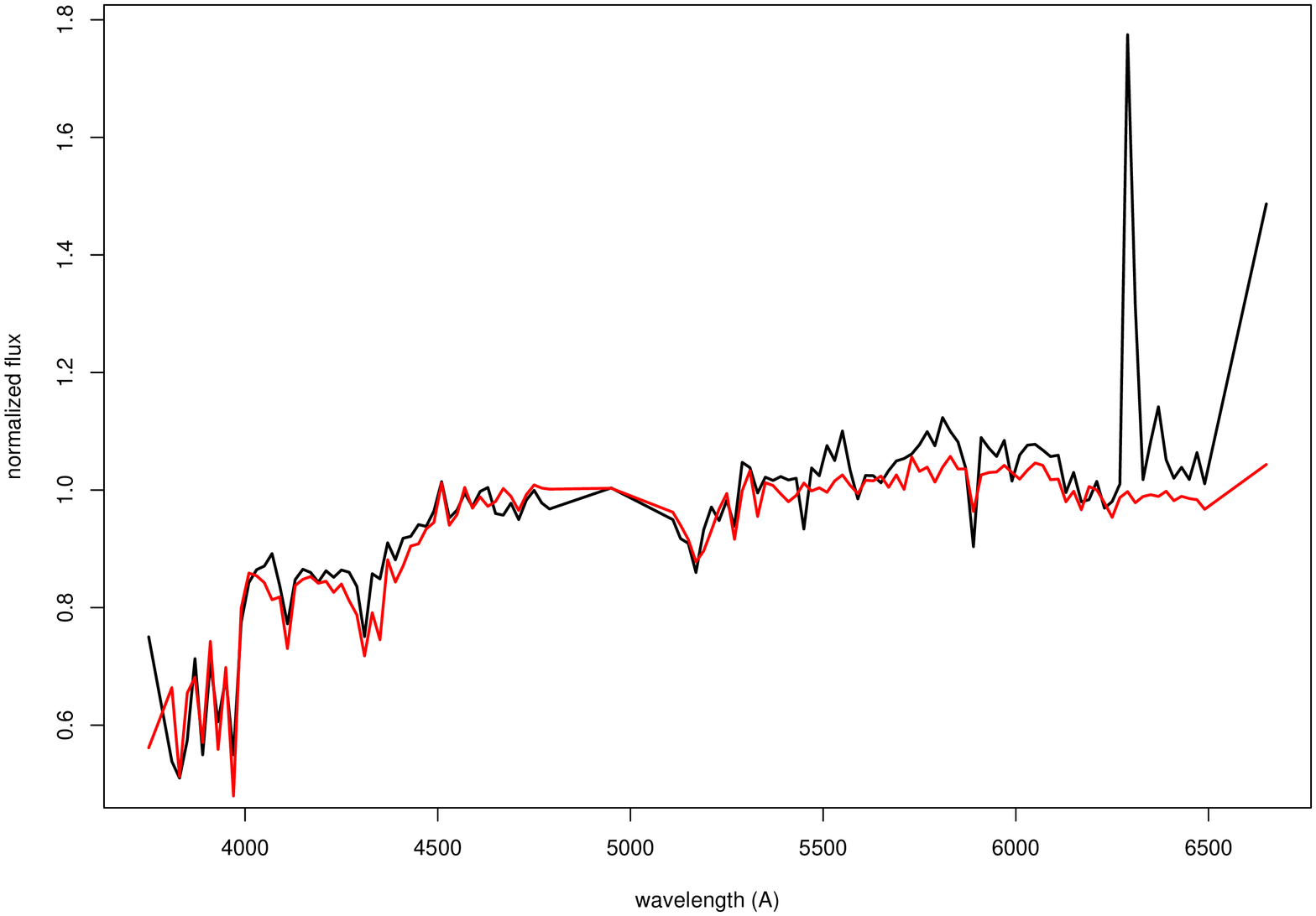}
\caption{The same with Fig \ref{a1} for the I0 irregular galaxies 
in Kennicutt's atlas. Here we present: NGC3034, NGC5195, NGC3077, 
and NGC6240.}
\label{a4}
\end{figure*}

\begin{figure*}[h]
\center
\includegraphics[angle=-90,width=0.225\columnwidth]{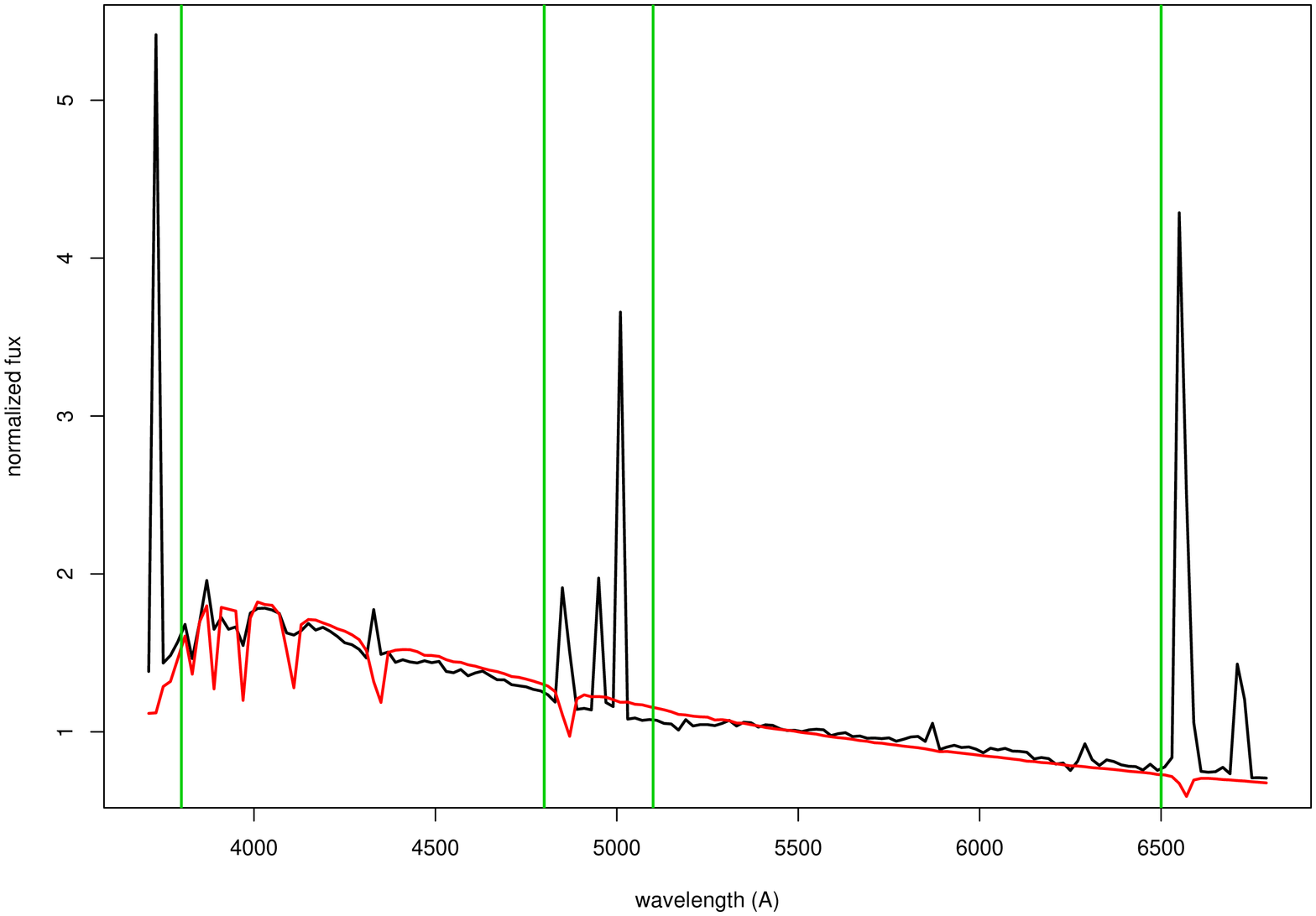}
\includegraphics[angle=-90,width=0.225\columnwidth]{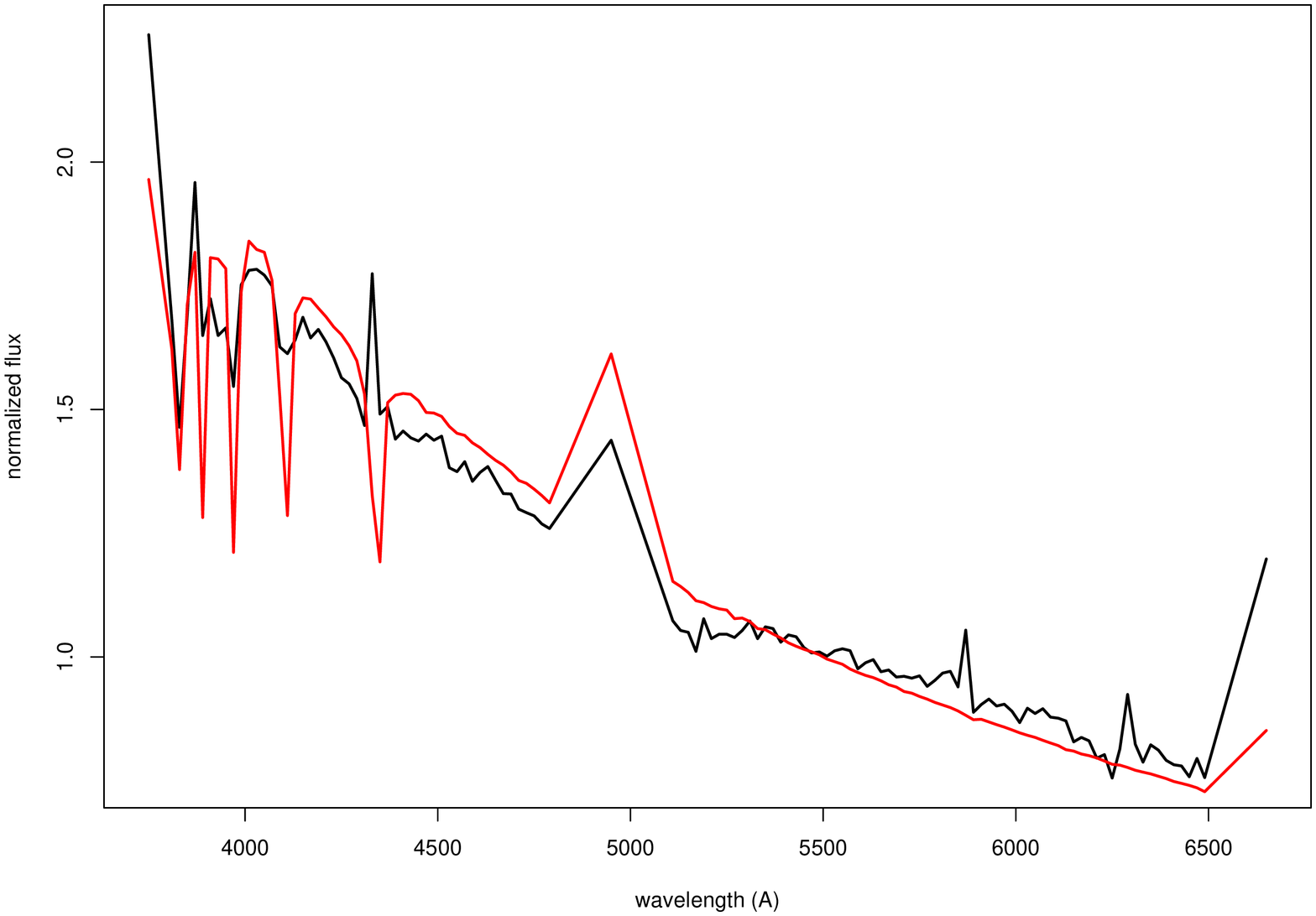}
\includegraphics[angle=-90,width=0.225\columnwidth]{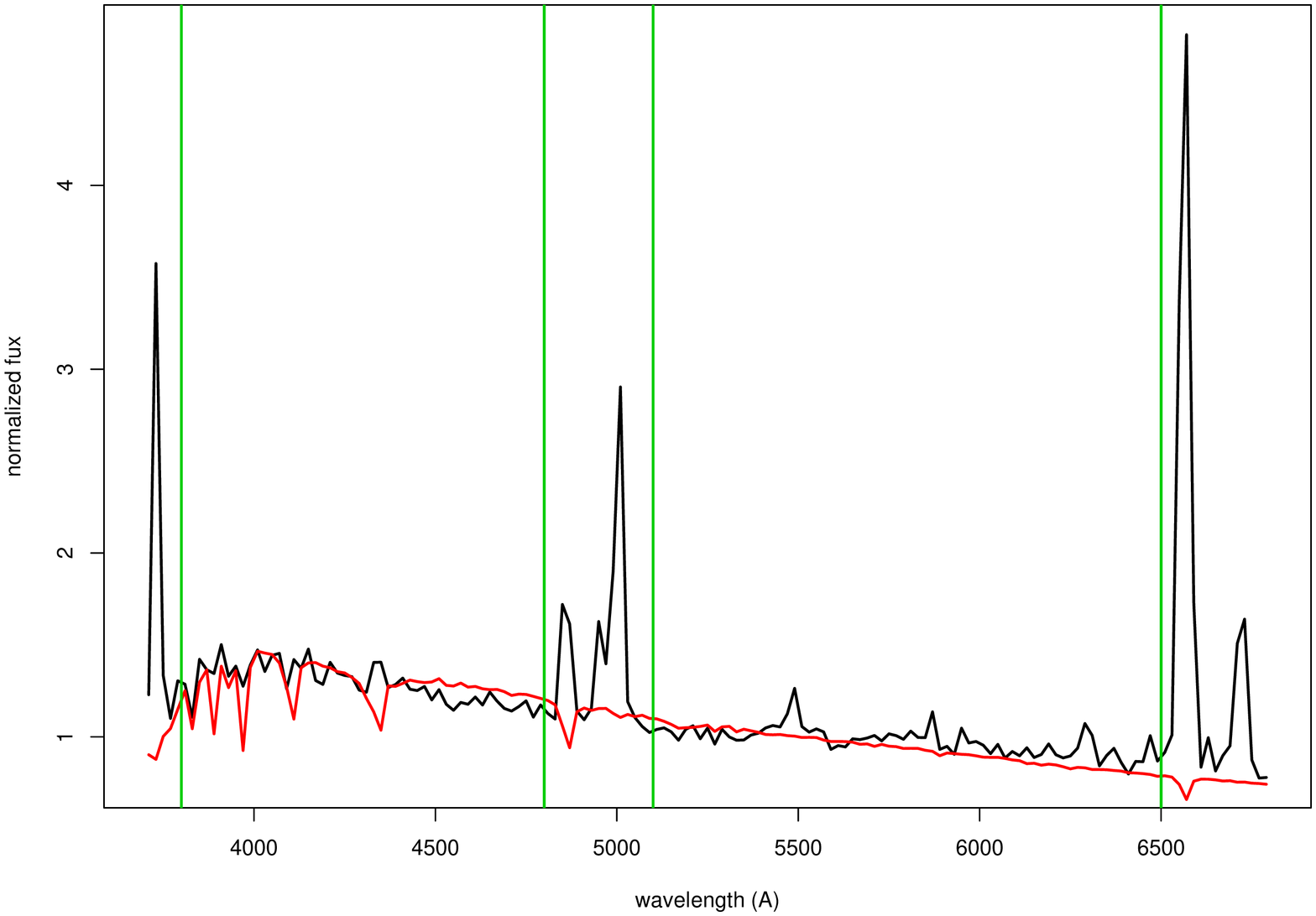}
\includegraphics[angle=-90,width=0.225\columnwidth]{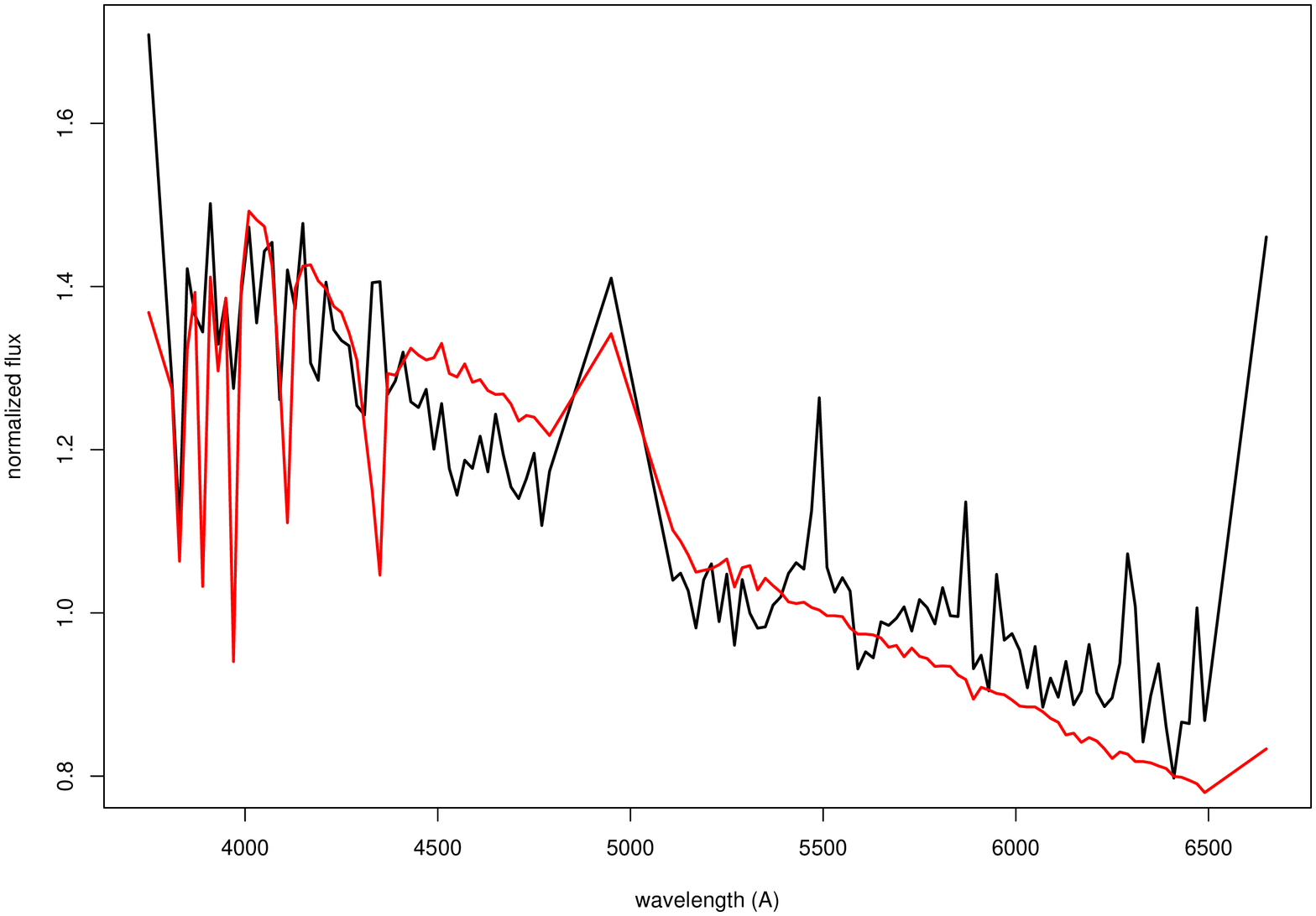}
\includegraphics[angle=-90,width=0.225\columnwidth]{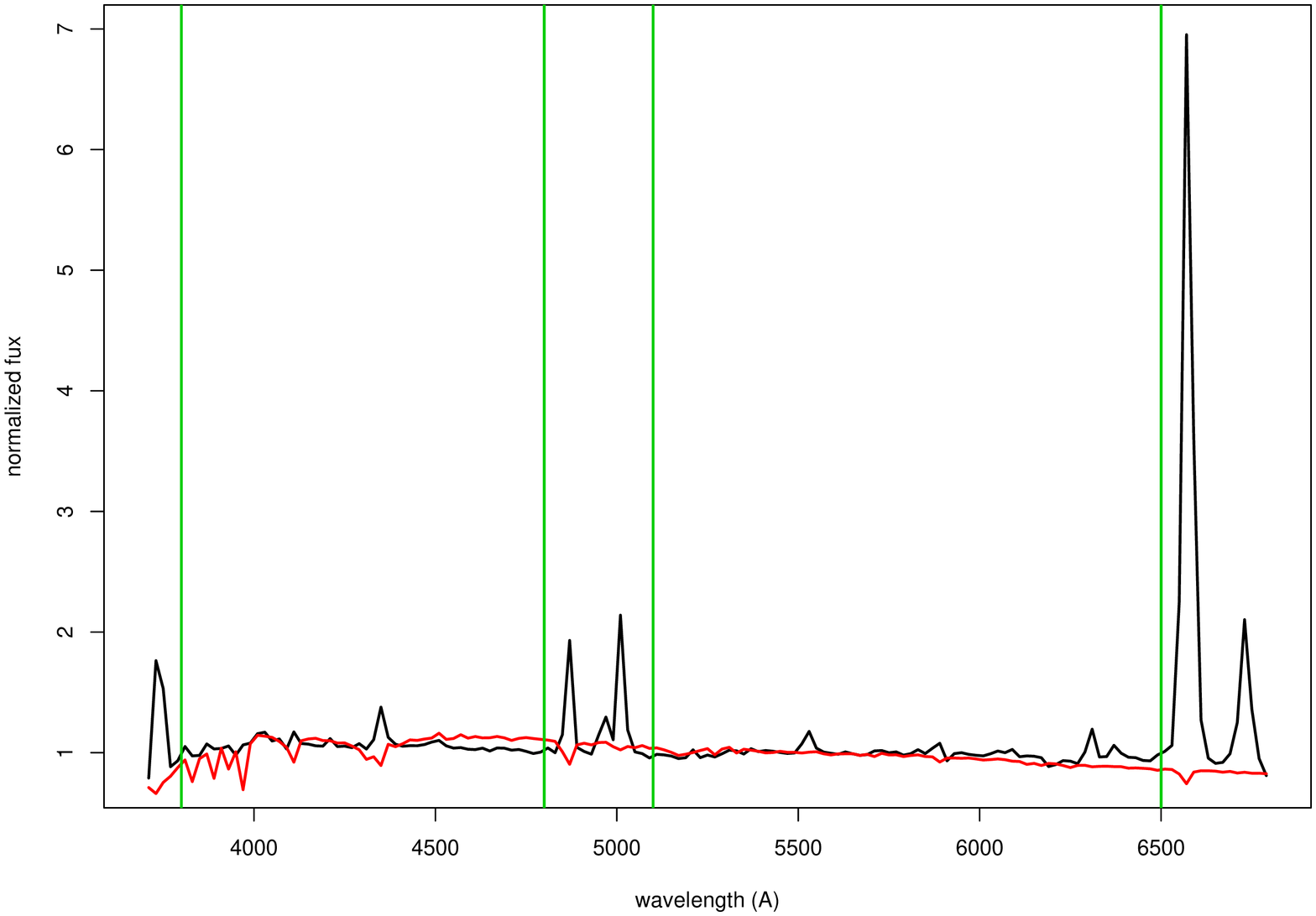}
\includegraphics[angle=-90,width=0.225\columnwidth]{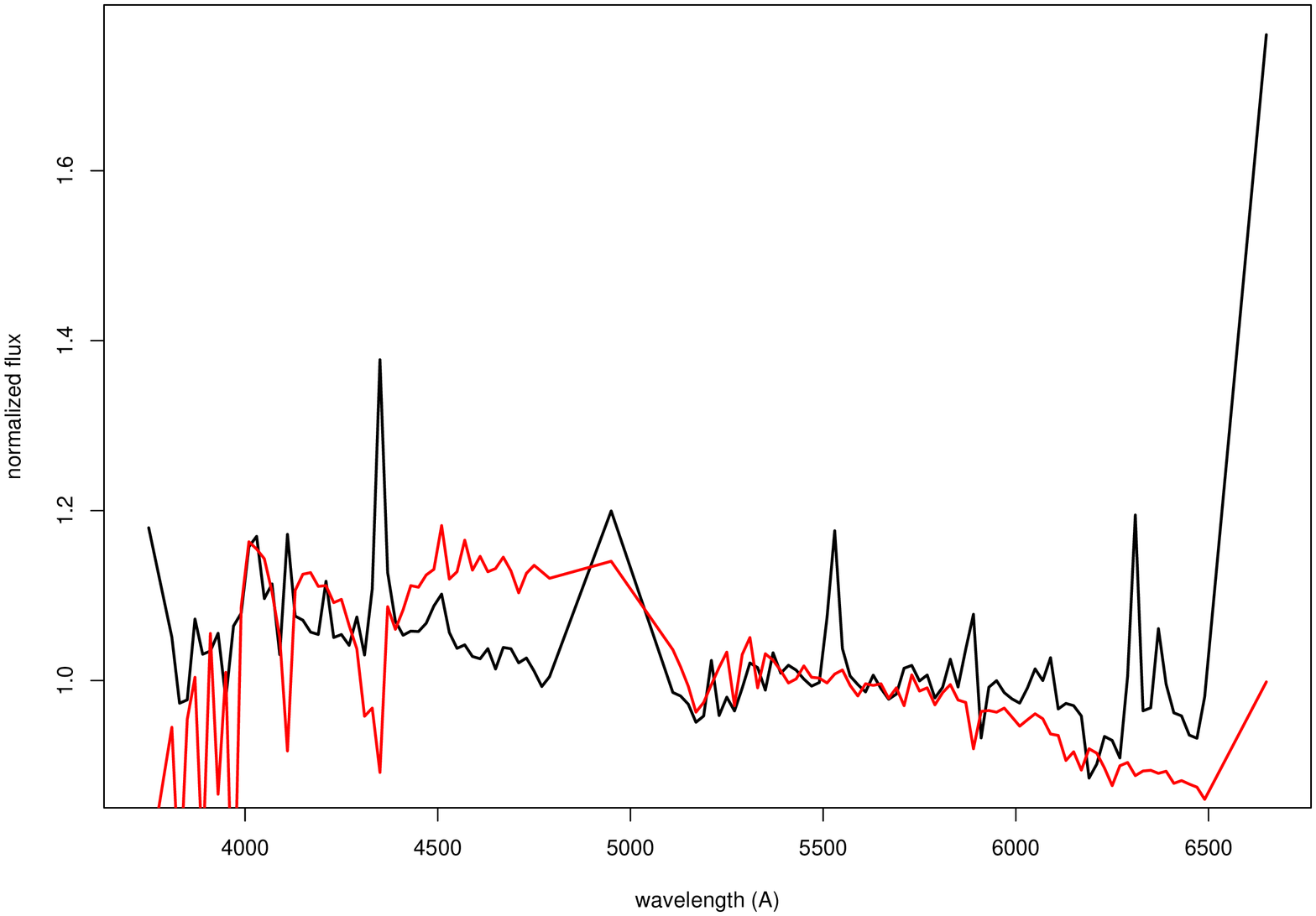}
\includegraphics[angle=-90,width=0.225\columnwidth]{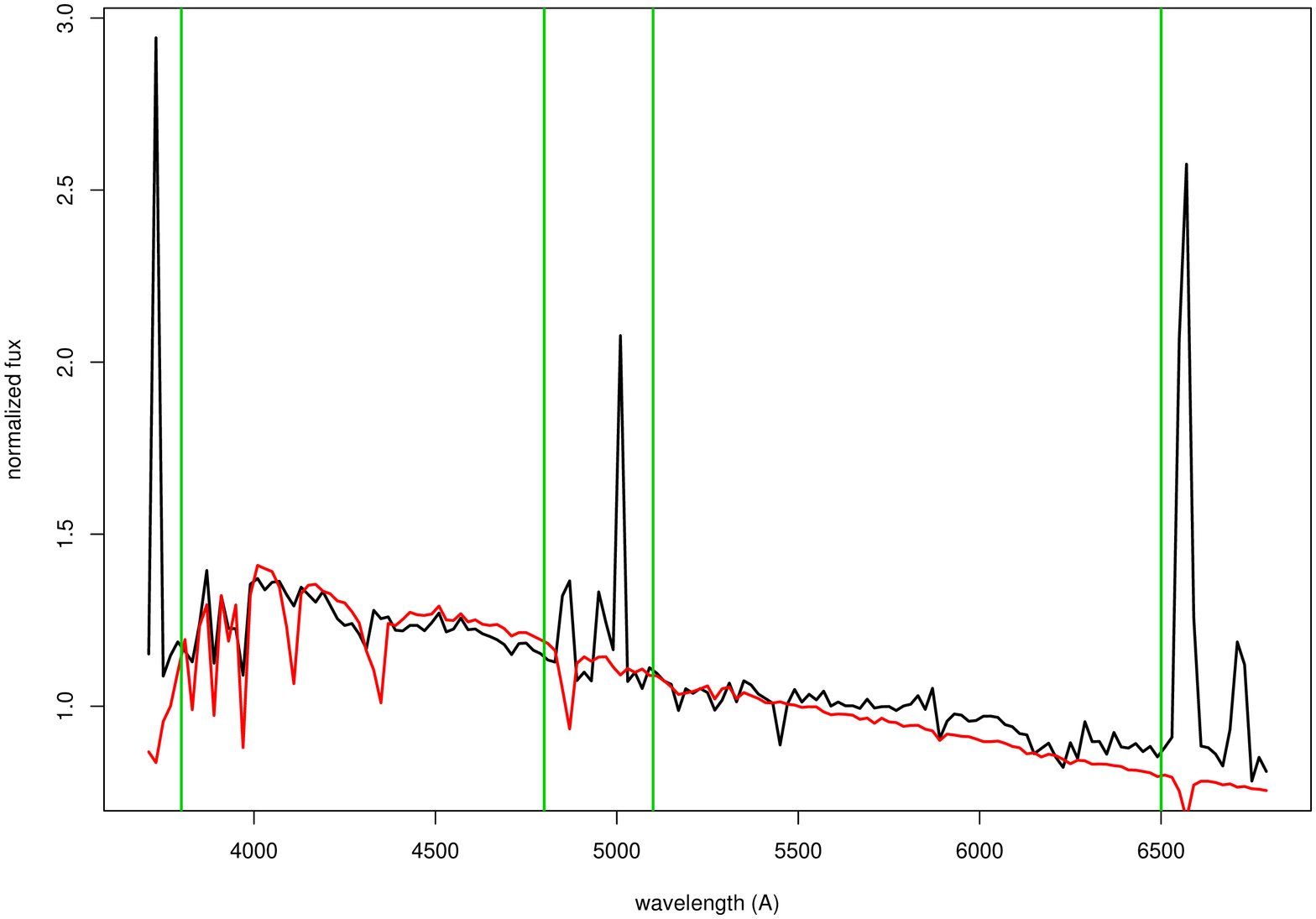}
\includegraphics[angle=-90,width=0.225\columnwidth]{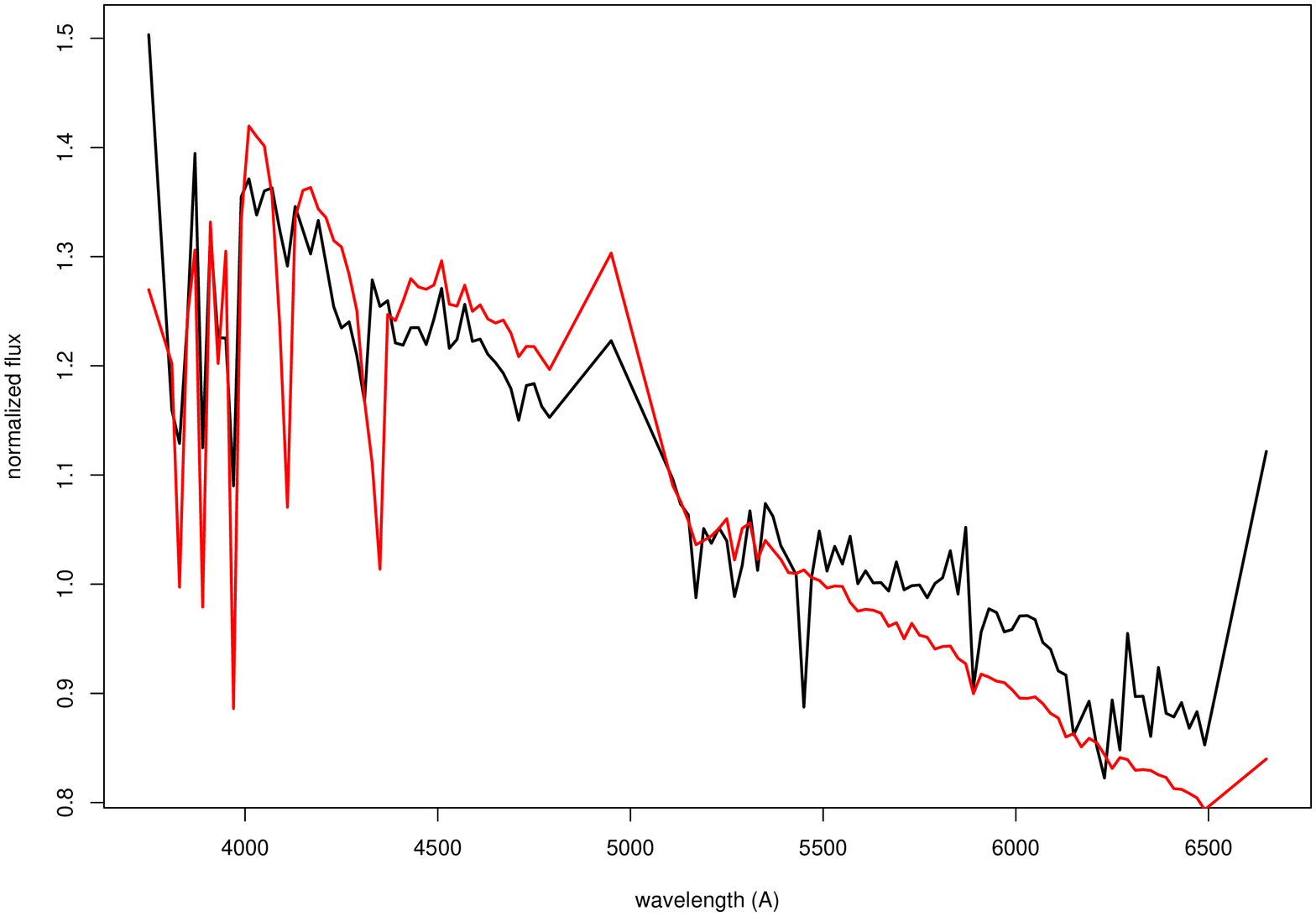}
\caption{The same with Fig \ref{a1} for the quenched star-forming 
galaxies with global starbursts in Kennicutt's atlas. Here we 
present: NGC3310, NGC6052, NGC3690, and UGC6697.}
\label{a5}
\end{figure*}

\begin{figure*}[h]
\center
\includegraphics[angle=-90,width=0.225\columnwidth]{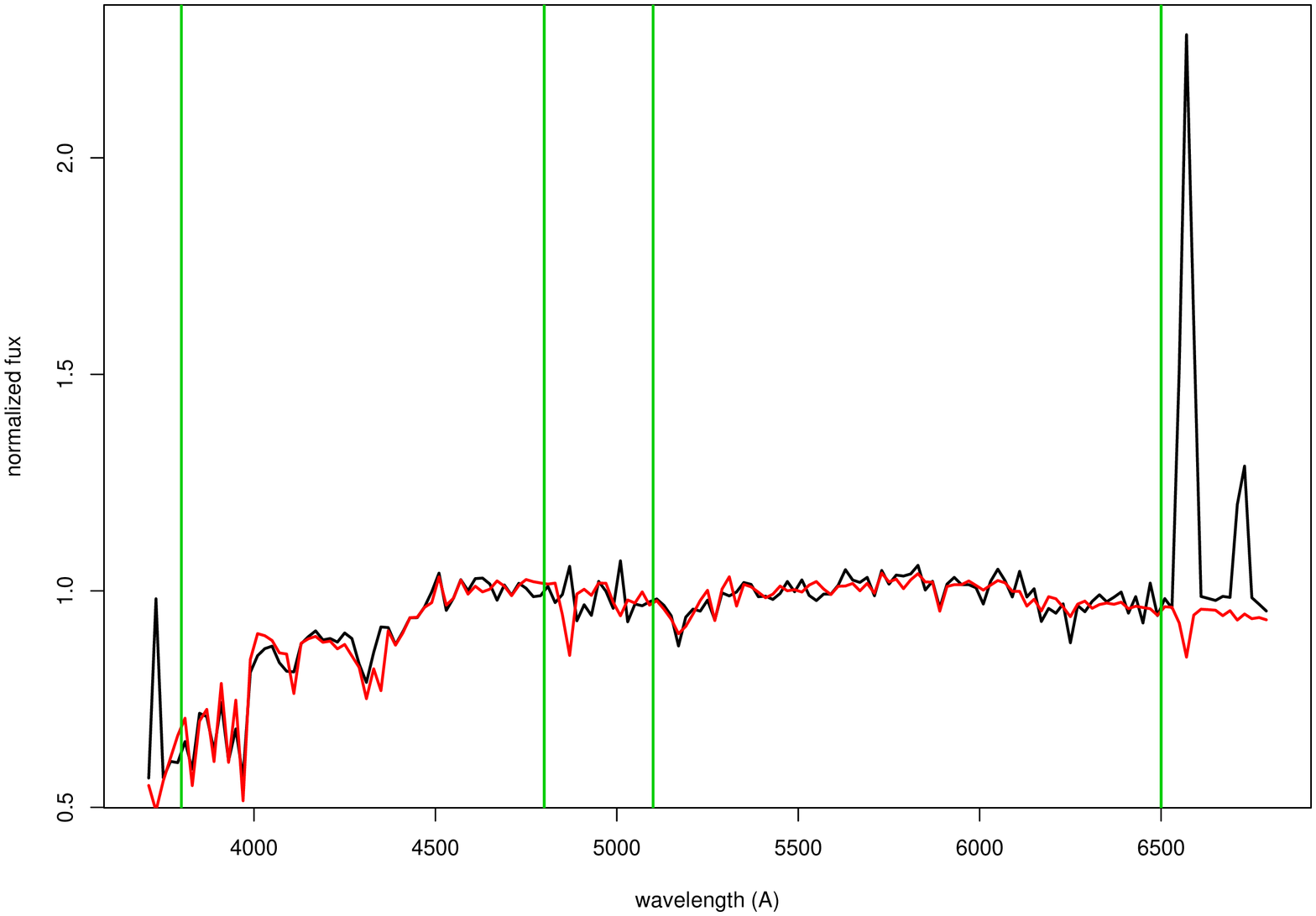}
\includegraphics[angle=-90,width=0.225\columnwidth]{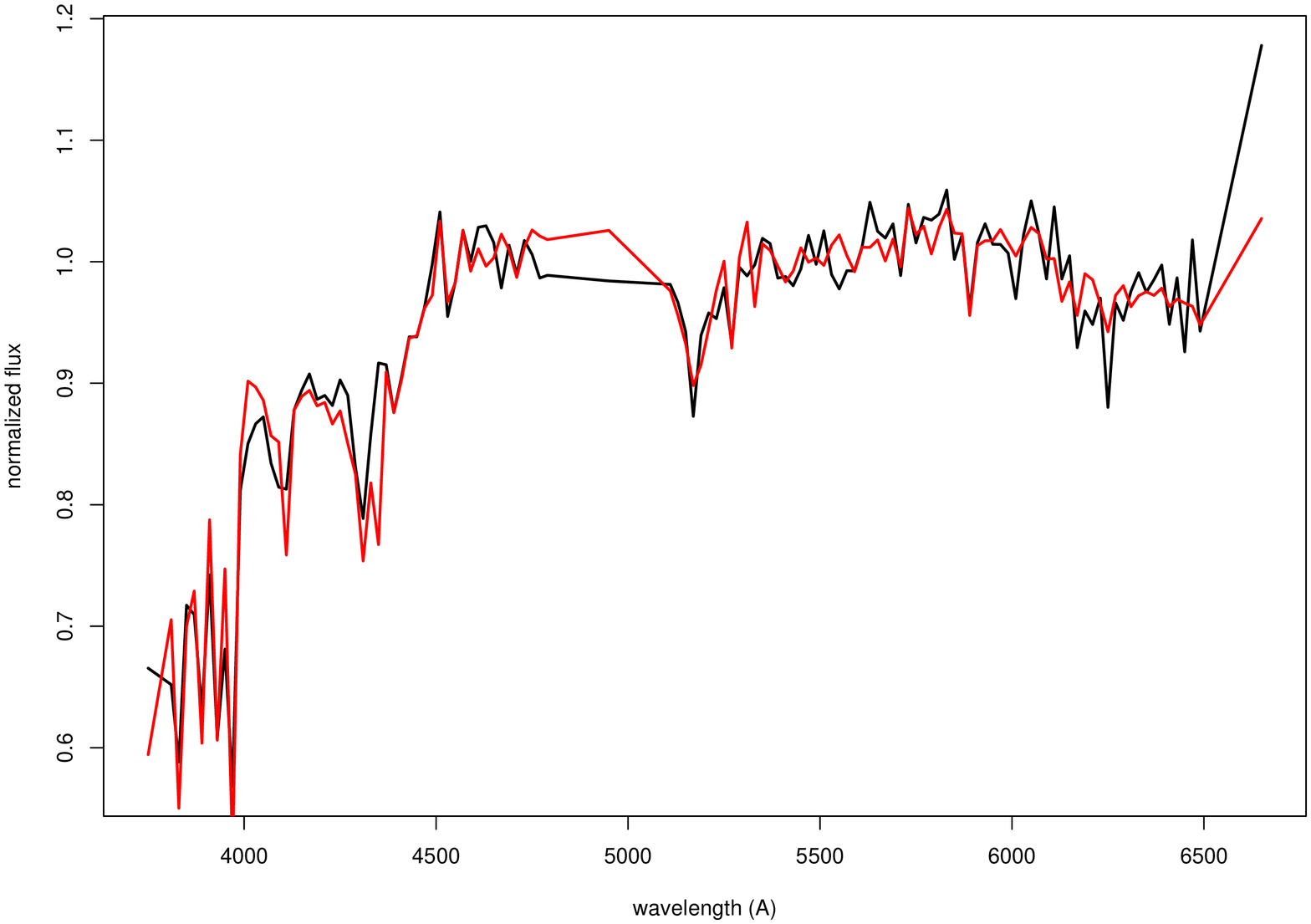}
\includegraphics[angle=-90,width=0.225\columnwidth]{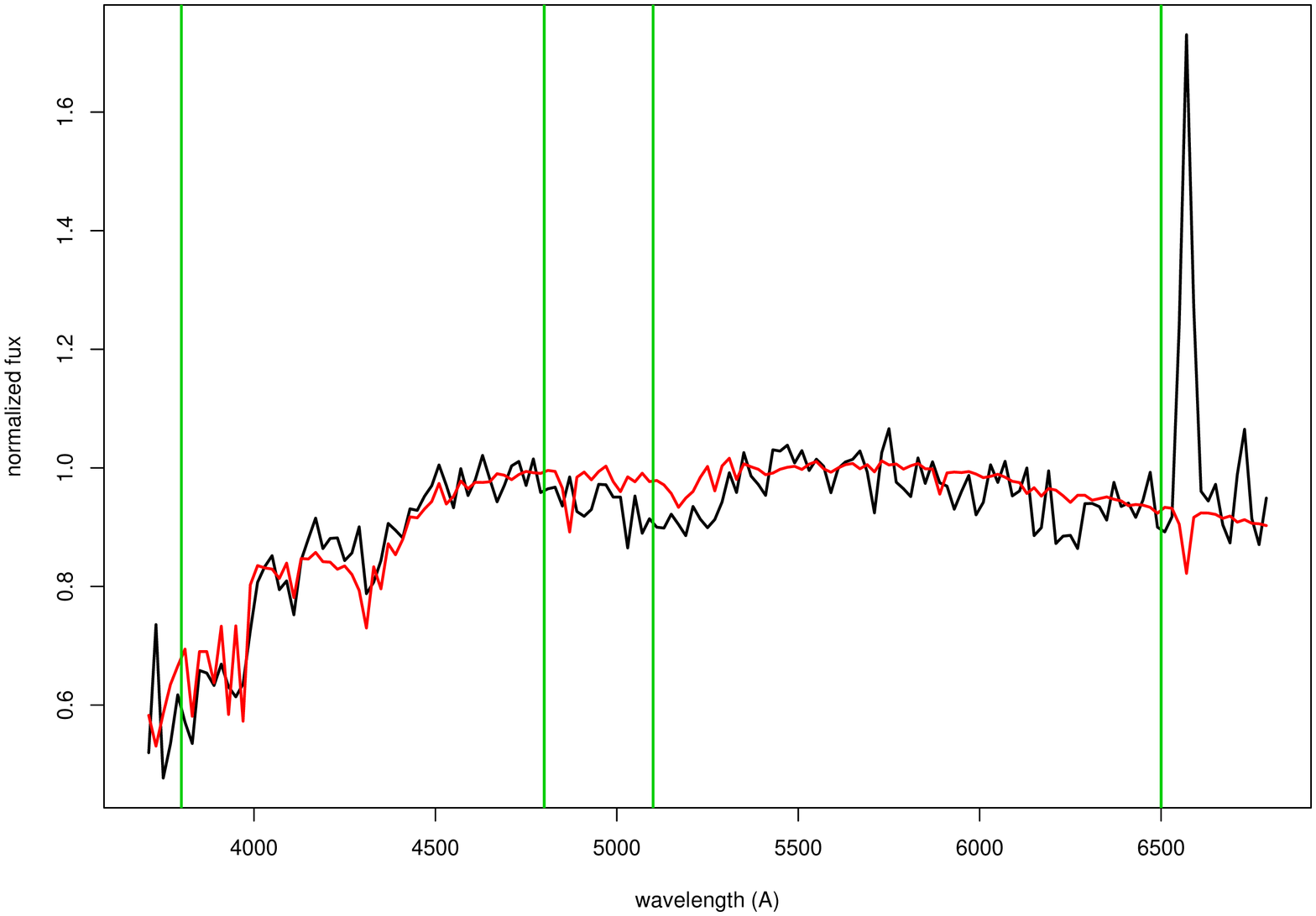}
\includegraphics[angle=-90,width=0.225\columnwidth]{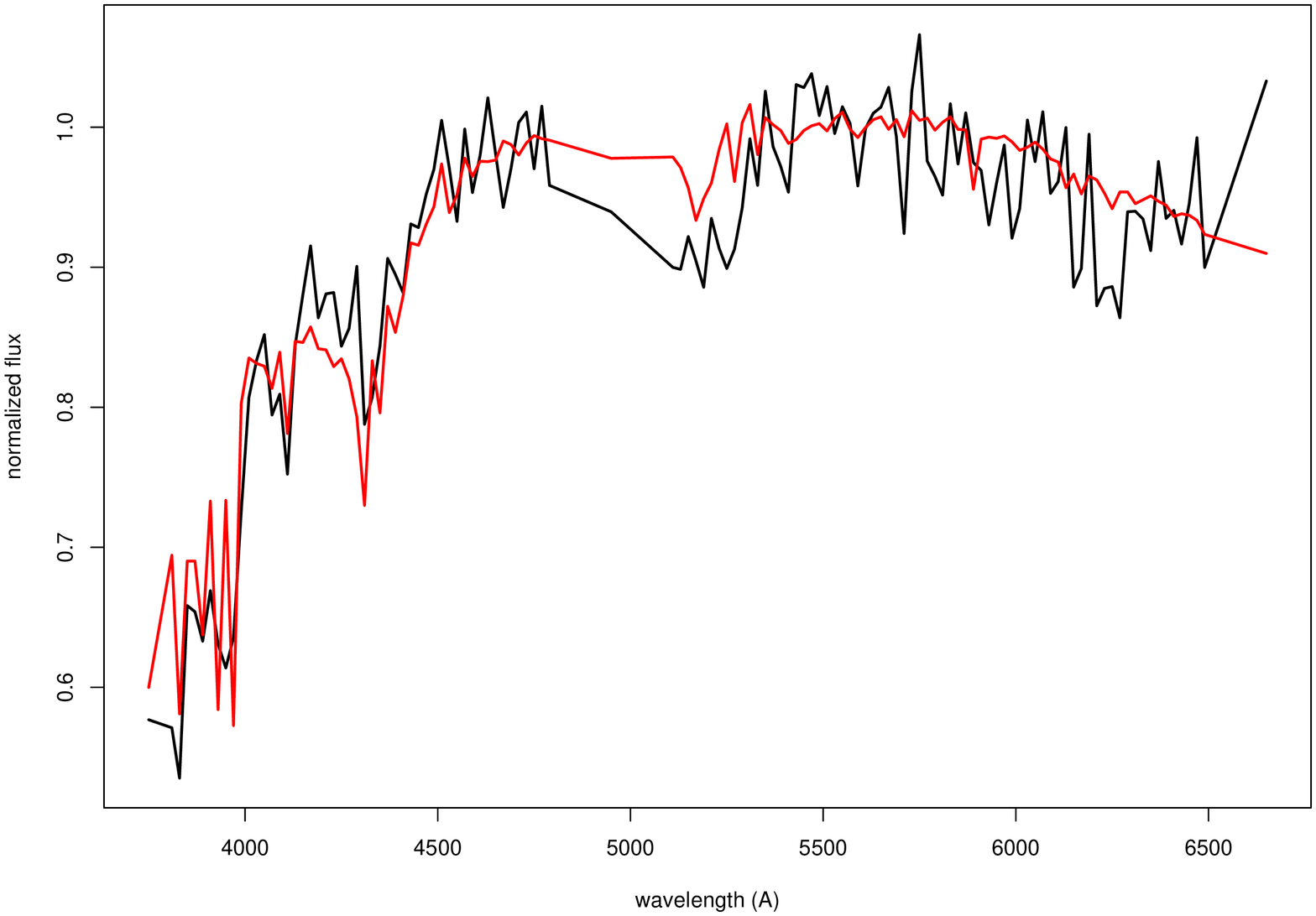}
\includegraphics[angle=-90,width=0.225\columnwidth]{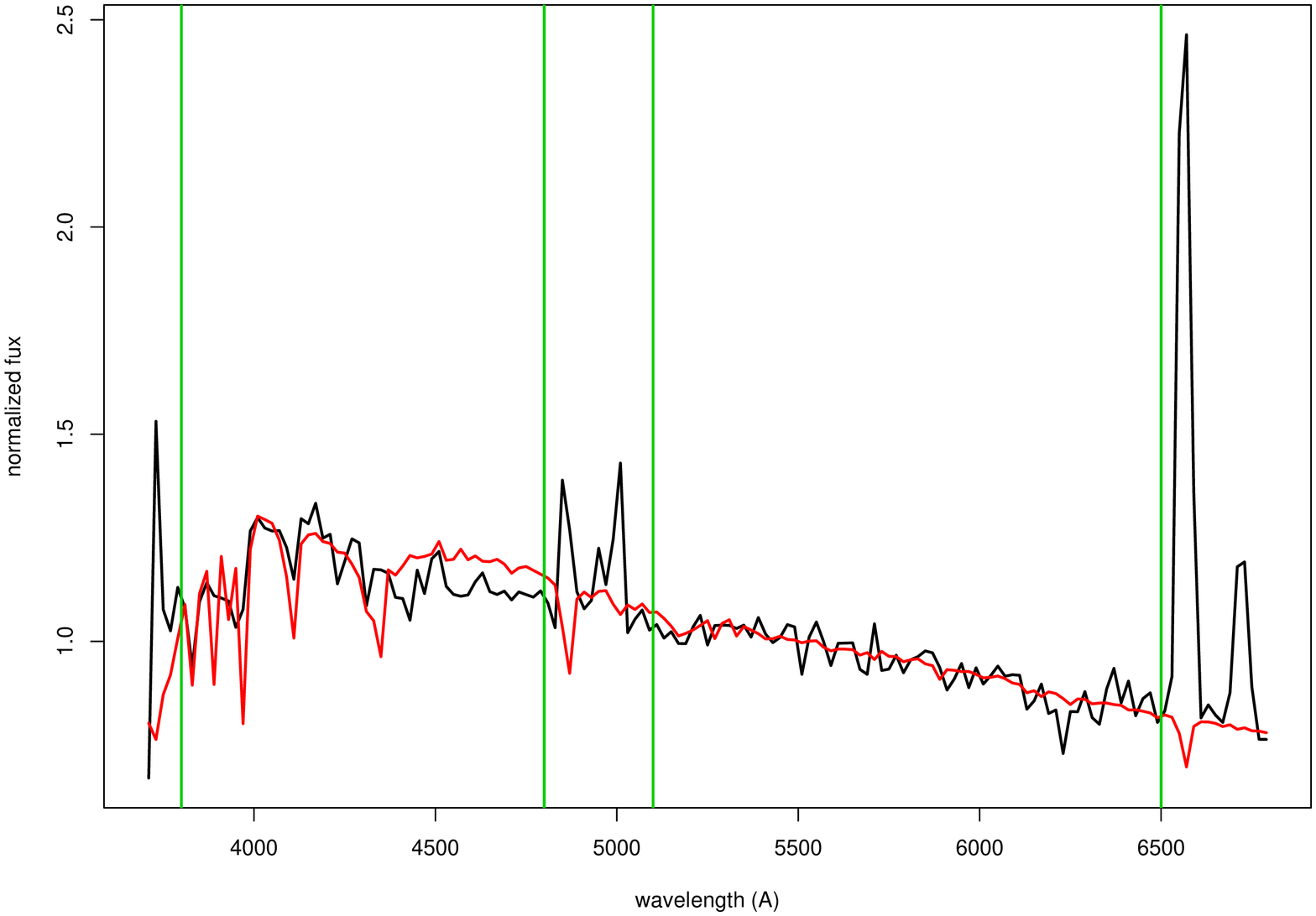}
\includegraphics[angle=-90,width=0.225\columnwidth]{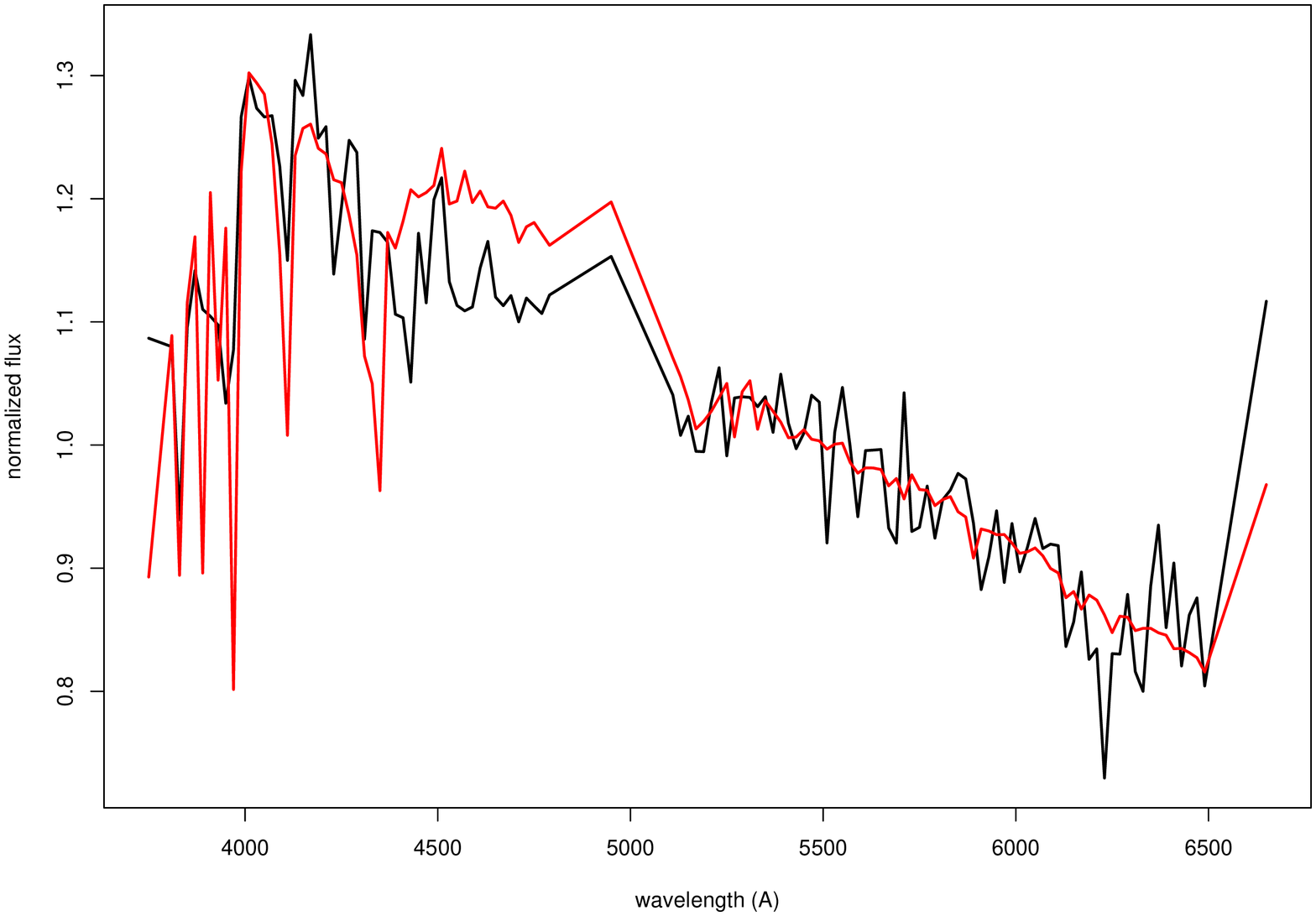}
\includegraphics[angle=-90,width=0.225\columnwidth]{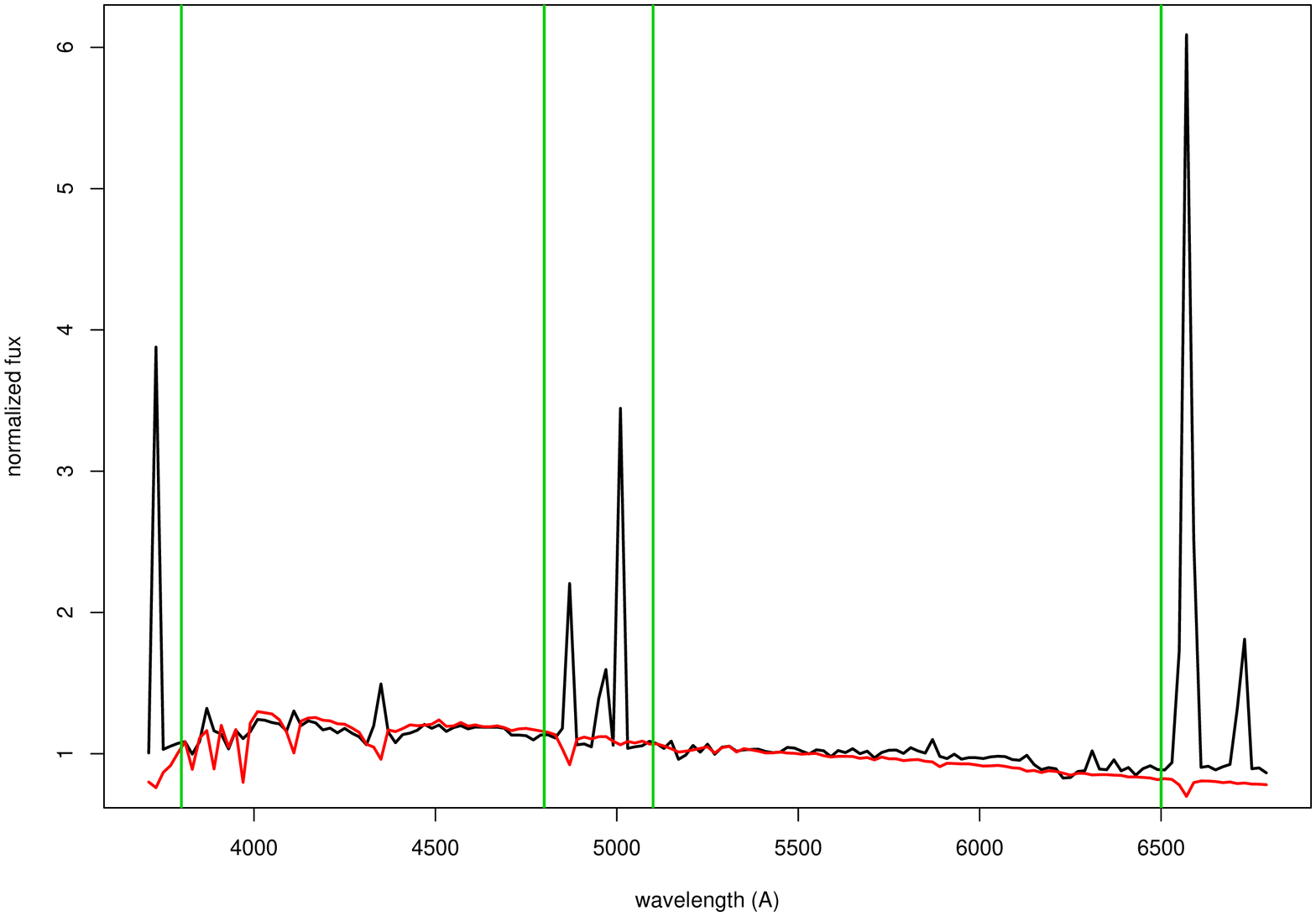}
\includegraphics[angle=-90,width=0.225\columnwidth]{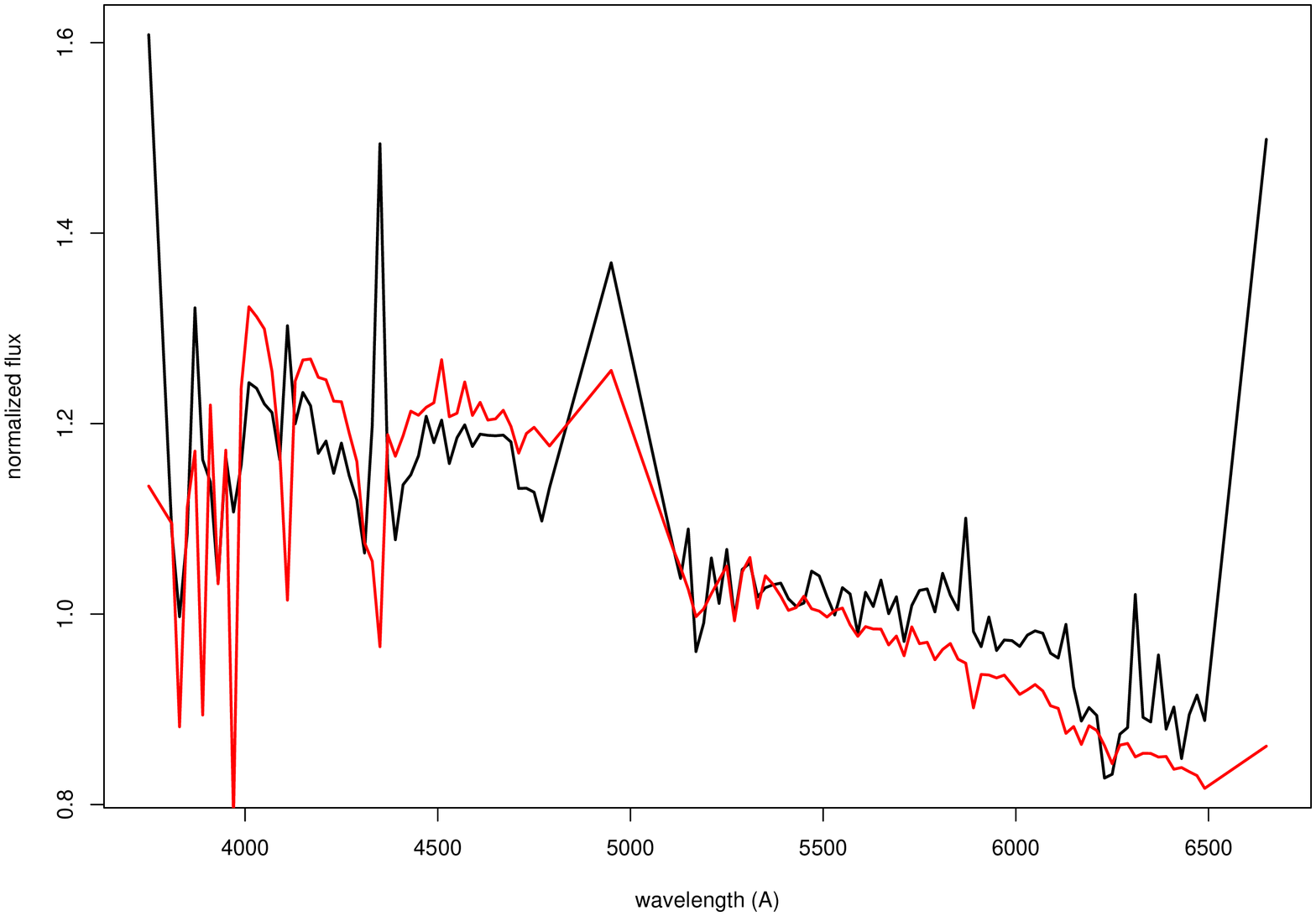}
\caption{The same with Fig \ref{a1} for the nuclear starburst 
galaxies in Kennicutt's atlas. Here we present: NGC2798, NGC3471, 
NGC5996, and NGC7714.}
\label{a6}
\end{figure*}

\end{document}